\documentclass[reqno]{amsart}

\usepackage{latexsym,pifont}
\usepackage{epsfig}

\usepackage{rotating}

\usepackage{ifpdf}
\ifpdf
\usepackage{epstopdf}
\usepackage{hyperref}
\else
\usepackage[hypertex]{hyperref}
\fi

\usepackage{amsfonts,amsmath,amssymb,mathrsfs,extarrows,MnSymbol}
\usepackage[all]{xy}

\usepackage{tikz}
\usetikzlibrary{matrix,arrows}

\newtheorem{Thm}{Theorem}
\newtheorem{Prop}[Thm]{Proposition}
\newtheorem{Lem}[Thm]{Lemma}
\newtheorem{Cor}[Thm]{Corollary}
\theoremstyle{definition}
\newtheorem{Rem}[Thm]{Remark}
\newtheorem{Def}[Thm]{Definition}
\newtheorem{Eg}[Thm]{Example}
\newtheorem{Conv}[Thm]{Convention}
\newtheorem{Quest}[Thm]{Question}
\newtheorem{Prob}[Thm]{Problem}

\newcommand{\alxydim}[2]{\begin{aligned}\xymatrix#1{#2}\end{aligned}}

\newcommand{\brem}{\begin{Rem}}
\newcommand{\erem}{\end{Rem}\medskip}
\newcommand{\beg}{\begin{Eg}}
\newcommand{\eeg}{\end{Eg}}
\newcommand{\bedef}{\begin{Def}}
\newcommand{\exdef}{
\end{Def}\vskip0.1cm}
\newcommand{\berop}{\begin{Prop}}
\newcommand{\eerop}{\end{Prop}}
\newcommand{\belem}{\begin{Lem}}
\newcommand{\elem}{\end{Lem}}
\newcommand{\bethe}{\begin{Thm}}
\newcommand{\ethe}{\end{Thm}}
\newcommand{\becor}{\begin{Cor}}
\newcommand{\ecor}{\end{Cor}}
\newcommand{\beroof}{\noindent\begin{proof}}
\newcommand{\eroof}{\end{proof}}
\newcommand{\becon}{\begin{Conv}}
\newcommand{\econ}{\begin{flushright}$\checkmark$\end{flushright}\end{Conv}}
\newcommand{\bequest}{\begin{Quest}}
\newcommand{\equest}{\end{Quest}}
\newcommand{\brob}{\begin{Prob}}
\newcommand{\erob}{\end{Prob}}

\newcommand{\barr}{\begin{array}}
\newcommand{\earr}{\end{array}}
\newcommand{\ben}{\begin{enumerate}}
\newcommand{\een}{\end{enumerate}}
\newcommand{\bit}{\begin{itemize}}
\newcommand{\eit}{\end{itemize}}

\newcommand{\qq}{\begin{eqnarray}}
\newcommand{\qqq}{\end{eqnarray}}

\newcommand{\nn}{\nonumber}

\newcommand{\ovl}[1]{\overline{#1}}
\newcommand{\unl}[1]{\underline{#1}}

\newcommand{\Reqref}[1]{Eq.\,\eqref{#1}}
\newcommand{\Rcite}[1]{Ref.\,\cite{#1}}
\newcommand{\Rxcite}[2]{Ref.\,\cite[#1]{#2}}

\newcommand\void[1]{}

\newcommand{\vref}[1]{\void{#1}}
\newcommand{\vReqref}[2]{\void{#1}Eq.\,(#2)}
\newcommand{\veqref}[2]{\void{#1}(#2)}

\newcommand{\tx}[1]{\textrm{#1}} 

\newcommand{\gt}[1]{\mathfrak{#1}}

\def\cA{\mathcal{A}}
\def\cB{\mathcal{B}}

\def\cD{\mathcal{D}}
\def\cE{\mathcal{E}}
\def\cF{\mathcal{F}}
\def\cG{\mathcal{G}}

\def\cJ{\mathcal{J}}

\def\ceL{\mathcal{L}}
\def\cM{\mathcal{M}}

\def\cO{\mathcal{O}}
\def\cP{\mathcal{P}}

\def\cT{\mathcal{T}}
\def\cU{\mathcal{U}}

\def\xcC{\mathscr{C}}
\def\xcD{\mathscr{D}}
\def\xcE{\mathscr{E}}
\def\xcF{\mathscr{F}}
\def\xcG{\mathscr{G}}

\def\xcI{\mathscr{I}}
\def\xcJ{\mathscr{J}}
\def\xcK{\mathscr{K}}
\def\xcL{\mathscr{L}}
\def\xcM{\mathscr{M}}
\def\xcN{\mathscr{N}}

\def\xcR{\mathscr{R}}

\def\xcU{\mathscr{U}}
\def\xcV{\mathscr{V}}
\def\xcW{\mathscr{W}}
\def\xcX{\mathscr{X}}

\def\bC{{\mathbb{C}}}

\def\bH{{\mathbb{H}}}

\def\bN{{\mathbb{N}}}

\def\bR{{\mathbb{R}}}
\def\bS{{\mathbb{S}}}

\def\bZ{{\mathbb{Z}}}

\def\a{\alpha}
\def\b{\beta}
\def\g{\gamma}
\def\G{\Gamma}
\def\d{\delta}
\def\D{\Delta}

\def\vep{\varepsilon}

\def\tht{\theta}
\def\Th{\Theta}

\def\la{\lambda}
\def\La{\Lambda}
\def\om{\omega}
\def\Om{\Omega}

\def\si{\sigma}
\def\Si{\Sigma}

\def\z{\zeta}

\def\Agt{\gt{A}}
\def\bgt{\gt{b}}
\def\Bgt{\gt{B}}

\def\Cgt{\gt{C}}

\def\Dgt{\gt{D}}

\def\Egt{\gt{E}}

\def\ggt{\gt{g}}

\def\hgt{\gt{h}}

\def\igt{\gt{i}}
\def\Igt{\gt{I}}
\def\jgt{\gt{j}}
\def\Jgt{\gt{J}}

\def\Kgt{\gt{K}}

\def\Lgt{\gt{L}}

\def\Mgt{\gt{M}}

\def\Rgt{\gt{R}}

\def\Sgt{\gt{S}}

\def\Tgt{\gt{T}}

\def\Vgt{\gt{V}}

\def\Wgt{\gt{W}}

\def\Xgt{\gt{X}}

\def\Ygt{\gt{Y}}

\def\Zgt{\gt{Z}}

\newcommand{\sfd}{{\mathsf d}}

\newcommand{\sfE}{{\mathsf E}}

\newcommand{\sfi}{{\mathsf i}}
\newcommand{\sfI}{{\mathsf I}}

\newcommand{\sfJ}{{\mathsf J}}
\newcommand{\sfk}{{\mathsf k}}

\newcommand{\sfL}{{\mathsf L}}

\newcommand{\sfN}{{\mathsf N}}

\newcommand{\sfp}{{\mathsf p}}
\newcommand{\sfP}{{\mathsf P}}

\newcommand{\sfT}{{\mathsf T}}

\newcommand{\txA}{{\rm A}}

\newcommand{\txB}{{\rm B}}

\newcommand{\txc}{{\rm c}}

\newcommand{\ee}{{\rm e}}

\newcommand{\txf}{{\rm f}}

\newcommand{\txg}{{\rm g}}
\newcommand{\txG}{{\rm G}}
\newcommand{\txh}{{\rm h}}
\newcommand{\txH}{{\rm H}}

\newcommand{\txK}{{\rm K}}

\newcommand{\txm}{{\rm m}}

\newcommand{\txp}{{\rm p}}

\def\vH{\check{H}}

\def\const{{\rm const}}

\def\id{{\rm id}}
\newcommand{\pr}{{\rm pr}}

\def\ev{{\rm ev}}

\def\obj{{\rm Ob}}
\def\mor{{\rm Hom}}
\def\morf{{\rm Mor}}
\def\dim{{\rm dim}}
\def\im{{\rm im}}
\def\ker{{\rm ker}}

\def\End{{\rm End}}

\newcommand{\Id}{{\rm Id}}
\def\Inv{{\rm Inv}}
\newcommand\Mod{{\rm Mod}}

\newcommand{\Gr}{{\rm Gr}}
\newcommand{\gtGr}{\mathfrak{Gr}}
\newcommand\Grbun[1]{\Gr\textrm{-}\gt{Bun}(#1)}
\newcommand\GMbun[1]{\txG\lx\xcM\textrm{-}\gt{Bun}(#1)}
\newcommand{\gtgr}{\mathfrak{gr}}

\newcommand\Gbun[1]{\txG\textrm{-}\gt{Bun}(#1)}

\def\Vol{{\rm Vol}}

\newcommand{\ic}{\imath}
\newcommand{\pLie}[1]{\,{-\hspace{-8pt}\xcL}_{#1}}
\def\p{\partial}

\def\con{\righthalfcup}
\newcommand{\Diff}{{\rm Diff}}

\def\emb{\hookrightarrow}

\def\curv{{\rm curv}}
\def\Hol{{\rm Hol}}

\newcommand{\cGk}{\cG_\sfk}

\def\bd1{{\boldsymbol{1}}}
\def\brd0{{\boldsymbol{0}}}

\def\diag{\textrm{diag}}

\def\Ad{{\rm Ad}}

\newcommand{\uj}{{\rm U}(1)}

\def\x{\times}
\def\ox{\otimes}

\def\lx{{\hspace{-0.04cm}\ltimes\hspace{-0.05cm}}}

\newcommand{\fibx}[2]{{}_{#1}\hspace{-3pt}\x_{#2}\hspace{-1pt}}

\def\lact{\vartriangleright}

\newcommand{\corr}[1]{\left\langle #1 \right\rangle}

\newcommand{\Vbra}[2]{\left[\,#1\,,\,#2\,\right]_{\rm V}}

\newcommand{\Vcon}[2]{\left(\,#1\,,\,#2\,\right)_{\con}}

\newcommand{\GBra}[2]{\lsem\,#1\,,\,#2\,\rsem}

\newcommand{\Mup}{{}^{\tx{\tiny $M$}}\hspace{-1pt}}

\newcommand{\xcMup}{{}^{\tx{\tiny $\xcM$}}\hspace{-1pt}}

\newcommand{\xcFup}{{}^{\tx{\tiny $\xcF$}}\hspace{-2pt}}
\newcommand{\txcFup}{{}^{\tx{\tiny $\widetilde\xcF$}}\hspace{-2pt}}
\newcommand{\Qup}{{}^{\tx{\tiny $Q$}}\hspace{-1pt}}

\newcommand{\Tnup}{{}^{\tx{\tiny $T_n$}}\hspace{-1pt}}

\newcommand{\ups}[1]{{}^{\tx{\tiny $#1$}}\hspace{-1pt}}
\newcommand{\dagu}{\textrm{\emph{${}^\dagger$}}}

\numberwithin{equation}{section} \numberwithin{Thm}{section}

\begin{document}

\begin{flushright}
ZMP-HH/10-3\\
Hamburger Beitr\"age zur Mathematik Nr.\,361\\
\end{flushright}
\vskip 5.0em

\title{\mbox{Defects, dualities and the geometry of strings via
gerbes} \\ II. Generalised geometries with a twist,\\ the gauge
anomaly and the gauge-symmetry defect}

\author{Rafa\l ~R.~Suszek${}^*$}
\address{\emph{Address:}
Katedra Metod Matematycznych Fizyki, Wydzia\l ~Fizyki Uniwersytetu
Warszawskiego, ul.\,Ho\.za 74, PL-00-682 Warszawa, Poland}
\email{suszek@fuw.edu.pl}\thanks{${}^*$ The author's work was done
partly under the EPSRC First Grant EP/E005047/1, the PPARC rolling
grant PP/C507145/1 and the Marie Curie network `Superstring Theory'
(MRTN-CT-2004-512194). He was also funded by the Collaborative
Research Centre 676 ``Particles, Strings and the Early Universe --
the Structure of Matter and Space-Time'', and subsequently from the
Polish Ministry of Science and Higher Education grant No.\,N N201
372736.}

\begin{abstract}
This is the second in a series of papers discussing, in the
framework of gerbe theory, canonical and geometric aspects of the
two-dimensional non-linear sigma model in the presence of conformal
defects in the world-sheet. Employing the formal tools worked out in
the first paper of the series, 1101.1126 [hep-th], a thorough
analysis of rigid symmetries of the sigma model is carried out, with
emphasis on algebraic structures on generalised tangent bundles over
the target space of the theory and over its state space that give
rise to a realisation of the symmetry algebra on states. The
analysis leads to a proposal for a novel differential-algebraic
construct extending the original definition of the (gerbe-twisted)
Courant algebroid on the generalised tangent bundles over the target
space in a manner co-determined by the structure of the 2-category
of abelian bundle gerbes with connection over it. The construct
admits a neat interpretation in terms of a relative Cartan calculus
associated with the hierarchy of manifolds that compose the target
space of the multi-phase sigma model. The paper also discusses at
length the gauge anomaly for the rigid symmetries, derived and
quantified cohomologically in a previous work of Gaw\c{e}dzki,
Waldorf and the author. The ensuing reinterpretation of the small
gauge anomaly in terms of the twisted relative Courant algebroid
modelling the Poisson algebra of Noether charges of the symmetries
is elucidated through an equivalence between a category built from
data of the gauged sigma model and that of principal bundles over
the world-sheet with a structural action groupoid based on the
target space. Finally, the large gauge anomaly is identified with
the obstruction to the existence of topological defect networks
implementing the action of the gauge group of the gauged sigma model
and those giving a local trivialisation of a gauge bundle of an
arbitrary topology over the world-sheet.
\end{abstract}


\maketitle

\tableofcontents

\section{Introduction}\label{sec:intro}

Ever since the seminal contributions by Noether and Wigner, precise
identification and subsequent investigation of symmetries of the
physical system, both in the classical and in the quantum r\'egime,
has become physicists' obsession as one of the most fundamental and
effective tools of a systematic construction and exploration of
mathematical models of physical phenomena. The numerous
manifestations of the Symmetry Principle include the structuring of
the state space of the physical theory in terms of the
representation theory of the relevant current algebra and the
constraining of the analytic form of correlation functions of the
quantised theory with the help of the Ward--Takahashi identities.
Within the framework of local field theory, the Symmetry Principle
is invariably accompanied by the Gauge Principle which stipulates
that global (or rigid) symmetries of the theory be rendered local,
whereupon the theory be descended (or reduced) to the `physical'
space of orbits of the action of the thus engendered gauge group.
This gauging procedure can meet with obstructions -- the so-called
gauge anomalies -- whose analysis has served to restrict the range
of admissible models of quantum field theory, working as a
super-selection rule for interaction schemes consistent with the
assumed gauge invariance.

The concept of symmetry develops novel geometric and cohomological
aspects in the context of multi-phase non-linear $\si$-models, with
the structure of a metric manifold on the fibre -- termed the target
space -- of the covariant configuration bundle\footnote{Recall that
the bundle is defined as the fibre bundle over the spacetime of the
field theory under consideration whose sections are precisely the
lagrangean fields of the theory.} extended, upon incorporation of
the so-called (topological) Wess--Zumino interaction term in the
action functional, to include a geometric realisation of a
distinguished class in an appropriate (relative) real Deligne
hypercohomology group of the target space. The coexistence of
distinct phases of the field theory is marked by embedding in its
space-time codimension-1 loci of field discontinuity -- termed
domain walls or defects -- carrying cohomological data, pulled back
from the target space, that ensure invariance of the multi-phase
$\si$-model under those space-time diffeomorphisms which preserve
the defect. The presence of a smooth structure on the target space
prompts questions as to the existence of a geometric (that is
algebroidal resp.\ groupoidal) target-space model, understood as a
pre-image under a structure-preserving map, of the canonical
presentation of rigid symmetries of the $\si$-model on the state
space of the latter, be it in their infinitesimal form (through
Noether hamiltonians) or in the finite form (through automorphisms
of the space of states). The obvious measure of naturalness of such
a symmetry model is its compatibility with the hypercohomological
structure over the target space necessitated by a rigorous
definition of the Wess--Zumino term, as well as a simple
interpretation of the gauge anomaly furnished by it. The multi-phase
character of the field theories of interest, and -- in particular --
the defect-duality correspondence established in
\Rcite{Suszek:2011hg}, impose further coherence constraints on an
admissible symmetry model as they suggest the emergence of natural
relations (or morphisms, in an appropriate category), of an
intrinsically cohomological quality, between symmetry models
assigned to the phases of the field theory that are mapped to one
another across those special defects -- termed symmetric -- which
are transmissive to the symmetry currents of the respective phases.
These relations are -- in turn -- subject to secondary constraints
at defect self-intersections, expressing compatibility of their
definition with trans-defect splitting-joining interactions
(represented by non-trivial space-time topologies). The said
compatibility conditions correspond, in the canonical description,
to the requirement that there exist an intertwiner, induced from the
data pulled back to the intersections from the target space, between
representations of the symmetry algebra resp.\ group carried by the
phases converging at the defect intersection.\medskip

A methodical derivation of the target-space symmetry model and
verification of its naturalness (in the two-dimensional setting) is
the main objective of the present paper. It is attained through
elaboration and essential extension, to the multi-phase setting of
interest, of the earlier results -- obtained by Alekseev and Strobl
in \Rcite{Alekseev:2004np} -- on the (Courant-)algebroidal nature of
the model for infinitesimal symmetries of the mono-phase field
theory, with the underlying hypercohomological structure encoded by
the Hitchin isomorphisms of \Rcite{Hitchin:2004ut}. Instrumental in
the construction is the canonical description of the multi-phase
$\si$-model set up in the companion work \cite{Suszek:2011hg}.
Further structural background and guiding insights are provided by
the works \cite{Gawedzki:2010rn,Gawedzki:2012fu} of Gaw\c{e}dzki,
Waldorf and the author on the geometry and cohomology of the gauge
anomaly of the two-dimensional non-linear $\si$-model with the
Wess--Zumino term, in which a proposal was advanced, and
subsequently backed up by ample evidence in its favour, for the
target-space model of finite symmetries amenable to gauging. It is
given by an equivariant structure on the string background of the
$\si$-model, composed of a self-coherent collection of 0-, 1- and
2-cells of the 2-category of abelian bundle gerbes with connection
over the nerve of the (symmetry) action groupoid based on the target
space of the $\si$-model. In this language, the gauge anomaly is to
be understood as the topological obstruction to the existence of
some such equivariant structure. Its careful reappraisal from the
vantage point offered by the target-space models for infinitesimal
and finite rigid symmetries developed in the present paper, in
conjunction with a correspondence (also worked out hereunder)
between a category naturally associated with data of the gauged
$\si$-model\footnote{An object of the category of interest is a
principal bundle over the space-time of the $\si$-model with the
structure group given by the group $\,\txG_\si\,$ under gauging that
has the following extra property: the bundle associated to it
through the action of $\,\txG_\si\,$ on the target space of the
$\si$-model (whose existence is assumed in the first place) admits a
global section, interpreted as a lagrangean field of the gauged
$\si$-model.} and the category of principal bundles over the
$\si$-model space-time with a distinguished structural Lie groupoid,
demonstrates the necessity of coupling gauge fields of arbitrary
topology to the string background of the $\si$-model and yields a
conclusive corroboration of the proposal of
Refs.\,\cite{Gawedzki:2010rn,Gawedzki:2012fu}, mentioned above, for
the necessary and sufficient structure with which to endow the
string background when gauging its (finite) rigid symmetries. The
basic idea employed in the proof of the proposal consists in
reinterpreting the gauge anomaly in terms of the obstruction to the
existence of a local trivialisation of a gauge bundle of arbitrary
topology over the space-time of the multi-phase $\si$-model. A minor
(technical) variation on the same idea permits to approach and
understand the gauge anomaly from yet another angle, to wit, as an
obstacle to implementing -- in the spirit of the defect-duality
correspondence -- the local (\textit{i.e.}\ gauged) action of the
symmetry group in the gauged $\si$-model through topological defect
networks with data carried by defect junctions of arbitrary valence
canonically induced, in the manner discussed in
Refs.\,\cite{Runkel:2008gr} and \cite{Suszek:2011hg}, from those
carried by the elementary 3-valent ones. The latter construction is
to be seen as an explicit realisation of the concept, put forward in
\Rcite{Suszek:2011hg}, of a simplicial duality background,
consistent with the definition, extracted from the categorial
quantisation scheme in \Rcite{Frohlich:2009gb}, of the conformal
field theory reduced to the orbit space of the action of the
symmetry group on the target space of the parent $\si$-model.
\bigskip

We conclude this section with an outline of the contents of the
present paper. Thus, in Section \ref{sec:gen-geom-mono}, some basic
generalised-geometric constructs are introduced that capture the
algebra of infinitesimal rigid symmetries of the mono-phase
$\si$-model (in an arbitrary space-time dimension), and the
underlying hypercohomological structure is discussed. In Section
\ref{sec:gen-geom-mono-2d-spec}, the formalism from the previous
section is specialised to the two-dimensional setting of immediate
interest and interpreted as a target-space model of the Poisson
(resp.\ commutator) algebra of Noether hamiltonians of the rigid
symmetry on the state space of the mono-phase $\si$-model. Section
\ref{sec:Vin-morph} examines the issue of identification of the
Noether hamiltonians for the two phases of the $\si$-model set in
correspondence by a conformal defect in the algebraic and canonical
frameworks set up in the preceding sections. Section
\ref{sec:ext-non-intersect} gives an extension of the target-space
model for infinitesimal symmetries valid in the presence of
non-intersecting (symmetric) defect lines. In Section
\ref{sec:intertwiner}, circumstances are examined in which Noether
charges of the rigid symmetry are additively conserved in the
cross-defect splitting-joining interactions. In Section
\ref{sec:rel-cohom}, the construction of the target-space model for
(infinitesimal) symmetries of an arbitrary multi-phase $\si$-model
is completed and subsequently reinterpreted in the framework of a
relative differential (Cartan) calculus for the hierarchy of smooth
manifolds that compose the target space of the $\si$-model. Section
\ref{sec:anomaly} contains a comprehensive analysis of the various
canonical and geometric facets of the gauge anomaly, culminating in
a hands-on construction of the topological defect network
implementing the local action of the symmetry group in the gauged
multi-phase $\si$-model, as well as a local (space-time) description
of that $\si$-model coupled to a gauge field of an arbitrary
topology. Finally, Section \ref{sec:con-&-out} recapitulates the
main result of the paper and lists some outstanding related problems
that deserve, in the author's opinion, to be addressed in near
future.\medskip

The present paper is to be viewed as a direct continuation of the
companion work \cite{Suszek:2011hg} to which it makes frequent
reference, borrowing the notation, invoking the definitions, and
making explicit use -- without additional preparations -- of some of
the constructions. In view of this intimate relation between the two
papers, and for the reader's convenience, detailed references to
\cite{Suszek:2011hg} have been distinguished by attaching the Roman
numeral ``I'' to the relevant reference labels, as in ``Section
I.2'', ``Figure I.7'', ``Theorem I.5.8'' and ``Eq.\,(I.4.10)''.
\bigskip

\noindent{\bf Acknowledgements:}  The author is much beholden to
Ingo Runkel for pertinent comments on the manuscript, and to Nils
Carqueville for his sustained interest in the project reported in
the present paper. He also gratefully acknowledges the kind
hospitality of Laboratoire de Physique de l'\'Ecole Normale
Sup\'erieure de Lyon, Albert-Einstein-Institut in Potsdam, and the
B{\c{e}}dlewo Mathematical Research and Conference Center, as well
as the inspiring atmosphere of the XXXI Workshop on Geometric
Methods in Physics in Bia\l owie\.za and of the XXIX International
Colloquium on Group-Theoretical Methods in Physics in Tianjin, where
various parts of this work were carried out.

\section{A differential-algebraic structure for the $\si$-model}
\label{sec:gen-geom-mono}

The canonical description of the two-dimensional $\si$-model shares
many important structural properties with that of a charged
point-like particle in the background of an abelian gauge field
coupling to the particle's charge, with the free-loop space of the
target space of the $\si$-model replacing the particle's target
space, and the transgression bundle induced by the gerbe playing the
r\^ole of the circle gauge bundle of the point-particle model. From
this vantage point, it proves instructive to first generalise the
field-theoretic and geometric concepts introduced in Section
\vref{sec:lagr}I.2, thereby gaining insights into certain natural
algebraic structures associated with $\si$-models at large and some
interesting interrelations between those structures on the target
space and -- upon transgression -- on the state space of the theory.
\bedef\label{def:sigmod-n}
Let $\,(\xcM,\txg)\,$ be a metric manifold, termed -- as in
\Rcite{Suszek:2011hg} -- the \textbf{target space}, with a closed
$(n+2)$-form $\,\txH_{(n)},\ n\in\bN\,$ with periods from $\,2\pi
\bZ\,$ and an \textbf{$n$-gerbe $\,\cG_{(n)}\,$} of curvature
$\,\curv(\cG_{(n)})=\txH_{(n)}\,$ over it, the latter being
understood in the sense of \Rcite{Chatterjee:1998},\ \textit{i.e.}\
as a differential-geometric structure representing a class in the
Deligne hypercohomology group
$\,\bH^{n+1}(\xcM,\cD(n+1)_\xcM^\bullet)$.\ Thus, for a choice
$\,\xcMup\cO\,$ of an open cover of $\,\xcM$,\ $\,\cG_{(n)}\,$ is
defined by its \textbf{local presentation} in terms of a \v
Cech--Deligne $(n+1)$-cochain
\qq\nn
\cG_{(n)}\xrightarrow{\rm loc.}(B_i,A_{ij},\ldots,g_{i_1 i_2\ldots
i_{n+2}})=:b_{(n)}\in\cA^{n+2,n+1}(\xcMup\cO)\,,
\qqq
satisfying the cohomological identity
\qq\label{eq:DGn-is-Hn}
\xcMup D_{(n+1)}b_{(n)}=(\txH_{(n)}\vert_{\xcMup\cO_i},0,0,\ldots,1)
\,,
\qqq
and determined up to \textbf{gauge transformations}
\qq\nn
b_{(n)}\mapsto b_{(n)}+\xcMup D_{(n)}\pi_{(n)}\,,\qquad\pi_{(n)}:=(
\Pi_i,\D_{ij},\ldots,\chi_{i_i i_2\ldots i_{n+1}})\in\cA^{n+2,n}(
\xcMup\cO)\,,
\qqq
all in the conventions set up in Definition \vref{def:loco}I.2.2.
The triple will be denoted jointly as
$\,(\xcM,\txg,\cG_{(n)})=:\cM_{(n )}\,$ and termed the
\textbf{$n$-target}. Furthermore, let $\,( \Om_{n+1},\eta)\,$ be a
closed oriented $(n+1)$-dimensional manifold with an intrinsic
minkowskian\footnote{The definition could readily be extended so as
to allow for generic intrinsic metrics of a lorentzian signature. As
the ensuing structure of a reparametrisation-invariant $\si$-model
is irrelevant to our considerations, we simply assume that the
minkowskian gauge for the intrinsic metric has been fixed in the
classical theory. \textit{Cf.}\ the footnote on p.\,9 of
\Rcite{Suszek:2011hg}.} metric $\,\eta=\diag(-1,+1, +1,\ldots,+1)$,\
termed the \textbf{world-volume} and embedded in $\,\xcM\,$ by a
continuously differentiable map $\,X:\Om_{n+1}\to \xcM$,\ to be
called the \textbf{embedding field}. The \textbf{$(n+1)$-dimensional
non-linear $\si$-model for embedding fields $\,X\,$ with $n$-target
$\,\cM_{(n)}\,$ on world-volume $\,(\Om_{n+1},\eta)\,$} is a theory
of continuously differentiable maps $\,X:\Om_{n+1}\to \xcM\,$
determined by the principle of least action applied to the action
functional
\qq\label{eq:sigma}
S^{(n+1)}_\si[X]=-\tfrac{1}{2}\,\int_{\Om_{n+1}}\,\txg_X(\sfd X
\overset{\wedge}{,}\star_\eta\sfd X)+S^{(n+1)}_{\rm top}[X]\,,
\qqq
in which
\bit
\item $\sfd X(\si)=\p_a X^\mu\,\sfd\si^a\ox\p_\mu\vert_{X(\si)}$,\
in local coordinates $\,\{\si^a\}^{a\in\ovl{1,n+1}}\,$ on
$\,\Om_{n+1}\,$ and $\{X^\mu \}^{\mu\in\ovl{1,\dim\,\xcM}}\,$ on
$\,\xcM$,\ and the target-space metric is assumed to act on the
second factor of the tensor product;
\item $\star_\eta\,$ is the Hodge operator on $\,\Om^\bullet(\Om_{n+
1})\,$ determined by $\,\eta$;
\item the topological term
\qq\nn
S^{(n+1)}_{\rm top}[X]=-\sfi\,\log\Hol_{\cG_{(n)}}(X)
\qqq
is defined by the hypersurface holonomy $\,\Hol_{\cG_{(n)}}(X)\,$ of
the $n$-gerbe, which is an obvious generalisation of the (2-)surface
holonomy of the (1-)gerbe $\,\cG\,$ from Definition
\vref{def:sigmod}I.2.7 (with defect contributions dropped),
\textit{i.e.}\ as a Cheeger--Simons differential character for
$(n+1)$-dimensional hypersurfaces, determined by a trivialisation of
the pullback $n$-gerbe $\,X^*\cG_{(n)}\,$ as
\qq\nn
\log\Hol_{\cG_{(n)}}(X)=[X^*\cG_{(n)}]\in\vH^n\left(\Om_{n+1},\uj
\right)\cong\uj\,.
\qqq
\eit
\exdef \noindent The last property of the topological term
immediately leads to
\berop\label{prop:var-sigmod-n}
Let $\,\cM_{(n)}=(\xcM,\txg,\cG_{(n)})\,$ be an $n$-target with
$n$-gerbe $\,\cG_{(n)}\,$ of curvature $\,\txH_{(n)}\in Z^{n+2}(\xcM
)$,\ and let $\,\xcV\,$ be a vector field on $\,\xcM\,$ with a
(local) flow $\,\xi_t:\xcM\to\xcM\,$ (assumed to exist). The
variation of the action functional $\,S^{(n+1)}_\si[X]\,$ of
\Reqref{eq:sigma} along $\,\xi_t\,$ is then given by
\qq\label{eq:var-sigmod-n}
\tfrac{\sfd\ }{\sfd t}\big\vert_{t=0}S^{(n+1)}_\si[\xi_t\circ X]=-
\tfrac{1}{2}\,\int_{\Om_{n+1}}\,(\pLie{\xcV}\txg)_X(\sfd X
\overset{\wedge}{,}\star_\eta\sfd X)
+\int_{\Om_{n+1}}\,X^*(\xcV\con\txH_{(n)})\,,
\qqq
where $\,\pLie{\xcV}\,$ is the Lie derivative in the direction of
the vector field $\,\xcV$.
\eerop
\beroof Obvious, through inspection. \textit{Cf.}\
\Rxcite{App.\,A.2}{Runkel:2008gr}. \eroof\medskip

\noindent From \Reqref{eq:var-sigmod-n}, we can immediately read off
internal (\textit{i.e.}\ rigid) symmetries of the
$(n+1)$-dimensional $\si$-model.
\becor\cite[App.\,A2]{Runkel:2008gr}\cite[Cor.\,2.2]{Gawedzki:2010rn}
In the notation of Proposition \ref{prop:var-sigmod-n}, internal
symmetries of the $(n+1)$-dimensional non-linear $\si$-model for
embedding fields $\,X\,$ with $n$-target $\,\cM_{(n )}\,$ on
world-volume $\,(\Om_{n+1},\eta)\,$ correspond to pairs
$\,(\xcV,\upsilon)\,$ composed of a vector field $\,\xcV\in\G(\sfT
M)\,$ that is \textbf{Killing} for $\,\txg$,
\qq\nn
\pLie{\xcV}\txg=0\,,
\qqq
and an $n$-form $\,\upsilon\in\Om^n(M)\,$ subject to the constraint
\qq\nn
\sfd\upsilon+\xcV\con\txH_{(n)}=0\,.
\qqq
\ecor The last observation points towards a distinguished and
natural r\^ole played by the bundle $\,\sfT M\oplus\wedge^n\sfT^*M
\to M\,$ over the fibre of the covariant configuration bundle of the
$\si$-model in the description of (infinitesimal internal)
symmetries of the latter. We shall, next, study the relevant
geometric and algebraic constructs in some detail with view to
elaborating this issue.
\bedef\label{def:Vin-str}
Let $\,\xcM\,$ be a smooth manifold of dimension $\,\dim\, \xcM\geq
n\in\bN$,\ with tangent bundle $\,\sfT\xcM\to\xcM\,$ and cotangent
bundle $\,\sfT^*\xcM\to\xcM$.\ The \textbf{generalised tangent
bundle of type $\,(1,n)\,$ over $\,\xcM\,$} is the Whitney sum
\qq\nn
\sfE^{(1,n)}\xcM:=\sfT\xcM\oplus\wedge^n\sfT^*\xcM\to\xcM \,.
\qqq
The vector bundle
\qq\nn
\sfE^{(n,1)}\xcM:=\wedge^n\sfT\xcM\oplus\sfT^*\xcM\to\xcM\,,\qquad
n\in\bN_{>0}\,,
\qqq
dual to $\,\sfE^{(1,n)}\xcM\,$ through the non-degenerate pairing of
sections
\qq\nn
\corr{\cdot,\cdot}\ :\ \G\bigl(\sfE^{(1,n)}\xcM\bigr)\x\G\bigl(
\sfE^{(n,1)}\xcM\bigr)\to C^\infty(\xcM,\bR)\ : \ (\xcV\oplus\nu,
\xcW\oplus\varpi)\mapsto\xcV\con\varpi+\xcW\con\nu\,,
\qqq
will be termed the \textbf{generalised cotangent bundle of type
$\,(n,1)\,$ over $\,\xcM$}. In the distinguished case of $\,n=0$,\
we define
\qq\nn
\sfE^{(0,1)}\xcM:=(\xcM\x\bR)\oplus\sfT^*\xcM\to\xcM\,,
\qqq
which is, again, dual to $\,\sfE^{(1,0)}\,$ by the non-degenerate
pairing of sections
\qq\nn
\corr{\cdot,\cdot}\ :\ \G\bigl(\sfE^{(1,0)}\xcM\bigr)\x\G\bigl(
\sfE^{(0,1)}\xcM\bigr)\to C^\infty(\xcM,\bR)\ : \ (\xcV\oplus f,g
\oplus\varpi)\mapsto\xcV\con\varpi+f\cdot g\,.
\qqq
The space of smooth sections of the generalised tangent bundle of
type $\,(1,n)\,$ is equipped with a natural antisymmetric bilinear
operation
\qq\label{eq:Vin-bra-1n}
\Vbra{\xcV\oplus\upsilon}{\xcW\oplus\varpi}:=[\xcV,\xcW]\oplus
\bigl(\pLie{\xcV}\varpi-\pLie{\xcW}\upsilon-\tfrac{1}{2}\,\sfd(\xcV
\con\varpi-\xcW\con\upsilon)\bigr)
\qqq
termed the \textbf{Vinogradov bracket} and introduced in
Refs.\,\cite{Vinogradov:1990un,Cabras:1992ex}. In the formula, the
bracket in the vector-field component of the right-hand side is the
standard Lie bracket of vector fields on $\,\xcM$,\ and we have, in
particular,
\qq\label{eq:Vin-bra-10}
\Vbra{\xcU\oplus f}{\xcV\oplus g}=[\xcU,\xcV]\oplus(\xcU(g)
-\xcV(f))
\qqq
for sections $\,\xcU\oplus f,\xcV\oplus g\in\G(\sfE^{(1,0)}\xcM)$.\
The quadruple $\,(\sfE^{(1,n)}\xcM,\Vbra{\cdot}{\cdot},\Vcon{\cdot}{
\cdot},\a_{\sfT\xcM})=:\Vgt^{(n)}\xcM$,\ containing, in addition to
the previously described elements, also the symmetric
\textbf{canonical contraction}
\qq
\barr{l} \Vcon{\cdot}{\cdot}\ :\ \G\bigl(\sfE^{(1,n)}\xcM\bigr)\x\G
\bigl(\sfE^{(1,n)}\xcM\bigr)\to\Om^{n-1}(\xcM)\ :\ (\xcV\oplus
\upsilon,\xcW\oplus\varpi)\mapsto\tfrac{1}{2}\,(\xcV\con\varpi+\xcW
\con\upsilon)\,,\quad n>0\cr\cr \Vcon{\cdot}{\cdot}\ :\ \G\bigl(
\sfE^{(1,0)}\xcM\bigr)\x\G\bigl(\sfE^{(1,0)}\xcM\bigr)\to\{0\}\ :\
(\xcV\oplus f,\xcW\oplus g)\mapsto 0\,,\earr\cr\cr
\label{eq:Vcon-def}
\qqq
and the \textbf{anchor} $\,\a_{\sfT\xcM}:\sfE^{(1,n)}\xcM\to\sfT
\xcM\,$ given by the canonical projection, will be called the
\textbf{canonical Vinogradov structure on $\,\sfE^{(1,n)}\xcM$}.
\exdef \noindent We readily establish the important property
\berop\label{prop:Vin-str-auts}
In the notation of Definition \ref{def:Vin-str},\ let $\,(f,F)\,$ be
an automorphism of $\,\sfE^{(1,n)}\xcM\,$ composed of a
diffeomorphism $\,f\in\Diff(\xcM)\,$ and a (fibre-wise) linear map
$\,F:\sfE^{(1,n)}\xcM\to\sfE^{(1,n)}\xcM\,$ covering $\,f\,$ in the
sense expressed by the commutative diagram
\qq\nn
\alxydim{@C=1.cm@R=1.cm}{\sfE^{(1,n)}\xcM \ar[r]^{F}
\ar[d]_{\pi_{\sfE^{(1,n)}\xcM}} & \sfE^{(1,n)}\xcM
\ar[d]^{\pi_{\sfE^{(1,n)}\xcM}}
\cr \xcM \ar[r]^{f} & \xcM}\,.
\qqq
Suppose also that $\,F\,$ is an automorphism of the Vinogradov
structure $\,\Vgt^{(n)}\xcM$,\ \textit{i.e.}\footnote{By a slight
abuse of the notation, we denote the map on sections by the same
symbol as the one used for the bundle map.}
\qq
\Vbra{\cdot}{\cdot}\circ(F,F)&=&F\circ\Vbra{\cdot}{\cdot}\,,
\label{eq:auto-Vin-bra}\\\cr \Vcon{\cdot}{\cdot}\circ(F,F)&=&(f^{-
1})^*\circ\Vcon{\cdot}{\cdot}\,.\label{eq:auto-can-contr}
\qqq
Then, the condition
\qq\nn
\a_{\sfT\xcM}\circ F=f_*\circ\a_{\sfT\xcM}
\qqq
follows automatically, and $\,F\,$ is necessarily of the form
\qq\nn
F=\widehat f\circ\ee^\txB\circ\widehat c_n\,,
\qqq
with
\qq\nn
\widehat f:=\left(\barr{cc} f_* & 0 \cr\cr 0 & (f^{-1})^* \earr
\right)
\qqq
acting on sections $\,\xcV\oplus\upsilon\in\G\bigl(\sfE^{(1,n)}\xcM
\bigr)\,$ as
\qq\nn
\widehat f\lact(\xcV\oplus\upsilon):=f_*\xcV\oplus(f^{-1})^*
\upsilon\,,
\qqq
with
\qq\label{eq:eeB}
\ee^\txB:=\left(\barr{cc} \id_{\G(\sfT\xcM)} & 0 \cr\cr \txB &
\id_{\Om^n(\xcM)}\earr\right)\,,\qquad\qquad\txB\in Z^{n+1}(\xcM)
\qqq
acting as
\qq\nn
\ee^\txB\lact(\xcV\oplus\upsilon):=\xcV\oplus(\upsilon+\xcV\con\txB)
\,,
\qqq
and with
\qq\nn
\widehat c_n=\left(\barr{cc} \id_{\G(\sfT\xcM)} & 0 \cr\cr 0 &
c^{\d_{n,0}}\,\id_{\Om^n(\xcM)} \earr \right)\,,\qquad c\in\bR^\x
\qqq
acting as
\qq\nn
\widehat c_n\lact(\xcV\oplus\upsilon):=\xcV\oplus c^{\d_{n,0}}\cdot
\upsilon\,.
\qqq
\eerop
\beroof
First of all, note that $\,\widehat f$,\ in which $\,f_*\,$ is the
covering map for $\,f\,$ on the total space $\,\sfT\xcM\,$ (and its
tensor powers) and $\,(f^{-1})^*\,$ has the same interpretation for
$\,\sfT^*\xcM\,$ (whence also its appearance on the right-hand side
of \Reqref{eq:auto-can-contr}), is an automorphism of
$\,\Vgt^{(n)}\xcM$.\ This follows immediately from the identities
\qq\nn
[\cdot,\cdot]\circ(f_*,f_*)=f_*\circ[\cdot,\cdot]\,,
\qqq
written in terms of the Lie bracket $\,[\cdot,\cdot]\,$ of vector
fields, and from
\qq\nn
f_*\xcV\con(f^{-1})^*\upsilon=(f^{-1})^*(\xcV\con\upsilon)\,,
\qqq
the latter being satisfied for arbitrary $\,\xcV\in\G(\sfT\xcM)\,$
and $\,\upsilon\in\Om^n(\xcM)$.

We may, next, consider the automorphism $\,(\id_\xcM,G):=\widehat
f^{-1}\circ F\,$ of $\,\Vgt^{(n)}\xcM$,\ covering the identity
diffeomorphism on $\,\xcM$.\ Let us begin with the case of $\,n>0$.\
Take an arbitrary $\,g\in C^\infty(\xcM,\bR)\,$ and compute, for any
$\,\Vgt,\Wgt\in\G\bigl(\sfE^{(1,n)} \xcM\bigr)$,\ the expression
\qq\nn
\Vbra{g\cdot\Vgt}{\Wgt}=g\cdot\Vbra{\Vgt}{\Wgt}-\a_{\sfT\xcM}(\Wgt)
(g)\cdot\Vgt+0\oplus\sfd g\wedge\Vcon{\Vgt}{\Wgt}\,.
\qqq
The assumption that $\,G\,$ is an automorphism of $\,\Vgt^{(n)}\xcM
\,$ covering the identity diffeomorphism gives
\qq\label{eq:Vin-aut-onf}\qquad\qquad
\bigl[\a_{\sfT\xcM}(\Wgt)(g)-\a_{\sfT\xcM}\bigl(G(\Wgt)\bigr)(g)
\bigr]\cdot G(\Vgt)=G\bigl(0\oplus\sfd g\wedge\Vcon{\Vgt}{\Wgt}
\bigr)-0\oplus\sfd g\wedge\Vcon{\Vgt}{\Wgt}\,.
\qqq
Upon choosing $\,\Vgt=\xcV\oplus 0\,$ and $\,\Wgt=\xcW\oplus 0\,$
for arbitrary vector fields $\,\xcV,\xcW$,\ so that $\,
\Vcon{\Vgt}{\Wgt}=0$,\ the above reduces to
\qq\label{eq:Vin-aut-aux1}
\bigl[\xcW(g)-\a_{\sfT\xcM}\bigr(G(\xcW\oplus 0)\bigr)(g)\bigr]
\cdot G(\xcV\oplus 0)=0\,.
\qqq
Clearly, $\,G\vert_{\G(\sfT\xcM)\oplus\{0\}}\not\equiv 0\,$ (as an
automorphism). Using this, in conjunction with the arbitrariness of
$\,g\,$ in \Reqref{eq:Vin-aut-aux1}, we conclude that the identity
\qq\nn
\a_{\sfT\xcM}\bigr(G(\xcW\oplus 0)\bigr)=\xcW
\qqq
has to hold true for all $\,\xcW\in\G(\sfT\xcM)$,\ whence
\qq\nn
G=\left(\barr{cc} \id_{\G(\sfT\xcM)} & G_{1,2} \cr\cr G_{2,1} &
G_{2,2} \earr\right)
\qqq
for some linear operators
\qq\nn
&G_{1,2}\ :\ \Om^n(\xcM)\to\G(\sfT\xcM)\,,\qquad\qquad G_{2,1}\ :\
\G(\sfT\xcM)\to\Om^n(\xcM)\,,&\cr\cr
&G_{2,2}\ :\ \Om^n(\xcM)\to\Om^n(\xcM)\,.&
\qqq
The second of the three, $\,G_{2,1}$,\ is a section of $\,\sfT^*
\xcM\ox\wedge^n\sfT^*\xcM\,$ which is readily seen, via
\qq\nn
0\equiv\Vcon{\xcV\oplus 0}{\xcW\oplus 0}=\Vcon{G(\xcV\oplus 0)}{G(
\xcW\oplus 0)}=\tfrac{1}{2}\,\left(G_{2,1}(\xcV,\xcW,\ldots)+G_{2,1}
(\xcW,\xcV,\ldots)\right)\,,
\qqq
valid for arbitrary vector fields $\,\xcV,\xcW$,\ to be an
$(n+1)$-form,
\qq\nn
G_{2,1}=:\txB\in\Om^{n+1}(\xcM)\,.
\qqq
Having established that, take $\,\Vgt=\xcV\oplus\upsilon\,$
arbitrary and set $\,\Wgt=(-\xcV)\oplus\upsilon$,\ so that, again,
$\,\Vcon{\Vgt}{\Wgt}=0\,$ and \Reqref{eq:Vin-aut-onf} yields
\qq\nn
G_{1,2}(\upsilon)(g)\cdot\bigl[\bigl(\xcV+G_{1,2}(\upsilon)\bigr)
\oplus\bigl(\xcV\con\txB+G_{2,2}(\upsilon)\bigr)\bigr]=0\,.
\qqq
The vanishing of the vector-field component on the left-hand side of
the above identity implies -- in virtue of the arbitrariness of $\,
\xcV\,$ and $\,g\,$ --
\qq\nn
G_{1,2}(\upsilon)\equiv 0\,.
\qqq
This ensures that the identity
\qq\nn
\a_{\sfT\xcM}\circ G=\a_{\sfT\xcM}
\qqq
obtains, and so we can rewrite \Reqref{eq:Vin-aut-onf} as
\qq\nn
\sfd g\wedge\Vcon{\Vgt}{\Wgt}=G_{2,2}\bigl(\sfd g\wedge
\Vcon{\Vgt}{\Wgt}\bigr)\,.
\qqq
We conclude that
\qq\nn
G_{2,2}=\id_{\Om^n(\xcM)}\,.
\qqq
At this stage, it remains to check Eqs.\,\eqref{eq:auto-Vin-bra} and
\eqref{eq:auto-can-contr} for the operator
\qq\nn
G=\left(\barr{cc} \id_{\G(\sfT\xcM)} & 0 \cr\cr \txB & \id_{\Om^n(
\xcM)}\earr\right)
\qqq
derived above. We find, for arbitrary sections $\,\Vgt=\xcV\oplus
\upsilon\,$ and $\,\Wgt=\xcW\oplus\varpi\,$ of $\,\sfE^{(1,n)}\xcM$,
\qq\nn
\Vcon{G(\Vgt)}{G(\Wgt)}=\tfrac{1}{2}\,(\xcV\con\xcW\con\txB+\xcV
\con\varpi+\xcW\con\xcV\con\txB+\xcW\con\upsilon)=\Vcon{\Vgt}{\Wgt}\,,
\qqq
which is the desired result, and
\qq\nn
\Vbra{G(\Vgt)}{G(\Wgt)}&=&\Vbra{\Vgt}{\Wgt}+0\oplus\bigl(
\pLie{\xcV}(\xcW\con\txB)-\pLie{\xcW}(\xcV\con\txB)-\tfrac{1}{2}\,
\sfd(\xcV\con \xcW\con\txB-\xcW\con\xcV\con\txB)\bigr)\cr\cr
&=&\Vbra{\Vgt}{\Wgt}+0\oplus\bigl([\xcV,\xcW]\con\txB-\xcV\con\xcW
\con\sfd\txB\bigr)\equiv G\bigl(\Vbra{\Vgt}{\Wgt}\bigr)-0\oplus\xcV
\con\xcW\con\sfd\txB\,,
\qqq
from which the thesis of the proposition follows for $\,n>0$.

Passing to the case of $\,n=0$,\ we note that, owing to the
triviality of the canonical contraction, \Reqref{eq:auto-can-contr}
is satisfied automatically, and \Reqref{eq:Vin-aut-onf} now
simplifies as
\qq\label{eq:gVW-Vbra}
\bigl[\a_{\sfT\xcM}(\Wgt)(g)-\a_{\sfT\xcM}\bigl(G(\Wgt)\bigr)(g)
\bigr]\cdot G(\Vgt)=0\,.
\qqq
Invoking the assumed automorphicity of $\,G$,\ we infer, due to the
arbitrariness of $\,g,\Vgt\,$ and $\,\Wgt$,\ that
\qq\nn
G=\left(\barr{cc} \id_{\G(\sfT\xcM)} & 0 \cr\cr \txB & C
\earr\right)
\qqq
for some $\,\txB\in\G(\sfT^*\xcM)\,$ and $\,C\in C^\infty(\xcM,\bR
)$.\ Upon substitution of the above into \Reqref{eq:auto-Vin-bra},
the latter being evaluated on $\,\Vgt=\xcV\oplus f\,$ and
$\,\Wgt=\xcW\oplus g$,\ we obtain the condition
\qq\nn
\xcW\con\xcV\con\sfd\txB+g\cdot\xcV(C)-f\cdot\xcW(C)=0\,,
\qqq
which leads to the result
\qq\label{eq:G-n0}
G=\left(\barr{cc} \id_{\G(\sfT\xcM)} & 0 \cr\cr \txB & c
\earr\right)
\qqq
with $\,\txB\in Z^1(\xcM)\,$ and $\,c\in\bR$.\ The requirement of
invertibility of $\,G\,$ ultimately fixes the range of $\,c\,$ as
$\,\bR\setminus\{0\}\,$ and thereby completes the proof. \eroof
\brem In the distinguished case of $\,n=1$,\ the generalised tangent
bundle of type $\,(1,n)\,$ becomes self-dual, the canonical
contraction coincides with the duality, and the canonical Vinogradov
structure is equivalent to the canonical Courant algebroid of
Refs.\,\cite{Courant:1990,Dorfman:1993,Liu:1997}, with $\,
\Vbra{\cdot}{\cdot}\,$ the canonical Courant bracket. Proposition
\ref{prop:Vin-str-auts} then reproduces the classification result of
\Rxcite{Prop.\,3.24}{Gualtieri:2003dx}.\erem

In order to put the distinguished case $\,n=0\,$ on equal footing
with the remaining cases, and -- more importantly -- with view to
subsequent applications of the formalism developed in the context of
the two-dimensional $\si$-model, we specialise the previous
definition as
\bedef
In the notation of Definition \ref{def:Vin-str} and of Proposition
\ref{prop:Vin-str-auts}, a \textbf{unital automorphism of
generalised tangent bundle $\,\sfE^{(1,n)}\xcM\,$} is an
automorphism $\,(f,F)\,$ of $\,\sfE^{(1,n)}\xcM\,$ with the
additional property that
\qq\nn
\pr_{\Om^n(\xcM)}\circ F=(f^{-1})^*\circ\pr_{\Om^n(\xcM)}\,.
\qqq
\exdef
\becon
From now onwards, all morphisms between generalised tangent bundles
will be assumed unital. Whenever possible, this will be explicitly
marked by a subscript $\,\textrm{u}\,$ on the symbols of the
relevant morphism sets. \econ

In the presence of an $n$-gerbe over $\,\xcM$,\ there arises a
natural notion of a topological twist of the bundle $\,\sfE^{(1,n)}
\xcM\,$ and of the algebraic structure $\,\Vgt^{(n)}\xcM\,$ on it.
In general,
\bedef\label{def:tw-gen-tan-bun}
Adopt the notation of Definition \ref{def:Vin-str}, and let
$\,\xcMup\cO=\{\xcMup\cO_i\}_{i\in \xcI_\xcM}\,$ be an open cover of
$\,\xcM\,$ with an index set $\,\xcI_\xcM$,\.\ A \textbf{twisted
generalised tangent bundle of type $\,(1,n)\,$ over $\,\xcM\,$} is a
vector bundle $\,\sfE^{(1,n)}_{\{\ggt_{ij}\}}\xcM\to\xcM\,$ with a
total space locally isomorphic to $\,\sfE^{(1,n)}\xcM\,$ and
determined by a collection $\,(\ggt_{ij})_{i,j\in\xcI_\xcM}\,$ of
\textbf{transition maps} $\,\ggt_{ij}\in\End_{\rm u}\bigl(\sfE^{(1,
n)}\xcM(\xcMup\cO_{ij})\bigr),\ \xcMup\cO_{ij}=\xcMup\cO_i\cap\xcMup
\cO_j$.\ The maps are required to cover the identity diffeomorphism
on $\,\xcM\,$ and to satisfy the usual cocycle condition
\qq\label{eq:gen-tw-cocyc}
(\ggt_{jk}\circ\ggt_{ij})\vert_{\xcMup\cO_{ijk}}=\ggt_{ik}
\vert_{\xcMup\cO_{ijk}}
\qqq
on non-empty triple intersections $\,\xcMup\cO_{ijk}=\xcMup\cO_i\cap
\xcMup\cO_j\cap\xcMup\cO_k$.\ Thus, the bundle has local sections
$\,\Vgt_i\in\sfE^{( 1,n)}\xcM(\xcMup\cO_i)\,$ related as per
\qq\nn
\Vgt_j\vert_{\xcMup\cO_{ij}}=\ggt_{ij}\lact\Vgt_i\vert_{\xcMup\cO_{i
j}}
\qqq
on non-empty double intersections $\,\xcMup\cO_{ij}$.\ An
\textbf{isomorphism between a pair $\,\sfE^{(1,n)}_{\{\ggt^\a_{ij}
\}}\xcM,\ \a\in\{1,2\} \,$} is a collection $\,\xcMup\chi=(\hgt_i
)_{i\in\xcI_\xcM}\,$ of local bundle maps $\,\hgt_i\in\End_{\rm u}
\bigl(\sfE^{(1,n)}\xcM(\xcMup\cO_i)\bigr)\,$ covering the identity
diffeomorphism on $\,\xcM\,$ and such that
\qq\nn
\ggt_{ij}^2=\hgt_j\circ\ggt_{ij}^1\circ\hgt_i^{-1}\,.
\qqq
These induce maps
\qq\label{eq:gen-tw-equiv-sec}
\Vgt_i^2=\hgt_i\lact\Vgt_i^1
\qqq
between the respective local sections $\Vgt_i^\a\in\sfE^{(1,n)}_{\{
\ggt^\a_{ij}\}}\xcM(\xcMup\cO_i)$.\exdef \brem Twisted generalised
tangent bundles (of type $\,(1,1)$) were first introduced in
\Rcite{Hitchin:2004ut}, \textit{cf.}\ also \Rcite{Baraglia:2007}, in
the restricted form in which the twist was determined by local data
of a gerbe, \textit{cf.}\ Corollary \ref{cor:tw-gen-tan-vs-ngerb}
for a generalisation of that result.\erem \brem We could also
consider more general isomorphisms covering diffeomorphisms between
different bases. That, however, while completely straightforward in
itself, would necessitate -- at least in the present (local)
formulation -- the introduction of \v Cech-extended manifold maps
(in the sense of \Rcite{Runkel:2008gr}), a complication that we
choose to avoid here.\erem \noindent We augment the above definition
with
\bedef\label{def:loc-glob-Vin-str}
In the notation of Definitions \ref{def:Vin-str} and
\ref{def:tw-gen-tan-bun}, a \textbf{local Vinogradov structure on
$\,\sfE^{(1,n)}_{\{\ggt_{ij}\}} \xcM\,$} is a collection of
Vinogradov structures $\,\left(\sfE^{(1,n)}(\xcMup\cO_i),
\Vbra{\cdot}{\cdot},\Vcon{\cdot}{\cdot},\a_{\sfT\xcM}\right)\,$ over
components $\,\xcMup\cO_i\,$ of $\,\xcMup\cO$.\ We say that there
exists a \textbf{global Vinogradov structure on $\,\sfE^{(1,n)}_{\{
\ggt_{ij}\}}\xcM\,$} iff the transition maps $\,\ggt_{ij}\,$ map the
local sections \emph{homomorphically} into one another, so that the
Vinogradov bracket of local sections $\,\Vgt_i,\Wgt_i\in\sfE^{(1,n
)}_{\{\ggt_{ij}\}}(\xcMup\cO_i)\,$ is also a local section over
$\,\xcMup\cO_i$,
\qq\nn
\ggt_{ij}\lact\Vbra{\Vgt_i}{\Wgt_i}=\Vbra{\Vgt_j}{\Wgt_j}\,.
\qqq

An \textbf{isomorphism between global Vinogradov structures
$\,\Vgt^{(n)}_{\{\ggt_{ij}\}}\xcM:=\left(\sfE^{(1,n)}_{\{\ggt^\a_{i
j}\}}\xcM,\Vbra{\cdot}{\cdot}^\a,\Vcon{\cdot}{\cdot}^\a,\a_{\sfT
\xcM}\right)\,$} is an isomorphism $\,\xcMup \chi\ :\ \sfE^{(1,n
)}_{\{\ggt^1_{ij}\}}\xcM\xrightarrow{\cong}\sfE^{(1,n)}_{\{
\ggt^2_{ij}\}}\xcM\,$ that lifts to a homomorphism of the respective
local Vinogradov structures as per
\qq\nn
\Vbra{\cdot}{\cdot}^2\circ(\hgt_i,\hgt_i)&=&\hgt_i\circ
\Vbra{\cdot}{\cdot}^1\,,\cr\cr
\Vcon{\cdot}{\cdot}^2\circ(\hgt_i,\hgt_i)&=&\Vcon{\cdot}{\cdot}^1
\,.
\qqq
\exdef \noindent We readily establish
\berop\label{prop:Vin-str-glob-tw} In the notation of Definitions
\ref{def:tw-gen-tan-bun} and \ref{def:loc-glob-Vin-str}, with
$\,\xcMup\cO\,$ a \emph{good} open cover of $\,\xcM$,\ there exists
a global Vinogradov structure $\,\Vgt^{(n)}_{\{\ggt_{ij}\}}\xcM\,$
on $\,\sfE^{(1,n )}_{\{\ggt_{ij}\}}\xcM\to\xcM\,$ iff the transition
maps of the bundle can be written as
\qq\label{eq:Vin-glob-on-gen-tw-trans}
\ggt_{ij}:=\ee^{(-1)^n\,\sfd A_{ij}}
\qqq
for some $\,A_{ij}\in\Om^n(\xcMup\cO_{ij})\,$ such that
\qq\label{eq:Vin-glob-on-gen-tw-etc}
(A_{jk}-A_{ik}+A_{ij})\vert_{\xcMup\cO_{ijk}}=(-1)^n\,\sfd^{(n)}
C_{ijk}
\qqq
for some $\,C_{ijk}\in\Om^{n-1}(\xcMup\cO_{ijk})$.\ Here, $\,\sfd^{(
n)}=\sfd\,$ for all $\,n\neq 0$,\ and $\,\sfd^{(0)}\,$ is the
trivial embedding of the sheaf of locally constant $\bR$-valued
functions into the sheaf of locally smooth $\bR$-valued functions,
\textit{cf.}\ Section \ref{sub:tan-sheaf}. There exists an
isomorphism between a pair $\,\Vgt^{(n)}_{\{\ggt_{ij}^\a\}}\xcM,\ \a
\in\{1,2\}\,$ of global Vinogradov structures iff there is an
isomorphism
\qq\nn
\xcMup\chi=(\hgt_i)_{i\in\xcI_\xcM}\ :\ \sfE^{(1,n)}_{\{\ggt_{ij}^1
\}}\xcM\xrightarrow{\cong}\sfE^{(1,n)}_{\{\ggt_{ij}^2\}}\xcM\,,
\qqq
understood as in Definition \ref{def:tw-gen-tan-bun}, with local
data of the form
\qq\label{eq:gauge-trans-gen-tw}
\hgt_i=\ee^{-\sfd\Pi_i}\,,
\qqq
for some $\,\Pi_i\in\Om^n(\xcMup\cO_i)$.
\eerop
\beroof
A simple consequence of the assumed goodness of the open cover and
of Proposition \ref{prop:Vin-str-auts}.\eroof\medskip \noindent We
can also twist the Vinogradov bracket itself, to wit,
\bedef\label{def:Om-tw-Vin}
Assume the notation of Definition \ref{def:Vin-str}. The
\textbf{$\Om_{(n+2)}$-twisted Vinogradov structure on $\,\sfE^{(1,n
)}\xcM\,$} is the quadruple $\,(\sfE^{(1,n)}\xcM,
\Vbra{\cdot}{\cdot}^{\Om_{(n+2)}},\Vcon{\cdot}{\cdot},\a_{\sfT\xcM}
)=:\Vgt^{(n),\Om_{(n+2)}}\xcM\,$ in which the antisymmetric bilinear
operation $\, \Vbra{\cdot}{\cdot}^{\Om_{(n+2)}}\,$ on sections of
$\,\sfE^{(1,n)} \xcM$,\ to be termed the \textbf{$\Om_{(n+2
)}$-twisted Vinogradov bracket}, is given by the formula
\qq\nn
\Vbra{\Vgt}{\Wgt}^{\Om_{(n+2)}}:=\Vbra{\Vgt}{\Wgt}+0\oplus\a_{\sfT
\xcM}(\Vgt)\con\a_{\sfT\xcM}(\Wgt)\con\Om_{(n+2)}\,,
\qqq
valid for all $\,\Vgt,\Wgt\in\G\bigl(\sfE^{(1,n)}\xcM\bigr)$,\ and
in which all the remaining components are the same as those of the
canonical Vinogradov structure $\,\Vgt^{(n)}\xcM$.

An \textbf{isomorphism between a pair $\,\Vgt^{(n),\Om_{(n+2)}^\a}
\xcM_\a,\ \a\in\{1,2\}\,$} is a vector-bundle isomorphism $\,\chi_{1
,2}:\sfE^{(1,n)}\xcM_1\xrightarrow{\cong}\sfE^{(1,n)}\xcM_2\,$
covering a diffeomorphism $\,h_{1,2}:\xcM_1\to\xcM_2\,$ that
satisfies the identities
\qq\nn
\Vbra{\cdot}{\cdot}^{\Om_{(n+2)}^2}\circ(\chi_{1,2},\chi_{1,2})&=&
\chi_{1,2}\circ\Vbra{\cdot}{\cdot}^{\Om_{(n+2)}^1}\,,\cr\cr
\Vcon{\cdot}{\cdot}\circ(\chi_{1,2},\chi_{1,2})&=&(h_{1,2}^{-1})^*
\circ\Vcon{\cdot}{\cdot}\,,\cr\cr
\a_{\sfT\xcM_2}\circ\chi_{1,2}&=&h_{1,2\,*}\circ\a_{\sfT\xcM_1}\,.
\qqq
\exdef \brem On specialisation to $\,n=1$,\ the last definition
reproduces the Courant bracket on the canonical generalised tangent
bundle twisted by a 3-form, as introduced in
\Rcite{Severa:2001qm}.\erem \noindent The two twisted structures are
related by the following
\berop\label{prop:tw-gen-tan-triv}
In the notation of Definitions \ref{def:Vin-str},
\ref{def:tw-gen-tan-bun} and \ref{def:loc-glob-Vin-str}, and
assuming that $\,\sfE^{(1,n)}_{\{\ggt_{ij}\}}\xcM\,$ carries a
global Vinogradov structure $\,\Vgt^{(n)}_{\{\ggt_{ij}\}}\xcM$,\ the
former admits a global trivialisation with local data
\qq\label{eq:hi-triv}
\hgt_i=\ee^{B_i}\,,\qquad\qquad B_i\in\Om^{n+1}(\xcMup\cO_i)
\qqq
iff there exists a homomorphism $\,\xcMup\chi=(\hgt_i)_{i\in
\xcI_\xcM}\ :\ \Vgt^{(n)}_{\{\ggt_{ij}\}}\xcM\to\Vgt^{(n),\Om_{(n+2
)}}\xcM\,$ between $\,\Vgt^{(n)}_{\{\ggt_{ij}\}}\xcM\,$ and the
$\Om_{(n+2)}$-twisted Vinogradov structure $\,\Vgt^{(n),\Om_{(n+2)}}
\xcM\,$ on $\,\sfE^{(1,n)}\xcM\,$ with the twist given by the global
closed $(n+2)$-form with restrictions
\qq\label{eq:Vin-twist-loc}
\Om_{(n+2)}\vert_{\xcMup\cO_i}:=\sfd B_i\,,
\qqq
\textit{i.e.}\ iff for any two local sections
$\,\Vgt_i,\Wgt_i\in\sfE^{(1,n )}_{\{\ggt_{ij}\}}\xcM(\xcMup\cO_i)\,$
and the corresponding sections
$\,\Vgt\vert_{\xcMup\cO_i}=\hgt_i\lact\Vgt_i\,$ and $\,\Wgt
\vert_{\xcMup\cO_i}=\hgt_i\lact\Wgt_i\,$ from $\,\sfE^{(1,n)}\xcM(
\xcMup\cO_i)\,$,\ we obtain
\qq
\Vbra{\Vgt}{\Wgt}^{\Om_{(n+2)}}&=&\hgt_i\lact\Vbra{\Vgt_i}{\Wgt_i}
\,,\label{eq:Hitch-Vbra}\\\cr
\Vcon{\Vgt}{\Wgt}&=&\Vcon{\Vgt_i}{\Wgt_i}\label{eq:Hitch-Vcon}\\\cr
\a_{\sfT\xcM}(\Vgt)&=&\a_{\sfT\xcM}(\Vgt_i)\,.
\label{eq:Hitch-Vanch}
\qqq
\eerop
\beroof
\bit
\item[$\Rightarrow$]
The bundle $\,\sfE^{(1,n)}_{\{\ggt_{ij}\}}\xcM\to\xcM\,$ admits a
global Vinogradov structure, and so -- in virtue of Proposition
\ref{prop:Vin-str-glob-tw} -- its transition maps have the form
\eqref{eq:Vin-glob-on-gen-tw-trans}. Their trivialisation in terms
of the $\,\hgt_i\,$ given in \Reqref{eq:hi-triv} yields the
equalities
\qq\nn
\sfd A_{ij}=(-1)^{n+1}(B_j-B_i)\vert_{\xcMup\cO_{ij}}\,,
\qqq
and so, in particular, the $\,\sfd B_i\,$ define a global $(n+2
)$-form $\,\Om_{(n+2)}\,$ on $\,\xcM\,$ as per
\Reqref{eq:Vin-twist-loc}. Using the results from the proof of
Proposition \ref{prop:Vin-str-auts}, the trivialisation $\,\xcMup
\chi=(\hgt_i)_{i\in\xcI_\xcM}\,$ is readily checked to define the
desired homomorphism (note that the twist in the definition of
$\,\sfE^{(1,n)}_{\{\ggt_{i j}\}}\xcM\,$ is restricted to the
component $\,\wedge^n\sfT^*\xcM$).
\item[$\Leftarrow$] Adducing the same arguments as in the proof of
Proposition \ref{prop:Vin-str-auts} (this time for the Vinogradov
brackets $\,\Vbra{\cdot}{\cdot}\,$ and
$\,\Vbra{\cdot}{\cdot}^{\Om_{(n+2)}}$), we readily establish that
the (unital) homomorphism $\,\xcMup\chi$,\ whose existence is
assumed, is necessarily of the form
\qq\nn
\xcMup\chi\vert_{\xcMup\cO_i}=\ee^{B_i}\,.
\qqq
It is then clear that the local automorphisms $\,\hgt_i:=\xcMup\chi
\vert_{\xcMup\cO_i}\,$ of $\,\sfE^{(1,n)}\xcM\,$ determine a
trivialisation of $\,\sfE^{(1,n)}_{\{\ggt_{ij}\}}\xcM\,$ via
\qq\nn
\ggt_{ij}=\bigl(\hgt_j^{-1}\circ\hgt_i\bigr)\vert_{\xcMup\cO_{ij}}\,.
\qqq
\eit
\eroof\medskip \noindent We then immediately establish
\becor\label{cor:tw-gen-tan-vs-ngerb}
In the notation of Definitions \ref{def:sigmod-n}, \ref{def:Vin-str}
and \ref{def:tw-gen-tan-bun}, the $n$-gerbe $\,\cG_{(n)}\,$
canonically defines a twisted generalised tangent bundle
$\,\sfE^{(1,n)}_{\{\ggt_{ij}\}}\xcM\,$ over $\,\xcM\,$ with a global
Vinogradov structure, via
\qq\nn
\ggt_{ij}=\ee^{(-1)^n\,\sfd A_{ij}}\,.
\qqq
The latter structure is homomorphic to the $\txH_{(n)}$-twisted
Vinogradov structure on $\,\sfE^{(1,n)}\xcM\,$ as per
\qq\nn
\xcMup\chi\ :\ \Vgt^{(n)}_{\{\ggt_{ij}\}}\xcM\to\Vgt^{(n),\Om_{(n+2
)}}\xcM\,,\qquad\qquad\xcMup\chi\vert_{\xcMup\cO_i}=\ee^{B_i}\,.
\qqq
A (trivially) twisted generalised tangent bundle of the type
described will be denoted as $\,\sfE^{(1,n)}_{\cG_{(n)}}\xcM\,$ and
termed the \textbf{$\cG_{(n)}$-twisted generalised tangent bundle of
type $\,(1,n)\,$ over $\,\xcM$}. Analogously, the corresponding
global Vinogradov structure will be denoted by
$\,\Vgt^{(n)}_{\cG_{(n )}}\xcM$. \ecor \brem The statement of the
corollary clearly makes sense as the transition maps satisfy the
standard cocycle condition on triple intersections
$\,\xcMup\cO_{ijk}\,$ in consequence of \Reqref{eq:DGn-is-Hn},
\qq\nn
(\ggt_{jk}\circ\ggt_{ij})\vert_{\xcMup\cO_{ijk}}=\ee^{(-1)^n\,\sfd(
A_{ij}+A_{jk})\vert_{\xcMup\cO_{ijk}}}=\ee^{(-1)^n\,\sfd A_{ik}
\vert_{\xcMup\cO_{ijk}}}=\ggt_{ik}\vert_{\xcMup\cO_{i jk}}\,,
\qqq
and gauge-equivalent choices of a local presentation of $\,\cG_{(n
)}$,\ as described in Definition \ref{def:sigmod-n}, yield
isomorphic bundles,
\qq\nn
b_{(n)}\mapsto b_{(n)}+D_{(n)}\pi_{(n)}\qquad\Longrightarrow\qquad
(\ggt_{ij},\Vgt_i)\mapsto\bigl(\hgt_j\circ\ggt_{ij}\circ\hgt_i^{-1}
,\hgt_i\lact\Vgt_i\bigr)\,,
\qqq
with $\,\hgt_i\,$ as in \Reqref{eq:gauge-trans-gen-tw}.\erem
\bigskip

The structures introduced in the foregoing paragraphs have an
immediate physical realisation, which we state as
\berop\label{prop:sigmod-n-symm}
In the notation of Definitions \ref{def:sigmod-n}, \ref{def:Vin-str}
and \ref{def:Om-tw-Vin},
\bit
\item[i)] internal symmetries of the $(n+1)$-dimensional $\si$-model
of Definition \ref{def:sigmod-n} correspond to those smooth sections
$\,\Vgt\,$ of $\,\sfE^{(1,n)}\xcM\,$ which are \textbf{Killing} for
$\,\txg$,
\qq\nn
\pLie{\a_{\sfT\xcM}(\Vgt)}\txg=0\,,
\qqq
and belong to the kernel of the linear differential operator
\qq\nn
\sfd_{\txH_{(n)}}\ :\ \G\bigl(\sfE^{(1,n)}\xcM\bigr)\to\Om^{n+1}(
\xcM)\ :\ \xcV\oplus\upsilon\mapsto\sfd\upsilon+\xcV\con\txH_{(n)}
\,;
\qqq
we shall call these sections \textbf{$\si$-symmetric} and denote the
corresponding subset in $\,\G\bigl(\sfE^{(1,n)}\xcM\bigr)\,$
as\linebreak $\,\G_\si\bigl(\sfE^{(1,n)}\xcM\bigr)$;
\item[ii)] the $\txH_{(n)}$-twisted Vinogradov bracket $\,
\Vbra{\cdot}{\cdot}^{\txH_{(n)}}\,$ closes on $\,\G_\si\bigl(\sfE^{(
1,n)}\xcM\bigr)$,
\qq\nn
\Vgt,\Wgt\in\G_\si\bigl(\sfE^{(1,n)}\xcM\bigr)\qquad\Longrightarrow
\qquad\Vbra{\Vgt}{\Wgt}^{\txH_{(n)}}\in\G_\si\bigl(\sfE^{(1,n)}\xcM
\bigr)\,,
\qqq
and every other bracket $\,\lsem\cdot,\cdot\rsem_\si\,$ on $\,\G_\si
\bigl(\sfE^{(1,n)}\xcM\bigr)\,$ with this property and such that
\qq\label{eq:constr-bra-EMsi}
\a_{\sfT\xcM}\circ\lsem\cdot,\cdot\rsem_\si=[\cdot,\cdot]\circ(
\a_{\sfT\xcM},\a_{\sfT\xcM})
\qqq
differs from $\,\Vbra{\cdot}{\cdot}^{\txH_{(n)}}\,$ by a linear
operator $\,\D:\G_\si\bigl(\sfE^{(1,n)}\xcM\bigr)\wedge\G_\si\bigl(
\sfE^{(1,n)}\xcM\bigr)\to Z^n(\xcM)$.
\eit
\eerop
\beroof
\bit
\item[Ad i)] This is a corollary to Proposition
\ref{prop:var-sigmod-n}. Note that the $n$-form component of a
$\si$-symmetric section is determined up to a closed $n$-form.
\item[Ad ii)] First of all, note that $\,\a_{\sfT\xcM}(\ker\,
\sfd_{\txH_{(n)}})\,$ is a Lie subalgebra of the Lie algebra of
vector fields on $\,\xcM\,$ as
\qq\nn
\xcV_\a\con\txH_{(n)}=-\sfd\upsilon_\a\,,\quad\a\in\{1,2\}\qquad
\Longrightarrow\qquad[\xcV_1,\xcV_2]\con\txH_{(n)}=
\pLie{\xcV_1}(\xcV_2\con\txH_{(n)})=-\sfd(\pLie{\xcV_1}\upsilon_2)
\,.
\qqq
This demonstrates the naturalness of constraints
\eqref{eq:constr-bra-EMsi}. Having noted that, take an arbitrary
pair $\,\Vgt=\xcV\oplus\upsilon,\Wgt
=\xcW\oplus\varpi\in\G_\si\bigl(\sfE^{(1,n)}\xcM\bigr)$,\ so that
\qq\nn
\xcV\con\txH_{(n)}=-\sfd\upsilon\,,\qquad\qquad\xcW\con\txH_{(n)}=-
\sfd\varpi\,,
\qqq
and hence also
\qq\nn
\pLie{\xcV}\txH_{(n)}=0=\pLie{\xcW}\txH_{(n)}
\qqq
due to the closedness of $\,\txH_{(n)}$.\ The exterior derivative of
the $n$-form component of the $\txH_{(n)}$-twisted Vinogradov
bracket
\qq\label{eq:H-tw-Vbra}
\Vbra{\Vgt}{\Wgt}^{\txH_{(n)}}=[\xcV,\xcW]\oplus\bigl(\pLie{\xcV}
\varpi-\pLie{\xcW}\upsilon-\tfrac{1}{2}\,\sfd(\xcV\con\varpi-\xcW
\con\upsilon)+\xcV\con\xcW\con\txH_{(n)}\bigr)\,.
\qqq
reads
\qq\nn
\sfd(\pLie{\xcV}\varpi-\pLie{\xcW}\upsilon+\xcV\con\xcW\con\txH_{(n)})
&=&\pLie{\xcV}\sfd\varpi+\sfd(-\xcW\con\sfd\upsilon+\xcV\con\xcW\con
\txH_{(n)})\cr\cr
&=&-\pLie{\xcV}(\xcW\con\txH_{(n)})=-\a_{\sfT\xcM}\bigl(
\Vbra{\Vgt}{\Wgt}^{\txH_{(n)}}\bigr)\con\txH_{(n)}
\qqq
as claimed. Furthermore, by assumption,
\qq\nn
\Vbra{\Vgt}{\Wgt}^{\txH_{(n)}}-\lsem\Vgt,\Wgt\rsem_\si=0\oplus\D(
\Vgt\wedge\Wgt)
\qqq
for some $\,\D:\G_\si\bigl(\sfE^{(1,n)}\xcM\bigr)\wedge\G_\si\bigl(
\sfE^{(1,n)}\xcM\bigr)\to\Om^n(\xcM)$,\ and the previous result
implies
\qq\nn
\sfd\D(\Vgt\wedge\Wgt)=\bigl(\a_{\sfT\xcM}\bigl(\lsem\Vgt,\Wgt
\rsem_\si\bigr)-\a_{\sfT\xcM}\bigl(\Vbra{\Vgt}{\Wgt}\bigr)\bigr)
\con\txH_{(n)}=0\,,
\qqq
thereby completing the proof of statement ii).
\eit\eroof \brem It is completely straightforward, at least on the
formal level, to pass to the canonical or even pre-quantum
description of the $(n+1)$-dimensional non-linear $\si$-model, in
which the $n$-gerbe $\,\cG_{(n)}\,$ plays a r\^ole analogous to that
of the (1-)gerbe in the familiar two-dimensional case, that is, in
particular, it canonically defines -- via a higher-dimensional
variant of the transgression map -- a pre-quantum bundle of the
theory. There then ensues a natural transgression scheme between the
attendant Vinogradov structures on the target space of the
$\si$-model and on its state space, in which the canonical
contraction enters through the definition of Noether currents and
the corresponding hamiltonian functions, and which identifies the
$\txH_{(n)}$-twisted Vinogradov structure on the generalised tangent
bundle $\,\sfE^{(1,n)}M\,$ as the sought-after algebraic counterpart
of the canonical Vinogradov structure on the state space of the
$(n+1)$-dimensional $\si$-model mentioned in the Introduction.
Instead of pursuing this issue at the hitherto level of generality,
we specialise our analysis directly to the case of immediate
interest, that is to the two-dimensional $\si$-model, leaving the
generalisation as a simple exercise. \erem

\section{Symmetries of the two-dimensional $\si$-model -- the
untwisted sector}\label{sec:gen-geom-mono-2d-spec}

Having extracted the concept of the generalised tangent bundle from
the lagrangean analysis of infinitesimal rigid symmetries of the $(n
+1)$-dimensional $\si$-model (with a topological term), we shall
next pose the question as to the r\^ole played by that concept in
the canonical treatment of the symmetries, based on an explicit
reconstruction of the phase space of the $\si$-model, understood as
a (pre-)symplectic manifold, in the so-called first-order formalism
of
Refs.\,\cite{Gawedzki:1972ms,Kijowski:1973gi,Kijowski:1974mp,Kijowski:1976ze,Szczyrba:1976,Kijowski:1979dj}
reported in \Rcite{Suszek:2011hg}. Our discussion will enable us to
regard the structure of a twisted Courant algebroid on the set of
$\si$-symmetric sections as a homomorphic target-space preimage of
the structure of a Poisson algebra on the set of the associated
Noether hamiltonians on the phase space of the $\si$-model.

Our passage to the phase space of the $\si$-model will be seen to
serve yet another purpose, to wit, that of demystifying the
emergence of the generalised geometry in the field-theoretic setting
of interest. The underlying idea is laid out in
\berop\label{prop:ham-as-genom}
Adopt the notation of Definition \ref{def:Vin-str} and let
$\,\sfP\,$ be a smooth manifold endowed with the structure of a
symplectic manifold $\,(\sfP,\Om)\,$ by a closed non-degenerate
2-form $\,\Om$.\ To every \textbf{hamiltonian function} $\,h\in
C^\infty(\sfP,\bR)$,\ \textit{i.e.}\ a smooth function on $\,\sfP$,\
there is associated a smooth section
\qq\nn
\Xgt_h=\xcX_h\oplus h
\qqq
of the generalised tangent bundle $\,\sfE^{(1,0)}\sfP\,$ from the
kernel of the linear differential operator
\qq\nn
\sfd_\Om\ :\ \G\bigl(\sfE^{(1,0)}\sfP\bigr)\to\G(\sfT^*\sfP)\ :\
\xcX \oplus f\mapsto\sfd f+\xcX\con\Om\,.
\qqq
Elements of $\,\ker\,\sfd_\Om\,$ will be called \textbf{hamiltonian
sections of $\,\sfE^{(1,0)}\sfP$},\ and a smooth vector field
$\,\xcX_h\,$ associated to $\,h\,$ as indicated above is termed a
\textbf{globally hamiltonian vector field}. The linear map
\qq\nn
\Xgt\ :\ C^\infty(\sfP,\bR)\to\G\bigl(\sfE^{(1,0)}\sfP\bigr)\ :\ h
\mapsto\Xgt_h
\qqq
determines a homomorphism between the Lie algebra $\,(C^\infty(\sfP,
\bR),\{\cdot,\cdot\}_\Om)\,$ of hamiltonian functions with the Lie
bracket given by the Poisson bracket induced by $\,\Om$,
\qq\nn
\{h_1,h_2\}_\Om:=\xcX_{h_2}\con\xcX_{h_1}\con\Om\,,\qquad h_1,h_2\in
C^\infty(\sfP,\bR)\,,
\qqq
and the Lie algebra $\,\bigl(\ker\,\sfd_\Om,\Vbra{\cdot}{\cdot}^\Om
\bigr)\,$ of hamiltonian sections of $\,\sfE^{(1,0)}\sfP\,$ with the
Lie bracket given by the $\Om$-twisted Vinogradov bracket
\qq\label{eq:tw-Vinbra}
\Vbra{\Xgt_{h_1}}{\Xgt_{h_2}}^\Om:=[\xcX_{h_1},\xcX_{h_2}]\oplus
\bigl(\xcX_{h_1}\con\sfd h_2-\xcX_{h_2}\con\sfd h_1+\xcX_{h_1}\con
\xcX_{h_2}\con\Om\bigr)\,.
\qqq
The $\Om$-twisted Vinogradov bracket is a unique -- up to addition
of a linear map $\,\G\bigl(
\sfE^{(1,0)}\sfP\bigr)\wedge\G\bigl(\sfE^{(1,0)}\sfP\bigr)\to\ker\,
\sfd\subset C^\infty(\sfP,\bR)\,$ to the 0-form
component\footnote{Note that the ambiguity in the definition of a
bracket with the properties listed is consistent with the ambiguity
in the definition of the hamiltonian function.} -- bilinear
antisymmetric operation on $\,\G \bigl(\sfE^{(1,0)}\sfP\bigr)\,$
with the properties
\qq\nn
\a_{\sfT\sfP}\circ\Vbra{\cdot}{\cdot}^\Om=[\cdot,\cdot]\circ(
\a_{\sfT\sfP},\a_{\sfT\sfP})
\qqq
and
\qq\nn
\Xgt_1,\Xgt_2\in\ker\,\sfd_\Om\qquad\Longrightarrow\qquad
\Vbra{\Xgt_1}{\Xgt_2}^\Om\in\ker\,\sfd_\Om\,,
\qqq
written in terms of the anchor $\,\a_{\sfT\sfP}\ :\ \sfE^{(1,0)}
\sfP\to\sfT\sfP\,$ (given by the canonical projection). The triple
$\,(\sfE^{(1,0)}\sfP,\Vbra{\cdot}{\cdot},\a_{\sfT\sfP})\,$ will be
referred to as the \textbf{canonical Vinogradov struture on
$\,\sfE^{(1,0)}\sfP\,$} henceforth.
\eerop
\beroof
The statement of the proposition follows directly from the
definition of a hamiltonian function, and from the simple property
\qq\nn
[\xcX_{h_1},\xcX_{h_2}]=\xcX_{\{h_1,h_2\}_\Om}
\qqq
of hamiltonian vector fields. The Jacobi identity for the
$\Om$-twisted Vinogradov bracket is then a consequence of the same
identity for the Poisson bracket.\eroof

As a first step towards our goal, let us specialise our
considerations to two dimensions, extending them simultaneously so
as to account for the existence of world-sheet defects.
\bedef\label{def:sigmod-2d}
Adopt the notation of Definitions \vref{def:net-field}I.2.6 and
\vref{def:sigmod}I.2.7., and let $\,\Bgt=(\cM,\cB,\cJ)\,$ be a
string background with target $\,\cM=(M,\txg,\cG)$,\ $\cG$-bi-brane
$\,\cB =\bigl(Q,\iota_\a,\om,\Phi\ \vert\ \a\in\{1,2\}\bigr)\,$ and
$(\cG,\cB)$-inter-bi-brane $\,\cJ=\bigsqcup_{n\in \bN_{\geq 3}}\,
\bigl(T_n,\bigl(\vep^{k,k+1}_n,\pi^{k,k+1}_n\ \vert\ k\in\ovl{1,n}
\bigr),\varphi_n\bigr)$,\ supported over target space $\,\xcF:=M
\sqcup Q \sqcup T,\ T=\bigsqcup_{n\geq 3}\,T_n$,\ all as introduced
in Definition \vref{def:bckgrnd}I.2.1. Moreover, let $\,\G\,$ be a
defect quiver from Definition I.2.6. The two-dimensional non-linear
$\si$-model for network-field configurations $\,(X\,\vert\,\G)\,$ in
string background $\,\Bgt\,$ on world-sheet $\,(\Si,\g)\,$ with
defect quiver $\,\G\,$ is a theory of continuously differentiable
maps $\,X\ :\ \Si\to\xcF\,$ determined by the principle of least
action applied to the action functional
\qq\label{eq:2d-sigma-def}
S_\si[(X\,\vert\,\G);\g]:=-\tfrac{1}{2}\,\int_\Si\,\txg(\sfd X
\overset{\wedge}{,}\star_\g\sfd X)-\sfi\,\log\Hol_{\cG,\Phi,(
\varphi_n)}(X\,\vert\,\G)\,.
\qqq
\exdef
\berop\cite[Props.\,\vref{prop:sympl-form-si-untw}I.3.11 \&
\vref{prop:sympl-form-si-tw}I.3.12]{Suszek:2011hg}
\label{prop:sympl-form-twuntw} Let $\,\sfP_{\si,\emptyset}\,$ and
$\,\sfP_{\si,\cB\,\vert\,(\pi,\vep)}\,$ be the untwisted and
1-twisted state spaces of the two-dimensional non-linear $\si$-model
of Definition \ref{def:sigmod-2d}, as introduced in Definitions
\vref{def:untw-phspace}I.3.9 and \vref{def:tw-phspace}I.3.10,
respectively. A (pre-)symplectic form on $\,\sfP_{\si,\emptyset}\,$
can be written as
\qq\label{eq:sympl-form-untw}
\Om_{\si,\emptyset}[(X,\txp)]=\int_{\bS^1}\,\Vol\left(\bS^1\right)
\wedge\left[\d\txp_\mu\wedge\d X^\mu+3(X_*\widehat t)^\la\,
\txH_{\la\mu\nu}\d X^\mu\wedge\d X^\nu\right]
\qqq
in terms of the canonical coordinates $\,(X,\sfp)\,$ on $\,\sfP_{\si
,\emptyset}\,$ and of components $\,\txH_{\la\mu\nu}\,$ of the
curvature 3-form $\,\txH\,$ of the gerbe $\,\cG$.\ Similarly, a
(pre-)symplectic form on $\,\sfP_{\si,\cB\,\vert\,(\pi,\vep)}\,$ can
be written as
\qq\label{eq:sympl-form-1tw}
\Om_{\si,\cB\,\vert\,(\pi,\vep)}[(X,\txp,q,V)]=\int_{\bS^1_{\{\pi
\}}}\,\Vol\left(\bS^1_{\{\pi\}}\right)\wedge\left[\d\txp_\mu\wedge
\d X^\mu+3(X_*\widehat t)^\la\,\txH_{\la\mu\nu}\d X^\mu\wedge\d
X^\nu\right]+\vep\,\om(q)
\qqq
in terms of the canonical coordinates $\,(X,\txp,q,V)\,$ on
$\,\sfP_{\si,\cB\,\vert\,(\pi,\vep)}\,$ and of the curvature
$\,\om\,$ of the $\cG$-bi-brane $\,\cB$.
\eerop
\noindent We are now ready to study the canonical description of
internal symmetries of the two-dimensional $\si$-model.

We start by recapitulating the algebraic structure on the set of
sections of the generalised tangent bundle over the target space in
the absence of defects. As a specialisation of Corollary
\ref{cor:tw-gen-tan-vs-ngerb} and Proposition
\ref{prop:sigmod-n-symm} to the case $\,n=1$,\ we obtain
\becor\label{cor:sigmod-symm-E11}
Adopt the notation of Definitions \ref{def:Vin-str},
\ref{def:tw-gen-tan-bun}, \ref{def:loc-glob-Vin-str} and
\ref{def:Om-tw-Vin}. Let $\,\Bgt\,$ be a string background with
target $\,\cM=(M,\txg,\cG )$,\ as detailed in Definition
\vref{def:bckgrnd}I.2.1, and denote by $\,\txH\in Z^3(M)\,$ the
curvature of $\,\cG$.\ Infinitesimal rigid symmetries of the
two-dimensional non-linear $\si$-model for network-field
configurations $\,(X\,\vert\,\emptyset)\,$ in string background
$\,\Bgt\,$ on world-sheet $\,(\Si,\g)\,$ with an empty defect quiver
$\,\G=\emptyset$,\ as described in Definition
\vref{def:sigmod}I.2.7, correspond to $\si$-symmetric sections
$\,\Vgt\in\G_\si\bigl(\sfE^{(1,1)}M\bigr)$,
\qq\nn
\pLie{\a_{\sfT M}(\Vgt)}\txg=0\,,\qquad\qquad\sfd_\txH\Vgt=0\,.
\qqq
The $\txH$-twisted Vinogradov bracket (of the $\txH$-twisted
Vinogradov structure $\,\Vgt^{(1),\txH}M$) closes on $\,\G_\si\bigl(
\sfE^{(1,1)}M\bigr)$,
\qq\nn
\Vgt,\Wgt\in\G_\si\bigl(\sfE^{(1,1)}M\bigr)\qquad\Longrightarrow
\qquad\Vbra{\Vgt}{\Wgt}^\txH\in\G_\si\bigl(\sfE^{(1,1)}M\bigr)\,.
\qqq
Equivalently, given an open cover $\,\Mup\cO=\{\cO^M_i\}_{i\in
\xcI_M}\,$ of $\,M\,$ with an index set $\,\xcI_M$,\ together with
the associated local presentation $\,(B_i,A_{ij},g_{ijk})\in\cA^{3,
2}(\Mup\cO)\,$ of $\,\cG$,\ as described in Definition
\vref{def:loco}I.2.2, the symmetries can be represented by
$\si$-symmetric sections $\,(\Vgt_i)_{i\in\xcI_M}\in\G_\si\bigl(
\sfE^{(1,1)}_\cG M\bigr)$,
\qq\nn
\pLie{\a_{\sfT M}(\Vgt_i)}\txg=0\,,\qquad\qquad\sfd\pr_{\sfT^*M}(
\Vgt_i)+\pLie{\a_{\sfT M}(\Vgt_i)}B_i=0\,.
\qqq
The Vinogradov bracket (of the global Vinogradov structure
$\,\Vgt^{(1)}_\cG M$,\ homomorphic to $\,\Vgt^{(1),\txH}M$) closes
on $\,\G_\si\bigl(\sfE^{(1,1)}_\cG M \bigr)$,
\qq\nn
(\Vgt_i)_{i\in \xcI_M},(\Wgt_i)_{i\in \xcI_M}\in\G_\si\bigl(\sfE^{(1
,1)}_\cG M\bigr)\qquad\Longrightarrow\qquad\bigl(
\Vbra{\Vgt_i}{\Wgt_i}\bigr)_{i\in \xcI_M}\in\G_\si\bigl(\sfE^{(1,1
)}_\cG M\bigr)\,.
\qqq
\ecor

It was demonstrated in the proof of statement ii) of Proposition
\ref{prop:sigmod-n-symm} that $\,\a_{\sfT M}(\ker\,\sfd_\txH)\,$ is
a Lie subalgebra of the Lie algebra of vector fields, and the same
is true for Killing vector fields. Hence, we establish
\berop\label{prop:Htw-Vbra-symm}
Adopt the notation of Definitions \ref{def:Vin-str} and
\ref{def:Om-tw-Vin}, and of Proposition \ref{prop:sigmod-n-symm}.
Let $\,(\xcM,\txg)\,$ be a metric manifold with a smooth closed
3-form $\,\txH\in Z^3(\xcM)$.\ The subspace $\,\a_{\sfT\xcM}\bigl(
\G_\si\bigl(\sfE^{(1,1)}\xcM \bigr)\bigr)\,$ is a Lie subalgebra, to
be denoted as $\,\ggt_\si$,\ within the Lie algebra of Killing
vector fields on $\,(\xcM,\txg)$.\ Fix a basis
$\,\{\xcK_A\}_{A\in\ovl{1,K_\si}},\ K_\si=\dim\, \ggt_\si\,$ in
$\,\ggt_\si\,$ such that the defining commutation relations
\qq\nn
[\xcK_A,\xcK_B]=f_{ABC}\,\xcK_C
\qqq
hold true for some structure constants $\,f_{ABC}$.\ The
corresponding $\si$-symmetric sections
\qq\nn
\Kgt_A=\xcK_A\oplus\kappa_A\,,\qquad\qquad\qquad\pLie{\xcK_A}\txg=
0\,,\qquad\qquad\sfd_\txH\Kgt_A=0
\qqq
satisfy the relations
\qq\label{eq:Vinbra-Ka}\qquad\qquad
\Vbra{\Kgt_A}{\Kgt_B}^\txH=f_{ABC}\,\Kgt_C+0\oplus(\D_{AB}-\sfd
\txc_{(AB)})\,,\qquad\qquad\left\{ \barr{l} \D_{AB}=\pLie{\xcK_A}
\kappa_B-f_{ABC}\,\kappa_C\,,\cr\cr
\txc_{(AB)}=\Vcon{\Kgt_A}{\Kgt_B}\,. \earr \right.
\qqq
\eerop
\beroof
Obvious, through inspection. \eroof\medskip

Having presented the target-space aspect of internal symmetries of
the $\si$-model action functional in the absence of defects, we may
next consider their symplectic realisation on the state space of the
untwisted sector of the theory. To this end, we should lift our
previous analysis to the symplectic space $\,(\sfP_{\si,\emptyset}
\equiv\sfT^*\sfL M,\Om_{\si,\emptyset})$,\ whereupon it develops in
complete analogy to the geometric discussion.
\bedef\label{def:LM-lifts}
Let $\,\xcM\,$ be a smooth manifold, $\,\sfL\xcM=C^\infty(
\bS^1,\xcM)\,$ its free-loop space, and
\qq\nn
\ev_\xcM\ :\ \sfL\xcM\x\bS^1\to\xcM
\qqq
the canonical evaluation map. Given a vector field $\,\xcV\in\G(\sfT
\xcM)$,\ denote by $\,\xi_t:\xcM\to\xcM\,$ the flow of $\,\xcV$.\
The \textbf{loop-space lift of vector field from $\,\xcM\,$} is a
linear map
\qq\nn
\sfL_*\ :\ \G(\sfT\xcM)\to\G(\sfT\sfL\xcM)\ :\ \xcV\mapsto\sfL_*
\xcV\,,\qquad\qquad\sfL^*\xcV(F)[X]:=\tfrac{\sfd\ }{\sfd t}\big
\vert_{t=0}F[\xi_t\circ X]\,,
\qqq
defined for an arbitrary functional $\,F\,$ on $\,\sfL\xcM\,$ and
for a free loop $\,X$.\ The \textbf{loop-space lift of $n$-form from
$\,\xcM\,$} is the linear map
\qq\nn
\sfL^*\ &:&\ \Om^n(\xcM)\to\Om^{n-1}(\sfL\xcM)\ :\ \upsilon\mapsto
\int_{\bS^1}\,\ev_\xcM^*\,\upsilon\,,\quad n\in\bN_{>0}\,,
\qqq
extended to the case $\,n=0\,$ as per
\qq\label{eq:triv-can-con-10}
\sfL^*\ :\ C^\infty(\xcM,\bR)\to\{0\}\ :\ f\mapsto 0\,.
\qqq
\exdef \noindent Basic properties of the two lifts are summarised in
the following
\belem\label{lem:lift-minus}
Adopt the notation of Definition \ref{def:LM-lifts} and denote by
$\,\d\,$ the (functional) exterior derivative on $\,\Om^\bullet(\sfL
\xcM)$.\ Then, for arbitrary $\,\xcV,\xcW\in\G(\sfT\xcM)\,$ and
$\,\upsilon\in\Om^n(\xcM)$,
\qq\nn
&\d\sfL^*\upsilon=-\sfL^*\sfd\upsilon\,,\qquad\qquad\sfL_*\xcV
\con\sfL^*\upsilon=-\sfL^*(\xcV\con\upsilon)\,,&\cr\cr
&[\sfL_*\xcV,\sfL_*\xcW]=\sfL_*[\xcV,\xcW]\,.&
\qqq
\elem
\beroof Obvious, through inspection. \eroof\medskip
\noindent In the next, physically motivated step, we obtain
\belem\label{lem:can-cotan-lift}
Adopt the notation of Definition \ref{def:LM-lifts} and of Lemma
\ref{lem:lift-minus}. The lift $\,\sfL^*\,$ induces a lift
\qq\nn
\widetilde\sfL^*:=\pi_{\sfT^*\sfL\xcM}^*\circ\sfL^*\ :\ \Om^n(\xcM)
\to\Om^{n-1}(\sfL\xcM)
\qqq
of $n$-forms on $\,\xcM\,$ to $(n-1)$-forms on the cotangent bundle
$\,\pi_{\sfT^*\sfL\xcM}:\sfT^*\sfL\xcM\to\sfL \xcM$.\ Analogously,
$\,\sfL_*\,$ induces a canonical lift
\qq\nn
\widetilde\sfL_*\ :\ \G\bigl(\sfT\xcM\bigr)\to\G\bigl(\sfT(\sfT^*
\sfL\xcM)\bigr)
\qqq
of vector fields $\,\xcV\in\G\bigl(\sfT\xcM\bigr)\,$ to those on
$\,\sfT^*\sfL\xcM$,\ fixed by the relations
\qq
\pi_{\sfT^*\sfL\xcM\,*}\widetilde\sfL_*\xcV&=&\sfL_*\xcV\,,\cr\cr
\pLie{\widetilde\sfL_*\xcV}\theta_{\sfT^*\sfL\xcM}&=&0\,,
\label{eq:cot-lift-pres-p}
\qqq
expressed in terms of the canonical 1-form $\,\theta_{\sfT^*\sfL
\xcM}\,$ on $\,\sfT^*\sfL\xcM\,$ given in
\vReqref{eq:can-1-cot}{I.3.14}. Then, for arbitrary $\,\xcV,\xcW\in
\G(\sfT\xcM)\,$ and $\,\upsilon\in\Om^n(\xcM)$,
\qq
&\d\widetilde\sfL^*\upsilon=-\widetilde\sfL^*\sfd\upsilon\,,
\qquad\qquad\widetilde\sfL_*\xcV\con\widetilde\sfL^*\upsilon=-
\widetilde\sfL^*(\xcV\con\upsilon)\,,&\label{eq:tiL-minus}\\\cr
&[\widetilde\sfL_*\xcV,\widetilde\sfL_*\xcW]=\widetilde\sfL_*[\xcV,
\xcW]\,.&\label{eq:tiL-Liebra}
\qqq
\elem
\beroof
Obvious, through inspection. \eroof \brem Relation
\eqref{eq:cot-lift-pres-p} ensures that the fibre coordinate $\,
\sfp_\nu\,$ of a point $\,\psi=(X^\mu,\sfp_\nu)\in\sfT^*\sfL\xcM\,$
has the tensorial properties of a component of a 1-form on
$\,\xcM$.\ Explicitly, the canonical lift of a vector field
$\,\xcV=\xcV^\mu\,\tfrac{\p\ }{\p X^\mu}\in\G(\sfT \xcM)\,$ can be
written in the form
\qq\nn
\widetilde\sfL_*\xcV[\psi]=\int_{\bS^1}\,\Vol(\bS^1)\,\left[
\xcV^\mu\bigl(X(\cdot)\bigr)\,\tfrac{\d\quad\ }{\d X^\mu(\cdot)}-
\sfp_\mu(\cdot)\,\p_\nu\xcV^\mu\bigl(X(\cdot)\bigr)\,\tfrac{\d\quad
\ }{\d \sfp_\nu(\cdot)}\right]\,.
\qqq
\erem The lifts give rise to a simple algebraic structure, namely,
\belem\label{lem:quasi-morph-Vin}
Adopt the notation of Definitions \ref{def:Vin-str},
\ref{def:Om-tw-Vin} and \ref{def:LM-lifts}, and of Lemma
\ref{lem:can-cotan-lift}. The pair $\,(\sfL_*,\sfL^*)\,$ induces a
linear mapping
\qq\nn
\sfL_{(1,n)}=\left( \barr{cc} \sfL_* & 0 \cr\cr 0 & \sfL^* \earr
\right)\ :\ \G\bigl(\sfE^{(1,n)}\xcM\bigr)\to\G\bigl(\sfE^{(1,n-1)}
\sfL\xcM\bigr)\ : \ \xcV\oplus\upsilon\mapsto\sfL_*\xcV\oplus\sfL^*
\upsilon\,,\quad n\in\bN
\qqq
that relates elements of the respective twisted Vinogradov
structures $\,\Vgt^{(n),\Om_{n+2}}\xcM\,$ and $\,\Vgt^{(n-1),\sfL^*
\Om_{n+2}}\sfL\xcM\,$ as
\qq\nn
\a_{\sfT\sfL\xcM}\circ\sfL_{(1,n)}&=&\sfL_*\circ\a_{\sfT\xcM}\,,\cr
\cr
\Vbra{\cdot}{\cdot}^{\sfL^*\Om_{(n+2)}}\circ(\sfL_{(1,n)},\sfL_{(1,
n)})&=&\sfL_{(1,n)}\circ\Vbra{\cdot}{\cdot}^{\Om_{(n+2)}}\,,\cr\cr
\Vcon{\cdot}{\cdot}\circ(\sfL_{(1,n)},\sfL_{(1,n)})&=&-\sfL^*\circ
\Vcon{\cdot}{\cdot}\,.
\qqq
The mapping admits an obvious (canonical) extension
\qq\label{eq:tiL11}
\widetilde\sfL_{(1,n)}=\left( \barr{cc} \widetilde\sfL_* & 0 \cr\cr
0 & \widetilde\sfL^* \earr \right)\ :\ \G\bigl(\sfE^{(1,n)}\xcM
\bigr)\to\G \bigl(\sfE^{(1,n-1)}\sfT^*\sfL\xcM\bigr)\,,
\qqq
that relates elements of the respective twisted Vinogradov
structures $\,\Vgt^{(n),\Om_{n+2}}\xcM\,$ and $\,\Vgt^{(n-1),
\widetilde\sfL^*\Om_{(n+2)}}\sfT^*\sfL\xcM\,$ as
\qq
\a_{\sfT(\sfT^*\sfL\xcM)}\circ\widetilde\sfL_{(1,n)}&=&\sfL_*\circ
\a_{\sfT\xcM}\,,\label{eq:alla}\\\cr
\Vbra{\cdot}{\cdot}^{\widetilde\sfL^*\Om_{(n+2)}}\circ(\widetilde
\sfL_{(1,n)},\widetilde\sfL_{(1,n)})&=&\widetilde\sfL_{(1,n)}\circ
\Vbra{\cdot}{\cdot}^{\Om_{(n+2)}}\,,\label{eq:Vinbra-lll-Vinbra}\\
\cr
\Vcon{\cdot}{\cdot}\circ(\widetilde\sfL_{(1,n)},\widetilde\sfL_{(1,
n)})&=&-\widetilde\sfL^*\circ\Vcon{\cdot}{\cdot}\,.
\label{eq:Vcon-lll-Vcon}
\qqq
\elem
\beroof
An immediate corollary to Lemmata \ref{lem:lift-minus} and
\ref{lem:can-cotan-lift}. \eroof\medskip

Putting together various results obtained so far and those of
\Rcite{Suszek:2011hg}, we arrive at a conclusion of immediate
relevance to the two-dimensional field theory of interest, phrased
as
\bethe\label{thm:ind-quasi-morph-glob-Vin}
Adopt the notation of Definition \ref{def:sigmod-2d}, of Corollaries
\ref{cor:tw-gen-tan-vs-ngerb} and \ref{cor:sigmod-symm-E11}, of
Proposition \ref{prop:sympl-form-twuntw}, and of Lemma
\ref{lem:can-cotan-lift}. Let
$\,\ceL_{\si,\emptyset}\to\sfP_{\si,\emptyset}\,$ the pre-quantum
bundle of the untwisted sector of the $\si$-model from Corollary
\vref{cor:preqb-untw}I.3.17. The gerbe $\,\cG\,$ canonically induces
a linear mapping
\qq\nn
\phi_{\si,\emptyset}\ :\ \sfE^{(1,1)}_\cG M\to\sfE^{(1,0
)}_{\ceL_{\si,\emptyset}}\sfP_{\si,\emptyset}
\qqq
(with the codomain twisted by the 0-gerbe $\,\ceL_{\si,\emptyset}$)
that relates elements of the respective global Vinogradov structures
$\,\Vgt^{(1)}_\cG M\,$ and $\,\Vgt^{(0)}_{\ceL_{\si,\emptyset}}
\sfP_{\si,\emptyset}\,$ as
\qq
\a_{\sfT\sfP_{\si,\emptyset}}\circ\phi_{\si,\emptyset}&=&\widetilde
\sfL_*\circ\a_{\sfT M}\,,\label{eq:anchPsi-phisi}\\\cr
\Vbra{\cdot}{\cdot}\circ(\phi_{\si,\emptyset},\phi_{\si,\emptyset})
&=&\phi_{\si,\emptyset}\circ\Vbra{\cdot}{\cdot}\,,
\label{eq:Vbra-phisi}\\\cr
\Vcon{\cdot}{\cdot}\circ(\phi_{\si,\emptyset},\phi_{\si,\emptyset})
&=&\widetilde\sfL^*\circ\Vcon{\cdot}{\cdot}\,.\label{eq:Vcon-phisi}
\qqq
\ethe
\beroof
Choose an open cover $\,\Mup\cO=\{\cO^M_i\}_{i\in\xcI_M}\,$ of $\,
M\,$ (with an index set $\,\xcI_M$), and induce from it an open
cover $\,\cO_{\sfT^*\sfL M}=\{ \cO^*_\igt\}_{\igt\in\xcI_{\sfT^*\sfL
M}}\,$ of $\,\sfT^*\sfL M\,$ in the same manner as in Corollary
\vref{cor:preqb-untw}I.3.17. Fix local presentations:
$\,(B_i,A_{ij}, g_{ijk})\in\cA^{3,2}(\Mup\cO)\,$ of the gerbe
$\,\cG$,\ and $\,(
\theta_{\si,\emptyset\,\igt},\g_{\si,\emptyset\,\igt\jgt})\in\cA^{2
,1}(\cO_{\sfT^*\sfL M})\,$ of the pre-quantum bundle $\,\ceL_{\si,
\emptyset}$,\ the latter as in
\vReqref{eq:loc-dat-preq-untw}{I.3.18}. Denote by $\,\txH\,$ the
curvature of $\,\cG$,\ and by $\,\theta_{\sfT^*\sfL M}\,$ the
canonical 1-form on $\,\sfT^*\sfL M\,$ from Eq.\,(I.3.14).\ By
virtue of Corollary \ref{cor:tw-gen-tan-vs-ngerb}, there exist
homomorphisms of the Vinogradov structures:
\qq\nn
\Mup\chi\ :\ \Vgt^{(1)}_\cG M\to\Vgt^{(1),\txH}M\,,\qquad\qquad\Mup
\chi\vert_{\cO^M_i}=\ee^{B_i}
\qqq
and
\qq\nn
{}^{\sfP_{\si,\emptyset}}\hspace{-2pt}\chi\ :\ \Vgt^{(0)}_{\ceL_{\si
,\emptyset}}\sfP_{\si,\emptyset}\to\Vgt^{(0),\Om_{\si,\emptyset}}
\sfP_{\si,\emptyset}\,,\qquad\qquad{}^{\sfP_{\si,\emptyset}}
\hspace{-2pt}\chi\vert_{\cO^*_\igt}=\ee^{\theta_{\si,\emptyset\,
\igt}}\,.
\qqq
The linear mapping in question can now be explicitly defined as
\qq\nn
\phi_{\si,\emptyset}:={}^{\sfP_{\si,\emptyset}}\hspace{-2pt}\chi^{-
1}\circ\ee^{\theta_{\sfT^*\sfL M}}\circ\widetilde\sfL_{(1,1)}\circ
\Mup\chi
\qqq
in terms of the linear mapping $\,\widetilde\sfL_{(1,1)}\,$ from
Lemma \ref{lem:quasi-morph-Vin}. The desired algebraic properties of
$\,\phi_{\si,\emptyset}\,$ listed in the proposition can readily be
verified by combining the results from Proposition
\ref{prop:tw-gen-tan-triv} and Lemma \ref{lem:quasi-morph-Vin}.
Note, in particular, that the last of them, \eqref{eq:Vcon-phisi},
follows from triviality of the canonical contraction on
$\,\sfE^{(1,0)}_{\ceL_{\si,\emptyset}} \sfP_{\si,\emptyset}$,\
\textit{cf.}\ \Reqref{eq:triv-can-con-10}. \eroof\medskip \noindent
The last theorem provides a clear-cut answer to the general question
raised in the Introduction to this section as it demonstrates the
existence of a straightforward correspondence between the
gerbe-induced (Courant-)algebraic structure on the generalised
tangent bundle of type $\,(1,1)\,$ over the target space of the
$\si$-model and the canonical Vinogradov structure on its untwisted
state space. It will be seen to organise the canonical description
of internal symmetries of the $\si$-model under study, to which we
turn next.
\berop\label{prop:ham-vs-sisymsec}
Adopt the notation of Definition \ref{def:Vin-str}, of Corollaries
\ref{cor:tw-gen-tan-vs-ngerb} and \ref{cor:sigmod-symm-E11}, of
Proposition \ref{prop:sigmod-n-symm}, of Theorem
\ref{thm:ind-quasi-morph-glob-Vin}, and of Lemmata
\ref{lem:lift-minus}, \ref{lem:can-cotan-lift} and
\ref{lem:quasi-morph-Vin}. Let $\,\ceL_\cG\to\sfL M\,$ be the
transgression bundle of Theorem \vref{thm:trans-untw}I.3.16, the
latter having local data $\,(E_\igt,G_{\igt\jgt})$,\ as explicited
in the constructive proof of the theorem, written for the open cover
$\,\cO_{\sfL M}=\{\cO_\igt\}_{\igt\in\xcI_{\sfL M}}\,$ of the
free-loop space $\,\sfL M=C^\infty(\bS^1,M)\,$ from Proposition
\vref{prop:cover-untw}I.3.13. Write
\qq\nn
\Tgt:=1\oplus\theta_{\sfT^*\sfL M}\in\G\bigl(\sfE^{(0,1)}\sfP_{\si,
\emptyset}\bigr)\,,
\qqq
and call the latter object the \textbf{canonical section of
$\,\sfE^{(1,0)}\sfP_{\si,\emptyset}$}.\ To every smooth
$\si$-symmetric section $\,\Vgt\,$ of $\,\sfE^{(1,1)}M\,$ there is
associated a \textbf{hamiltonian function $\,h_\Vgt$},\
\textit{i.e.}\ a smooth function on $\,\sfP_{\si,\emptyset}\,$
satisfying the defining relation
\qq\nn
\a_{\sfT\sfP_{\si,\emptyset}}\bigl(\widetilde\sfL_{(1,1)}\Vgt\bigr)
\con\Om_{\si,\emptyset}=:-\d h_\Vgt\,.
\qqq
The hamiltonian function is given by the formula
\qq\nn
h_\Vgt=\corr{\widetilde\sfL_{(1,1)}\Vgt,\Tgt}\,.
\qqq
The \textbf{pre-quantum hamiltonian for $\,h_\Vgt$},\ as explicited
in Definition \vref{def:prequantise}I.3.4, is a linear operator
$\,\widehat\cO_{h_\Vgt}\,$ on
$\,\G\bigl(\ceL_{\si,\emptyset}\bigr)\,$ with restrictions
\qq\label{eq:preq-ham-gen}
\widehat\cO_{h_\Vgt}\vert_{\pi_{\sfT^*\sfL M}^{-1}(\cO_\igt)}=-\sfi
\,\pLie{\a_{\sfT\sfP_{\si,\emptyset}}\bigl(\ee^{-\theta_{\sfT^*\sfL
M}}\lact\widetilde\Vgt_\igt\bigr)}+\corr{\ee^{-\theta_{\sfT^*\sfL
M}}\lact\widetilde\Vgt_\igt,\Tgt}=:\widehat h_{\widetilde\Vgt_\igt}
\,,
\qqq
expressed in terms of local sections
\qq\label{eq:preq-ham-gen-ingr}
\widetilde\Vgt_\igt:=\ee^{-\pi_{\sfT^*\sfL M}^*E_\igt}\lact
\widetilde\sfL_{(1,1)}\Vgt\in\sfE^{(1,0)}_{\pi_{\sfT^*\sfL M}^*
\ceL_\cG}\sfP_{\si,\emptyset}\bigl(\pi_{\sfT^*\sfL M}^{-1}(\cO_\igt)
\bigr)\,.
\qqq
Given a pair $\,\Vgt,\Wgt\,$ of $\si$-symmetric sections of
$\,\sfE^{(1,1)}M$,\ the Poisson bracket of the associated
hamiltonian functions, determined by $\,\Om_{\si,\emptyset}\,$ in
the manner detailed in Remark \vref{rem:Mars-Wein}I.3.3, reads
\qq\label{eq:Poiss-bra-ham}
\{h_\Vgt,h_\Wgt\}_{\Om_{\si,\emptyset}}=
h_{\Vbra{\Vgt}{\Wgt}^\txH}\,.
\qqq
The commutator of the corresponding pre-quantum hamiltonians
satisfies (locally) the relation
\qq\label{eq:comm-preq-ham}
[\widehat h_{\widetilde\Vgt_\igt},\widehat h_{\widetilde\Wgt_\igt}]
=-\sfi\,\widehat h_{\Vbra{\widetilde\Vgt_\igt}{\widetilde\Wgt_\igt}}
\,,
\qqq
written in terms of the bracket of the global Vinogradov structure
$\,\Vgt^{(0)}_{\pi_{\sfT^*\sfL M}^*\ceL_\cG}\sfP_{\si,\emptyset}$.
\eerop
\beroof
Begin by noting that the symplectic form of
\Reqref{eq:sympl-form-untw} can be written as
\qq\nn
\Om_{\si,\emptyset}=\d\theta_{\sfT^*\sfL M}+\widetilde\sfL^*\txH
\equiv\d_{\widetilde\sfL^*\txH}\Tgt\,,
\qqq
and so, using Eqs.\,\eqref{eq:anchPsi-phisi},
\eqref{eq:cot-lift-pres-p} and \eqref{eq:tiL-minus}, as well as the
assumption $\,\Vgt\in \ker\,\sfd_\txH$,\ we find, for the canonical
projection $\,\pr_{\sfT^*M}:\sfE^{(1,1)}M\to\sfT^*M$,
\qq\nn
\a_{\sfT\sfP_{\si,\emptyset}}\bigl(\widetilde\sfL_{(1,1)}\Vgt\bigr)
\con\Om_{\si,\emptyset}&=&\widetilde\sfL_*\a_{\sfT M}(\Vgt)\con
\bigl(\d\theta_{\sfT^*\sfL M}+\widetilde\sfL^*\txH\bigr)=-\d\bigl(
\widetilde\sfL_*\a_{\sfT M}(\Vgt)\con\theta_{\sfT^*\sfL M}\bigr)-
\widetilde\sfL^*\bigl(\a_{\sfT M}(\Vgt)\con\txH\bigr)\cr\cr
&=&-\d\bigl(\widetilde\sfL_*\a_{\sfT M}(\Vgt)\con\theta_{\sfT^*\sfL
M}+\widetilde\sfL^*\pr_{\sfT^*M}(\Vgt)\bigr)\,,
\qqq
as claimed. The form of the pre-quantum hamiltonian then follows
directly from the general definition of
Eq.\,\veqref{eq:pre-ham-gen}{I.3.8}, and we readily see, through
direct inspection, that the local objects $\,\widetilde\Vgt_\igt\,$
are in the image of an isomorphism defined analogously to the
isomorphism $\,{}^{\sfP_{\si,\emptyset}}\hspace{-2pt}\chi^{-1}\,$
from the constructive proof of Theorem
\ref{thm:ind-quasi-morph-glob-Vin}.

The Poisson bracket of a pair of hamiltonian functions can be
computed directly but instead let us combine our observation from
Proposition \ref{prop:ham-as-genom} with the findings of Lemma
\ref{lem:quasi-morph-Vin} to render the algebraic structure that
underlies the calculation manifest. First, we write down the
hamiltonian section $\,\Xgt_{h_\Vgt}\equiv\widetilde\Vgt\,$ of
$\,\sfE^{(1,0)}\sfP_{\si,\emptyset}\,$ associated to $\,h_\Vgt$.\
Clearly,
\qq\nn
\widetilde\Vgt=\a_{\sfT\sfP_{\si,\emptyset}}\bigl(\widetilde\sfL_{(
1,1)}\Vgt\bigr)\oplus\corr{\widetilde\sfL_{(1,1)}\Vgt,\Tgt}=
\ee^{\theta_{\sfT^*\sfL M}}\lact\widetilde\sfL_{(1,1)}\Vgt\,,
\qqq
and so, exploiting the results from the proof of Proposition
\ref{prop:Vin-str-auts} and taking into account
\Reqref{eq:Vinbra-lll-Vinbra}, we find, for a pair $\,\Vgt,\Wgt\in
\G_\si\bigl(\sfE^{(1,1)}M\bigr)$,\
\qq\nn
\Vbra{\widetilde\Vgt}{\widetilde\Wgt}^{\Om_{\si,\emptyset}}&\equiv&
\Vbra{\ee^{\theta_{\sfT^*\sfL M}}\lact\widetilde\sfL_{(1,1)}
\Vgt}{\ee^{\theta_{\sfT^*\sfL M}}\lact\widetilde\sfL_{(1,1)}
\Wgt}^{\Om_{\si,\emptyset}}=\ee^{\theta_{\sfT^*\sfL M}}\lact
\Vbra{\widetilde\sfL_{(1,1)}\Vgt}{\widetilde\sfL_{(1,1)}
\Wgt}^{\widetilde\sfL^*\txH}\cr\cr
&=&\ee^{\theta_{\sfT^*\sfL M}}\lact\widetilde\sfL_{(1,1)}\bigl(
\Vbra{\Vgt}{\Wgt}^\txH\bigr)\equiv
\widetilde{\Vbra{\Vgt}{\Wgt}^\txH}\,.
\qqq
By virtue of Proposition \ref{prop:ham-as-genom}, this confirms
\Reqref{eq:Poiss-bra-ham}.

Passing, next, to the pre-quantum hamiltonians, we first note that
they satisfy -- in consequence of Eq.\,(I.3.9) -- the algebra
\qq\nn
[\widehat h_{\widetilde\Vgt_\igt},\widehat h_{\widetilde\Wgt_\igt}]=
-\sfi\,\widehat\cO_{\{h_\Vgt,h_\Wgt\}_{\Om_{\si,\emptyset}}}
\vert_{\cO^*_\igt}\,,
\qqq
which can be rewritten -- with the help of
Eqs.\,\eqref{eq:Poiss-bra-ham} and \eqref{eq:preq-ham-gen}, taken
together with \Reqref{eq:preq-ham-gen-ingr} -- as
\qq\nn
[\widehat h_{\widetilde\Vgt_\igt},\widehat h_{\widetilde\Wgt_\igt}]=
-\sfi\,\left(-\sfi\,\pLie{\a_{\sfT\sfP_{\si,\emptyset}}\bigl(\ee^{-
\theta_{\si,\emptyset\,\igt}}\lact\widetilde\sfL_{(1,1)}
\Vbra{\Vgt}{\Wgt}^\txH\bigr)}+\corr{\ee^{-\theta_{\si,\emptyset\,
\igt}}\lact\widetilde\sfL_{(1,1)}\Vbra{\Vgt}{\Wgt}^\txH,\Tgt}
\right)\,.
\qqq
Employing Eqs.\,\eqref{eq:Vinbra-lll-Vinbra} and
\eqref{eq:Hitch-Vbra} once more, we then find
\qq\nn
[\widehat h_{\widetilde\Vgt_\igt},\widehat h_{\widetilde\Wgt_\igt}]
&=&-\sfi\,\left(-\sfi\,\pLie{\a_{\sfT\sfP_{\si,\emptyset}}\bigl(
\ee^{-\theta_{\si,\emptyset\,\igt}}\lact\Vbra{\widetilde\sfL_{(1,1
)}\Vgt}{\widetilde\sfL_{(1,1)}\Wgt}^{\widetilde\sfL^*\txH}\bigr)}+
\corr{\ee^{-\theta_{\si,\emptyset\,\igt}}\lact\Vbra{\widetilde
\sfL_{(1,1)}\Vgt}{\widetilde\sfL_{(1,1)}\Wgt}^{\widetilde\sfL^*
\txH},\Tgt}\right)\cr\cr
&=&-\sfi\,\bigl(-\sfi\,\pLie{\a_{\sfT\sfP_{\si,\emptyset}}\bigl(
\ee^{-\theta_{\sfT^*\sfL M}}\lact\Vbra{\widetilde
\Vgt_\igt}{\widetilde\Wgt_\igt}\bigr)}+\corr{\ee^{-\theta_{\sfT^*
\sfL M}}\lact\Vbra{\widetilde\Vgt_\igt}{\widetilde\Wgt_\igt},\Tgt}
\bigr)\equiv-\sfi\,\widehat h_{\Vbra{\widetilde\Vgt_\igt}{\widetilde
\Wgt_\igt}}\,,
\qqq
as claimed. \eroof\medskip

We are now ready to discuss at length the realisation of internal
symmetries of the $\si$-model on the classical and pre-quantum state
space of the untwisted sector of the theory. Thus,
\berop\label{prop:sympl-goes-ham-untw}
In the notation of Corollary \ref{cor:sigmod-symm-E11}, of
Propositions \ref{prop:Htw-Vbra-symm} and
\ref{prop:ham-vs-sisymsec}, and of Theorem
\ref{thm:ind-quasi-morph-glob-Vin}, the $\si$-symmetric sections
$\,\Kgt_A\,$ determine a symplectic realisation of $\,\ggt_\si\,$ on
$\,C^\infty\bigl(\sfP_{\si, \emptyset},\bR\bigr)\,$ by the
hamiltonian functions $\,h_{\Kgt_A}\,$ and an operator realisation
on $\G\bigl(\ceL_{\si, \emptyset}\bigr)\,$ by the pre-quantum
hamiltonians $\,\widehat\cO_{h_{\Kgt_A}}\,$ with local restrictions
$\,\widehat h_{\widetilde\Kgt_A\,\igt}$.\ The former realisation is
hamiltonian,
\qq\label{eq:Poiss-class-ham-Ka-HS}
\{h_{\Kgt_A},h_{\Kgt_B}\}_{\Om_{\si,\emptyset}}=f_{ABC}\,
h_{\Kgt_C}\,,
\qqq
iff the $\,\Kgt_A\,$ can be chosen such that
\qq\label{eq:HS-2}
\pLie{\xcK_A}\kappa_B=f_{ABC}\,\kappa_C+\sfd D_{AB}
\qqq
for some $\,D_{AB}\in C^\infty(M,\bR)$,\ in which case also
\qq\label{eq:Vinbra-Ka-HS}
\Vbra{\Kgt_A}{\Kgt_B}^\txH=f_{ABC}\,\Kgt_C+0\oplus\tfrac{1}{2}\,
\sfd(D_{AB}-D_{BA})
\qqq
and
\qq\label{eq:tw-Vinbra-Kai-HS}
\Vbra{\widetilde\Kgt_{A\,\igt}}{\widetilde\Kgt_{B\,\igt}}=f_{ABC}\,
\widetilde\Kgt_{C\,\igt}\,,
\qqq
so that
\qq\label{eq:comm-quant-ham-Ka-HS}
[\widehat h_{\widetilde\Kgt_A\,\igt},\widehat h_{\widetilde
\Kgt_B\,\igt}]=-\sfi\,f_{ABC}\,\widehat h_{\widetilde\Kgt_C
\,\igt}\,.
\qqq
\eerop
\beroof
First, invoking \Reqref{eq:Vinbra-Ka} in conjunction with
\Reqref{eq:tiL-minus}, we rewrite \Reqref{eq:Poiss-bra-ham} in the
present setting as
\qq\nn
\{h_{\Kgt_A},h_{\Kgt_B}\}_{\Om_{\si,\emptyset}}=f_{ABC}\,
h_{\Kgt_C}+\widetilde\sfL^*(\D_{AB}-\sfd \txc_{(AB)})=f_{ABC}\,
h_{\Kgt_C}+\widetilde\sfL^*\D_{AB}
\qqq
in the notation of \Reqref{eq:Vinbra-Ka}. Clearly, the realisation
of $\,\ggt_\si\,$ is hamiltonian iff $\,\D_{AB}=\sfd D_{AB}\,$ for
some $\,D_{AB}\in C^\infty(M,\bR)$,\ which is, indeed, tantamount to
\Reqref{eq:HS-2}. Moreover, note that, in this case,
\qq\nn
\sfd
\txc_{(AB)}=\tfrac{1}{2}\,\bigl(\pLie{\xcK_A}\kappa_B+\xcK_A\con
\xcK_B\con\txH+\pLie{\xcK_B}\kappa_A+\xcK_B\con\xcK_A\con\txH\bigr)
=\tfrac{1}{2}\,\sfd(D_{AB}+D_{BA})\,,
\qqq
whence \Reqref{eq:Vinbra-Ka-HS} follows.

Passing to the pre-quantum hamiltonians, under the assumption that
\Reqref{eq:HS-2} holds true, we readily verify the identity
\qq\nn
\Vbra{\widetilde\Kgt_{A\,\igt}}{\widetilde\Kgt_{B\,\igt}}&=&\ee^{-
\pi_{\sfT^*M}^*E_\igt}\lact\widetilde\sfL_{(1,1)}
\Vbra{\Kgt_A}{\Kgt_B}^\txH=\ee^{-\pi_{\sfT^*M}^*E_\igt}\lact
\widetilde\sfL_{(1,1)}\bigl(f_{ABC}\,\Kgt_C+0\oplus\tfrac{1}{2}\,
\sfd(D_{AB}-D_{BA})\bigr)\cr\cr
&=&f_{ABC}\,\widetilde\Kgt_{C\,\igt}
\qqq
by reversing and repeating the manipulations carried out in the
proof of \Reqref{eq:comm-preq-ham}, which reproduces
\Reqref{eq:tw-Vinbra-Kai-HS} and thereby completes the proof. \eroof
\brem\label{rem:AS-res} Another piece of evidence in favour of the
relevance of the geometry of the generalised tangent bundle in the
canonical description of the two-dimensional $\si$-model comes from
the study of the Poisson algebra of the Noehter currents ($\widehat
t\,$ is the normalised tangent vector field on $\,\bS^1$)
\qq\label{eq:symm-curr-def}
J_{\Kgt_A}[\psi]=\xcK_A^\mu\,\sfp_\mu+(X_*\widehat t)^\mu\,
\kappa_{A\,\mu}\,,\qquad\qquad\psi=(X^\mu,\sfp_\nu)
\qqq
of the theory, furnishing an anomalous field-theoretic
representation of the algebra $\,\ggt_\si$.\ This is, in fact, the
structure originally examined in the pioneering
\Rcite{Alekseev:2004np} in which the link between the
current-symmetry algebra of the $\si$-model and the structure of a
Courant algebroid, twisted according to the standard prescription
first suggested in \Rcite{Severa:2001qm}, on the generalised tangent
bundle $\,\sfE^{(1,1)}M\,$ was established. A straightforward
computation, first carried out in \Rcite{Alekseev:2004np}, yields
the identity
\qq
\{J_{\Kgt_A}(t,\varphi),J_{\Kgt_B}(t,\varphi')\}_{\Om_{\si,
\emptyset}}=J_{\Vbra{\Kgt_A}{\Kgt_B}^\txH}(t,\varphi)\,\d(\varphi-
\varphi')-2\corr{\Kgt_A,\Kgt_B}\bigl(t,\tfrac{1}{2}(\varphi+
\varphi')\bigr)\,\d'(\varphi-\varphi')\cr\cr
\label{eq:curr-bra-anom}
\qqq
for the $\txH$-twisted Vinogradov bracket on $\,\G(\sfE^{(1,1)}M)\,$
(identical with the Courant bracket\footnote{There is, in fact, a
whole family of brackets on $\,\G(\sfE^{(1,1)}M)\,$ of different
skew-symmetry properties and jacobiators, including, in particular,
the Dorfman bracket of \Rcite{Dorfman:1987}. They can be obtained
from the above calculation by replacing $\,\tfrac{1}{2}\,(\varphi+
\varphi')\,$ in the anomalous second term on the right-hand side of
\Reqref{eq:curr-bra-anom} with a generic argument $\,\varphi_\la=\la
\,\varphi+(1-\la)\,\varphi',\ \la\in[0,1]\,$ and by changing the
first term accordingly.} in this special case).\erem \brem The
contents of the present section seem to indicate that it is natural,
in the context of the two-dimensional field theory of interest, to
separate the algebraic structure on the generalised tangent bundle
over the target space $\,M\,$ of the $\si$-model from
field-theoretic considerations of internal symmetries of the latter.
Although the presence of a gerbe over $\,M\,$ can affect this
structure, either via the topological twist of the bundle itself or,
equivalently, via the twist of the Vinogradov bracket on it, it is
not a priori clear how one could extract from the structure any
information on the geometry of the target. That one can actually do
so was shown in \Rcite{Hitchin:2005in}, and we pause briefly to
demonstrate what can be learnt from the original argument about the
transition undergone by the geometry of the tangent bundle as we
pass from an untwisted to an $\txH$-twisted Vinogradov structure on
the associated generalised tangent bundle of type $\,(1,1)$.

To these ends, we consider a target $\,\cM=(M,\txg,\cG)$,\ together
with the generalised tangent bundle $\,\sfE^{(1,1)}M\to M\,$ and the
$\txH$-twisted Vinogradov structure $\,\Vgt^{(1),\txH}M\,$ on it,
with $\,\txH=H_{\la \mu\nu}\,\sfd X^\mu\wedge\sfd X^\mu\wedge\sfd
X^\nu\in Z^3( M)\,$ the curvature of $\,\cG$.\ We represent -- after
Hitchin -- the metric $\,\txg\,$ on $\,\G(\sfT M)\,$ by its graph in
$\,\sfE^{(1,1 )}M$,\ \textit{i.e.}\ by the subbundle of rank
$\,\dim\,M\,$ with fibre
\qq\nn
{\rm graph}(\txg)\vert_m:=\{\ \xcV_+(m):=\xcV(m)\oplus\txg_m(\xcV(m)
,\cdot) \quad\vert\quad \xcV\in\G(\sfT M) \ \}\subset\sfE^{(1,1)}_m
M\,,\qquad m\in M
\qqq
on which the canonical contraction becomes positive definite (for
$\,\txg\,$ Riemannian),
\qq\nn
\Vcon{\xcV_+}{\xcV_+}=\txg(\xcV,\xcV)\,.
\qqq
Thus, we think of $\,\txg\,$ as the so-called \textbf{generalised
metric}, in the sense of \Rxcite{Def.\,3}{Hitchin:2005in}, defining
a splitting
\qq\nn
&\sfE^{(1,1)}M={\rm graph}(\txg)\oplus{\rm graph}(\txg
)^{\perp_{\Vcon{\cdot}{\cdot}}}\,,&\cr\cr
&\xcV\oplus\upsilon=\tfrac{1}{2}\,\left(\xcV+\txg^{-1}(\upsilon,
\cdot)\right)_+\oplus\tfrac{1}{2}\,\left(\xcV-\txg^{-1}(\upsilon,
\cdot)\right)_-\,,&
\qqq
where $\,\xcW_\pm=\xcW\oplus\txg(\pm\xcW,\cdot)$.\ Next, we readily
check, \textit{cf.}\ \Rxcite{Thm.\,2}{Hitchin:2005in}, that the
linear operator $\,\nabla_\xcV:\G(\sfT M)\to\G(\sfT M)\,$ given, for
arbitrary $\,\xcV,\xcW\in\G(\sfT M)$,\ by the formula
\qq\nn
0\oplus 2\txg(\nabla_\xcV\xcW,\cdot):=\Vbra{\xcV_-}{\xcW_+}^\txH-[
\xcV,\xcW ]_-
\qqq
defines a metric connection
\qq\nn
\nabla\ &:&\ \G(\sfT M)\to\G(\sfT^*M)\ox\G(\sfT M)
\qqq
on $\,\sfT M$,\ with
\qq\nn
\nabla_\xcV\xcW=\xcV^\la\,\bigl(\p_\la\xcW^\nu+\G_{\la\mu}^{\ \nu}
\,\xcW^\mu\bigr)\,\p_\nu\,,
\qqq
where
\qq\nn
\G_{\la\mu}^{\ \nu}=\bigl\{\begin{smallmatrix} \nu \\ \la\mu
\end{smallmatrix}\bigr\}-3\,(\txg^{-1})^{\nu\rho}\,\txH_{\rho\la
\mu}\,.
\qqq
Hence, the passage from the untwisted Vinogradov structure on
$\,\sfE^{(1,1)}M\,$ to the twisted one can be understood in terms of
induction of a torsion-full (Weitzenb\"ock) connection on the
tangent bundle of the base manifold $\,M\,$ that extends the
standard symmetric Levi-Civita connection. In this manner,
$\,\Vgt^{(1),\txH}M\,$ can encode non-trivial information on the
geometry (of the tangent bundle) of the $\si$-model target. It
deserves to be pointed out that the above identification of the
curvature of the gerbe with the torsion component of a metric
connection on $\,\sfT M\,$ arises independently in the framework of
spectral non-commutative geometry of (the supersymmetric extension
of) the CFT of the quantised $\si$-model, mentioned in Remark
\vref{rem:NCG}I.3.20, \textit{cf.}\
Refs.\,\cite{Frohlich:1993es,Recknagel:2006hp}.\erem

\section{Morphisms of Vinogradov structures and symmetry
transmission across defects}\label{sec:Vin-morph}

The observations made in the preceding section suggest that we begin
our study of structures induced by data carried by the defect in the
geometry of the generalised tangent bundle of the $\si$-model
background in abstraction from symmetries of the two-dimensional
field theory. Thus, we find
\bethe\label{thm:bib-as-morph}
Adopt the notation of Definitions \ref{def:Vin-str} and
\ref{def:Om-tw-Vin}, and of Corollary \ref{cor:tw-gen-tan-vs-ngerb}.
Let $\,\Bgt\,$ be a string background with target $\,\cM=(M,\txg,\cG
)\,$ and $\cG$-bi-brane $\,\cB=\bigl(Q,\iota_\a, \om,\Phi\ \vert\
\a\in\{1,2\}\bigr)\,$ as in Definition \vref{def:bckgrnd}I.2.1, and
let $\,\txH\in Z^3(M)\,$ be the curvature of $\,\cG$.\ Finally, let
$\,\Qup\chi_\a\ :\ \Vgt^{(1)}_{\iota_\a^*\cG}Q\to\Vgt^{(1),
\iota_\a^*\txH}Q,\ \a\in\{1,2\}\,$ be the canonical isomorphisms of
Corollary \ref{cor:tw-gen-tan-vs-ngerb}. Then, the following
statements hold true:
\bit
\item[i)] The curvature $\,\om\,$ of the $\cG$-bi-brane $\,\cB\,$
canonically determines an isomorphism
\qq\nn
\b_\om\ :\ \Vgt^{(1),\iota_1^*\txH}Q\xrightarrow{\ \cong\ }\Vgt^{(1
),\iota_2^*\txH}Q \,.
\qqq
\item[ii)] The 1-isomorphism $\,\Phi\,$ of the $\cG$-bi-brane
$\,\cB\,$ canonically induces an isomorphism
\qq\nn
\b_\Phi\ :\ \Vgt^{(1)}_{\iota_1^*\cG}Q\xrightarrow{\ \cong\ }\Vgt^{(
1)}_{\iota_2^*\cG}Q\,.
\qqq
\item[iii)] The above isomorphisms are intertwined by the
isomorphisms $\,\Qup\chi_\a\,$ in the sense expressed by the
commutative diagram
\qq\nn
\alxydim{@C=1.cm@R=1.cm}{\Vgt^{(1)}_{\iota_1^*\cG}Q \ar[r]^{\b_\Phi}
\ar[d]_{\Qup\chi_1} & \Vgt^{(1)}_{\iota_2^*\cG}Q \ar[d]^{\Qup\chi_2}
\cr \Vgt^{(1),\iota_1^*\txH}Q \ar[r]^{\b_\om} & \Vgt^{(1),\iota_2^*
\txH}Q}\,.
\qqq
\eit
\ethe
\beroof
\bit
\item[Ad i)] Consider the isomorphism
\qq\nn
\ee^{-\om}\ :\ \sfE^{(1,1)}Q\xrightarrow{\cong}\sfE^{(1,1)}Q
\qqq
covering the identity diffeomorphism on the base $\,Q$.\ Reasoning
as in the proof of Proposition \ref{prop:Vin-str-auts}, and using
\vReqref{eq:curv-constr}{I.3.21}, we establish the identities
\qq\nn
\Vbra{\cdot}{\cdot}^{\iota_2^*\txH}\circ(\ee^{-\om},\ee^{-\om})&=&
\ee^{-\om}\circ\Vbra{\cdot}{\cdot}^{\iota_1^*\txH}\,,\cr\cr
\Vcon{\cdot}{\cdot}\circ(\ee^{-\om},\ee^{-\om})&=&
\Vcon{\cdot}{\cdot}\,,\cr\cr \a_{\sfT Q}\circ\ee^{-\om}&=&\a_{\sfT
Q}\,.
\qqq
This permits to set
\qq\label{eq:betom}
\b_\om:=\ee^{-\om}\,.
\qqq
\item[Ad ii)] Choose open covers $\,\Mup\cO=\{\cO^M_i\}_{i\in
\xcI_M}\,$ and $\,\Qup\cO=\{\Qup\cO_a\}_{a\in\xcI_Q}\,$ of the
target space $\,M\,$ and of the $\cG$-bi-brane world-volume $\,Q$,\
respectively, for which there exist \v Cech-extended $\cG$-bi-brane
maps $\,(\iota_\a,\phi_\a),\ \a\in\{1,2\}$,\ and fix local
presentations, associated with these covers, for the gerbe, $\,(B_i
,A_{ij},g_{ijk})\in\cA^{3,2}(\Mup\cO)$,\ and for the $\cG$-bi-brane
1-isomorphism, $\,(P_a,K_{ab})\in\cA^{2,1}(\Qup\cO)$,\ all as
described in Definition \vref{def:loco}I.2.2. The transition maps
$\,\ggt_{ab}^\a\,$ of the $\,\sfE^{(1,1)}_{\iota_\a^*\cG}Q\,$ are
then given by the formula
\qq\nn
\ggt_{ab}^\a=\ee^{\iota_\a^*B_{\phi_\a(a)}-\iota_\a^*B_{\phi_\a(b
)}}\,.
\qqq
We readily check, with the help of the cohomological identity from
\veqref{eq:DPhi-is}{I.2.5}, that the isomorphism
\qq\nn
(\bgt_{\Phi\,a})_{a\in\xcI_Q}\ :\ \sfE^{(1,1)}_{\iota_1^*\cG}Q
\xrightarrow{\ \cong\ }\sfE^{(1,1)}_{\iota_2^*\cG}Q\,,\qquad\qquad
\bgt_{\Phi\,a}:=\ee^{-\sfd P_a}
\qqq
covering the identity diffeomorphism on the base $\,Q\,$ satisfies
the identity
\qq\nn
\ggt_{ab}^2=\bgt_{\Phi\,b}\circ\ggt_{ab}^1\circ\bgt_{\Phi\,a}^{-1}
\qqq
on $\,\cO_{ab}^Q$,\ and so -- by virtue of Proposition
\ref{prop:Vin-str-glob-tw} -- it is meaningful to define
\qq\label{eq:betPhi}
\b_\Phi:=(\bgt_{\Phi\,a})_{a\in\xcI_Q}\,.
\qqq
\item[Ad iii)] An immediate corollary to \vReqref{eq:DPhi-is}{I.2.4},
taking into account Eqs.\,\eqref{eq:betom} and \eqref{eq:betPhi}.
\eit
\eroof\bigskip

Having established an independent interpretation of the structure of
a $\cG$-bi-brane in the context of the geometry of the generalised
tangent bundle, we may next return to the main point of our
interest, that is the physics of the two-dimensional $\si$-model in
the presence of defects. In Section \vref{sec:def-as-iso}I.4, the
latter were straightforwardly related to dualities of the
$\si$-model. The following result attests, once more, the
naturalness of the algebraic structures introduced in this paper by
demonstrating their simple behaviour under dualities.
\berop\label{prop:dual-vs-morf}
Adopt the notation of Definitions \ref{def:Vin-str} and
\ref{def:Om-tw-Vin}, and of Theorems \ref{thm:bib-as-morph} and
\ref{thm:ind-quasi-morph-glob-Vin}. Let $\,\Igt_\si(\cB)\,$ be the
isotropic subspace in $\,\sfP_{\si, \emptyset}^{\x
2}=\sfP_{\si,\emptyset}\x\sfP_{\si, \emptyset}\,$ defined in
Proposition \vref{prop:DGC-as-iso}I.4.1. Suppose that $\,
\Igt_\si(\cB)\,$ is a graph of a symplectomorphism $\,\b_\cB:
\sfP_{\si,\emptyset}\to\sfP_{\si,\emptyset}$,\ which is the case, in
particular, if the $\cG$-bi-brane $\,\cB\,$ together with the Defect
Gluing Condition \veqref{eq:DGC}{I.2.7} define a pre-quantum duality
of the untwisted sector of the $\si$-model, understood in the sense
of Definition \vref{def:pqsymm}I.4.7. Then,
$\,(\b_\cB,\widehat\b_\cB)$,\ with the covering map
\qq\nn
\widehat\b_\cB=\left(\barr{cc} \b_{\cB\,*} & 0 \cr\cr 0 & (\b_\cB^{-
1})^* \earr \right)\,,
\qqq
is an automorphism of $\,\Vgt^{(0),\Om_{\si,\emptyset}}\sfP_{\si,
\emptyset}$.\ Conversely, every (unital) automorphism $\,(f,F)\,$ of
$\,\Vgt^{(0),\Om_{\si,\emptyset}}\sfP_{\si,\emptyset}\,$ is of the
form
\qq\nn
F=\widehat f\circ\ee^\txB
\qqq
for some $\,\widehat f\,$ and $\,\ee^\txB\,$ as in Proposition
\ref{prop:Vin-str-auts}, and for $\,\txB\,$ a unique 1-form on
$\,\sfP_{\si,\emptyset}\,$ such that
\qq\nn
f^*\Om_{\si,\emptyset}-\Om_{\si,\emptyset}=\d\txB\,.
\qqq
\eerop
\beroof
The first statement of the proposition is readily checked through
inspection, and so we pass immediately to the second one. As in the
proof of Proposition \ref{prop:Vin-str-auts}, we decompose the
bundle map
\qq\nn
F=\widehat f\circ G
\qqq
into the standard term $\,\widehat f\,$ that covers $\,f$,\ and the
completion $\,G\,$ covering the identity diffeomorphism on the base
$\,\sfP_{\si,\emptyset}$.\ Using the identity
\qq\nn
\widehat f\circ\Vbra{\cdot}{\cdot}^{f^*\Om_{\si,\emptyset}}=
\Vbra{\cdot}{\cdot}^{\Om_{\si,\emptyset}}\circ\left(\widehat f,
\widehat f\right)\,,
\qqq
we rewrite the (co)defining property
\qq\label{eq:F-def}
\Vbra{\cdot}{\cdot}^{\Om_{\si,\emptyset}}\circ(F,F)=F\circ
\Vbra{\cdot}{\cdot}^{\Om_{\si,\emptyset}}
\qqq
of $\,F\,$ in the form
\qq\label{eq:G-from-Fdef}
G\circ\Vbra{\cdot}{\cdot}^{\Om_{\si,\emptyset}}=
\Vbra{\cdot}{\cdot}^{f^*\Om_{\si,\emptyset}}\circ(G,G)\,.
\qqq
We subsequently evaluate both sides of the last relation on the pair
$\,(g \cdot\Vgt,\Wgt)$,\ whereby we arrive, once again, at the
consistency condition \eqref{eq:gVW-Vbra} that leads to the familiar
form \eqref{eq:G-n0} of $\,G$.\ The requirement that $\,F\,$ be
unital yields the desired result
\qq\nn
G=\ee^\txB\,,\qquad\txB\in\Om^1(\sfP_{\si,\emptyset})\,.
\qqq
Substituting the ensuing $\,F=\widehat f\circ\ee^\txB\,$ back into
relation \eqref{eq:F-def} and evaluating the latter on a pair $\,(
\xcV\oplus g,\xcW\oplus h)$,\ we obtain
\qq\nn
&&[f_*\xcV,f_*\xcW]\oplus\left[f_*\xcV\left(f^{-1\,*}(h+\xcW\con
\txB)\right)-f_*\xcW\left(f^{-1\,*}(g+\xcV\con\txB)\right)+f_*\xcV
\con f_*\xcW\con\Om_{\si,\emptyset}\right]\cr\cr
&=&f_*[\xcV,\xcW]\oplus f^{- 1\,*}\left(\xcV(h)-\xcW(g)+\xcV\con
\xcW\con\Om_{\si,\emptyset}+[\xcV,\xcW]\con\txB\right)\,,
\qqq
whence also the consistency constraint
\qq\nn
\xcV\con\xcW\con\left(f^*\Om_{\si,\emptyset}-\Om_{\si,\emptyset}
-\d\txB\right)=0\,,
\qqq
from which we recover the claim of the proposition. \eroof\medskip

We are now fully equipped for the study of mechanisms of
transmission of symmetries between phases of the $\si$-model across
world-sheet defects that separate them. The obvious starting point
of our analysis is a counterpart of Proposition
\ref{prop:var-sigmod-n}, readily extractable from the results of
\Rcite{Runkel:2008gr}, that holds true in the presence of an
embedded defect quiver for $\,n=1\,$ (a generalisation of this
result to higher-dimensional cases is completely straightforward).
\berop\cite[App.\,A2]{Runkel:2008gr}\label{prop:var-sigmod-def}
Adopt the notation of Definition \ref{def:sigmod-2d} and of
Proposition \ref{prop:sympl-form-twuntw}. Let $\,\xcV\,$ be a vector
field on the target space of the background $\,\xcF:=\xcM\sqcup
Q\sqcup \bigsqcup_{n\in\bN_{\geq 3}}\,T_n \,$ with a (local) flow
$\,\xi_t\ :\ \xcF\to\xcF\,$ and restrictions
$\,\xcMup\xcV:=\xcV\vert_\xcM,\ \xcM\in\{M,Q,T_n\}\,$ such that
\qq\label{eq:iota-align}
\iota_{\a\,*}\Qup\xcV=\Mup\xcV\vert_{\iota_\a(Q)}\,,
\qqq
and
\qq\nn
\pi_{n\,*}^{k,k+1}\Tnup\xcV=\Qup\xcV\vert_{\pi_n^{k,k+1}(T_n)}\,.
\qqq
The variation along $\,\xi_t\,$ of the action functional of
\Reqref{eq:2d-sigma-def} reads
\qq
\tfrac{\sfd\ }{\sfd t}\vert_{t=0}S_\si[(\xi_t\circ X\,\vert\,\G);\g]
=-\tfrac{1}{2}\, \int_\Si\,(\pLie{\Mup \xcV}\txg)_X(\sfd X
\overset{\wedge}{,}\star_\g\sfd X)+\int_\Si\,X^*(\Mup\xcV\con\txH)+
\int_\G\,(X\vert_\G)^*(\Qup\xcV\con\om)\,.\cr\cr
\label{eq:var-sigmod}
\qqq
\eerop

\noindent Combining the above result with the statement of Corollary
\ref{cor:sigmod-symm-E11}, we can give a compact algebraic
description of internal symmetries of the $\si$-model in the
presence of circular defects, which we formulate as
\berop\label{prop:sigmod-symm-def}
Adopt the notation of Definition \ref{def:Vin-str} and of Theorem
\ref{thm:bib-as-morph}, and write
\qq\nn
\D_Q:=\iota_2^*-\iota_1^*\,.
\qqq
Infinitesimal rigid symmetries of the two-dimensional non-linear
$\si$-model for network-field configurations $\,(X\,\vert \,\G)\,$
in string background $\,\Bgt\,$ on world-sheet $\,(\Si,\g )\,$ with
a defect quiver $\,\G\,$ composed of a finite number of
non-intersecting circular defect lines, as described in Definition
\vref{def:sigmod}I.2.7, correspond to pairs $\,(\Mup\Vgt,\Qup\Vgt
)\,$ consisting, each, of a $\si$-symmetric section $\,\Mup\Vgt\in
\G_\si\bigl(\sfE^{(1,1)}M \bigr)\,$ of $\,\sfE^{(1,1)}M$,
\qq\label{eq:sigmod-symm-bulk}
\pLie{\a_{\sfT M}(\Mup\Vgt)}\txg=0\,,\qquad\qquad\sfd_\txH\Mup\Vgt=
0\,,
\qqq
and of a $\Mup\Vgt$-twisted $\si$-symmetric section $\,\Qup\Vgt\in
\G\bigl(\sfE^{(1,0)}Q\bigr)\,$ of $\,\sfE^{(1,0)}Q$,
\qq\label{eq:sigmod-symm-def}
\sfd_\om\Qup\Vgt=-\D_Q\pr_{\sfT^*M}(\Mup\Vgt)\,,
\qqq
written in terms of the canonical projection $\,\pr_{\sfT^*M}:
\sfE^{(1,1)}M\to\sfT^*M$,\ and subject to the
\textbf{$\iota_\a$-alignment condition}:
\qq\label{eq:sigmod-symm-cons}
\a_{\sfT M}(\Mup\Vgt)\vert_{\iota_\a(Q)}=\iota_{\a\,*}\a_{\sfT Q}(
\Qup\Vgt)\,.
\qqq
\eerop
\beroof
The sufficiency of the conditions listed is a straightforward
corollary to Proposition \ref{prop:var-sigmod-def}. That they are
also necessary was demonstrated in the proof of Proposition 2.24 of
\Rcite{Gawedzki:2012fu}. \eroof
\medskip\noindent There are some important consequences of the
statement of symmetry of the $\si$-model in the presence of defects.
We begin with
\berop\label{prop:ham-cont-across}
In the notation of Definition \ref{def:Vin-str}, and of Theorems
\ref{thm:bib-as-morph} and \ref{thm:ind-quasi-morph-glob-Vin}, the
Poisson(-bracket) algebra of the hamiltonian functions on
$\,\sfP_{\si,\emptyset}\,$ assigned to those $\si$-symmetric
sections of $\,\sfE^{(1,1)}M\,$ which admit an extension to a pair
of $\si$-symmetric sections from $\,\sfE^{(1,1)}M\sqcup\sfE^{(1,0)}
Q\,$ subject to the $\iota_\a$-alignment condition
\eqref{eq:sigmod-symm-cons} is continuous across $\,\G$.
\eerop
\beroof
Take an arbitrary pair of $\iota_\a$-aligned $\si$-symmetric
sections $\,\bigl(\Mup\xcV\oplus\upsilon,\Qup\xcV\oplus\xi\bigr)=:(
\Mup\Vgt,\Qup\Vgt)$.\ The proof boils down to demonstrating the
equality of the values $\,h_{\Mup\Vgt}[\psi_1]\,$ and $\,h_{\Mup
\Vgt}[\psi_2]\,$ attained by the hamiltonian function $\,h_{\Mup
\Vgt}\,$ on a pair $\,(\psi_1,\psi_2)=\bigl((X_1,\sfp_1),(X_2,\sfp_2
)\bigr)\,$ of untwisted states from the isotropic subspace
$\,\Igt_\si(\cB)\,$ introduced in Proposition
\vref{prop:DGC-as-iso}I.4.1. We obtain, in the notation adopted from
the proof of that proposition, and using
Eqs.\,\veqref{eq:DGC}{I.2.7} and \eqref{eq:sigmod-symm-def},
\qq\nn
h_{\Mup\Vgt}[\psi_1]&=&\int_{\bS^1}\,\Vol(\bS^1)\,\bigl(\Mup\xcV(
X_1)\con\sfp_1+(X_{1\,*}\widehat t)\con\upsilon(X_1)\bigr)\cr\cr
&=&\int_{\bS^1}\,\Vol(\bS^1)\,\bigl(\Qup\xcV(X)\con(\sfp_1\circ
\iota_{1\,*})+(X_*\widehat t)\con\iota_1^*\upsilon(X)\bigr)\cr\cr
&=&\int_{\bS^1}\,\Vol(\bS^1)\,\bigl(\Qup\xcV(X)\con(\sfp_2\circ
\iota_{2\,*})+(X_*\widehat t)\con(\iota_1^*\upsilon-\Qup\xcV\con\om)
(X)\bigr)\cr\cr
&=&\int_{\bS^1}\,\Vol(\bS^1)\,\bigl(\Qup\xcV(X)\con(\sfp_2\circ
\iota_{2\,*})+(X_*\widehat t)\con(\iota_2^*\upsilon+\sfd\xi)(X)
\bigr)\cr\cr
&=&\int_{\bS^1}\,\Vol(\bS^1)\,\bigl(\Mup\xcV(X_2)\con\sfp_2+(X_{2\,
*}\widehat t)\con\upsilon(X_2)\bigr)\cr\cr
&=&h_{\Mup\Vgt}[\psi_2]\,.
\qqq
This is manifestly consistent with the structure of the Poisson
algebra on the space of hamiltonian functions. \eroof
\medskip\noindent The last result carries over directly to the
pre-quantum r\'egime, in which we have the analogous
\berop\label{prop:ham-simil-across}
Adopt the notation of Definition \ref{def:Vin-str}, and of Theorems
\ref{thm:bib-as-morph} and \ref{thm:ind-quasi-morph-glob-Vin}, and
assume that the isotropic submanifold
$\,\Igt_\si(\cB)\subset\sfP_{\si,\emptyset}\x\sfP_{\si,
\emptyset}\,$ defined in Proposition \vref{prop:DGC-as-iso}I.4.1 is
a graph of a symplectomorphism. The unitary similarity
transformation on the set of pre-quantum hamiltonians of the
untwisted sector of the $\si$-model defined by the bundle
isomorphism $\,\Dgt_\si(\cB)\,$ from the proof of Theorem
\vref{thm:def-dual}I.4.9 preserves (element-wise) the subalgebra
composed of those pre-quantum hamiltonians which are assigned to the
$\si$-symmetric sections of $\,\sfE^{(1,1)}M\,$ admitting an
extension to a pair of $\si$-symmetric sections from $\,\sfE^{(1,1)}
M\sqcup\sfE^{(1,0)}Q\,$ subject to the $\iota_\a$-alignment
condition \eqref{eq:sigmod-symm-cons}.
\eerop
\beroof
Fix an open cover $\,\cO_{\Igt_\si(\cB)}=\{\cO_{\igt^1}^*\x
\cO_{\igt^2}^*\}_{\xcI_{\Igt_\si(\cB)}}\,$ of $\,\Igt_\si(\cB)\,$ as
in the proof of Theorem \vref{thm:def-dual}I.4.9 and take the
associated data $\,\pr_\a^*(\theta_{\si,\emptyset\,\igt^\a},\g_{\si,
\emptyset\,\igt^\a\jgt^\a}),\ \a\in\{1,2\}\,$ of the pullbacks
$\,\pr_\a^*\ceL_{\si,\emptyset}\,$ of the pre-quantum bundle
$\,\ceL_{\si,\emptyset}\to\sfP_{\si,\emptyset}\,$ from the same
corollary, and those of the bundle isomorphism $\,\Dgt_\si(\cB)$,\
denoted by $\,f_{\si\,(\igt^1,\igt^2 )}\,$ and given in
\vReqref{eq:duality-iso-bib}{I.4.12}. The latter relate local
sections $\,\pr_\a^*s_{\igt^\a}:\cO_{\igt^1}^*\x\cO_{\igt^2}^*
\to\pr_\a^*\ceL_{\si,\emptyset}\,$ over $\,\Igt_\si(\cB)\ni(\psi_1,
\psi_2)\,$ as per
\qq\nn
s_{\igt^2}[\psi_2]=f_{\si,\cB\,(\igt^1,\igt^2)}[(\psi_1,\psi_2)]
\cdot s_{\igt^1}[\psi_1]\,.
\qqq
Take, next, a section $\,\Vgt=\xcV\oplus\upsilon\in\G_\si\bigl(
\sfE^{(1,1)}M\bigr)\,$ and consider the associated local pre-quantum
hamiltonians $\,\widehat h_{\widetilde\Vgt_\igt}\,$ from
\Reqref{eq:preq-ham-gen}, written out explicitly as
\qq\nn
\widehat h_{\widetilde\Vgt_\igt}=-\sfi\,\pLie{\widetilde\sfL_*
\xcV}-\widetilde\sfL_*\xcV\con\theta_{\si,\emptyset\,\igt}+h_\Vgt
\,,
\qqq
\textit{cf.}\ Eq.\,(I.3.8). Upon invoking continuity of the
hamiltonian function $\,h_\Vgt\,$ across the defect, demonstrated in
the proof of Proposition \ref{prop:ham-cont-across}, and using
relation \veqref{eq:f12-as-iso}{I.4.10}, we then find
\qq\nn
\widehat h_{\widetilde\Vgt_{\igt^2}}[\psi_2]\lact s_{\igt^2}[\psi_2
]&\equiv&\bigl(-\sfi\,\pLie{\widetilde\sfL_*\xcV[\psi_2]}
\vert_{\psi_1=\const}-\widetilde\sfL_*\xcV\con\theta_{\si,\emptyset
\,\igt^2}[\psi_2]+ h_\Vgt[\psi_2]\bigr)s_{\igt^2}[\psi_2]\cr\cr
&=&\bigl(-\sfi\,\pLie{\widetilde\sfL_*\xcV[\psi_2]}\vert_{\psi_1=
\const}-\sfi\,\pLie{\widetilde\sfL_*\xcV[\psi_1]}\vert_{\psi_2=
\const}-\widetilde\sfL_*\xcV\con\theta_{\si,\emptyset\,\igt^2}[
\psi_2]+h_\Vgt[\psi_2]\bigr)s_{\igt^2}[\psi_2]\cr\cr
&=&f_{\si,\cB\,(\igt^1,\igt^2)}[(\psi_1,\psi_2)]\cdot
\bigl(-\sfi\,\pLie{\widetilde\sfL_*\xcV[\psi_1]}\vert_{\psi_2=
\const}-\widetilde\sfL_*\xcV\con\theta_{\si,\emptyset\,\igt^2}[
\psi_2]+h_\Vgt[\psi_1]\cr\cr
&&-\bigl(\widetilde\sfL_*\xcV[\psi_2]\vert_{\psi_1=\const}+
\widetilde\sfL_*\xcV[\psi_1]\vert_{\psi_2=\const}\bigr)\con\sfi\,\d
\log f_{\si,\cB\,(\igt^1,\igt^2)}[(\psi_1,\psi_2)]\bigr)s_{\igt^1}[
\psi_1]\cr\cr
&=&f_{\si,\cB\,(\igt^1,\igt^2)}[(\psi_1,\psi_2)]\cdot\bigl(-\sfi\,
\pLie{\widetilde\sfL_*\xcV[\psi_1]}\vert_{\psi_2=\const}-\widetilde
\sfL_*\xcV\con\theta_{\si,\emptyset\,\igt^1}[\psi_1]+h_\Vgt[\psi_1]
\bigr)s_{\igt^1}[\psi_1]\cr\cr
&\equiv&f_{\si,\cB\,(\igt^1,\igt^2)}[(\psi_1,\psi_2)]\cdot\bigl(
\widehat h_{\widetilde\Vgt_{\igt^1}}[\psi_1]\lact s_{\igt^1}[\psi_1
]\bigr)\,,
\qqq
as claimed.\eroof\medskip

The present section rendered more precise the intuitively clear
assignment, to the geometric data carried by the defect, of
morphisms in the category of (twisted) Vinogradov structures on the
(twisted) generalised tangent bundle over the target space of the
background, the objects of the latter category being viewed as
target-space counterparts of the canonical Vinogradov structure on
the state space of the untwisted sector of the $\si$-model,
naturally associated with symmetries of that sector. It also
clarified the conditions under which symmetries of the theory are
mapped to one another across the defect on the level of the
corresponding hamiltonian functions and pre-quantum hamiltonians,
and -- in so doing -- pointed towards an extension of the previous
category that would accommodate the $\iota_\a$-aligned
$\si$-symmetric sections. The natural question as to the precise
nature of this extension becomes particularly pronounced when
discussing a realisation of the transmitted symmetries in the
twisted sector of the theory, which we examine closely in the next
section.

\section{Paired bracket structures and symmetries of the twisted
sector}\label{sec:ext-non-intersect}

The emergence of the distinguished gerbe bi-modules associated with
bi-branes follows a natural pattern of cohomological, or -- more
abstractly -- categorial descent, laid out in
\Rcite{Stevenson:2000wj} and further elaborated in
\Rcite{Fuchs:2009si} and similar in spirit to the one discussed in
Remark \vref{rem:duality-scheme}I.5.6, in which a lower-rank
cohomological structure arises from trivialisation of a
pullback-cohomology\footnote{\textit{Cf.}\ \Rcite{Murray:1994db}.}
coboundary obtained by pulling back a higher-rank structure to a
correspondence space along a number of smooth maps between the bases
of the geometric objects corresponding to the two structures. In the
process, the classifying cohomology for the lower-rank structure
inherits a twist, \textit{cf.}\ \vReqref{eq:DPhi-is}{I.2.4}, which
couples the two structures together. Drawing inspiration from the
intimate relationship between ($n$-)gerbes and bracket structures on
generalised tangent bundles, noted in
Ref.\,\cite{Hitchin:2004ut,Gualtieri:2003dx} and further elaborated
in the preceding sections, we propose to follow the same line of
reasoning in the algebraic setting of generalised geometry. In so
doing, we use the principle of compatibility with the symmetry
content of the two-dimensional $\si$-model as a natural measure of
naturalness of our constructions. We are thus led to the following
\bedef\label{def:pair-tw-bra-str}
Let $\,(M,Q)\,$ be a pair of smooth manifolds, equipped with a pair
of smooth maps $\,\iota_\a:Q\to M,\ \a\in\{1,2\}\,$ and a pair $\,(
\txH,\om)\in\Om^3(M)\x\Om^2(Q)\,$ of globally defined forms. Write
$\,\D_Q=\iota_2^*-\iota_1^*\,$ and assume
\qq\label{eq:dom-DelH}
\sfd\om+\D_Q\txH=0\,.
\qqq
Adopt the notation of Definitions \ref{def:Vin-str} and
\ref{def:Om-tw-Vin} and let $\,\pr_{\sfT^*M}:\sfE^{(1,1)}M\to
\sfT^*M\,$ and $\,\pr_{\sfT^*Q}:\sfE^{(1,0)}Q\to\sfT^*Q\,$ be the
canonical projections. The \textbf{$(\txH,\om;\D_Q)$-twisted bracket
structure on $\iota_\a$-paired generalised tangent bundles
$\,\sfE^{( 1,1)}M\sqcup\sfE^{(1,0)}Q\to M\sqcup Q\,$} is the
quadruple
\qq\nn
\bigl(\sfE^{(1,1)}M\sqcup\sfE^{(1,0)}Q,\GBra{\cdot}{\cdot}^{(\txH,
\om;\D_Q)},\Vcon{\cdot}{\cdot},\a_{\sfT(M\sqcup Q)}\bigr)=:\Mgt^{(1
,0),(\txH,\om;\D_Q)}(M\sqcup Q)
\qqq
in which $\,\Vcon{\cdot}{\cdot}\,$ and $\,\a_{\sfT(M\sqcup Q)}\,$
restrict to the respective canonical contractions and anchors on the
component generalised tangent bundles, and in which
$\,\GBra{\cdot}{\cdot}^{(\txH,\om;\D_Q)}\,$ is the antisymmetric
bilinear operation on smooth sections of $\,\sfE^{(1,1)}M\sqcup
\sfE^{(1,0)}Q\,$ that assigns to a pair $\,\Vgt,\Wgt\in\G\bigl(
\sfE^{(1,1)}M\sqcup\sfE^{( 1,0)}Q\bigr)\,$ of sections, with
restrictions $\,\Vgt\vert_\xcM=\xcMup\Vgt,\ \Wgt\vert_\xcM=\xcMup
\Wgt,\ \xcM\in\{M,Q\}$,\ another section with restrictions
\qq\nn
\GBra{\Vgt}{\Wgt}^{(\txH,\om;\D_Q)}\vert_M&=&\Vbra{\Mup\Vgt}{\Mup
\Wgt}^\txH\,,\cr\cr
\GBra{\Vgt}{\Wgt}^{(\txH,\om;\D_Q)}\vert_Q&=&\Vbra{\Qup\Vgt}{\Qup
\Wgt}^\om+0\oplus\tfrac{1}{2}\,\bigl(\a_{\sfT Q}(\Qup\Vgt)\con\D_Q
\pr_{\sfT^*M}(\Mup\Wgt)-\a_{\sfT Q}(\Qup\Wgt)\con\D_Q\pr_{\sfT^*M}(
\Mup\Vgt)\bigr)\,.
\qqq
Given two such structures, $\,\Mgt^{(1,0),(\txH_n,\om_n;\D_{Q_n})}(
M_n\sqcup Q_n),\ n\in\{1,2\}$,\ on the respective
$\iota_\a^n$-paired generalised tangent bundles $\,\sfE^{(1,1)}M_n
\sqcup\sfE^{(1,0)}Q_n\to M_n\sqcup Q_n$,\ a
\textbf{(factorised\footnote{We could, in principle, contemplate
more general mappings, mixing sections of the two pairs of
generalised tangent bundles involved.}) homomorphism between twisted
bracket structures on paired generalised tangent bundles} is a
quadruple $\, \bigl(f^{(1)},F^{(1,1)},f^{(0)},F^{(1,0)}\bigr)\,$
which consists of a pair of diffeomorphisms\footnote{Clearly, one
could relax the requirement that the base maps be diffeomorphisms,
whereupon a notion of a morphism of the two brackets would be
obtained. Here, we consider the more rigid structure with view to
the subsequent physical applications.}
\qq\nn
f^{(1)}\ :\ M_1\to M_2\,,\qquad\qquad f^{(0)}\ :\ Q_1\to Q_2
\qqq
compatible with the $\,\iota_\a^n\,$ in the sense expressed by the
commutative diagram
\qq\nn
\alxydim{@C=1.cm@R=1.cm}{Q_1 \ar[r]^{f^{(0)}} \ar[d]_{\iota_\a^1} &
Q_2 \ar[d]^{\iota_\a^2} \cr M_1 \ar[r]^{f^{(1)}} & M_2}\,,
\qqq
together with the vector-bundle maps
\qq\nn
F^{(1,1)}\ :\ \sfE^{(1,1)}M_1\to\sfE^{(1,1)}M_2\,,\qquad\qquad F^{(1
,0)}\ :\ \sfE^{(1,0)}Q_1\to\sfE^{(1,0)}Q_2
\qqq
that cover $\,f^{(1)}\,$ and $\,f^{(0)}$,\ respectively, in the
sense expressed by the commutative diagrams
\qq\nn
\alxydim{@C=1.cm@R=1.cm}{\sfE^{(1,1)}M_1 \ar[r]^{F^{(1,1)}}
\ar[d]_{\pi_{\sfT M_1}\circ\a_{\sfT M_1}} & \sfE^{(1,1)}M_2
\ar[d]^{\pi_{\sfT M_2}\circ\a_{\sfT M_2}} \cr M_1 \ar[r]^{f^{(1)}} &
M_2}\,,\qquad\qquad \alxydim{@C=1.cm@R=1.cm}{\sfE^{(1,0)}Q_1
\ar[r]^{F^{(1,0)}} \ar[d]_{\pi_{\sfT Q_1}\circ\a_{\sfT Q_1}} &
\sfE^{(1,0)}Q_2 \ar[d]^{\pi_{\sfT Q_2}\circ\a_{\sfT Q_2}}
\cr Q_1 \ar[r]^{f^{(0)}} & Q_2}\,,
\qqq
and such that the following identities hold true for $\,F^{(1,1
\sqcup 0)}=F^{(1,1)}\sqcup F^{(1,0)}$:
\qq
\GBra{\cdot}{\cdot}^{(\txH_2,\om_2;\D_{Q_2})}\circ(F^{(1,1\sqcup 0
)},F^{(1,1\sqcup 0)})&=&F^{(1,1\sqcup 0)}\circ\GBra{\cdot}{\cdot}^{(
\txH_1,\om_1;\D_{Q_1})}\,,\label{eq:pair-tw-bra-aut-1}\\\cr
\Vcon{\cdot}{\cdot}\circ(F^{(1,1\sqcup 0)},F^{(1,1\sqcup 0)})&=&
\bigl(\bigl(f^{(1)\,-1}\bigr)^*\sqcup\bigl(f^{(0)\,-1}\bigr)^*\bigr)
\circ\Vcon{\cdot}{\cdot}\,,\label{eq:pair-tw-bra-aut-2}\\\cr
\a_{\sfT(M_2\sqcup Q_2)}\circ F^{(1,1 \sqcup 0)}&=&\bigl(f^{(1)}_*
\sqcup f^{(0)}_*\bigr)\circ\a_{\sfT(M_1\sqcup Q_1)}\,.
\label{eq:pair-tw-bra-aut-3}
\qqq
\exdef \noindent In analogy with Proposition
\ref{prop:Vin-str-auts}, we readily prove
\berop\label{prop:aut-pair-tw-bra-str}
Adopt the notation of Definitions \ref{def:Vin-str} and
\ref{def:pair-tw-bra-str}, and suppose that $\,\bigl(f^{(1)},F^{(1,
1)},f^{(0)},F^{(1,0)}\bigr)\,$ is an automorphism of the $(\txH,\om;
\D_Q)$-twisted bracket structure on $\iota_\a$-paired generalised
tangent bundles $\,\sfE^{(1,1)}M\sqcup\sfE^{(1,0)}Q$.\ Then,
$\,\bigl(f^{(1)},F^{(1,1)},f^{(0)},F^{(1,0)}\bigr)\,$ necessarily
has the following properties:
\bit
\item[i)] the base maps $\,f^{(1)}\,$ and $\,f^{(0)}\,$ are
diffeomorphisms such that
\qq\label{eq:base-maps-aut-paired}
f^{(1)\,*}\txH-\txH=\sfd\txB^{(1)}\,,\qquad\qquad f^{(0)\,*}\om-\om
=\sfd\txB^{(0)}
\qqq
for some $\,\txB^{(1)}\in\Om^2(M)\,$ and $\,\txB^{(0)}\in\Om^1(Q
)$,\ of which the former is further constrained by the condition
\qq\label{eq:DQ-MupV-B1}
\D_Q\bigr(\Mup\xcV\con\txB^{(1)}\bigl)=0\,,
\qqq
to be satisfied for an arbitrary vector field $\,\Mup\xcV\,$ on
$\,M$;
\item[ii)] the bundle maps take the form
\qq\label{eq:aut-pair-tw-bra-str-bunmap}
F^{(1,1)}\sqcup F^{(1,0)}=\left(\widehat f^{(1)}\circ\ee^{\txB^{(1
)}}\right)\sqcup\left(\widehat f^{(0)}\circ\ee^{\txB^{(0)}}\right)
\,.
\qqq
\eit
Upon restriction to the subspace $\,\G_{\iota_\a}\bigl(\sfE^{(1,1)}M
\sqcup\sfE^{(1,0)}Q\bigr)\subset\G\bigl(\sfE^{(1,1)}M\sqcup\sfE^{(1,
0)}Q\bigr)\,$ composed of those sections, to be termed
\textbf{$\iota_\a$-aligned}, which satisfy the additional condition
\qq\label{eq:iota-pairing}
\iota_{\a\,*}\circ\a_{\sfT Q}=\a_{\sfT M}\vert_{\iota_\a(Q)}\,,
\qqq
the set of automorphisms extends to include those with base maps
constrained as in the first of
Eqs.\,\eqref{eq:base-maps-aut-paired}, and with bundle maps as in
\Reqref{eq:aut-pair-tw-bra-str-bunmap} but now written for forms
$\,\txB^{(1)}\,$ and $\,\txB^{(0)}\,$ subject to the constraint
\qq\label{eq:aut-pair-tw-bra-str-paired}
\sfd\txB^{(0)}=f^{(0)\,*}\om-\om+\D_Q\txB^{(1)}\,.
\qqq
\eerop
\beroof The proof goes along similar lines as that of Proposition
\ref{prop:dual-vs-morf}, which is also how the form of the bundle
map $\,F^{(1,1)}\,$ is established. Only now one considers an
automorphism $\,G:=\widehat{f^{(1)}}^{-1}\circ F^{(1,1)}\,$ of
$\,\sfE^{(1,1)}M\,$ satisfying the analogon of relation
\eqref{eq:G-from-Fdef}.

The sole non-trivial statement that has to be verified is the one
concerning the explicit form of the bundle map $\,F^{(1,0)}$.\ We
begin by noting that condition \eqref{eq:pair-tw-bra-aut-3} fixes
the map in the form
\qq\nn
F^{(1,0)}=\left( \barr{cc} \id_{\sfT Q} & 0 \cr\cr \txB^{(0)} & C
\earr \right)
\qqq
for some $\,\txB^{(0)}\in\G(\sfT^*Q)\,$ and $\,C\in C^\infty(Q,\bR
)$.\ Due to the triviality of condition
\eqref{eq:pair-tw-bra-aut-2}, we are left with condition
\eqref{eq:pair-tw-bra-aut-1} to be imposed. Take arbitrary sections
$\,\Vgt,\Wgt\in\G\bigl(\sfE^{(1,1)} M\sqcup\sfE^{(1,0)}Q\bigr)\,$
with restrictions $\,(\Vgt,\Wgt)\vert_M=(\Mup\xcV\oplus\upsilon,
\Mup\xcW\oplus\varpi)\,$ and $\,(\Vgt,\Wgt)\vert_Q=(\Qup\xcV\oplus
\xi,\Qup\xcW\oplus\z)$.\ The condition now boils down to the
identity
\qq\nn
&&[\Qup\xcV,\Qup\xcW]\con\txB^{(0)}+C\cdot\bigl(\Qup\xcV\con\sfd\z-
\Qup\xcW\con\sfd\xi+\Qup\xcV\con\Qup\xcW\con\om+\tfrac{1}{2}\,
\bigl(\Qup\xcV\con\D_Q\varpi-\Qup\xcW\con\D_Q\upsilon\bigr)\bigr)
\cr\cr
&=&\Qup\xcV\con\sfd\bigl(\Qup\xcW\con\txB^{(0)}+C\cdot\z\bigr)-\Qup
\xcW\con\sfd\bigl(\Qup\xcV\con\txB^{(0)}+C\cdot\xi\bigr)+\Qup\xcV
\con\Qup\xcW\con f^{(0)\,*}\om\cr\cr
&&+\tfrac{1}{2}\,\bigl(\Qup\xcV\con\D_Q\bigl(\varpi+\Mup\xcW\con
\txB^{(1)}\bigr)-\Qup\xcW\con\D_Q\bigl(\upsilon+\Mup\xcV\con\txB^{(
1)}\bigr)\bigr)\,,
\qqq
or -- after obvious cancellations --
\qq\nn
\Qup\xcW\con\Qup\xcV\con\bigl(\sfd\txB^{(0)}+(C-f^{(0)\,*})\,\om
\bigr)+(\z\,\Qup\xcV-\xi\,\Qup\xcW)\con\sfd C\cr\cr
=\tfrac{1}{2}\,\bigl[\Qup\xcW\con\bigl(\D_Q\bigl(\upsilon+\Mup\xcV
\con\txB^{(1)}\bigr)-C\,\D_Q\upsilon\bigr)-\Qup\xcV\con\bigl(\D_Q
\bigl(\varpi+\Mup\xcW\con\txB^{(1)}\bigr)-C\,\D_Q\varpi\bigr)\bigr]
\,.
\qqq
On setting $\,\Qup\xcW=-\Qup\xcV,\ \Mup\xcW=-\Mup\xcV\,$ and
$\,\varpi=- \upsilon$,\ the above simplifies as
\qq\nn
(\xi+\z)\,\Qup\xcV\con\sfd C=0\,,
\qqq
whence $\,C\in\bR$.\ Keeping the same relation between the vector
components but letting $\,\upsilon\,$ and $\,\varpi\,$ vary
independently, we fix the value of the constant as $\,C=1\,$ (for
$\,\iota_1\not\equiv\iota_2$,\ which we assume). This leaves us with
the condition
\qq\label{eq:penult}
\Qup\xcW\con\Qup\xcV\con\left(\sfd\txB^{(0)}+(1-f^{(0)\,*})\om
\right)+\tfrac{1}{2}\,\bigl(\Qup\xcV\con\D_Q\bigl(\Mup\xcW\con
\txB^{(1)}\bigr)-\Qup\xcW\con\D_Q\bigl(\Mup\xcV\con\txB^{(1)}\bigr)
\bigr)=0\,.
\qqq
Up to now, the special choices made along the way were always
consistent with the additional constraint \eqref{eq:iota-pairing},
and so differentiation between generic automorphisms and the
extended ones for the restricted bracket structure starts at this
point.

In order to ultimately constrain the former, set $\,\Mup\xcV=0=\Mup
\xcW\,$ to obtain the second of
Eqs.\,\eqref{eq:base-maps-aut-paired}. The ensuing constraint
\qq\nn
\Qup\xcV\con\D_Q\bigl(\Mup\xcW\con\txB^{(1)}\bigr)-\Qup\xcW\con\D_Q
\bigl(\Mup\xcV\con\txB^{(1)}\bigr)=0
\qqq
is then tantamount to \Reqref{eq:DQ-MupV-B1}, which proves the first
part of the proposition.

As for $\iota_\a$-aligned sections, note, first of all, that the
restriction makes sense as
\qq\nn
\a_{\sfT M}\bigl(\GBra{\Vgt}{\Wgt}^{(\txH,\om;\D_Q)}\bigr)
\vert_{\iota_\a(Q)}&=&[\a_{\sfT M}(\Vgt),\a_{\sfT M}(\Wgt)]
\vert_{\iota_\a(Q)}=[\iota_{\a\,*}\circ\a_{\sfT Q}(\Vgt),\iota_{\a
\,*}\circ\a_{\sfT Q}(\Wgt)]\cr\cr
&=&\iota_{\a\,*}\circ[\a_{\sfT Q}(\Vgt),\a_{\sfT
Q}(\Wgt)]=\iota_{\a\,*}\circ\a_{\sfT
Q}\bigl(\GBra{\Vgt}{\Wgt}^{(\txH,\om ;\D_Q)}\bigr)\,.
\qqq
Upon restriction, \Reqref{eq:penult} rewrites as
\qq\nn
\Qup\xcV\con\Qup\xcW\con\left(\sfd\txB^{(0)}+(1-f^{(0)\,*})\om-\D_Q
\txB^{(1)}\right)=0\,,
\qqq
and so \Reqref{eq:aut-pair-tw-bra-str-paired} is reproduced. This
completes the proof of the proposition.\eroof\medskip

The constraint \eqref{eq:iota-pairing} is completely natural in the
physical context of our analysis as it is directly built into the
structure of the $\si$-model for world-sheets with an embedded
defect quiver, \textit{cf.}\ \Reqref{eq:iota-align}. That it is also
distinguished from a purely geometric point of view is shown in the
following
\berop\label{prop:pair-tw-gen-tan-bun}
Adopt the notation of Definitions \ref{def:Vin-str} and
\ref{def:pair-tw-bra-str}, and of Theorem \ref{thm:bib-as-morph}.
Choose open covers $\,\Mup\cO=\{\Mup \cO_i\}_{i\in\xcI_M}\,$ and
$\,\Qup\cO=\{\Qup\cO_a\}_{a\in\xcI_Q}\,$ such that there exist \v
Cech extensions $\,\check\iota_\a=(\iota_\a ,\phi_\a)\,$ of the
$\cG$-bi-brane maps as in Definition I.2.2. Let $\,\sfE^{(1,1)}_\cG
M\to M\,$ be the $\cG$-twisted generalised tangent bundle associated
with $\,\Mup\cO$,\ and let $\,\sfE^{(1,0 )}_\cB
Q:=\sfE^{(1,0)}_{\{\ggt_{ab}\}}Q\to Q\,$ be the generalised tangent
bundle twisted by a local presentation $\,(P_a,K_{ab})\in
\cA^{2,1}(\Qup\cO)\,$ of $\,\Phi\,$ associated, in the manner
specified in Definition \vref{def:loco}I.2.2, with $\,\Qup\cO$,\
with the twist determined by the transition maps
\qq\nn
\ggt_{ab}=\ee^{(P_a-P_b)\vert_{\Qup\cO_{ab}}}\,.
\qqq
Write
\qq\nn
\check\D_Q:=\check\iota_2^*-\check\iota_1^*\,.
\qqq
A \textbf{global (twisted) bracket structure
\qq\nn
\Mgt^{(1,0),(\cdot,\cdot;\D_Q)}_{(\cG,\cB)}(M\sqcup Q)=\bigl(\sfE^{(
1,1)}_\cG M\sqcup\sfE^{(1,0)}_\cB Q,\GBra{\cdot}{\cdot}^{(\cdot,
\cdot;\check\D_Q)},\Vcon{\cdot}{\cdot},\a_{\sfT(M\sqcup Q)}\bigr)
\qqq
on $(\cG,\cB)$-twisted $\iota_\a$-paired generalised tangent bundles
$\,\sfE^{(1,1)}_\cG M\sqcup\sfE^{(1,0)}_\cB Q\,$} (understood in
analogy with the global Vinogradov structure of Definition
\ref{def:loc-glob-Vin-str}) exists, in general, exclusively on the
subspace of $\iota_\a$-aligned sections
$\,\G_{\iota_\a}\bigl(\sfE^{(1,1)}_\cG M\sqcup\sfE^{(1,0)}_\cB Q
\bigr)\subset\G\bigl(\sfE^{(1,1)}_\cG M\sqcup \sfE^{(1,0)}_\cB Q
\bigr)$.\ The restricted bracket structure
$\,\Mgt^{(1,0),(0,0;\check\D_Q)}(M \sqcup
Q)\vert_{\G_{\iota_\a}\bigl(\sfE^{(1,1)}_\cG M\sqcup\sfE^{(1
,0)}_\cB Q\bigr)}=:\Mgt^{(1,0),(0,0;\check\D_Q)}_{(\cG,\cB),
\iota_\a}(M\sqcup Q)\,$ is homomorphic with the restricted $(\txH,
\om;\D_Q)$-twisted bracket structure $\,\Mgt^{(1,0),(\txH,\om;\D_Q
)}(M\sqcup Q)\vert_{\G_{\iota_\a}\bigl(\sfE^{(1,1)}M\sqcup\sfE^{(1,
0)}Q \bigr)}=:\Mgt^{(1,0),(\txH,\om;\D_Q)}_{\iota_\a}(M\sqcup Q)$,\
and the homomorphism
\qq\nn
{}^{\tx{\tiny $M\sqcup Q$}}\hspace{-2pt}\chi\ :\ \Mgt^{(1,0),(0,0;
\check\D_Q)}_{(\cG,\cB),\iota_\a}(M\sqcup Q)\to\Mgt^{(1,0),(\txH,\om
;\D_Q)}_{\iota_\a}(M\sqcup Q)
\qqq
restricts as
\qq\nn
{}^{\tx{\tiny $M\sqcup Q$}}\hspace{-2pt}\chi\ :\ \sfE^{(1,1)}_\cG M
\xrightarrow{\cong}\sfE^{(1,1)}M\,,\qquad\qquad{}^{\tx{\tiny
$M\sqcup Q$}}\hspace{-2pt}\chi\ :\ \sfE^{(1,0 )}_\cB Q
\xrightarrow{\cong}\sfE^{(1,0)}Q
\qqq
with local data
\qq\nn
{}^{\tx{\tiny $M\sqcup Q$}}\hspace{-2pt}\chi\vert_{\Mup\cO_i}=
\ee^{B_i}\,,\qquad\qquad{}^{\tx{\tiny $M\sqcup Q$}}\hspace{-2pt}
\chi\vert_{\Qup\cO_a}=\ee^{P_a}
\qqq
determined by a local presentation of $\,\cB\,$ as above and that of
the gerbe, $\,(B_i,A_{ij},g_{ijk})\in\cA^{3,2}(\Mup\cO)$.
\eerop
\beroof
In virtue of Corollary \ref{cor:tw-gen-tan-vs-ngerb}, and due to the
triviality of the canonical contraction on $\,\sfE^{(1,0)}Q$,\ the
proof of the existence of a global (twisted) bracket structure on
$\,\sfE^{(1,1)}_\cG M\sqcup\sfE^{(1,0)}_\cB Q\,$ reduces to checking
the required properties of the bracket $\,\GBra{\cdot}{\cdot}^{(
\cdot,\cdot;\check\D_Q)}\,$ restricted to $\,Q$.\ Choose open covers
$\,\Mup\cO\,$ and $\,\Qup\cO\,$ as described, and take the
associated local presentation of $\,\Bgt$.\ Given a pair $\,\Vgt,
\Wgt\,$ of sections of $\,\sfE^{(1,1)}_\cG M\sqcup\sfE^{(1,0)}_\cB
Q$,\ with restrictions $\,(\Vgt,\Wgt)\vert_{\Mup\cO_i}=(\Mup\xcV
\oplus\upsilon_i,\Mup\xcW\oplus \varpi_i)\,$ and $\,(\Vgt,\Wgt)
\vert_{\Qup\cO_a}=(\Qup\xcV\oplus\xi_a,\Qup\xcW\oplus\z_a)$,\ we
readily compute, using \vReqref{eq:DPhi-is}{I.2.4} and for
$\,\GBra{\Vgt}{\Wgt}^{(\cdot,\cdot;\check\D_Q)}_a=
\GBra{\Vgt}{\Wgt}^{(\cdot,\cdot;\check\D_Q)}\vert_{\Qup\cO_a}$,
\qq\nn
\bigl(\GBra{\Vgt}{\Wgt}^{(\cdot,\cdot;\check\D_Q)}_b-
\GBra{\Vgt}{\Wgt}^{(\cdot,\cdot;\check\D_Q)}_a\bigr)_{\Qup\cO_{ab}}=0
\oplus\D_{ab}
\qqq
with
\qq\nn
\D_{ab}&=&\bigl[\Qup\xcV\con\sfd(\z_b-\z_a)-\Qup\xcW\con\sfd(\xi_b-
\xi_a)+\tfrac{1}{2}\,\Qup\xcV\con\bigl(\iota_2^*(\varpi_{\phi_2(b)}
-\varpi_{\phi_2(a)})-\iota_1^*(\varpi_{\phi_1(b)}-\varpi_{\phi_1(i
)})\bigr)\cr\cr
&&-\tfrac{1}{2}\,\Qup\xcW\con\bigl(\iota_2^*(\upsilon_{\phi_2(b)}-
\upsilon_{\phi_2(a)})-\iota_1^*(\upsilon_{\phi_1(b)}-
\upsilon_{\phi_1(a)})\bigr)\bigr]\vert_{\Qup\cO_{ab}}\cr\cr
&=&\bigl\{\Qup\xcV\con\sfd\bigl(\Qup\xcW\con(P_a-P_b)\bigr)-\Qup
\xcW\con\sfd\bigl(\Qup\xcV\con(P_a-P_b)\bigr)\cr\cr
&&+\tfrac{1}{2}\,\Qup\xcV\con\bigl[\iota_2^*\bigl(\Mup\xcW\con(
B_{\phi_2(a)}-B_{\phi_2(b)})\bigr)-\iota_1^*\bigl(\Mup\xcW\con(
B_{\phi_1(a)}-B_{\phi_1(b)})\bigr)\bigr]\cr\cr
&&-\tfrac{1}{2}\,\Qup\xcW\con\bigl[\iota_2^*\bigl(\Mup\xcV\con(
B_{\phi_2(a)}-B_{\phi_2(b)})\bigr)-\iota_1^*\bigl(\Mup\xcV\con(
B_{\phi_1(a)}-B_{\phi_1(b)})\bigr)\bigr]\bigr\}\vert_{\Qup\cO_{ab}}
\cr\cr
&=&[\Qup\xcV,\Qup\xcW]\con(P_a-P_b)\vert_{\Qup\cO_{ab}}+\Qup
\xcV\con\Qup\xcW\con\bigl(\iota_2^*\sfd A_{\phi_2(a)\phi_2(b)}-
\iota_1^*\sfd A_{\phi_1(a)\phi_1(b)}\bigr)\cr\cr
&&-\tfrac{1}{2}\,\bigl\{\Qup\xcV\con\bigl[\iota_2^*\bigl(\Mup\xcW
\con\sfd A_{\phi_2(a)\phi_2(b)}\bigr)-\iota_1^*\bigl(\Mup\xcW\con
\sfd A_{\phi_1(a)\phi_1(b)}\bigr)\bigr]\cr\cr
&&-\Qup\xcW\con\bigl[\iota_2^*\bigl(\Mup\xcV\con\sfd A_{\phi_2(a)
\phi_2(b)}\bigr)-\iota_1^*\bigl(\Mup\xcV\con\sfd A_{\phi_1(a)\phi_1
(b)}\bigr)\bigr]\bigr\}\,.
\qqq
The first term in the above expression has the desired form, and it
is immediately clear that the condition for the other terms to
cancel out (generically) coincides with the defining relation
\eqref{eq:iota-pairing}.

Passing to the second statement of the proposition, we see once more
that it remains to prove it for the bracket restricted to $\,Q$.\
Thus, we have to show, for any two sections $\,\Vgt,\Wgt\,$ of
$\,\sfE^{(1,1)}M\sqcup\sfE^{(1,0)}Q$,\ with restrictions
$\,(\Vgt,\Wgt)\vert_\xcM=(\xcMup\Vgt,\xcMup\Wgt),\ \xcM\in\{M,Q\}$,\
the equality
\qq\nn
\Vbra{\ee^{-P_a}\lact\Qup\Vgt}{\ee^{-P_a}\lact\Qup\Wgt}+0\oplus
\tfrac{1}{2}\,\bigl[\a_{\sfT Q}\bigl(\ee^{-P_a}\lact\Qup\Vgt\bigr)
\con\bigl(\iota_2^*\pr_{\sfT^*M}\bigl(\ee^{-B_{\phi_2(a)}}\lact\Mup
\Wgt\bigr)-\iota_1^*\pr_{\sfT^*M}\bigl(\ee^{-B_{\phi_1(a)}}\lact\Mup
\Wgt\bigr)\bigr)\cr\cr -\a_{\sfT Q}\bigl(\ee^{-P_a}\lact\Qup\Wgt
\bigr)\con\bigl(\iota_2^*\pr_{\sfT^*M}\bigl(\ee^{-B_{\phi_2(a)}}
\lact\Mup\Vgt\bigr)-\iota_1^*\pr_{\sfT^*M}\bigl(\ee^{-B_{\phi_1(a)}}
\lact\Mup\Vgt\bigr)\bigr)\bigr]\cr\cr =\ee^{-P_a}\lact\bigl[
\Vbra{\Qup\Vgt}{\Qup\Wgt}^\om+0\oplus\tfrac{1}{2}\,\bigl(\a_{\sfT Q}
\bigl(\Qup\Vgt\bigr)\con\D_Q\pr_{\sfT^*M}\bigl(\Mup\Wgt\bigr)-
\a_{\sfT Q}\bigl(\Qup\Wgt\bigr)\con\D_Q\pr_{\sfT^*M}\bigl(\Mup\Vgt
\bigr)\bigr)\bigr]\,.
\qqq
It is verified through a straightforward calculation employing
exactly the same arguments as those invoked in the proof of the
first part of the proposition.\eroof \brem It is perhaps worth
emphasising at this stage that the very definition of the
$(\cG,\cB)$-twisted $\iota_\a$-paired generalised tangent bundles
$\,\sfE^{(1,1)}_\cG M\sqcup\sfE^{(1,0 )}_\cB Q\,$ ensures the
existence of an isomorphism between any two such bundles twisted by
(gauge-)equivalent local presentations of $\,\Bgt$,\ and this
property is inherited by the global (twisted) bracket structure
under the restriction.\erem

The distinguished $\iota_\a$-aligned sections reappear in the
algebraic description of symmetries of the $\si$-model, which we
give in
\berop\label{prop:sisym-iotalign}
Adopt the notation of Definitions \ref{def:Vin-str} and
\ref{def:pair-tw-bra-str}, of Theorem \ref{thm:bib-as-morph}, and of
Propositions \ref{prop:sigmod-symm-def} and
\ref{prop:pair-tw-gen-tan-bun}. Infinitesimal rigid symmetries of
the two-dimensional non-linear $\si$-model for network-field
configurations $\,(X\,\vert \,\G)\,$ in string background $\,\Bgt\,$
on world-sheet $\,(\Si,\g )\,$ with a defect quiver $\,\G\,$
composed of a finite number of non-intersecting circular defect
lines, as described in Definition \vref{def:sigmod}I.2.7, correspond
to those $\iota_\a$-aligned sections of
$\,\sfE^{(1,1)}M\sqcup\sfE^{(1,0)}Q\,$ which satisfy conditions
\eqref{eq:sigmod-symm-bulk} and \eqref{eq:sigmod-symm-def}. We shall
call them \textbf{$\si$-symmetric $\iota_\a$-aligned sections of
$\,\sfE^{(1,1)}M\sqcup\sfE^{(1,0)}Q$},\ and denote the corresponding
subset in $\,\G\bigl(\sfE^{(1,1)}M\sqcup\sfE^{(1,0)}Q \bigr)\,$ as
$\,\G_{\iota_\a,\si}\bigl(\sfE^{(1,1)}M\sqcup\sfE^{(1,0 )}Q\bigr)$.\
The bracket $\,\GBra{\cdot}{\cdot}^{(\txH,\om;\D_Q)}\,$ closes on
$\,\G_{\iota_\a,\si}\bigl(\sfE^{(1,1)}M\sqcup\sfE^{(1,0)}Q \bigr)$,
\qq\nn
\Vgt,\Wgt\in\G_{\iota_\a,\si}\bigl(\sfE^{(1,1)}M\sqcup\sfE^{(1,0)}Q
\bigr)\qquad\Longrightarrow\qquad\GBra{\Vgt}{\Wgt}^{(\txH,\om;\D_Q)}
\in\G_{\iota_\a,\si}\bigl(\sfE^{(1,1)}M\sqcup\sfE^{(1,0)}Q\bigr)\,,
\qqq
and every bracket with this property differs from
$\,\GBra{\cdot}{\cdot}^{(\txH,\om;\D_Q)}\,$ by a linear map on
$\,\G_{\iota_\a,\si}\bigl(\sfE^{(1,1)}M\sqcup\sfE^{(1,0)}Q\bigr)
\wedge\G_{\iota_\a,\si}\bigl(\sfE^{(1,1)}M\sqcup\sfE^{(1,0)}Q
\bigr)\,$ with values given by pairs $\,(\txB_1,\txB_0)\in\Om^1(M)
\x C^\infty(Q,\bR)\,$ subject to the constraints
\qq\label{eq:corr-Vin}
\sfd\txB_1=0\,,\qquad\qquad\sfd\txB_0+\D_Q\txB_1=0\,.
\qqq
Equivalently, the symmetries can be represented by $\si$-symmetric
$\iota_\a$-aligned sections $\,\Vgt\in\G_{\iota_\a,\si}\bigl(
\sfE^{(1,1)}_\cG M\sqcup\sfE^{(1,0)}_\cB Q \bigr)\,$ of $\,\sfE^{(1
,1)}_\cG M\sqcup\sfE^{(1,0)}_\cB Q$,\ with restrictions $\,\Vgt
\vert_{\xcMup\cO_i}=\xcMup\Vgt_i,\ i\in\xcI_\xcM,\ \xcM\in\{M,Q\}$,
\qq\nn
\pLie{\a_{\sfT M}(\Mup\Vgt_i)}\txg=0\,,\qquad\qquad\left\{ \barr{l}
\sfd\pr_{\sfT^*M}(\Mup\Vgt_i)+\pLie{\a_{\sfT M}(\Mup\Vgt_i)}B_i=0
\cr\cr \sfd\pr_{\sfT^*Q}(\Qup\Vgt_a)+\pLie{\a_{\sfT Q}(\Qup\Vgt_a)}
P_a=-\check\D_Q\pr_{\sfT^*M}(\Mup\Vgt_\bullet)_a\earr\right.\,.
\qqq
The bracket $\,\GBra{\cdot}{\cdot}^{(0,0;\check\D_Q)}\,$ closes on
$\,\G_{\iota_\a,\si}\bigl(\sfE^{(1,1)}_\cG M\sqcup\sfE^{(1,0)}_\cB Q
\bigr)$,
\qq\nn
\Vgt,\Wgt\in\G_{\iota_\a,\si}\bigl(\sfE^{(1,1)}_\cG M\sqcup\sfE^{(1,
0)}_\cB Q\bigr)\qquad\Longrightarrow\qquad\GBra{\Vgt}{\Wgt}^{(0,0;
\check\D_Q)}\in\G_{\iota_\a,\si}\bigl(\sfE^{(1,1)}_\cG M\sqcup
\sfE^{(1,0)}_\cB Q\bigr)\,,
\qqq
and every bracket with this property differs from
$\,\GBra{\cdot}{\cdot}^{(0,0;\check\D_Q)}\,$ by a linear map on
$\,\G_{\iota_\a,\si}\bigl(\sfE^{(1,1)}_\cG M\sqcup\sfE^{(1,0)}_\cB Q
\bigr)\wedge\G_{\iota_\a,\si}\bigl(\sfE^{(1,1)}_\cG M\sqcup\sfE^{(1,
0)}_\cB Q\bigr)\,$ with local values $\,(\txB_{1,i},\txB_{0,a})\in
\Om^1(\Mup\cO_i)\x C^\infty(\Qup\cO_a,\bR)\,$ constrained as in
\Reqref{eq:corr-Vin}.
\eerop
\beroof
The correspondence between symmetries and $\si$-symmetric
$\iota_\a$-aligned sections of $\,\sfE^{(1,1)}M\sqcup\sfE^{(1,0)}Q
\,$ was established in Proposition \ref{prop:sigmod-symm-def}. In
the light of Corollary \ref{cor:sigmod-symm-E11}, it remains to
verify the relation
\qq\nn
\sfd_\om\GBra{\Vgt}{\Wgt}^{(\txH,\om;\D_Q)}\vert_Q+\D_Q\pr_{\sfT^*
M}\bigl(\GBra{\Vgt}{\Wgt}^{(\txH,\om;\D_Q)}\vert_M\bigr)=0\,.
\qqq
Write $\,(\Mup\Vgt,\Mup\Wgt)=(\Mup\xcV\oplus\upsilon,\Mup\xcW\oplus
\varpi)\,$ and$\,(\Qup\Vgt,\Qup\Wgt)=(\Qup\xcV\oplus\xi,\Qup\xcW
\oplus\z)$.\ Using condition \eqref{eq:iota-align} alongside the
assumption $\,\Vgt,\Wgt\in\G_{\iota_\a,\si}\bigl(\sfE^{(1,1)}M
\sqcup\sfE^{(1,0)}Q\bigr)$,\ we obtain
\qq\nn
&&\sfd\bigl(\Qup\xcV\con\sfd\z-\Qup\xcW\con\sfd\xi+\Qup\xcV\con\Qup
\xcW\con\om+\tfrac{1}{2}\,\bigl(\Qup\xcV\con\D_Q\varpi-\Qup\xcW\con
\D_Q\upsilon\bigr)\bigr)+[\Qup\xcV,\Qup\xcW]\con\om\cr\cr
&&+\D_Q\bigl(\pLie{\Mup\xcV}\varpi-\pLie{\Mup\xcW}\upsilon-
\tfrac{1}{2}\,\sfd\bigl(\Mup\xcV\con\varpi-\Mup\xcW\con\upsilon
\bigr)+\Mup\xcV\con\Mup\xcW\con\txH\bigr)\cr\cr
&=&-\pLie{\Qup\xcV}\bigl(\Qup\xcW\con\om\bigr)+\pLie{\Qup\xcW}
\bigl(\Qup\xcV\con\om\bigr)+\sfd\bigl(\Qup\xcV\con\Qup\xcW\con\om
\bigr)+[\Qup\xcV,\Qup\xcW]\con\om-\Qup\xcV\con\Qup\xcW\con
\sfd\om=0\,,
\qqq
as claimed. The uniqueness of the bracket up to a bilinear map on
$\,\G_{\iota_\a,\si}\bigl(\sfE^{(1,1)}M\sqcup\sfE^{(1,0)}Q\bigr)
\wedge\G_{\iota_\a,\si}\bigl(\sfE^{(1,1)}M\sqcup\sfE^{(1,0)}Q
\bigr)\,$ with values described in the thesis of the proposition is
obvious. Finally, the relations defining $\si$-symmetric
$\iota_\a$-aligned sections of $\,\sfE^{(1,1)}_\cG
M\sqcup\sfE^{(1,0)}_\cB Q\,$ rephrase those defining $\si$-symmetric
$\iota_\a$-aligned sections of $\,\sfE^{(1
,1)}M\sqcup\sfE^{(1,0)}Q$,\ and the closure of the corresponding
bracket follows from Proposition
\ref{prop:pair-tw-gen-tan-bun}.\eroof
\medskip

The reconstruction of the algebraic structure present on the set of
those symmetries of the untwisted sector of the $\si$-model that are
transmitted across a conformal defect is an obvious prerequisite for
understanding their symplectic realisation on the \emph{twisted}
sector of the theory, with the twist determined by the geometric
data carried by the defect. In order to attain this goal, we should
first lift the structures obtained hitherto on the target space and
the bi-brane world-volume to the twisted loop spaces
$\,\sfL_{Q\vert\{(P_k,\vep_k)\}}\,$ and to the respective cotangent
bundles. For the sake of concretness and brevity, we restrict our
analysis to the 1-twisted case.
\bedef\label{def:tw-loop-sp-lifts}
Let $\,(M,Q)\,$ be a pair of smooth manifolds, equipped with a pair
of smooth maps $\,\iota_\a:Q\to M,\ \a\in\{1,2\}$,\ and let
$\,\sfL_{Q\vert(\pi,\vep)}M\,$ be the 1-twisted loop space with
coordinates $\,(X,q)$,\ as introduced in Definition
\vref{def:tw-phspace}I.3.10. Write $\,\bS^1_\pi:=\bS^1\setminus\{
\pi\}\,$ for $\,\pi\in\bS^1$,\ and denote by $\,\ev_{M,\pi}\ :\
\sfL_{Q\vert(\pi,\vep)}M\x\bS^1_\pi\to M\,$ the canonical evaluation
map. A pair $\,(\Mup\xcV,\Qup\xcV)\in\G(\sfT M)\x\G(\sfT Q)\,$ of
vector fields will be called \textbf{$\iota_\a$-aligned} iff
\qq\nn
\iota_{\a*}\Qup\xcV=\Mup\xcV\vert_{\iota_\a(Q)}\,,
\qqq
and the corresponding subset in $\,\G(\sfT M)\x\G(\sfT Q)\,$ will be
denoted as $\,\G_{\iota_\a}(\sfT M\sqcup\sfT Q)$.\ The (local) flow
$\,\xi_t=\Mup\xi_t\sqcup\Qup\xi_t:\xcM\sqcup Q\to\xcM\sqcup Q\,$ of
a $\iota_\a$-aligned pair $\,(\Mup\xcV,\Qup\xcV)\,$ (assumed to
exist) satisfies the condition
\qq\nn
\iota_\a\circ\Qup\xi_t=\Mup\xi_t\vert_{\iota_\a(Q)}\,.
\qqq
The \textbf{1-twisted loop-space lift of $\iota_\a$-aligned pair of
vector fields on $\,M\sqcup Q\,$} is a linear map
\qq
\sfL^{Q\vert(\pi,\vep)}_{\iota_\a\,*}\ :\ \G_{\iota_\a}(\sfT M\sqcup
\sfT Q)\to\G(\sfT\sfL_{Q\vert(\pi,\vep)}M)\ :\ (\Mup\xcV,\Qup\xcV)
\mapsto\bigl(\sfL_*^\pi\Mup\xcV,\Qup\xcV\circ\pr_Q\bigr)=:\sfL^{Q
\vert(\pi,\vep)}_*(\Mup\xcV,\Qup\xcV)\,,\cr\label{eq:Liota-vec}
\qqq
written in terms of a loop-space lift $\,\sfL^\pi_*\,$ determined
just as the loop-space lift $\,\sfL_*\,$ in Definition
\ref{def:LM-lifts} (\textit{i.e.}\ through action on functionals of
1-twisted loops) and of the canonical projection $\,\pr_Q\ :\
\sfL_{Q\vert(\pi ,\vep)}M\to Q$,\ so that
\qq\nn
\Qup\xcV\circ\pr_Q[(X,q)]=\Qup\xcV(q)\,.
\qqq
The \textbf{1-twisted loop-space lift of $n$-form from $\,M\,$} is a
linear map
\qq\nn
\sfL^{Q\vert(\pi,\vep)\,*}\ :\ \Om^n(M)\to\Om^{n-1}(\sfL_{Q\vert(\pi
,\vep)}M)\ :\ \upsilon\mapsto\int_{\bS^1_\pi}\,\ev_{M,\pi}^*
\upsilon=:\sfL^{Q\vert(\pi,\vep)\,*}\upsilon\,,\quad n\in\bN_{>0}\,,
\qqq
extended to the case of $\,n=0\,$ as per
\qq\nn
\sfL^{Q\vert(\pi,\vep)\,*}\ :\ C^\infty(M,\bR)\to\{0\}\ :\ f\mapsto
0\,.
\qqq
Similarly, the \textbf{1-twisted loop-space lift of $n$-form from
$\,Q\,$} is a linear map
\qq\nn
\unl\sfL^{Q\vert(\pi,\vep)\,*}\ :\ \Om^n(Q)\to\Om^n(\sfL_{Q\vert(
\pi,\vep)}M)\ :\ \upsilon\mapsto\pr_Q^*\upsilon\,,\quad n\in\bN\,.
\qqq
The lifts thus defined can, in turn, be combined into a lift
\qq\nn
\sfL^{Q\vert(\pi,\vep)}_{(1,1\sqcup 0)}\ :\ \G_{\iota_\a}\bigl(
\sfE^{(1,1)}M\sqcup\sfE^{(1,0)}Q\bigr)\to\G\bigl(\sfE^{(1,0)}
\sfL_{Q\vert(\pi,\vep)}M\bigr)
\qqq
with restrictions
\qq\nn
\sfL^{Q\vert(\pi,\vep)}_{(1,1\sqcup 0)}\vert_{\G_{\iota_\a}(\sfT M
\sqcup\sfT Q)}:=\sfL^{Q\vert(\pi,\vep)}_{\iota_\a\,*}
\qqq
and
\qq\nn
\sfL^{Q\vert(\pi,\vep)}_{(1,1\sqcup 0)}\vert_{\G\left(\sfT^*M\sqcup
(Q\x\bR)\right)}:=\sfL^{Q\vert(\pi,\vep)\,*}\circ\pr_{\G(\sfT^*M)}+
\vep\,\unl\sfL^{Q\vert(\pi,\vep)\,*}\circ\pr_{C^\infty(Q,\bR)}\,,
\qqq
written in terms of the canonical projections $\,\pr_{\G(\sfT^*M)}\
:\ \G\left(\sfT^*M\sqcup(Q\x\bR)\right)\to\G(\sfT^*M)\,$ and
$\,\pr_{\G(\sfT^*M)}\ :\ \G\left(\sfT^*M\sqcup(Q\x\bR)\right)\to
C^\infty(Q,\bR)$.\exdef \noindent By way of preparation for the
subsequent discussion, we give
\belem\label{lem:tw-paired-lifts}
Adopt the notation of Definitions \ref{def:Vin-str} and
\ref{def:tw-loop-sp-lifts}, Theorem \ref{thm:bib-as-morph} and
Proposition \ref{prop:aut-pair-tw-bra-str}. Let $\,\sfP_{\si,\cB\,
\vert\,(\pi,\vep)}\,$ be the 1-twisted state space with the
canonical projections $\,\pr_{\sfL_{Q\vert(\pi,\vep)}M}\ :\
\sfP_{\si,\cB\,\vert\,(\pi,\vep)}\to\sfL_{Q\vert(\pi,\vep)}M\,$ and
$\,\pr_{\sfT^*
\sfL_{Q\vert(\pi,\vep)}M}:\sfP_{\si,\cB\,\vert\,(\pi,\vep)}\to\sfT^*
\sfL_{Q\vert(\pi,\vep)} M$,\ all as introduced in Definition
\vref{def:tw-phspace}I.3.10. Denote by $\,\theta_{\sfT^*\sfL_{Q\vert
(\pi,\vep)}M}\,$ the canonical 1-form on $\,\sfT^*\sfL_{Q\vert(\pi,
\vep)}M\,$ given in that definition. The lifts $\,\sfL^{Q\vert(\pi,
\vep)\,*}\,$ and $\,\unl\sfL^{Q\vert(\pi,\vep)\,*}\,$ induce the
respective lifts
\qq\nn
\widetilde\sfL^{Q\vert(\pi,\vep)\,*}:=\pr_{\sfL_{Q\vert(\pi,\vep)}
M}^*\circ\sfL^{Q\vert(\pi,\vep)\,*}\,,\qquad\qquad\widetilde{\unl
\sfL}^{Q\vert(\pi,\vep)\,*}:=\pr_{\sfL_{Q\vert(\pi,\vep)}M}^*\circ
\unl\sfL^{Q\vert(\pi,\vep)\,*}\,,
\qqq
and, analogously, the lift $\,\unl\sfL^{Q\vert(\pi,\vep)}_{\iota_\a
\,*}\,$ induces a canonical lift
\qq\nn
\widetilde\sfL^{Q\vert(\pi,\vep)}_{\iota_\a\,*}\ :\ \G_{\iota_\a}(
\sfT M\sqcup\sfT Q)\to\G(\sfT\sfP_{\si,\cB\,\vert\,(\pi,\vep)})
\qqq
fixed by the relations
\qq
\pr_{\sfL_{Q\vert(\pi,\vep)}M\,*}\circ\widetilde\sfL^{Q\vert(\pi,
\vep)}_{\iota_\a\,*}&=&\sfL^{Q\vert(\pi,\vep)}_{\iota_\a\,*}\,,
\label{eq:tiL-sqcup-can-1}\\\cr
\pLie{\widetilde\sfL^{Q\vert(\pi,\vep)}_{\iota_\a\,*}(\Mup\xcV,\Qup
\xcV)}\pr_{\sfT^*\sfL_{Q\vert(\pi,\vep)}M}^*\theta_{\sfT^*\sfL_{Q
\vert(\pi,\vep)}M}&=&0\,,\label{eq:tiL-sqcup-can-2}
\qqq
to be satisfied for any $\,(\Mup\xcV,\Qup \xcV)\in\G_{\iota_\a}(
\sfT M\sqcup\sfT Q)$.\ The above can, in turn, be combined into a
lift
\qq\nn
\widetilde\sfL^{Q\vert(\pi,\vep)}_{(1,1\sqcup 0)}\ :\ \G_{\iota_\a}
\bigl(\sfE^{(1,1)}M\sqcup\sfE^{(1,0)}Q\bigr)\to\G\bigl(\sfE^{(1,0)}
\sfP_{\si,\cB\,\vert\,(\pi,\vep)}\bigr)
\qqq
with restrictions
\qq\label{eq:restr-tan-tiLcup}
\widetilde\sfL^{Q\vert(\pi,\vep)}_{(1,1\sqcup 0)}
\vert_{\G_{\iota_\a}(\sfT M\sqcup\sfT Q)}:=\widetilde\sfL^{Q\vert(
\pi,\vep)}_{\iota_\a\,*}
\qqq
and
\qq\label{eq:restr-cotan-tiLcup}
\widetilde\sfL^{Q\vert(\pi,\vep)}_{(1,1\sqcup 0)}\vert_{\G\left(
\sfT^*M\sqcup(Q\x\bR)\right)}:=\widetilde\sfL^{Q\vert(\pi,\vep)\,*}
\circ\pr_{\Om^1(M)}+\vep\,\widetilde{\unl\sfL}^{Q\vert(\pi,\vep)\,
*}\circ\pr_{C^\infty(Q,\bR)}\,.
\qqq
The various lifts have the following properties
\qq
\d\widetilde\sfL^{Q\vert(\pi,\vep)\,*}\upsilon&=&-\widetilde\sfL^{Q
\vert(\pi,\vep)\,*}\sfd\upsilon+\vep\,\widetilde{\unl\sfL}^{Q\vert(
\pi,\vep)\,*}\D_Q\upsilon\,,\label{eq:tiL-minus-DQ-tw}\\\cr
\d\widetilde{\unl\sfL}^{Q\vert(\pi,\vep)\,*}\xi&=&\widetilde{\unl
\sfL}^{Q\vert(\pi,\vep)\,*}\sfd\xi\,,\label{eq:unL-plus1-tw}\\\cr
\cr
\widetilde\sfL^{Q\vert(\pi,\vep)}_{\iota_\a\,*}\bigl(\Mup\xcV,\Qup
\xcV\bigr)\con\widetilde\sfL^{Q\vert(\pi,\vep)\,*}\upsilon&=&-
\widetilde\sfL^{Q\vert(\pi,\vep)\,*}\bigl(\Mup\xcV\con\upsilon
\bigr)\,,\label{eq:tiL-minus-tw}\\\cr
\widetilde\sfL^{Q\vert(\pi,\vep)}_{\iota_\a\,*}\bigl(\Mup\xcV,\Qup
\xcV\bigr)\con\widetilde{\unl\sfL}^{Q\vert(\pi,\vep)\,*}\xi&=&
\widetilde{\unl\sfL}^{Q\vert(\pi,\vep)\,*}\bigl(\Qup\xcV\con\xi
\bigr)\,,\label{eq:tiL-plus-tw}
\qqq
written for arbitrary $\,(\Mup\xcV,\Qup\xcV)\in\G_{\iota_\a}(\sfT M
\sqcup\sfT Q),\upsilon\in\Om^n(M)\,$ and $\,\xi\in\Om^n(Q)$. \elem
\beroof
Obvious, through inspection. \eroof\medskip

The next result establishes the sought-after connection between the
(twisted) bracket structure on $(\cG,\cB)$-twisted $\iota_\a$-paired
generalised tangent bundles $\,\sfE^{(1,1)}_\cG
M\sqcup\sfE^{(1,0)}_\cB Q\,$ and the canonical Vinogradov structure
on the (1-)twisted state space of the $\si$-model in the presence of
defects, thus realising the general correspondence scheme
anticipated in the Introduction.
\bethe\label{thm:ind-quasi-morph-glob-Vin-tw}
Adopt the notation of Corollary \ref{cor:tw-gen-tan-vs-ngerb}, of
Theorem \ref{thm:bib-as-morph}, of Proposition
\ref{prop:pair-tw-gen-tan-bun}, and of Lemma
\ref{lem:tw-paired-lifts}. Let
$\,\ceL_{\si,\cB\,\vert\,(\pi,\vep)}\to
\sfP_{\si,\cB\,\vert\,(\pi,\vep)}\,$ be the pre-quantum bundle from
Corollary \vref{cor:preqb-tw}I.3.19. The pair $\,(\cG,\cB)\,$
canonically induces a linear mapping
\qq\nn
\phi_{\si,\cB\,\vert\,(\pi,\vep)}\ :\ \G_{\iota_\a}\bigl(\sfE^{(1,1
)}_\cG M\sqcup\sfE^{(1,0)}_\cB Q\bigr)\to\G\bigl(\sfE^{(1,0
)}_{\ceL_{\si,\cB\,\vert\,(\pi,\vep)}}\sfP_{\si,\cB\,\vert\,(\pi,\vep)}
\bigr)
\qqq
that relates elements of the respective global structures $\,\Mgt^{(
1,0),(0,0;\check\D_Q)}_{(\cG,\cB),\iota_\a}(M\sqcup Q)\,$ and
$\,\Vgt^{(0)}_{\ceL_{\si,\cB\,\vert\,(\pi,\vep)}}\sfP_{\si,\cB\,\vert\,(\pi,
\vep)}\,$ as
\qq
\a_{\sfT\sfP_{\si,\cB\,\vert\,(\pi,\vep)}}\circ\phi_{\si,\cB\,\vert\,(\pi,
\vep)}&=&\widetilde\sfL^{Q\vert(\pi,\vep)}_{\iota_\a\,*}\circ
\a_{\sfT(M\sqcup Q)}\,,\label{eq:anchPsi-phisi-tw}\\\cr
\Vbra{\cdot}{\cdot}\circ(\phi_{\si,\cB\,\vert\,(\pi,\vep)},\phi_{\si,
\cB\,\vert\,(\pi,\vep)})&=&\phi_{\si,\cB\,\vert\,(\pi,\vep)}\circ
\GBra{\cdot}{\cdot}^{(0,0;\check\D_Q)}\,,\label{eq:Vbra-phisi-tw}
\\\cr
\Vcon{\cdot}{\cdot}\circ(\phi_{\si,\cB\,\vert\,(\pi,\vep)},\phi_{\si,
\cB\,\vert\,(\pi,\vep)})&=&\widetilde\sfL^{Q\vert(\pi,\vep)\,*}\circ
\pr_{\Om^0(M)}\circ\Vcon{\cdot}{\cdot}\equiv 0\,.
\label{eq:Vcon-phisi-tw}
\qqq
\ethe
\beroof
Consider the generalised tangent bundles $\,\sfE^{(1,1)}M\,$ and
$\,\sfE^{(1,0)}Q\,$ in keeping with Definition \ref{def:Vin-str}.
Denote by $\,\txH\in Z^3(M)\,$ the curvature of $\,\cG$,\ and let
$\,\Mgt^{(1,0),(\txH,\om;\D_Q)}_{\iota_\a}(M\sqcup Q)\,$ be the
$(\txH,\om;\D_Q)$-twisted bracket structure on $\iota_\a$-paired
generalised tangent bundles $\,\sfE^{(1,1)} M\sqcup \sfE^{(1,0)}Q$,\
as described in Definition \ref{def:pair-tw-bra-str}, under
restriction to the subset
$\,\G_{\iota_\a}\bigl(\sfE^{(1,1)}M\sqcup\sfE^{(1,0)}Q\bigr)\,$ of
$\iota_\a$-aligned sections of $\,\sfE^{(1,1)}M\sqcup\sfE^{(1,0)}Q$,
\ introduced in Proposition \ref{prop:aut-pair-tw-bra-str}. In
virtue of Proposition \ref{prop:pair-tw-gen-tan-bun}, there exists a
homomorphism of bracket structures
\qq\nn
{}^{\tx{\tiny $M\sqcup Q$}}\hspace{-2pt}\chi\ :\ \Mgt^{(1,0),(0,0;
\check\D_Q)}_{(\cG,\cB),\iota_\a}(M\sqcup Q)\to\Mgt^{(1,0),(\txH,\om
;\D_Q)}_{\iota_\a}(M\sqcup Q)\,.
\qqq
Consider, next, the generalised tangent bundle $\,\sfE^{(1,0)}
\sfP_{\si,\cB\,\vert\,(\pi,\vep)}\,$ equipped with the $\Om_{\si,\cB
\vert(\pi,\vep)}$-twisted Vinogradov structure $\,\Vgt^{(0),
\Om_{\si,\cB\,\vert\,(\pi,\vep)}}\sfP_{\si,\cB\,\vert\,(\pi,\vep)}$,\
detailed in Definition \ref{def:Om-tw-Vin}. Corollary
\ref{cor:tw-gen-tan-vs-ngerb} states the existence of a
homomorphisms of the Vinogradov structures
\qq\nn
{}^{\sfP_{\si,\cB\,\vert\,(\pi,\vep)}}\hspace{-2pt}\chi\ :\ \Vgt^{(0
)}_{\ceL_{\si,\cB\,\vert\,(\pi,\vep)}}\sfP_{\si,\cB\,\vert\,(\pi,\vep)}\to
\Vgt^{(0),\Om_{\si,\cB\,\vert\,(\pi,\vep)}}\sfP_{\si,\cB\,\vert\,(\pi,\vep
)}\,,
\qqq
given in terms of local data of
$\,\ceL_{\si,\cB\,\vert\,(\pi,\vep)}$,\ \textit{cf.}\ the proof of
Theorem \ref{thm:ind-quasi-morph-glob-Vin}. The linear mapping
announced in the theorem is now explicitly defined as
\qq\nn
\phi_{\si,\cB\,\vert\,(\pi,\vep)}:={}^{\sfP_{\si,\cB\,\vert\,(\pi,\vep)}}
\hspace{-2pt}\chi^{-1}\circ\ee^{\theta_{\sfT^*\sfL_{Q\vert(\pi,\vep
)}M}}\circ\widetilde\sfL^{Q\vert(\pi,\vep)}_{(1,1\sqcup 0)}\circ
{}^{\tx{\tiny $M\sqcup Q$}}\hspace{-2pt}\chi
\qqq
in terms of the lift $\,\widetilde\sfL^{Q\vert(\pi,\vep)}_{(1,1
\sqcup 0)}\,$ from Lemma \ref{lem:tw-paired-lifts}. The linearity of
the mapping thus defined follows immediately from condition
\eqref{eq:iota-pairing} of $\iota_\a$-alignment as the latter
enforces a common scaling of the two restrictions (to $\,M\,$ and to
$\,Q$) of a section from $\,\G_{\iota_\a}\bigl(\sfE^{(1,1)}_\cG M
\sqcup\sfE^{(1,0)}_\cB Q\bigr)$.\ Moreover, relation
\eqref{eq:Vcon-phisi-tw} is satisfied automatically due to -- on one
hand -- the identity
\qq\nn
\widetilde\sfL^{Q\vert(\pi,\vep)\,*}\circ\pr_{\Om^0(M)}\equiv 0\,,
\qqq
\textit{cf.}\ \Reqref{eq:triv-can-con-10}, and -- on the other hand
-- the triviality of the canonical contraction on $\,\sfE^{(1,0
)}_{\ceL_{\si,\cB\,\vert\,(\pi,\vep)}}\sfP_{\si,\cB\,\vert\,(\pi,\vep)}$.\
This leaves us with the other two relations to check.

The first of the two, \Reqref{eq:anchPsi-phisi-tw}, derives directly
from the definition of $\,\phi_{\si,\cB\,\vert\,(\pi,\vep)}\,$ in
which all mappings except for the lift leave the vector-field
components unchanged, and in which the lift itself restricts to
vector-field components as in \Reqref{eq:restr-tan-tiLcup}. In order
to prove the other one, \Reqref{eq:Vbra-phisi-tw}, it suffices to
demonstrate the identity
\qq\label{eq:pull-tiL-out}\qquad\qquad
\Vbra{\cdot}{\cdot}^{\widetilde\sfL^{Q\vert(\pi,\vep)\,*}\txH+
\widetilde{\unl\sfL}^{Q\vert(\pi,\vep)\,*}\om}\circ\bigl(\widetilde
\sfL^{Q\vert(\pi,\vep)}_{(1,1\sqcup 0)},\widetilde\sfL^{Q\vert(\pi,
\vep)}_{(1,1\sqcup 0)}\bigr)=\widetilde\sfL^{Q\vert(\pi,\vep)}_{(1,1
\sqcup 0)}\circ\GBra{\cdot}{\cdot}^{(\txH,\om;\D_Q)}\,.
\qqq
Take a pair of sections $\,\Vgt,\Wgt\in\G_{\iota_\a}\bigl(
\sfE^{(1,1)}M\sqcup\sfE^{(1,0)}Q\bigr)\,$ and denote the respective
restrictions to $\,M\,$ and $\,Q\,$  as $\,\Vgt\vert_M=\Mup\xcV
\oplus\upsilon,\ \Vgt\vert_Q=\Qup\xcV\oplus\xi\,$ and $\,\Wgt\vert_M
=\Mup \xcW\oplus\varpi,\ \Wgt\vert_Q=\Qup\xcW\oplus\z$.\
Furthermore, for the sake of transparency, represent $\,\Vgt\,$ as
$\,(\Mup\xcV,\Qup \xcV)\oplus(\upsilon,\xi)$,\ and $\,\Wgt\,$ as
$\,(\Mup\xcW,\Qup \xcW)\oplus(\varpi,\z)$,\ and similarly for their
bracket. Upon invoking conditions \eqref{eq:tiL-sqcup-can-1} and
\eqref{eq:tiL-sqcup-can-2} in conjunction with
\Reqref{eq:Liota-vec}, condition \eqref{eq:iota-pairing}, and
Eqs.\,\eqref{eq:tiL-minus-tw} together with
\eqref{eq:tiL-minus-DQ-tw}, and \eqref{eq:tiL-plus-tw} together with
\eqref{eq:unL-plus1-tw}, this yields
\qq\nn
\widetilde\sfL^{Q\vert(\pi,\vep)}_{(1,1\sqcup 0)}\circ
\GBra{\Vgt}{\Wgt}^{(\txH,\om;\D_Q)}&=&\widetilde\sfL^{Q\vert(\pi,
\vep)}_{\iota_\a\,*}\bigl([\Mup\xcV,\Mup\xcW],[\Qup\xcV\,
,\,\Qup\xcW]\bigr)\cr\cr
&&\oplus\bigl[\widetilde\sfL^{Q\vert(\pi,\vep)\,*}\bigl(\pLie{\Mup
\xcV}\varpi-\pLie{\Mup\xcW}\upsilon-\tfrac{1}{2}\,\sfd\bigl(\Mup
\xcV\con\varpi-\Mup\xcW\con\upsilon\bigr)+\Mup\xcV\con\Mup\xcW\con
\txH\bigr)\cr\cr
&&\hspace{-1cm}+\vep\,\widetilde{\unl\sfL}^{Q\vert(\pi,\vep)\,*}
\bigl(\Qup\xcV\con\sfd\z-\Qup\xcW\con\sfd\xi+\Qup\xcV\con\Qup\xcW
\con\om+\tfrac{1}{2}\,\bigl(\Qup\xcV\con\D_Q\varpi-\Qup\xcW\con\D_Q
\upsilon\bigr)\bigr)\bigr]\cr\cr
&=&[\widetilde\sfL^{Q\vert(\pi,\vep)}_{\iota_\a\,*}\bigl(\Mup\xcV
,\Qup\xcV\bigr),\widetilde\sfL^{Q\vert(\pi,\vep)}_{\iota_\a\,*}
\bigl(\Mup\xcW,\Qup\xcW\bigr)]\cr\cr
&&\oplus\bigl(\widetilde\sfL^{Q\vert(\pi,\vep)}_{\iota_\a\,*}\bigl(
\Mup\xcV,\Qup\xcV\bigr)\con\d\widetilde\sfL^{Q\vert(\pi,\vep)\,*}
\varpi-\tfrac{\vep}{2}\,\widetilde\sfL^{Q\vert(\pi,\vep)}_{\iota_\a
\,*}\bigl(\Mup\xcV,\Qup\xcV\bigr)\con\widetilde{\unl\sfL}^{Q\vert(
\pi,\vep)\,*}\D_Q\varpi\cr\cr
&&-\widetilde\sfL^{Q\vert(\pi,\vep)}_{\iota_\a\,*}\bigl(\Mup\xcW,
\Qup\xcW\bigr)\con\d\widetilde\sfL^{Q\vert(\pi,\vep)\,*}\upsilon+
\tfrac{\vep}{2}\,\widetilde\sfL^{Q\vert(\pi,\vep)}_{\iota_\a\,*}
\bigl(\Mup\xcW,\Qup\xcW\bigr)\con\widetilde{\unl\sfL}^{Q\vert(\pi,
\vep)\,*}\D_Q\upsilon\cr\cr
&&+\widetilde\sfL^{Q\vert(\pi,\vep)}_{\iota_\a\,*}\bigl(\Mup\xcV,
\Qup\xcV\bigr)\con\widetilde\sfL^{Q\vert(\pi,\vep)}_{\iota_\a\,*}
\bigl(\Mup\xcW,\Qup\xcW\bigr)\con\widetilde\sfL^{Q\vert(\pi,\vep)\,
*}\txH\cr\cr
&&+\vep\,\bigl(\widetilde\sfL^{Q\vert(\pi,\vep)}_{\iota_\a\,*}
\bigl(\Mup\xcV,\Qup\xcV\bigr)\con\d\widetilde{\unl\sfL}^{Q\vert(\pi
, \vep)\,*}\z-\widetilde\sfL^{Q\vert(\pi,\vep)}_{\iota_\a\,*}\bigl(
\Mup\xcW,\Qup\xcW\bigr)\con\d\widetilde{\unl\sfL}^{Q\vert(\pi,\vep)
\,*}\xi\bigr)\cr\cr
&&+\widetilde\sfL^{Q\vert(\pi,\vep)}_{\iota_\a\,*}\bigl(\Mup\xcV,
\Qup\xcV\bigr)\con\widetilde\sfL^{Q\vert(\pi,\vep)}_{\iota_\a\,*}
\bigl(\Mup\xcW,\Qup\xcW\bigr)\con\widetilde{\unl\sfL}^{Q\vert(\pi,
\vep)\,*}\om\cr\cr
&&+\tfrac{\vep}{2}\,\widetilde{\unl\sfL}^{Q\vert(\pi,\vep)\,*}
\bigl(\Qup\xcV\con\D_Q\varpi-\Qup\xcW\con\D_Q \upsilon\bigr)\bigr)
\cr\cr
&=&[\widetilde\sfL^{Q\vert(\pi,\vep)}_{\iota_\a\,*}\bigl(\Mup\xcV
,\Qup\xcV\bigr),\widetilde\sfL^{Q\vert(\pi,\vep)}_{\iota_\a\,*}
\bigl(\Mup\xcW,\Qup\xcW\bigr)]\cr\cr
&&\oplus\bigl(\widetilde\sfL^{Q\vert(\pi,\vep)}_{\iota_\a\,*}\bigl(
\Mup\xcV,\Qup\xcV\bigr)\con\d\bigl(\widetilde\sfL^{Q\vert(\pi,\vep)
\,*}\varpi+\vep\,\widetilde{\unl\sfL}^{Q\vert(\pi,\vep)\,*}\z\bigr)
\cr\cr
&&-\widetilde\sfL^{Q\vert(\pi,\vep)}_{\iota_\a\,*}\bigl(\Mup\xcW,
\Qup\xcW\bigr)\con\d\bigl(\widetilde\sfL^{Q\vert(\pi,\vep)\,*}
\upsilon+\vep\,\widetilde{\unl\sfL}^{Q\vert(\pi,\vep)\,*}\xi\bigr)
\cr\cr
&&+\widetilde\sfL^{Q\vert(\pi,\vep)}_{\iota_\a\,*}\bigl(\Mup\xcV,
\Qup\xcV\bigr)\con\widetilde\sfL^{Q\vert(\pi,\vep)}_{\iota_\a\,*}
\bigl(\Mup\xcW,\Qup\xcW\bigr)\con\bigl(\widetilde\sfL^{Q\vert(\pi,
\vep)\,*}\txH+\vep\,\widetilde{\unl\sfL}^{Q\vert(\pi,\vep)\,*}\om
\bigr)\,.
\qqq
Comparison with \Reqref{eq:Vin-bra-10} and, subsequently, with
Eqs.\,\eqref{eq:restr-cotan-tiLcup} and \eqref{eq:restr-tan-tiLcup}
permits to rewrite the above concisely as
\qq\nn
\widetilde\sfL^{Q\vert(\pi,\vep)}_{(1,1\sqcup 0)}\circ
\GBra{\Vgt}{\Wgt}^{(\txH,\om;\D_Q)}&=&\bigl[\widetilde\sfL^{Q
\vert(\pi,\vep)}_{\iota_\a\,*}\bigl(\Mup\xcV,\Qup\xcV\bigr)\oplus
\bigl(\widetilde\sfL^{Q\vert(\pi,\vep)\,*}\upsilon+\vep\,
\widetilde{\unl\sfL}^{Q\vert(\pi,\vep)\,*}\xi\bigr)\,,\cr\cr
&&\,\widetilde\sfL^{Q\vert(\pi,\vep)}_{\iota_\a\,*}\bigl(\Mup\xcW,
\Qup\xcW\bigr)\oplus\bigl(\widetilde\sfL^{Q\vert(\pi,\vep)\,*}
\varpi+\vep\,\widetilde{\unl\sfL}^{Q\vert(\pi,\vep)\,*}\z\bigr)\,
\bigr]^{\widetilde\sfL^{Q\vert(\pi,\vep)\,*}\txH+\vep\,
\widetilde{\unl\sfL}^{Q\vert(\pi,\vep)\,*}\om}_{\rm V}\cr\cr
&=&\Vbra{\widetilde\sfL^{Q\vert(\pi,\vep)}_{1,1\sqcup 0}
\Vgt}{{\widetilde\sfL^{Q\vert(\pi,\vep)}_{1,1\sqcup 0}
\Wgt}}^{\widetilde\sfL^{Q\vert(\pi,\vep)\,*}\txH+\vep\,
\widetilde{\unl\sfL}^{Q\vert(\pi,\vep)\,*}\om}\,,
\qqq
which concludes the proof. \eroof\medskip

We are now fully equipped to discuss, in the algebraic framework
elaborated above, a realisation of symmetries in the twisted sector
of the theory.
\berop\label{prop:ham-vs-sisymsec-tw} Adopt the notation of Definitions
\ref{def:Vin-str} and \ref{def:tw-loop-sp-lifts}, of Corollary
\ref{cor:tw-gen-tan-vs-ngerb}, of Theorems \ref{thm:bib-as-morph}
and \ref{thm:ind-quasi-morph-glob-Vin-tw}, of Propositions
\ref{prop:aut-pair-tw-bra-str} and \ref{prop:pair-tw-gen-tan-bun},
and of Lemma \ref{lem:tw-paired-lifts}. Let $\,\ceL_{(\cG,\cB)\vert(
\pi,\vep)}\to\sfL_{Q\vert(\pi,\vep)}M\,$ be the transgression bundle
of Theorem \vref{thm:trans-tw}I.3.18, with local data $\,(E_{(\pi,
\vep)\,\igt},G_{(\pi,\vep)\,\igt\jgt})$,\ as explicited in the
constructive proof of the theorem, written for the open cover
$\,\cO_{\sfL_{Q\vert(\pi,\vep)}M}=\{\cO_{(\pi,\vep)\,\igt}\}_{\igt
\in\xcI_{\sfL_{Q\vert(\pi,\vep)} M}}\,$ of $\,\sfL_{Q\vert(\pi,\vep
)}\,$ from Proposition \vref{prop:cover-tw}I.3.14. Write
\qq\nn
\Tgt_{\cB\,\vert\,(\pi,\vep)}:=1\oplus\pr_{\sfT^*\sfL_{Q\vert(\pi,\vep
)}M}^*\theta_{\sfT^*\sfL_{Q\vert(\pi,\vep)}M}\in\G\bigl(\sfE^{(0,1)}
\sfP_{\si,\cB\,\vert\,(\pi,\vep)}\bigr)\,,
\qqq
and call the latter object the \textbf{canonical section of
$\,\sfE^{(1,0)}\sfP_{\si,\cB\,\vert\,(\pi,\vep)}$}.\ To every
$\si$-symmetric $\iota_\a$-aligned section $\,\Vgt\in\G_{\iota_\a,
\si}\bigl(\sfE^{(1,1)}M\sqcup\sfE^{(1,0)}Q\bigr)\,$ there is
associated a \textbf{hamiltonian function $\,h^{\cB\,\vert\,
\vep}_\Vgt$},\ \textit{i.e.}\ a smooth function on $\,\sfP_{\si,\cB
\vert(\pi,\vep)}\,$ satisfying the defining relation
\qq\nn
\a_{\sfT\sfP_{\si,\cB\,\vert\,(\pi,\vep)}}\bigl(\widetilde\sfL^{Q\vert(
\pi,\vep)}_{(1,1\sqcup
0)}\Vgt\bigr)\con\Om_{\si,\cB\,\vert\,(\pi,\vep )}=:-\d
h^{\cB\,\vert\,\vep}_\Vgt\,.
\qqq
The hamiltonian function is given by the formula
\qq\label{eq:class-ham-tw}
h^{\cB\,\vert\,\vep}_\Vgt=\corr{\widetilde\sfL^{Q\vert(\pi,\vep)}_{(1,1
\sqcup 0)}\Vgt,\Tgt_{\cB\,\vert\,(\pi,\vep)}}\,.
\qqq
The \textbf{pre-quantum hamiltonian for
$\,h^{\cB\,\vert\,\vep}_\Vgt$},\ as explicited in Definition
\vref{def:prequantise}I.3.4, is the linear operator $\,
\widehat\cO_{h^{\cB\,\vert\,\vep}_\Vgt}\,$ on
$\,\G\bigl(\ceL_{\si,\cB \vert(\pi,\vep)}\bigr)\,$ with restrictions
\qq\label{eq:preq-ham-gen-tw}
\widehat\cO_{h^{\cB\,\vert\,\vep}_\Vgt}\vert_{\pr_{\sfL_{Q\vert(\pi,
\vep)}M}^{-1}\bigl(\cO_{(\pi,\vep)\,\igt}\bigr)}\\\cr
=-\sfi\,\pLie{\a_{\sfT\sfP_{\si,\cB\,\vert\,(\pi,\vep)}}\bigl(\ee^{-
\pr_{\sfT^*\sfL_{Q\vert(\pi,\vep)}M}^*\theta_{\sfT^*\sfL_{Q\vert(
\pi,\vep)}M}}\lact\widetilde\Vgt_\igt\bigr)}+\corr{\ee^{-
\pr_{\sfT^*\sfL_{Q\vert(\pi,\vep)}M}^*\theta_{\sfT^*\sfL_{Q\vert(
\pi,\vep)}M}}\lact\widetilde\Vgt_\igt,\Tgt_{\cB\,\vert\,(\pi,\vep)}}=:
\widehat h^{\cB\,\vert\,\vep}_{\widetilde\Vgt_\igt}\,,\nonumber
\qqq
expressed in terms of local sections
\qq\label{eq:preq-ham-gen-ingr-tw}\qquad\qquad
\widetilde\Vgt_\igt:=\ee^{-E_{(\pi,\vep)\,\igt}}\lact\widetilde
\sfL^{Q\vert(\pi,\vep)}_{(1,1\sqcup 0)}\Vgt\in\bigl(\sfE^{(1,0
)}_{\pr_{\sfL_{Q\vert(\pi,\vep)}M}^*\ceL_{(\cG,\cB)\vert(\pi,\vep
)}}\sfP_{\si,\cB\,\vert\,(\pi,\vep)}\bigr)\bigl(\pr_{\sfL_{Q\vert(\pi,
\vep)}M}^{-1}\bigl(\cO_{(\pi,\vep)\,\igt}\bigr)\bigr)\,.
\qqq
Given two $\si$-symmetric $\iota_\a$-aligned sections $\,\Vgt,
\Wgt$,\ the Poisson bracket of the associated hamiltonian functions,
determined by $\,\Om_{\si,\cB\,\vert\,(\pi,\vep)}\,$ in the manner
detailed in Remark \vref{rem:Mars-Wein}I.3.3, reads
\qq\label{eq:Poiss-bra-ham-tw}
\{h^{\cB\,\vert\,\vep}_\Vgt,h^{\cB\,\vert\,\vep}_\Wgt\}_{\Om_{\si,
\cB\,\vert\,(\pi,\vep)}}=h^{\cB\,\vert\,\vep}_{\GBra{\Vgt}{\Wgt}^{(\txH,\om
;\D_Q)}}\,.
\qqq
The commutator of the corresponding pre-quantum hamiltonians is
(locally) given by
\qq\label{eq:comm-preq-ham-tw}
[\widehat h^{\cB\,\vert\,\vep}_{\widetilde\Vgt_\igt},\widehat
h^{\cB\,\vert\,\vep}_{\widetilde \Wgt_\igt}]=-\sfi\,\widehat h^{\cB
\vert\vep}_{\Vbra{\widetilde\Vgt_\igt}{\widetilde\Wgt_\igt}}\,.
\qqq
\eerop
\beroof
The proof proceeds along the same lines as for Proposition
\ref{prop:ham-vs-sisymsec}. Thus, we first rewrite the symplectic
form of the 1-twisted sector from \Reqref{eq:sympl-form-1tw} as
\qq\nn
\Om_{\si,\cB\,\vert\,(\pi,\vep)}=\d_{\widetilde\sfL^{Q\vert(\pi,\vep)\,
*}\txH+\vep\,\widetilde{\unl\sfL}^{Q\vert(\pi,\vep)\,*}\om}
\Tgt_{\cB\,\vert\,(\pi,\vep)}\,.
\qqq
Take an arbitrary $\,\Vgt\in\G_{\iota_\a}\bigl(\sfE^{(1,1)}M\sqcup
\sfE^{(1,0)}Q\bigr)\,$ and denote its restrictions to $\,M\,$ and
$\,Q\,$ as $\,\Vgt\vert_M=\Mup\xcV\oplus\upsilon\,$ and $\,\Vgt
\vert_Q=\Qup\xcV\oplus\xi$,\ respectively, representing $\,\Vgt\,$
as $\,(\Mup\xcV,\Qup \xcV)\oplus(\upsilon,\xi)$.\ Then, using
conditions \eqref{eq:tiL-sqcup-can-1} and \eqref{eq:tiL-sqcup-can-2}
together with Eqs.\,\eqref{eq:Liota-vec}, \eqref{eq:tiL-minus-tw}
and \eqref{eq:tiL-plus-tw}, the conditions of $\si$-symmetricity and
$\iota_\a$-alignment of $\,\Vgt$,\ and, finally,
Eqs.\,\eqref{eq:unL-plus1-tw} and \eqref{eq:tiL-minus-DQ-tw}, we
obtain,
\qq\nn
&&\a_{\sfT\sfP_{\si,\cB\,\vert\,(\pi,\vep)}}\bigl(\widetilde\sfL^{Q
\vert(\pi,\vep)}_{(1,1\sqcup
0)}\Vgt\bigr)\con\Om_{\si,\cB\,\vert\,(\pi, \vep)}\cr\cr
&=&\widetilde\sfL^{Q\vert(\pi,\vep)}_{\iota_\a\,*}\bigl(\Mup\xcV,
\Qup\xcV\bigr)\con\bigl(\d\pr_{\sfT^*\sfL_{Q\vert(\pi,\vep)}M}^*
\theta_{\sfT^*\sfL_{Q\vert(\pi,\vep)}M}+\widetilde\sfL^{Q\vert(\pi,
\vep)\,*}\txH+\vep\,\widetilde{\unl\sfL}^{Q\vert(\pi,\vep)\,*}\om
\bigr)\cr\cr
&=&-\d\bigl(\widetilde\sfL^{Q\vert(\pi,\vep)}_{\iota_\a\,*}\bigl(
\Mup\xcV,\Qup\xcV\bigr)\con\pr_{\sfT^*\sfL_{Q\vert(\pi,\vep)}M}^*
\theta_{\sfT^*\sfL_{Q\vert(\pi,\vep)}M}\bigr)-\widetilde\sfL^{Q
\vert(\pi,\vep)\,*}\bigl(\Mup\xcV\con\txH\bigr)+\vep\,
\widetilde{\unl\sfL}^{Q\vert(\pi,\vep)\,*}\bigl(\Qup\xcV\con\om
\bigr)\cr\cr
&=&-\d\bigl(\widetilde\sfL^{Q\vert(\pi,\vep)}_{\iota_\a\,*}\bigl(
\Mup\xcV,\Qup\xcV\bigr)\con\pr_{\sfT^*\sfL_{Q\vert(\pi,\vep)}M}^*
\theta_{\sfT^*\sfL_{Q\vert(\pi,\vep)}M}+\widetilde\sfL^{Q\vert(\pi,
\vep)\,*}\upsilon\bigr)+\vep\,\widetilde{\unl\sfL}^{Q\vert(\pi,\vep
)\,*}\bigl(\D_Q\upsilon+\Qup\xcV\con\om\bigr)\cr\cr
&=&-\d\corr{\widetilde\sfL^{Q\vert(\pi,\vep)}_{\iota_\a\,*}\bigl(
\Mup\xcV,\Qup\xcV\bigr)\oplus\bigl(\widetilde\sfL^{Q\vert(\pi,\vep)
\,*}\upsilon+\vep\,\widetilde{\unl\sfL}^{Q\vert(\pi,\vep)\,*}\xi
\bigr),\Tgt_{\cB\,\vert\,(\pi,\vep)}}\,,
\qqq
whence \Reqref{eq:class-ham-tw} ensues upon invoking the definition
of the lift $\,\widetilde \sfL^{Q\vert(\pi,\vep)}_{(1,1\sqcup 0)}$.\
The pre-quantum hamiltonian can then be reproduced, in the form
stipulated, by specialisation of the general definition (I.3.8), and
we easily see, through direct inspection, that the local objects
$\,\widetilde\Vgt_\igt\,$ are in the image of an isomorphism defined
analogously to the isomorphism
$\,{}^{\sfP_{\si,\cB\,\vert\,(\pi,\vep)}} \hspace{-2pt}\chi^{-1}\,$
from the constructive proof of Theorem
\ref{thm:ind-quasi-morph-glob-Vin-tw}.

As a corollary to the above, we obtain a hamiltonian section
$\,\widetilde\Vgt\in\G\bigl(\sfE^{(1,0)}\sfP_{\si,\cB\,\vert\,(\pi,\vep
)}\bigr)\,$ for every $\si$-symmetric $\iota_\a$-aligned section
$\,\Vgt\in\G_{\iota_\a,\si}\bigl(\sfE^{(1,1)}M\sqcup\sfE^{(1,0)}Q
\bigr)$,\ given by
\qq\nn
\widetilde\Vgt=\ee^{\pr_{\sfT^*\sfL_{Q\vert(\pi,\vep)}M}^*
\theta_{\sfT^*\sfL_{Q\vert(\pi,\vep)}M}}\lact\widetilde\sfL^{Q\vert
(\pi,\vep)}_{(1,1\sqcup 0)}\Vgt\,.
\qqq
Consider a pair of sections $\,\Vgt,\Wgt\in\G_{\iota_\a,\si}\bigl(
\sfE^{(1,1)}M\sqcup\sfE^{(1,0)}Q\bigr)\,$ and the respective
hamiltonian sections $\,\widetilde\Vgt\,$ and $\,\widetilde\Wgt$.\
The Poisson bracket of the corresponding hamiltonian functions can
be extracted from the canonical Vinogradov bracket
\qq\nn
\Vbra{\widetilde\Vgt}{\widetilde\Wgt}^{\Om_{\si,\cB\,\vert\,(\pi,\vep
)}}&\equiv&\Vbra{\ee^{\pr_{\sfT^*\sfL_{Q\vert(\pi,\vep)}M}^*
\theta_{\sfT^*\sfL_{Q\vert(\pi,\vep)}M}}\lact\widetilde\sfL^{Q\vert
(\pi,\vep)}_{(1,1\sqcup 0)}\Vgt}{\ee^{\pr_{\sfT^*\sfL_{Q\vert(\pi,
\vep)}M}^*\theta_{\sfT^*\sfL_{Q\vert(\pi,\vep)}M}}\lact\widetilde
\sfL^{Q\vert(\pi,\vep)}_{(1,1\sqcup
0)}\Wgt}^{\Om_{\si,\cB\,\vert\,(\pi, \vep)}}\cr\cr
&=&\ee^{\pr_{\sfT^*\sfL_{Q\vert(\pi,\vep)}M}^*\theta_{\sfT^*\sfL_{Q
\vert(\pi,\vep)}M}}\lact\Vbra{\widetilde\sfL^{Q\vert(\pi,\vep)}_{(1
,1\sqcup 0)}\Vgt}{\widetilde\sfL^{Q\vert(\pi,\vep)}_{(1,1\sqcup 0)}
\Wgt}^{\widetilde\sfL^{Q\vert(\pi,\vep)\,*}\txH+\vep\,
\widetilde{\unl\sfL}^{Q\vert(\pi,\vep)\,*}\om}\cr\cr
&=&\ee^{\pr_{\sfT^*\sfL_{Q\vert(\pi,\vep)}M}^*\theta_{\sfT^*\sfL_{Q
\vert(\pi,\vep)}M}}\lact\widetilde\sfL^{Q\vert(\pi,\vep)}_{(1,1
\sqcup 0)}\bigl(\GBra{\Vgt}{\Wgt}^{(\txH,\om;\D_Q)}\bigr)\equiv
\widetilde{\GBra{\Vgt}{\Wgt}^{(\txH,\om;\D_Q)}}\,,
\qqq
calculated with the help of the results from the proof of
Proposition \ref{prop:Vin-str-auts} and \Reqref{eq:pull-tiL-out}.
This proves \Reqref{eq:Poiss-bra-ham-tw}.

Upon (partially) reversing the last chain of equalities, using
Eq.\,(I.3.9) and introducing the local data
$\,(\theta_{\si,\cB\,\vert\,(\pi,\vep)\,\igt}
,\g_{\si,\cB\,\vert\,(\pi,\vep)\,\igt\jgt})\,$ of the pre-quantum
bundle from Corollary \vref{cor:preqb-tw}I.3.19 (associated with the
open cover
$\,\cO_{\sfP_{\si,\cB\,\vert\,(\pi,\vep)}}=\{\pr_{\sfL_{Q\vert(\pi,\vep
)}M}^{-1}\bigl(\cO_{(\pi,\vep)\,\igt}\bigr)\}_{\igt\in\xcI_{\sfL_{Q
\vert(\pi,\vep)}M}}$), we readily derive the commutator of the
(local) pre-quantum hamiltonians,
\qq\nn
[\widehat h^{\cB\,\vert\,\vep}_{\widetilde\Vgt_\igt},\widehat
h^{\cB\,\vert\,\vep}_{\widetilde\Wgt_\igt}]&=&-\sfi\,\widehat\cO_{\{
\,h^{\cB\,\vert\,\vep}_\Vgt,h^{\cB\,\vert\,\vep}_\Wgt\}_{\Om_{\si,\cB
\vert(\pi,\vep)}}}\big\vert_{\pr_{\sfL_{Q\vert(\pi,\vep)}M}^{-1}\bigl(
\cO_{(\pi,\vep)\,\igt}\bigr)}\cr\cr
&=&-\sfi\,\bigg(-\sfi\,\pLie{\a_{\sfT\sfP_{\si,\cB\,\vert\,(\pi,\vep)}}
\bigl(\ee^{-\theta_{\si,\cB\,\vert\,(\pi,\vep)\,\igt}}\lact\widetilde
\sfL^{Q\vert(\pi,\vep)}_{(1,1\sqcup 0)}\GBra{\Vgt}{\Wgt}^{(\txH,\om;
\D_Q)}\bigr)}\cr\cr
&&+\corr{\ee^{-\theta_{\si,\cB\,\vert\,(\pi,\vep)\,\igt}}\lact
\widetilde\sfL^{Q\vert(\pi,\vep)}_{(1,1\sqcup 0)}
\GBra{\Vgt}{\Wgt}^{(\txH,\om;\D_Q)},\Tgt_{\cB\,\vert\,(\pi,\vep)}}\bigg)
\cr\cr
&=&-\sfi\,\bigg(-\sfi\,\pLie{\a_{\sfT\sfP_{\si,\cB\,\vert\,(\pi,\vep)}}
\bigl(\ee^{-\theta_{\si,\cB\,\vert\,(\pi,\vep)\,\igt}}\lact
\Vbra{\widetilde\sfL^{Q\vert(\pi,\vep)}_{(1,1\sqcup 0)}
\Vgt}{\widetilde\sfL^{Q\vert(\pi,\vep)}_{(1,1\sqcup
0)}\Wgt}^{\widetilde\sfL^{Q\vert(\pi,\vep)\,*}\txH+\vep\,
\widetilde{\unl\sfL}^{Q\vert(\pi,\vep)\,*}\om}\bigr)}\cr\cr
&&+\corr{\ee^{-\theta_{\si,\cB\,\vert\,(\pi,\vep)\,\igt}}\lact
\Vbra{\widetilde\sfL^{Q\vert(\pi,\vep)}_{(1,1\sqcup 0)}
\Vgt}{\widetilde\sfL^{Q\vert(\pi,\vep)}_{(1,1\sqcup 0)}
\Wgt}^{\widetilde\sfL^{Q\vert(\pi,\vep)\,*}\txH+\vep\,
\widetilde{\unl\sfL}^{Q\vert(\pi,\vep)\,*}\om},\Tgt_{\cB\,\vert\,(
\pi,\vep)}}\bigg)\cr\cr
&=&-\sfi\,\bigg(-\sfi\,\pLie{\a_{\sfT\sfP_{\si,\cB\,\vert\,(\pi,\vep)}}
\bigl(\ee^{-\pr_{\sfT^*\sfL_{Q\vert(\pi,\vep)}M}^*\theta_{\sfT^*
\sfL_{Q\vert(\pi,\vep)}M}}\lact\Vbra{\widetilde
\Vgt_\igt}{\widetilde\Wgt_\igt}\bigr)}\cr\cr
&&+\corr{\ee^{-\pr_{\sfT^*\sfL_{Q\vert(\pi,\vep)}M}^*\theta_{\sfT^*
\sfL_{Q\vert(\pi,\vep)}M}}\lact\Vbra{\widetilde
\Vgt_\igt}{\widetilde\Wgt_\igt},\Tgt_{\cB\,\vert\,(
\pi,\vep)}}\bigg)\,,
\qqq
in conformity with \Reqref{eq:comm-preq-ham-tw}. This completes the
proof of the proposition. \eroof\medskip

It is natural to ask about the conditions under which the symplectic
realisation of the internal symmetries of the $\si$-model on the
twisted sector of the theory becomes hamiltonian. A clear-cut answer
is best phrased upon organising the symplectic data in hand in a
manner similar to the untwisted case. Thus,
\berop\label{prop:Htw-Vbra-symm-tw}
Adopt the notation of Definitions \ref{def:Vin-str} and
\ref{def:pair-tw-bra-str}, and of Proposition
\ref{prop:sisym-iotalign}. The subspace $\,\a_{\sfT(M\sqcup Q)}
\bigl(\G_{\iota_\a,\si}\bigl(\sfE^{(1,1)}M\sqcup\sfE^{(1,0)}Q\bigr)
\bigr)\,$ is a Lie subalgebra, to be denoted as $\,\ggt_\si$,\
within the Lie algebra of vector fields on $\,M\sqcup Q\,$ with a
Killing restriction to $\,M$.\ Fix a basis $\,\{\xcK_A\}_{A\in
\ovl{1,\dim\,\ggt_\si}}$,\ with restrictions $\,\xcK_A\vert_M=\Mup
\xcK_A\,$ and $\,\xcK_A\vert_Q=\Qup\xcK_A\,$ such that the defining
commutation relations
\qq\nn
[\xcK_A,\xcK_B]=f_{ABC}\,\xcK_C
\qqq
hold true for some structure constants $\,f_{ABC}$.\ Assuming that
condition \ref{eq:dom-DelH} is satisfied, the corresponding
$\si$-symmetric $\iota_\a$-aligned sections $\,\Kgt_A\,$ with
restrictions
\qq\nn
&\Kgt_A\vert_M=\Mup\xcK_A\oplus\kappa_A=:\Mup\Kgt_A\,,\qquad\qquad
\Kgt_A\vert_Q=\Qup\xcK_A\oplus k_A=:\Qup\Kgt_A\,,&\cr\cr\cr
&\left\{ \barr{l} \pLie{\Mup\xcK_A}\txg=0\cr\cr
\sfd_\txH\Mup\Kgt_A=0 \earr \right.\,,\qquad\qquad\qquad \left\{
\barr{l} \iota_{\a\,*}\Qup\xcK_A=\Mup\xcK_A\vert_{\iota_\a(Q)}\cr\cr
\sfd_\om\Qup\Kgt_A=-\D_Q\kappa_A \earr \right.&
\qqq
and the canonical contraction (with a trivial restriction to $\,Q$)
\qq\nn
\txc_{(AB)}=\Vcon{\Kgt_A}{\Kgt_B}\vert_M
\qqq
satisfy the relations
\qq\label{eq:Vinbra-Ka-tw}
\GBra{\Kgt_A}{\Kgt_B}^{(\txH,\om;\D_Q)}=f_{ABC}\,\Kgt_C+0
\oplus\a_{AB}
\qqq
with
\qq\nn
\a_{AB}\vert_\xcM=\left\{ \barr{ll}
\pLie{\Mup\xcK_A}\kappa_B-f_{ABC}\,\kappa_C-\sfd \txc_{(AB)} \quad &
\tx{on} \quad \xcM=M \cr\cr \pLie{\Qup\xcK_A}k_B-f_{ABC} \,k_C+\D_Q
\txc_{(AB)} \quad & \tx{on} \quad \xcM=Q \earr \right.\,.
\cr\cr
\qqq
\eerop
\beroof
Obvious, through inspection. \eroof \noindent The symplectic
realisation of the symmetries is further characterised in
\berop\label{prop:sympl-goes-ham-tw}
In the notation of Theorem \ref{thm:bib-as-morph}, of Lemma
\ref{lem:tw-paired-lifts}, of Theorem
\ref{thm:ind-quasi-morph-glob-Vin-tw}, and of Propositions
\ref{prop:ham-vs-sisymsec-tw} and \ref{prop:Htw-Vbra-symm-tw}, the
sections $\,\Kgt_A\,$ determine a symplectic realisation of
$\,\ggt_\si\,$ on $\,C^\infty\bigl(\sfP_{\si,\cB\,\vert\,(
\pi,\vep)},\bR\bigr)\,$ by hamiltonian functions $\,h^{\cB\,\vert\,
\vep}_{\Kgt_A}$,\ and an operator realisation of $\,\ggt_\si\,$ on
$\G\bigl(\ceL_{\si,\cB\,\vert\,(\pi,\vep)}\bigr)\,$ by pre-quantum
hamiltonians $\,\widehat\cO_{h^{\cB\,\vert\, \vep}_{\Kgt_A}}\,$ with
local restrictions $\,\widehat h^{\cB\,\vert\,
\vep}_{\widetilde\Kgt_A\,\igt}$.\ The former realisation is
hamiltonian,
\qq\label{eq:Poiss-class-ham-Ka-FM}
\{h^{\cB\,\vert\,\vep}_{\Kgt_A},h^{\cB\,\vert\,\vep}_{\Kgt_B}\,
\}_{\Om_{\si,\cB\,\vert\,(\pi,\vep)}}=f_{ABC}\,h^{\cB\,\vert\,
\vep}_{\Kgt_C}\,,
\qqq
iff the $\,\Kgt_A\,$ can be chosen such that
\qq\label{eq:FM-2}
\pLie{\Mup\xcK_A}\kappa_B=f_{ABC}\,\kappa_C+\sfd\Mup D_{AB}\,,\qquad
\qquad\pLie{\Qup\xcK_A}k_B=f_{ABC}\,k_C-\D_Q\Mup D_{AB}-\Qup D_{AB}
\qqq
for some $\,\Mup D_{AB}\in C^\infty(M,\bR)\,$ and (local) constants
$\,\Qup D_{AB}$.\ In this case also
\qq\label{eq:GBra-Ka-FM}
\GBra{\Kgt_A}{\Kgt_B}^{(\txH,\om;\D_Q)}=f_{ABC}\,\Kgt_C+0\oplus
\bigl(\sfd(\Mup D_{AB}- \txc_{(AB)}),-\D_Q(\Mup D_{AB}- \txc_{(AB)})
-\Qup D_{AB}\bigr)
\qqq
and
\qq\label{eq:tw-Vinbra-Kai-FM}
\Vbra{\widetilde\Kgt_{A\,\igt}}{\widetilde\Kgt_{B\,\igt}}=f_{ABC}\,
\widetilde\Kgt_{C\,\igt}\,.
\qqq
The latter identity then implies
\qq\label{eq:comm-quant-ham-Ka-FM}
[\widehat h^{\cB\,\vert\,\vep}_{\widetilde\Kgt_A\,\igt},\widehat
h^{\cB\,\vert\,\vep}_{\widetilde \Kgt_B\,\igt}]=-\sfi\,f_{ABC}
\,\widehat h^{\cB\,\vert\,\vep}_{\widetilde\Kgt_C\,\igt}\,.
\qqq
\eerop
\beroof
A proof is given in Section \ref{sub:proof}. It invokes some
elementary facts from the theory of singular and differential
(co)homology of paired manifolds $\,\alxydim{}{Q
\ar@<.5ex>[r]^{\iota_1} \ar@<-.5ex>[r]_{\iota_2} & M}\,$ of the kind
discussed in the last two sections. The relevant formalism will be
set up in Section \ref{sec:rel-cohom}. \eroof \noindent We have
established a physically motivated algebraic structure on the space
of (distinguished) sections of generalised tangent bundles over the
composite target space of the non-linear $\si$-model in the presence
of circular defects in the world-sheet. The structure can be
understood as a target-space model of the Poisson algebra of Noether
charges of rigid symmetries of the $\si$-model. Prior to giving it
an interpretation independent of the physical context of interest,
we pause to complete the canonical description of the symmetries for
a generic multi-phase $\si$-model, admitting the possibility of
self-intersecting defects.

\section{Intertwiners of the symmetry algebra from inter-bi-brane
data}\label{sec:intertwiner}

The physical Leitmotiv of the analysis carried out in the foregoing
sections was to understand mechanisms of symmetry transmission
across conformal defects, and -- in this manner -- to pave the way
to adding more structure to the correspondence between defects and
$\si$-model dualities worked out in Section \vref{sec:def-as-iso}I.4
by deriving constraints under which not merely the Virasoro modules
in the state space of the (quantum) theory but also their submodules
closed under the action of an extended current symmetry algebra are
mapped into one another by the symplectomorphism (resp. by the
endomorphism of the pre-quantum bundle) defined by the data of the
defect. In the present section, we bring this line of thought to its
logical conclusion and restate the questions concerning the fate of
the internal symmetries at the defect quiver in the setting of
Section \vref{sec:fusion}I.5, that is for state spaces under fusion.
Based on the findings of that section, it is well-justified to
expect that the data carried by defect junctions of those defect
quivers whose defect lines are transmissive to some internal
symmetries of the untwisted sector of the theory give rise to
intertwiners between representations of the symmetry algebra
furnished by the state spaces under fusion. This expectation will be
rendered rigorous and then proven below. For the sake of
transparency of the discussion, we shall restrict it to the simplest
non-trivial configurations of state spaces under fusion, to wit,
those studied in Section \vref{sec:fusion}I.5. For the same reason,
we shall also extract the physically relevant structures from the
extensive algebraic framework set up earlier in the paper and
proceed with our reasoning in a completely explicit fashion, leaving
a more abstract formulation of the results as an exercise for the
interested reader.\medskip

Cross-defect fusion processes generically involve non-trivial defect
junctions. Therefore, a prerequisite for our subsequent discussion
is a geometric description of infinitesimal (rigid) symmetries of
the $\si$-model on world-sheets with arbitrary embedded defect
quivers, which we can infer from Proposition
\ref{prop:var-sigmod-def}. Proposition \ref{prop:sigmod-symm-def},
valid for circular defects, is now generalised to
\berop\label{prop:sigmod-symm-def-junct}
Adopt the notation of Definitions \ref{def:Vin-str},
\ref{def:Om-tw-Vin} and \ref{def:sigmod-2d}. Denote by $\,\txH\in
Z^3(M)\,$ the curvature of the gerbe $\,\cG$,\ and write
\qq\label{eq:DelQ-and-DelTn}
\D_Q:=\iota_2^*-\iota_1^*\,,\qquad\qquad\D_{T_n}:=\sum_{k=1}^n\,
\vep_n^{k,k+1}\,\pi_n^{k,k+1\,*}\,.
\qqq
Infinitesimal rigid symmetries of the two-dimensional non-linear
$\si$-model for network-field configurations $\,(X\,\vert\,\G)\,$ in
string background $\,\Bgt\,$ on world-sheet $\,(\Si,\g)\,$ with a
defect quiver $\,\G$, as described in Definition
\vref{def:sigmod}I.2.7, correspond to triples $\,(\Mup\Vgt,\Qup
\Vgt,\Tnup\xcV)\,$ consisting of a $\si$-symmetric section $\,\Mup
\Vgt\in\G_\si\bigl(\sfE^{(1,1)}M\bigr)\,$ of $\,\sfE^{(1,1)}M$,\ as
defined by \Reqref{eq:sigmod-symm-bulk}, of a $\Mup\Vgt$-twisted
$\si$-symmetric section $\,\Qup\Vgt\in\G\bigl(
\sfE^{(1,0)}Q\bigr)\,$ of $\,\sfE^{(1,0)}Q$,\ as defined by
\Reqref{eq:sigmod-symm-def} and relations
\qq\label{eq:sigmod-symm-def-junct}
\D_{T_n}\pr_{C^\infty(Q,\bR)}(\Qup\Vgt)=0\,,
\qqq
written in terms of the canonical projection $\,\pr_{C^\infty(Q,\bR
)}:\sfE^{(1,0)}Q\to C^\infty(Q,\bR)$,\ and of a family of vector
fields $\,\Tnup \xcV\,$ on the respective manifolds $\,T_n$.\ These
are subject to the \textbf{$(\iota_\a,\pi_n^{k,k+1})$-alignment
conditions}
\qq\label{eq:iotapi-align}
\a_{\sfT M}(\Mup\Vgt)\vert_{\iota_\a(Q)}=\iota_{\a\,*}\a_{\sfT Q}(
\Qup\Vgt)\,,\qquad\qquad\a_{\sfT Q}(\Qup\Vgt)\vert_{\pi_n^{k,k+1}(Q
)}=\pi_{n\,*}^{k,k+1}\Tnup\xcV\,.
\qqq
\eerop
\noindent We are now ready to study at length the issue of charge
conservation at generic interaction vertices in the canonical
framework developed earlier.

As the first configuration of state spaces under fusion, we treat
the situation illustrated in Figure \vref{fig:fusion}I.5, that is we
consider pairs of states from the untwisted sector of the
$\si$-model fused across the defect. The first obvious issue is the
definition of the hamiltonian functions and of the corresponding
pre-quantum hamiltonians for those symmetries of the untwisted
sector which are transmitted across the defect, in the sense of
Proposition \ref{prop:sigmod-symm-def}.
\becor\label{cor:ham-fus-sub-untw}
Adopt the notation of Definitions \ref{def:Vin-str},
\ref{def:sigmod-2d} and \ref{def:pair-tw-bra-str}, of Propositions
\ref{prop:sympl-form-twuntw}, \ref{prop:ham-vs-sisymsec} and
\ref{prop:sisym-iotalign}, and of Theorem
\ref{thm:ind-quasi-morph-glob-Vin}. Let
$\,\sfP^{\circledast\cB}_{\si, \emptyset}\,$ be the $\cB$-fusion
subspace of the untwisted string from Definition
\vref{def:int-sub-untw}I.5.4, with the choice
$\,\cO_{\sfP^{\circledast\cB}_{\si,\emptyset}}=\{
\cO^{\circledast\cB}_{(\igt^1,\igt^2)}\}_{\igt^1,\igt^2\in
\xcI_{\sfL M}}\,$ of an open cover induced from the (sufficiently
fine) open cover $\,\cO_{\sfL M}\,$ in the manner detailed in the
proof of Theorem \vref{thm:cross-def-int-untw}I.5.5. Take an
arbitrary $\si$-symmetric $\iota_\a$-aligned section $\,\Vgt\in
\G_{\iota_\a,\si}\bigl(\sfE^{(1,1)}M\sqcup\sfE^{(1,0)}Q\bigr)\,$
with restrictions $\,\Vgt\vert_M=\Mup\xcV\oplus\upsilon\,$ and
$\,\Vgt \vert_Q=\Qup\xcV\oplus\xi$.\ Write $\,\sfI=[0,\pi]\,$ and
let $\,\tau:\bS^1\to\bS^1\,$ be the $\pi$-shift map from
\vReqref{eq:prev-shift-id}{I.5.2}. The hamiltonian function
$\,h^{\circledast\cB}_\Vgt\,$ on $\,\sfP_{\si,\emptyset}^{\x 2}\,$
associated to $\,\Vgt\,$ restricts to $\,\sfP^{\circledast\cB}_{\si
,\emptyset}\,$ as
\qq
h^{\circledast\cB}_\Vgt[(\psi_1,\psi_2)]&=&\int_\sfI\,\Vol(\sfI)\,
\bigl[\Mup\xcV\bigl(X_2(\cdot)\bigr)\con\sfp_2+(X_{2\,*}\widehat t)
\con\upsilon\bigl(X_2(\cdot)\bigr)\bigr]
\label{eq:class-ham-fus-untw}\\\cr
&&+\int_{\tau(\sfI)}\,\Vol\bigl(\tau(\sfI)\bigr)\,\bigl[\Mup\xcV
\bigl(X_1(\cdot)\bigr)\con\sfp_1+(X_{1\,*}\widehat t')\con\upsilon
\bigl(X_1(\cdot)\bigr)\bigr]+Y_{1,2}^*\xi(\pi)-Y_{1,2}^*\xi(0)\,,
\nonumber
\qqq
written in terms of the tangent vector field $\,\widehat t\,$ and
the volume form $\,\Vol(\sfI)\,$ on $\,\sfI$,\ the tangent vector
field $\,\widehat t'\,$ and the volume form $\,\Vol\bigl(\tau(
\sfI)\bigr)\,$ on $\,\tau(\sfI)$,\ and for an arbitrary pair $\,(
\psi_1,\psi_2)\in\sfP^{\circledast\cB}_{\si,\emptyset}\,$ of states,
represented by the respective Cauchy data $\,\psi_\a=(X_\a,\sfp_\a)
,\ \a\in\{1,2 \}\,$ and glued along the open path $\,Y_{1,2}\in\sfI
Q$.\ The corresponding pre-quantum hamiltonian for
$\,h^{\circledast\cB}_\Vgt$,\ constructed in conformity with
Definition \vref{def:prequantise}I.3.4, has local restrictions
\qq\label{eq:preq-ham-fus-untw}
\widehat\cO_{h^{\circledast\cB}_\Vgt}\vert_{\cO^{\circledast\cB}_{(\igt^1,\igt^2)}}=
-\sfi\,\pLie{(\Mup\widetilde\xcV,\Qup\widetilde\xcV)}-\bigl(\Mup\widetilde\xcV,\Qup\widetilde\xcV\bigr)\con\theta_{\si,
\circledast\cB\,(\igt^1,\igt^2)}+h^{\circledast\cB}_\Vgt=:\widehat
h^{\circledast \cB}_{\Vgt\,(\igt^1,\igt^2)}\,,
\qqq
expressed in terms of the restrictions
\qq\nn
\Mup\widetilde\xcV[(\psi_1,\psi_2)]&=&\int_\sfI\,\Vol(\sfI)\,\bigl[
\Mup\xcV^\mu\bigl(X_2(\cdot)\bigr)\,\tfrac{\d\quad\ }{\d X_2^\mu(
\cdot)}-\sfp_{2\,\mu}(\cdot)\,\p_\nu\Mup\xcV^\mu\bigl(X_2(\cdot)
\bigr)\,\tfrac{\d\quad\ }{\d\sfp_{2\,\nu}(\cdot)}\bigr]\cr\cr
&&+\int_{\tau(\sfI)}\,\Vol\bigl(\tau(\sfI)\bigr)\,\bigl[
\Mup\xcV^\mu\bigl(X_1(\cdot)\bigr)\,\tfrac{\d\quad\ }{\d X_1^\mu(
\cdot)}-\sfp_{1\,\mu}(\cdot)\,\p_\nu\Mup\xcV^\mu\bigl(X_1(\cdot)
\bigr)\,\tfrac{\d\quad\ }{\d\sfp_{1\,\nu}(\cdot)}\bigr]\cr\cr\cr
\Qup\widetilde\xcV[(\psi_1,\psi_2)]&=&\int_\sfI\,\Vol(\sfI)\,\Qup
\xcV^A\bigl(Y_{1,2}(\cdot)\bigr)\,\tfrac{\d\quad\ \ \ }{\d
Y_{1,2}^A(\cdot)}
\qqq
of the lift of the vector-field component of $\,\Vgt\,$ to
$\,\sfP_{\si, \emptyset}^{\x 2}$,\ and of the local data
$\,(\theta_{\si,
\circledast\cB\,(\igt^1,\igt^2)},\g_{\si,\circledast\cB\,(\igt^1,
\igt^2)(\jgt^1,\jgt^2)})$,\ derived in the proof of Theorem
\vref{thm:cross-def-int-untw}I.5.5, of the restriction $\,\ceL_{\si,
\circledast\cB}=\bigl(\pr_1^*\ceL_{\si,\emptyset}\ox\pr_2^*
\ceL_{\si,\emptyset}\bigr)\vert_{\sfP^{\circledast\cB}_{\si,
\emptyset}}\,$ to $\,\sfP^{\circledast\cB}_{\si,\emptyset}\,$ of the
tensor product of pullbacks of $\,\ceL_{\si,\emptyset}\,$ along the
canonical projections $\,\pr_\a:\sfP_{\si,\emptyset}^{\x 2}\to
\sfP_{\si,\emptyset}$. \ecor \beroof The formula for $\,h_\Vgt\,$
readily follows from the expression for the restricted symplectic
form $\,\ovl\Om^+_{\si,\emptyset}$,\ given in
\vReqref{eq:Om+-restr-fus}{I.D.1} and taken in conjunction with the
condition of the $\si$-symmetricity of $\,\Vgt$,\ whereas
\Reqref{eq:preq-ham-fus-untw} is a specialisation of the general
definition \veqref{eq:pre-ham-gen}{I.3.8}. \eroof\medskip

\noindent The last corollary forms the basis of the following
important result:
\bethe\label{thm:DJI-aug-intertw-untw}
Adopt the notation of Definitions \ref{def:Vin-str} and
\ref{def:sigmod-2d}, of Propositions \ref{prop:sympl-form-twuntw},
\ref{prop:ham-vs-sisymsec} and \ref{prop:sisym-iotalign}, and of
Corollary \ref{cor:ham-fus-sub-untw}. Let
$\,\Igt_\si(\circledast\cB:\cJ:\cB )\,$ be the $2\to 1$
cross-$(\cB,\cJ)$ interaction subspace of the untwisted string
within $\,\sfP_{\si,\emptyset}^{\x 3}=\sfP_{\si,
\emptyset}\x\sfP_{\si,\emptyset}\x\sfP_{\si,\emptyset}\,$ described
in that definition. Finally, let $\,\Vgt\,$ be a $\si$-symmetric
section from $\,\G_{\iota_\a,\si}\bigl(\sfE^{(1,1)}M\sqcup\sfE^{(1,0
)}Q\bigr)\,$ to which there are associated hamiltonian functions:
$\,h_\Vgt\,$ on $\,\sfP_{\si,\emptyset}\,$ and $\,h^{\circledast
\cB}_\Vgt\,$ on $\,\sfP^{\circledast\cB}_{\si,\emptyset}$.\ The
values attained by the pullbacks $\,\pr_3^*h_\Vgt\,$ and
$\,(\pr_1\x\pr_2)^*h^{\circledast\cB}_\Vgt \,$ along the canonical
projections $\,\pr_n:\sfP_{\si, \emptyset}^{\x
3}\to\sfP_{\si,\emptyset},\ n\in\{1,2,3\}\,$ coincide on
$\,\Igt_\si(\circledast\cB:\cJ:\cB)\,$ iff the condition
\qq\label{eq:DJI-for-xi}
\bigl(\pi^{1,2\,*}_3+\pi^{2,3\,*}_3-\pi^{3,1\,*}_3\bigr)
\pr_{C^\infty(Q,\bR)}(\Vgt)=0
\qqq
is satisfied on $\,T_3\,$ for the canonical projection
$\,\pr_{C^\infty(Q,\bR)}:\sfE^{(1,1)}M\sqcup\sfE^{(1,0)}Q\to
C^\infty(Q,\bR)$.\ Furthermore, assuming that $\,\Igt_\si(
\circledast\cB:\cJ:\cB)\,$ projects (canonically) onto each of the
three cartesian factors in $\,\sfP_{\si,\emptyset}^{\x 3}$,\ the
unitary similarity transformation between the set of pre-quantum
hamiltonians on $\,\sfP^{\circledast\cB}_{\si,\emptyset}\,$ and
those on $\,\sfP_{\si,\emptyset}\,$ defined by the bundle
isomorphism $\,\Jgt_{\si,(\circledast\cB:\cJ:\cB)}\,$ from Theorem
\vref{thm:cross-def-int-untw}I.5.5 preserves (element-wise) the
respective subalgebras composed of those pre-quantum hamiltonians
which are assigned, in the manner explicited in Proposition
\ref{prop:ham-vs-sisymsec} and Corollary \ref{cor:ham-fus-sub-untw},
respectively, to the $\iota_\a$-aligned $\si$-symmetric sections of
$\,\sfE^{(1,1)}M \sqcup\sfE^{(1,0)}Q\,$ iff the same condition holds
true. \ethe
\beroof
The proof goes along the same lines as those of Propositions
\ref{prop:ham-cont-across} and \ref{prop:ham-simil-across}. Take a
$\iota_\a$-aligned $\si$-symmetric section $\,\Vgt\,$ with
restrictions $\,\Vgt\vert_M=\Mup\xcV\oplus\upsilon\,$ and $\,\Vgt
\vert_Q=\Qup\xcV\oplus\xi$.\ In the classical setting, we compute,
substituting the defining relations
\veqref{eq:nfix-half}{I.5.9}-\veqref{eq:loop-nfix-half}{I.5.11} of
$\,\Igt_\si( \circledast\cB:\cJ:\cB)\ni(\psi_1,\psi_2,\psi_3),\
\psi_n=(X_n, \sfp_n),\ n\in\{1,2,3\}\,$ in
\Reqref{eq:class-ham-fus-untw},
\qq\nn
h^{\circledast\cB}_\Vgt[(\psi_1,\psi_2)]&=&\int_\sfI\,\Vol(\sfI)\,\bigl[\Qup\xcV
\bigl(Y_{2,3}(\cdot)\bigr)\con(\sfp_2\circ\iota_{1\,*})+(Y_{2,3\,*}
\widehat t)\con\iota_1^*\upsilon\bigl(Y_{2,3}(\cdot)\bigr)\bigr]\cr
\cr
&&\hspace{-.5cm}+\int_{\tau(\sfI)}\,\Vol\bigl(\tau(\sfI)\bigr)\,
\bigl[\Qup\xcV\bigl(Y_{1,3}(\cdot)\bigr)\con(\sfp_1\circ\iota_{1\,
*})+(Y_{1,3\,*}\widehat t')\con\iota_1^*\upsilon\bigl(Y_{1,3}(\cdot)
\bigr)\bigr]+Y_{1,2}^*\xi(\pi)-Y_{1,2}^*\xi(0)\cr\cr
&=&\int_\sfI\,\Vol(\sfI)\,\bigl[\Qup\xcV\bigl(Y_{2,3}(\cdot)\bigr)
\con(\sfp_3\circ\iota_{2\,*})+(Y_{2,3\,*}\widehat t)\con\bigl(
\iota_1^*\upsilon-\Qup\xcV\con\om\bigr)\bigl(Y_{2,3}(\cdot)\bigr)
\bigr]\cr\cr
&&+\int_{\tau(\sfI)}\,\Vol\bigl(\tau(\sfI)\bigr)\,
\bigl[\Qup\xcV\bigl(Y_{1,3}(\cdot)\bigr)\con(\sfp_3\circ\iota_{2\,
*})+(Y_{1,3\,*}\widehat t')\con\bigl(\iota_2^*\upsilon-\Qup\xcV\con
\om\bigr)\bigl(Y_{1,3}(\cdot)\bigr)\bigr]\cr\cr
&&+Y_{1,2}^*\xi(\pi)-Y_{1,2}^*\xi(0)\cr\cr
&=&h_\Vgt[\psi_3]+Y_{2,3}^*\xi(\pi)-Y_{2,3}^*\xi(0)+Y_{1,3}^*\xi(0)
-Y_{1,3}^*\xi(\pi)+Y_{1,2}^*\xi( \pi)-Y_{1,2}^*\xi(0)\cr\cr
&=&h_\Vgt[\psi_3]+Z^*\bigl(\pi^{1,2\,*}_3+\pi^{2,3\,*}_3-\pi^{3,1\,
*}_3\bigr)\bigl(\xi(\pi)-\xi(0)\bigr)\,,
\qqq
whence the first statement of the theorem follows.

Passing to the pre-quantum r\'egime, fix an open cover
$\,\cO_{\Igt_\si(\circledast\cB:\cJ:\cB)}=\{\cO^*_{\igt^1}\x
\cO^*_{\igt^2}\x\cO^*_{\igt^3}\}_{\igt^1,\igt^2,\igt^3\in\xcI_{\sfL
M}}\,$ as in the proof of Theorem \vref{thm:cross-def-int-untw}I.5.5
and take the associated data
$\,\pr_n^*(\theta_{\si,\emptyset\,\igt^n},
\g_{\si,\emptyset\,\igt^n\jgt^n}),\ n\in\{1,2,3\}\,$ of the pullback
bundles $\,\pr_n^*\ceL_{\si,\emptyset}$,\ and those of the bundle
isomorphism $\,\Jgt_{\si,(\circledast\cB:\cJ:\cB)}\,$,\ denoted by
$\,f^{+-}_{\si\,(\igt^1,\igt^2,\igt^3)}\,$ and given in
\vReqref{eq:f+-iso}{I.D.4}. The latter determine a relation between
local sections $\,s_{(\igt^1,\igt^2)}:=\pr_1^*s_{\igt^1}\ox\pr_2^*
s_{\igt^2}:\cO_{\igt^1}^*\x\cO_{\igt^2}^*\x\cO^*_{\igt^3}\to\pr_1^*
\ceL_{\si,\emptyset}\ox\pr_2^*\ceL_{\si,\emptyset}\,$ and $\,\pr_3^*
s_{\igt^3}:\cO_{\igt^1}^*\x\cO_{\igt^2}^*\x\cO^*_{\igt^3}\to\pr_3^*
\ceL_{\si,\emptyset}\,$ over $\,\Igt_\si(\circledast\cB:\cJ:\cB)\,$
of the form
\qq\nn
s_{(\igt^1,\igt^2)}[(\psi_1,\psi_2)]=f^{+-}_{\si\,(\igt^1,\igt^2,
\igt^3)}[(\psi_1,\psi_2,\psi_3)]\cdot s_{\igt^3}[\psi_3]\,.
\qqq
Taking into account Eqs.\,\eqref{eq:preq-ham-fus-untw} and
\veqref{eq:f+-dlog}{I.D.3} and using the explicit formula
\eqref{eq:preq-ham-gen} for $\,\widehat h_{\widetilde\Vgt_{\igt^3}}[
\psi_3]$,\ we obtain
\qq\nn
&&\widehat h^{\circledast\cB}_{\Vgt\,(\igt^1,\igt^2)}[(\psi_1,
\psi_2)]\lact s_{(\igt^1,\igt^2)}[(\psi_1,\psi_2)]\cr\cr
&=&\bigl(-\sfi\,\pLie{(\Mup\widetilde\xcV,\Qup\widetilde\xcV)[(\psi_1,\psi_2)]}
\vert_{\psi_3=\const}-\sfi\,\pLie{\widetilde\sfL_*\Mup\xcV[\psi_3]}
\vert_{\psi_1,\psi_2=\const}-\bigl(\Mup\widetilde\xcV,\Qup\widetilde\xcV\bigr)\con\theta_{\si,
\circledast\cB\,(\igt^1,\igt^2)}[(\psi_1,\psi_2)]\cr\cr
&&+h^{\circledast\cB}_\Vgt[(\psi_1,\psi_2)]\bigr)f^{+-}_{\si\,(\igt^1,\igt^2,
\igt^3)}[(\psi_1,\psi_2,\psi_3)]\cdot s_{\igt^3}[\psi_3]\cr\cr
&=&f^{+-}_{\si\,(\igt^1,\igt^2,\igt^3)}[(\psi_1,\psi_2,\psi_3)]
\cdot\bigl[-\sfi\,\pLie{\widetilde\sfL_*\Mup\xcV[\psi_3]}
\vert_{\psi_1,\psi_2=\const}-\bigl(\Mup\widetilde\xcV,\Qup\widetilde\xcV\bigr)\con\theta_{\si,
\circledast\cB\,(\igt^1,\igt^2)}[(\psi_1,\psi_2)]\cr\cr
&&-\bigl(\bigl(\Mup\widetilde\xcV,\Qup\widetilde\xcV\bigr)[(\psi_1,\psi_2)]\vert_{\psi_3=\const}+
\widetilde\sfL_*\Mup\xcV[\psi_3]\vert_{\psi_1,\psi_2=\const}\bigr)
\con\sfi\,\d\log f^{+-}_{\si\,(\igt^1,\igt^2,\igt^3)}[(\psi_1,\psi_2
,\psi_3)]\cr\cr
&&+h^{\circledast\cB}_\Vgt[(\psi_1,\psi_2)]\bigr]s_{\igt^3}[\psi_3]\cr\cr
&=&f^{+-}_{\si\,(\igt^1,\igt^2,\igt^3)}[(\psi_1,\psi_2,\psi_3)]
\cdot\bigl(-\sfi\,\pLie{\widetilde\sfL_*\Mup\xcV[\psi_3]}
\vert_{\psi_1,\psi_2=\const}-\widetilde\sfL_*\Mup\xcV\con\theta_{\si,
\emptyset\,\igt^3}[\psi_3]+h_\Vgt[\psi_3]\cr\cr
&&+Z^*\bigl(\pi^{1,2\,*}_3+\pi^{2,3\,*}_3-\pi^{3,1\,
*}_3\bigr)\bigl(\xi(\pi)-\xi(0)\bigr)\bigr)s_{\igt^3}[\psi_3]
\qqq
The above simply restates, in the setting in hand, the general rule:
in the presence of an isomorphism of pre-quantum bundles, the only
obstruction to having pre-quantum hamiltonians preserved by a
similarity transformation induced from the isomorphism can come from
non-equality of the corresponding hamiltonian functions pulled back
to the graph of the underlying symplectomorphism. Thus, upon
imposing \Reqref{eq:DJI-for-xi}, and in that case only, we find
\qq\nn
\widehat h^{\circledast\cB}_{\Vgt\,(\igt^1,\igt^2)}[(\psi_1,\psi_2)
]\lact s_{(\igt^1,\igt^2)}[(\psi_1,\psi_2)]=f^{+-}_{\si\,(\igt^1,
\igt^2,\igt^3)}[(\psi_1,\psi_2,\psi_3)]\cdot\bigl(\widehat
h_{\widetilde\Vgt_{\igt^3}}[\psi_3]\lact s_{\igt^3}[\psi_3]\bigr)\,,
\qqq
as claimed. \eroof\medskip \brem The last result is the first
rigorous statement concerning the anticipated relation between the
geometric data carried by defect junctions of a defect quiver with
defect lines that are transmissive to some internal symmetries of
the untwisted sector of the $\si$-model and intertwiners of the
algebra of those symmetries realised on the multi-string state
space. It is also easily generalised to more complex interaction
schemes for untwisted states -- in particular, in the case of an
$n$-string analogon of the process considered, in which untwisted
states pass through a defect junction of valence $n$ the condition
of equality of the hamiltonian functions associated to a
$\iota_\a$-aligned $\si$-symmetric section $\,\Vgt\,$ from the above
proof (resp.\ of similarity of the corresponding pre-quantum
hamiltonians) for the incoming and outgoing states takes the form
\qq\label{eq:DJI-for-xi-gen}
\D_{T_n}\pr_{C^\infty(Q,\bR)}(\Vgt)=0\,.
\qqq
Comparing the latter with the characterisation of $\si$-model
symmetries in the presence of self-intersecting defects, given in
Proposition \ref{prop:sigmod-symm-def-junct}, we conclude that the
charges of a $\si$-model symmetry that is preserved in the presence
of a defect quiver are automatically additively conserved in the
processes of a cross-defect splitting-joining interaction.\erem

The conclusive piece of evidence in favour of the interpretation of
the inter-bi-brane data and the the associated DJI for transmissive
defects in terms of intertwiners of the algebra of symmetries of the
$\si$-model comes from the twisted sector, in which we consider (for
the sake of concreteness) the simple fusion pattern depicted in
Figure \vref{fig:tw-fusion}I.7, in which two 1-twisted states are
fused, whereupon a single 1-twisted state is produced. We have
\bethe\label{thm:ham-cons-tw}
Adopt the notation of Definitions \ref{def:Vin-str} and
\ref{def:sigmod-2d}, and of Propositions
\ref{prop:sympl-form-twuntw}, \ref{prop:sisym-iotalign} and
\ref{prop:ham-vs-sisymsec-tw}. Let $\,\sfP^{\circledast\cB_{\rm
triv}}_{\si,\cB\,\vert\,(\vep_1,\vep_2)}\,$ be the $\cB_{\rm
triv}$-fusion subspace of the 1-twisted string from Definition
\vref{def:int-sub-tw}I.5.7, and let $\,\Igt_\si(\circledast\cB_{\rm
triv}:\cJ:\cB_{\rm triv})^{\cB\,\vert\,( \vep_1,\vep_2,\vep_3)}\,$
be the $2\to 1$ cross-$(\cB_{\rm triv}, \cJ)$ interaction subspace
of the 1-twisted string within
$\,\sfP_{\si,\cB\,\vert\,(\vep_1,\vep_2,
\vep_3)}^{+-}=\sfP_{\si,\cB\,\vert\,(\pi,\vep_1)}\x\sfP_{\si,\cB\,\vert\,(
\pi,\vep_2)}\x\sfP_{\si,\cB\,\vert\,(\pi,\vep_2)}\,$ described in
that definition. Finally, let $\,\Vgt\,$ be a $\si$-symmetric
section from
$\,\G_{\iota_\a,\si}\bigl(\sfE^{(1,1)}M\sqcup\sfE^{(1,0)}Q \bigr)\,$
with restrictions $\,\Vgt\vert_M=\Mup\xcV\oplus \upsilon\,$ and
$\,\Vgt\vert_Q=\Qup\xcV\oplus\xi\,$ to which there are associated
hamiltonian functions: $\,h^{\cB\,\vert\,\vep_3}_\Vgt\,$ on
$\,\sfP_{\si,\cB\,\vert\,(\pi,\vep_3)}\,$ and
\qq\nn
h^{\cB\,\vert\,(\vep_1,\vep_2)}_\Vgt[(\psi_1,\psi_2)]&=&\int_\sfI\,
\Vol(\sfI)\,\bigl[\Mup\xcV\bigl(X_2(\cdot)\bigr)\con\sfp_2+(X_{2\,
*}\widehat t)\con\upsilon\bigl(X_2(\cdot)\bigr)\bigr]\cr\cr
&&+\int_{\tau(\sfI)}\,\Vol\bigl(\tau(\sfI)\bigr)\,\bigl[\Mup\xcV
\bigl(X_1(\cdot)\bigr)\con\sfp_1+(X_{1\,*}\widehat t')\con\upsilon
\bigl(X_1(\cdot)\bigr)\bigr]+\vep_1\,\xi(q_1)+\vep_2\,\xi(q_2)\,,
\qqq
on $\,\sfP^{\circledast\cB_{\rm
triv}}_{\si,\cB\,\vert\,(\vep_1,\vep_2 )}\ni(\psi_1,\psi_2),\
\psi_\a=(X_\a,\sfp_\a,q_\a,V_\a),\ \a\in\{1, 2\}$,\ the latter being
written in terms of the tangent vector field $\,\widehat t\,$ and
the volume form $\,\Vol(\sfI)\,$ on $\,\sfI$,\ as well as the
tangent vector field $\,\widehat t'\,$ and the volume form
$\,\Vol\bigl(\tau( \sfI)\bigr)\,$ on $\,\tau(\sfI)$.\ The values
attained by the pullbacks $\,\pr_3^*h^{\cB\,\vert\, \vep_3}_\Vgt\,$
and $\,(\pr_1,\pr_2)^*h^{\cB\,\vert\,(\vep_1,\vep_2 )}_\Vgt\,$ along
the canonical projections $\,\pr_3:\sfP_{\si,\cB
\vert(\vep_1,\vep_2,\vep_3)}^{+-}\to\sfP_{\si,\cB\,\vert\,(\pi,\vep_3
)}\,$ and
$\,(\pr_1,\pr_2):\sfP_{\si,\cB\,\vert\,(\vep_1,\vep_2,\vep_3
)}^{+-}\to\sfP^{\circledast\cB_{\rm triv}}_{\si,\cB\,\vert\,(\vep_1,
\vep_2)}\,$ coincide on $\,\Igt_\si(\circledast\cB_{\rm triv}:\cJ:
\cB_{\rm triv})^{\cB\,\vert\,(\vep_1,\vep_2,\vep_3)}\,$ iff
condition \eqref{eq:DJI-for-xi} is satisfied on $\,T_3$.\
Furthermore, assuming that $\,\Igt_\si(\circledast\cB:\cJ:\cB) \,$
projects (canonically) onto each of the three cartesian factors
$\,\sfP_{\si, \cB\,\vert\,(\pi,\vep_n)}\,$ in
$\,\sfP_{\si,\cB\,\vert\,(\vep_1,\vep_2, \vep_3)}^{+-}$,\ the
unitary similarity transformation between the set of pre-quantum
hamiltonians on $\,\sfP^{\circledast\cB_{\rm
triv}}_{\si,\cB\,\vert\,(\vep_1,\vep_2)}\,$ and those on
$\,\sfP_{\si, \cB\,\vert\,(\pi,\vep_3)}\,$ defined by the bundle
isomorphism $\,\Jgt_{\si,(\circledast\cB_{\rm triv}:\cJ:\cB_{\rm
triv})}^{\cB \vert(\vep_1,\vep_2,\vep_3)}\,$ from Theorem
\vref{thm:cross-def-int-tw}I.5.8 preserves (element-wise) the
respective subalgebras composed of those pre-quantum hamiltonians
which are assigned, in the canonical manner, to the
$\iota_\a$-aligned $\si$-symmetric sections of $\,\sfE^{(1,1)}M
\sqcup\sfE^{(1,0)}Q\,$ iff the same condition holds true. \ethe
\beroof The proof is a straightforward variation of the proof of
Theorem \ref{thm:DJI-aug-intertw-untw}. \eroof

\beg\label{eg:WZW-def}\textbf{Symmetry transmission across the
maximally symmetric WZW defects}\\[-8pt]

\noindent In order to prepare the ground for subsequent analysis of
the maximally symmetric WZW defects, described in
\Rcite{Runkel:2010} (\textit{cf.}\ also Example I.2.13 for the
notation used), let us first note that the sections of the
generalised tangent bundle $\,\sfE^{(1,1)}\xcG\,$ over the group
manifold of a Lie group $\,\xcG\,$ which define the isometries of
the Cartan--Killing metric and preserve the Cartan 3-form are given
by
\qq\nn
\Lgt_A=L_A\oplus\bigl(-\tfrac{\sfk}{8\pi}\,\tht^A_L\bigr)\,,\qquad
\qquad\Rgt_A=R_A\oplus\tfrac{\sfk}{8\pi}\,\tht^A_R
\qqq
in terms of the components $\,\theta_L^A\,$ (resp.\ $\,\theta_R^A$)
of the left-invariant (resp.\ right-invariant) Maurer--Cartan 1-form
$\,\theta_L=\theta_L^A\ox t_A\,$ (resp.\ $\,\theta_R=\theta_R^A\ox
t_A$) and of the standard left-invariant (resp.\ right-invariant)
vector fields $\,L_A\,$ (resp.\ $\,R_A$) dual to them. Here, the
$\,t_A\,$ are the generators of the Lie algebra $\,\ggt\,$ of
$\,\xcG\,$ obeying the structure relations
\qq\nn
[t_A,t_B]=\txf_{ABC}\,t_C\,,
\qqq
with $\,\txf_{ABC}\in\bC\,$ the structure constants of $\,\ggt$.\
The sections satisfy the simple $\txH_\sfk$-twisted
Vinogradov-bracket algebra
\qq\nn
\Vbra{\Lgt_A}{\Lgt_B}^{\txH_\sfk}=\txf_{ABC}\,\Lgt_C\,,\qquad
\qquad\Vbra{\Rgt_A}{\Rgt_B}^{\txH_\sfk}=\txf_{ABC}\,\Rgt_C\,,
\qquad\qquad\Vbra{\Lgt_A}{\Rgt_B}^{\txH_\sfk}=0\,.
\qqq
They generate the right and left regular translations on the group,
and so yield, through definition \Reqref{eq:symm-curr-def}, the
right and left Ka\v c--Moody currents $\,J_H=J_H^A\ox t_A,\ H\in\{L
,R\}$,\ respectively,
\qq\nn
J_{\Lgt_A}=-\tfrac{1}{4}\,J_R^A\,,\qquad\qquad J_{\Rgt_A}=-
\tfrac{1}{4}\,J_L^A\,.
\qqq
In virtue of Proposition \ref{prop:sympl-goes-ham-untw}, the
associated hamiltonian functions and pre-quantum hamiltonians
furnish two independent representations of the Lie algebra
$\,\ggt\,$ of the Lie group $\,\xcG$.\ Having thus made contact with
our previous considerations from Example I.2.13, we may now discuss
the reduction of the bulk symmetry in the presence of the
defects.\medskip

\noindent \emph{\textbf{Symmetries preserved by the boundary
$\cGk$-bi-brane.}} We begin with the boundary defect and the
attendant bi-brane $\,\cB^\p_\sfk$,\ for which the analysis
simplifies enormously: the tangent space to a conjugacy class
$\,\xcC_\la\subset Q^\p_\sfk\,$ is spanned by the axial combinations
$\,R_A-L_A\,$ of the basic right- and left-invariant vector fields
on the group, and so we should look for $\iota_\la$-aligned
$\si$-symmetric sections of $\,\sfE^{(1,1)}\xcG\sqcup\sfE^{(1,0)}
Q^\p_\sfk\,$ amidst those obtained from the corresponding
combinations
\qq\nn
\Rgt_A-\Lgt_A=(R_A-L_A)\oplus\tfrac{\sfk}{8\pi}\,(\tht^A_R+\tht^A_L
)
\qqq
in the bulk. The latter are readily checked to satisfy the
$\si$-symmetricity condition in the form
\qq\nn
(R_A-L_A)\con\om^\p_{\sfk,\la}-\tfrac{\sfk}{8\pi}\,\iota_\la^*(
\tht^A_L+\tht^A_R)=0\,.
\qqq
Thus, we obtain a basis of $\iota_\la$-aligned $\si$-symmetric
sections
\qq\nn
\Agt_A=(R_A-L_A,R_A-L_A)\oplus\tfrac{\sfk}{8\pi}\,(\tht^A_R+
\tht^A_L,0)\,,
\qqq
with the $(\txH_\sfk,\om^\p_\sfk;-\iota_{Q^\p_\sfk}^*)$-twisted
brackets
\qq\nn
\GBra{\Agt_A}{\Agt_B}^{(\txH_\sfk,\om^\p_\sfk;-\iota_{Q^\p_\sfk}^*
)}=\txf_{ABC}\,\Agt_C\,.
\qqq
We conclude that the symmetry preserved by the boundary maximally
symmetric WZW defect is the adjoint (axial) component of the
left-right symmetry of the defect-free theory, generated by the
currents
\qq\nn
J_{\Lgt_A}-J_{\Rgt_A}=\tfrac{1}{4}\,(J_L^A-J_R^A)\,,
\qqq
and that the hamiltonian functions and pre-quantum hamiltonians
assigned to the sections $\,\Agt_A\,$ furnish a representation of a
single copy of $\,\ggt$.\medskip

\noindent \emph{\textbf{Symmetries preserved by the non-boundary
$\cGk$-bi-brane.}} In the non-boundary case, the geometry of the
world-volume $\,Q_\sfk\,$ of the $\cGk$-bi-brane $\,\cB_\sfk$,\ in
conjunction with the choice of the maps $\,\iota_\a\,$ detailed in
Example I.2.13, offer -- via the tangent maps $\,\iota_{\a \,*}\,$
-- an unrestrained choice of linear combinations of the basic left-
and right-invariant vector fields on the target space. Indeed, one
easily verifies that the vector fields $\,{}^{\tx{\tiny
$Q_\sfk$}}\hspace{-2pt}L_A\,$ and $\,{}^{\tx{\tiny
$Q_\sfk$}}\hspace{-2pt}R_A\,$ with values
\qq\nn
{}^{\tx{\tiny $Q_\sfk$}}\hspace{-2pt}L_A(g,h)=L_A(g)+(L_A-R_A)(h)
\,,\qquad\qquad{}^{\tx{\tiny $Q_\sfk$}}\hspace{-2pt}R_A(g,h)=R_A(g)
\qqq
on the bi-brane world-volume $\,Q_\sfk\ni(g,h)\,$ push forward to
the vector fields $\,L_A\,$ and $\,R_A$,\ respectively,
\qq\nn
\iota_{\a\,*}{}^{\tx{\tiny
$Q_\sfk$}}\hspace{-2pt}L_A=L_A\,,\qquad\qquad\iota_{\a\,*}{}^{\tx{\tiny
$Q_\sfk$}}\hspace{-2pt}R_A=R_A\,.
\qqq
At this stage, it remains to calculate
\qq\nn
{}^{\tx{\tiny
$Q_\sfk$}}\hspace{-2pt}L_A\con\om_\sfk+\D_{Q_\sfk}\bigl(-\tfrac{\sfk}{8\pi}\,\tht^A_L\bigr)=0\,,\qquad
\qquad{}^{\tx{\tiny $Q_\sfk$}}\hspace{-2pt}R_A\con\om_\sfk+\D_{Q_\sfk}\bigl(\tfrac{\sfk}{8\pi}\,\tht^A_R\bigr)=0
\qqq
over $\,Q_\sfk$,\ whereupon a basis can be chosen in $\,\G_{\iota_A,
\si}\bigl(\sfE^{(1,1)}\xcG\sqcup\sfE^{(1,0)}Q_\sfk\bigr)\,$ with
elements
\qq\nn
\Lgt_A=\bigl(L_A,{}^{\tx{\tiny $Q_\sfk$}}\hspace{-2pt}L_A\bigr)
\oplus\bigl(-\tfrac{\sfk}{8\pi}\,\tht^A_L,0\bigr)\,,\qquad\qquad
\Rgt_A=\bigl(R_A,{}^{\tx{\tiny $Q_\sfk$}}\hspace{-2pt}R_A\bigr)
\oplus\bigl(\tfrac{\sfk}{8\pi}\,\tht^A_R,0\bigr)\,.
\qqq
In this basis, we find the $(\txH_\sfk,\om_\sfk;\D_{Q_\sfk}
)$-twisted brackets
\qq\nn
&\GBra{\Lgt_A}{\Lgt_B}^{(\txH_\sfk,\om_\sfk;\D_{Q_\sfk})}=\txf_{ABC}\,
\Lgt_C\,,\qquad\qquad\GBra{\Rgt_A}{\Rgt_B}^{(\txH_\sfk,
\om_\sfk;\D_{Q_\sfk})}=\txf_{ABC}\,\Rgt_C\,,&\cr\cr
&\GBra{\Lgt_A}{\Rgt_B}^{(\txH_\sfk,\om_\sfk;\D_{Q_\sfk})}=0\,.&
\qqq
We are thus led to conclude that the full left-right symmetry of the
defect-free theory is preserved by the defect. As the 0-form
components of the $\iota_\a$-aligned $\si$-symmetric sections are
trivial, \Reqref{eq:DJI-for-xi-gen} is satisfied, and so we have a
non-anomalous realisation of the symmetry algebra on multi-string
state spaces.
\eeg

\brem There is an important conclusion that can be drawn from our
presentation of the symmetries preserved by the non-boundary
$\cGk$-bi-brane, to wit, it transpires that whatever the
world-volume of the corresponding $(\cGk,\cB_\sfk)$-inter-bi-brane,
charges of the full $\ggt\oplus\ggt$-symmetry are going to be
additively conserved in arbitrary cross-defect interaction
processes. In the light of the world-sheet interpretation of such
processes, as illustrated, \textit{e.g.}, in Figure I.7, this
observation points to the existence of a straightforward
correspondence between junctions of the maximally symmetric WZW
defects and spaces of intertwiners of the action of the symmetry
group $\,\xcG\,$ of the (bulk) WZW model. This seems to fit nicely
with the classificatory results of \Rcite{Frohlich:2006ch}, where,
in particular, the defect junctions (of valence $n$) in the
\emph{quantised} WZW model were related to the so-called conformal
blocks for the ($n$-)punctured Riemann sphere. The remarkable
consistency between these results, derived in the rigorous
categorial quantisation scheme for the WZW $\si$-model, and our
conclusions, based on the canonical analysis conveyed entirely in
geometric terms, hinges on the identification, detailed in
\Rxcite{Sec.\,5}{Gawedzki:1999bq}, between the said conformal blocks
and certain distinguished $\xcG$-invariant tensors. Further evidence
of an apparent correspondence between classical and quantum
maximally symmetric WZW defect junctions is presented in
\Rcite{Runkel:2010}.\erem

\section{The complete twisted bracket structure, and relative
cohomology}\label{sec:rel-cohom}

In the preceding sections, we have amassed ample evidence in favour
of the identification of twisted bracket structures on (twisted)
generalised tangent bundles over the target space $\,M\,$ and the
bi-brane world-volume $\,Q\,$ of the $\si$-model for a world-sheet
with circular non-intersecting defect lines as the right
differential-algebraic constructs that carry complete information on
(infinitesimal) rigid symmetries of the physical theory of interest.
Below, we shall complete our description of the generalised geometry
of the target space of the multi-phase $\si$-model for world-sheets
with generic defect quivers by adjoining an appropriate structure on
the inter-bi-brane world-volume and thus defining an extension of
the previously introduced twisted bracket structure to (the
generalised tangent bundle over) the composite target space
$\,M\sqcup Q\sqcup T$.

The said extension, while well-justified from the physical vantage
point adopted in this paper, may still seem somewhat \emph{ad hoc}
to a more mathematically oriented reader. We shall attempt to amend
this situation in the second part of the present section by
reinterpreting the complete twisted bracket structure in terms of
the relative cohomology of the field space of the multi-phase
$\si$-model.

\subsection{The twisted bracket structure for the full string
background}\label{sub:tan-sheaf}

It proves helpful to begin the search for a \emph{natural}
completion of the hitherto construction by reappraising the
correspondence between generalised tangent bundles equipped with a
twisted bracket and generalised tangent bundles twisted by local
data of a geometric object (such as, \textit{e.g.}, a gerbe or a
circle bundle), equipped with an untwisted bracket. The existence of
Hitchin-type isomorphisms between the two structures strongly
suggests to regard the underlying geometry as that of a
\emph{sheaf}-theoretic extension of the tangent bundle, or -- to
enable a uniform treatment -- of the tangent sheaf $\,\cT\xcM\,$ of
a given manifold $\,\xcM$,\ \textit{cf.}, \textit{e.g.},
\Rcite{Ramanan:2004}. The choice of the sheaves to work with is
immediately indicated by the cohomological description of the
geometric objects entering the definitions of twisted generalised
tangent bundles encountered earlier. Thus, we are led to consider
the following differential complex:
\qq\label{eq:Del-comp}
\cT^*_\bullet\xcM\ :\ 0\xrightarrow{\ \sfd^{(-1)}\ }\cT^*_0\xcM
\xrightarrow{\ \sfd^{(0)}\ }\cT^*_1\xcM\xrightarrow{\ \sfd^{(1)}\ }
\cT^*_2\xcM\xrightarrow{\ \sfd^{(2)}\ }\cdots
\qqq
of differential sheaves:
\bit
\item $\cT^*_0\xcM:=\unl\bR$,\ the sheaf of locally constant
real-valued functions on $\,\xcM$;
\item $\cT^*_{q+1}\xcM:=\unl\Om^q(\xcM),\ q\in\bN$,\ the sheaf of
locally smooth $q$-forms on $\,\xcM$,
\eit
with the coboundary operators given by the zero map\footnote{We
introduce this map for the sake of consistency of the notation to be
used in the remainder of the paper.} $\,\sfd^{(-1)}$, \ the
canonical embedding $\,\sfd^{(0)}\ :\ \unl\bR\emb\unl\Om^0(\xcM)\,$
and the de Rham differentials $\,\sfd^{(q+1)}:=\sfd$.\ The above
complex contains a distinguished subcomplex:
\qq\label{eq:sub-Del-comp}
\sfT^*_\bullet\xcM\ :\ 0\xrightarrow{\ \sfd^{(-1)}\ }\sfT^*_0\xcM
\xrightarrow{\ \sfd^{(0)}\ } \sfT^*_1\xcM\xrightarrow{\ \sfd^{(1)}\
}\sfT^*_2\xcM\xrightarrow{\ \sfd^{(2)}\ }\cdots
\qqq
composed of
\bit
\item $\sfT^*_0\xcM:=\bR^{\pi_0(\xcM)}$,\ the bundle of real-valued
functions on $\,\xcM\,$ constant on its connected components (the
latter forming the set $\,\pi_0(\xcM)$);
\item $\sfT^*_{q+1}\xcM:=\Om^q(\xcM),\ q\in\bN$,\ the bundle of
smooth $q$-forms on $\,\xcM$.
\eit
Using these, and the tangent sheaf $\,\cT\xcM\,$ of $\,\xcM$,\ we
next introduce
\bedef\label{def:gen-tan-sheaf}
In the above notation, the \textbf{generalised tangent sheaf of type
$(1,q)$} is the direct sum
\qq\nn
\cE^{(1,q)}\xcM:=\cT\xcM\oplus\cT^*_q\xcM\,.
\qqq
It comes with the obvious \textbf{anchor (map)}
\qq\nn
\a_{\cT\xcM}\ :\ \cE^{(1,q)}\xcM\to\cT\xcM
\qqq
and the \textbf{canonical contraction}
\qq\nn
\Vcon{\cdot}{\cdot}\ &:&\ \G\bigl(\cE^{(1,q+1)}\xcM\bigr)\x\G\bigl(
\cE^{(1,q)}\xcM\bigr)\to\G\bigl(\cT^*_q\xcM\bigr)\ :\ (\xcV\oplus
\upsilon_i,\xcW\oplus\varpi_i)\mapsto\tfrac{1}{2}\,(\xcV\con
\varpi_i+\xcW\con\upsilon_i)\,,\cr\cr \Vcon{\cdot}{\cdot}\ &:&\
\G\bigl(\cE^{(1,m)}\xcM\bigr)\x\G\bigl(\cE^{(1,m)}\xcM\bigr)\to
\{0\}\ :\ (\xcV\oplus\upsilon_i,\xcW\oplus\varpi_i)\mapsto 0\,,
\qquad m\in\{0,1\}\,.
\qqq
The sheaf $\,\cE^{(1,q)}\xcM\,$ can be endowed with the
\textbf{Vinogradov bracket} $\,\Vbra{\cdot}{\cdot}^{(q)}\,$ defined
for $\,q>1\,$ and $\,q=1\,$ as in Eqs.\,\eqref{eq:Vin-bra-1n} and
\eqref{eq:Vin-bra-10}, respectively, and extended to $\,\cE^{(1,0)}
\xcM\,$ by embedding the latter in $\,\xcE^{(1,1)}\xcM$,\
\textit{i.e.} as per
\qq\nn
\Vbra{\xcV\oplus c_i}{\xcW\oplus d_i}^{(0)}=[\xcV,\xcW]\oplus 0\,.
\qqq
The bracket for $\,q\geq 1\,$ can be twisted by an arbitrary $(q+2
)$-form $\,\txH_{(q+2)}\in\Om^{q+2}(\xcM)\,$ as in
\Reqref{eq:tw-Vinbra}.

Upon restriction of the components of $\,\cE^{(1,q)}\xcM\,$ to the
respective smooth subsheaves, we obtain the \textbf{restricted
generalised tangent sheaf of type $\,(1,q)$}
\qq\nn
\widehat\sfE^{(1,q)}\xcM:=\sfT\xcM\oplus\sfT^*_q\xcM
\qqq
with the structure inherited from that on $\,\cE^{(1,q)}\xcM$.\ On
the latter, we may also induce the \textbf{$\txH_{(q+2)}$-twisted
Vinogradov structure}
\qq\nn
\widehat\Vgt^{(q),\txH_{(q+2)}}\xcM=\bigl(\widehat\sfE^{(1,q)}\xcM,
\Vbra{\cdot}{\cdot}^{\txH_{(q+2)}},\Vcon{\cdot}{\cdot},\a_{\sfT
\xcM}\bigr)\,.
\qqq
\exdef Clearly, whenever $\,\cE^{(1,q)}\,$ gives rise to a twisted
generalised tangent bundle (in the sense of Definition
\ref{def:tw-gen-tan-bun}), we may require that $\,\a_{\sfT\xcM}\,$
and $\,\Vcon{\cdot}{\cdot}\,$ be globally defined, and that
$\,\Vbra{\cdot}{\cdot}^{(q)}\,$ map pairs of sections into sections,
whereby we retrieve the familiar statements of Propositions
\ref{prop:Vin-str-glob-tw} and \ref{prop:tw-gen-tan-triv}. We shall
not pursue this issue further. Instead, we consider
\bedef\label{def:tw-bra-str-sheaf}
Adopt the notation of Definition \ref{def:gen-tan-sheaf}. Let $\,\{M
,Q,T_n\,\vert\,n\in\bN_{\geq 3}\}\,$ be a family of smooth
manifolds, equipped with a collection of smooth maps
\qq\nn
\iota_\a\ :\ Q\to M\,,\qquad\a\in\{1,2\}\qquad\qquad\pi_n^{k,k+1}\
:\ T_n\to Q\,,\qquad k\in\ovl{1,n}\,,
\qqq
satisfying the identity
\qq\label{eq:DelTn-DelQ}
\D_{T_n}\circ\D_Q=0
\qqq
for $\,\D_Q\,$ and $\,\D_{T_n}\,$ as in \Reqref{eq:DelQ-and-DelTn}
(for some fixed collection of signs $\,\vep_n^{k,k+1},\ k\in\ovl{1,
n}$), and with a collection of smooth differential forms $\,\txH_{(3
)}\in\Om^3(M),\ \txH_{(2)}\in\Om^2(Q)\,$ and $\,\txH_{(1)}^n\in
\Om^1(T_n)$.\ Write
\qq\nn
\xcF:=M\sqcup Q\sqcup\bigsqcup_{n\geq 3}\,T_n\,.
\qqq
Assume that the forms satisfy the \textbf{curvature descent
relations}
\qq\nn
\D_Q\txH_{(3)}=-\sfd\txH_{(2)}\,,\qquad\qquad\D_{T_n}\txH_{(2)}=-
\sfd\txH_{(1)}^n\,.
\qqq
The \textbf{$(\iota_\a,\pi_n^{k,k+1})$-paired restricted generalised
tangent sheaves} are defined as
\qq\nn
\widehat\sfE^{(1,2\sqcup 1\sqcup 0)}\xcF:=\widehat\sfE^{(1,2)}M
\sqcup\widehat\sfE^{(1,1)}Q\sqcup\bigsqcup_{n\geq 3}\,\widehat
\sfE^{(1,0)}T_n\to\xcF\,.
\qqq
We restrict to those sections $\,\Vgt=(\Mup\xcV,\Qup\xcV,\Tnup\xcV)
\oplus(\upsilon,\xi,c)\,$ thereof which are \textbf{$(\iota_\a,
\pi_n^{k,k+1})$-aligned}, \textit{i.e.}\ those obeying the
conditions
\qq\nn
\iota_{\a\,*}\Qup\xcV=\Mup\xcV\vert_{\iota_\a(Q)}\,,\qquad\qquad
\pi^{k,k+1}_{n\,*}\Tnup\xcV=\Qup\xcV\vert_{\pi_n^{k,k+1}(T_n)}\,,
\qqq
and which are subject to \textbf{section descent equations}
\qq\label{eq:sde}
\sfd^{(2)}_{\txH_{(3)}}(\Mup\xcV\oplus\upsilon)=0\,,\qquad\qquad
\sfd^{(1)}_{\txH_{(2)}}(\Qup\xcV\oplus\xi)=-\D_Q\upsilon\,,\qquad
\qquad\sfd^{(0)}_{\txH_{(1)}}(\Tnup\xcV\oplus c)=-\D_{T_n}\xi\,,
\qquad\qquad
\qqq
written in terms of the twisted differentials
\qq\nn
\sfd^{(q)}_{\txH_{(q+1)}}(\xcV\oplus v)\equiv\sfd^{(q)}v+\xcV\con
\txH_{(q+1)}\,,
\qqq
where, in particular, $\,\sfd^{(0)}c\equiv c$.\ We denote the set of
all these sections as $\,\G_{(\iota_\a,\pi_n^{k,k+1}),\sfd}\left(
\widehat\sfE^{(1,2\sqcup 1\sqcup 0)}\xcF\right)$.

The
\textbf{$(\txH_{(3)},\txH_{(2)},\txH_{(1)};\D_Q,\D_{T_n})$-twisted
bracket structure} on $\,\widehat\sfE^{(1,2\sqcup 1\sqcup 0)}\xcF\,$
is the quadruple
\qq\nn
\widehat\Mgt_{(\iota_\a,\pi_n^{k,k+1})}^{(2,1,0),(\txH_{(3)},
\txH_{(2)},\txH_{(1)};\D_Q,\D_{T_n})}:=\bigl(\widehat\sfE^{(1,2\sqcup
1\sqcup 0)}\xcF,\GBra{\cdot}{\cdot}^{(\txH_{(3)},\txH_{(2)},\txH_{(1
)};\D_Q,\D_{T_n})},\Vcon{\cdot}{\cdot},\a_{\sfT\xcF}\bigr)\,,
\qqq
with the anchor map and the canonical contraction restricting to the
anchor maps and canonical contractions of the component $\txH_{(q+2
)}$-twisted Vinogradov structures, and with the $(\txH_{(3)},\txH_{(
2)},\txH_{(1)};\D_Q,\D_{T_n})$-twisted bracket restricting as
\qq
\GBra{\Vgt}{\Wgt}^{(\txH_{(3)},\txH_{(2)},\txH_{(1)};\D_Q,\D_{T_n}
)}\vert_{M\sqcup Q}&=&\GBra{\Vgt\vert_{M\sqcup Q}}{\Wgt\vert_{M
\sqcup Q}}^{(\txH_{(3)},\txH_{(2)};\D_Q)}\,,\cr
&&\label{eq:tw-brac-2cat}\\
\GBra{\Vgt}{\Wgt}^{(\txH_{(3)},\txH_{(2)},\txH_{(1)};\D_Q,\D_{T_n}
)}\vert_{T_n}&=&\Vbra{\Vgt\vert_{T_n}}{\Wgt\vert_{T_n}}^{(0)}\,,
\nonumber
\qqq
where $\,\GBra{\cdot}{\cdot}^{(\txH_{(3)},\txH_{(2)};\D_Q)}\,$ is
the twisted bracket structure from Definition
\ref{def:pair-tw-bra-str}. \exdef \noindent We find
\berop\label{prop:2tw-bra-close}
In the notation of Definition \ref{def:tw-bra-str-sheaf}, the $(
\txH_{(3)},\txH_{(2)},\txH_{(1)};\D_Q,\D_{T_n})$-twisted bracket
closes on the set $\,\G_{(\iota_\a,\pi_n^{k,k+1}),\sfd}\left(
\widehat\sfE^{(1,2\sqcup 1\sqcup 0)}\xcF\right)\,$ of $(\iota_\a,
\pi_n^{k,k+1})$-aligned sections of the restricted generalised
tangent sheaves $\,\widehat\sfE^{(1,2\sqcup 1\sqcup 0)}\xcF\,$
subject to the section descent equations \eqref{eq:sde}.
\eerop
\beroof
The only thing that has to be demonstrated is the identity
\qq\nn
\sfd^{(0)}_{\txH_{(1)}}\GBra{\Vgt}{\Wgt}^{(\txH_{(3)},\txH_{(2)},
\txH_{(1)};\D_Q,\D_{T_n})}\vert_{T_n}=-\D_{T_n}\pr_{\sfT^*_1 Q}
\bigl(\GBra{\Vgt}{\Wgt}^{(\txH_{(3)},\txH_{(2)},\txH_{(1)};\D_Q,
\D_{T_n})}\vert_{M\sqcup Q}\bigr)\,,
\qqq
which, for $\,\Vgt=(\Mup\xcV,\Qup\xcV,\Tnup\xcV)\oplus(\upsilon,\xi,
c)\,$ and $\,\Wgt=(\Mup\xcW,\Qup\xcW,\Tnup\xcW)\oplus(\varpi,\z,d)$,
\ follows from
\qq\nn
&&-\D_{T_n}\bigl(\Qup\xcV\con\sfd\z-\Qup\xcW\con\sfd\xi+\Qup\xcV
\con\Qup\xcW\con\txH_{(2)}+\tfrac{1}{2}\,\bigl(\Qup\xcV\con\D_Q
\varpi-\Qup\xcW\con\D_Q\upsilon\bigr)\bigr)\cr\cr
&=&\Tnup\xcV\con\sfd(\Tnup\xcW\con\txH_{(1)})-\Tnup\xcW\con\sfd(
\Tnup\xcV\con\txH_{(1)})+\Tnup\xcV\con\Tnup\xcW\con\sfd\txH_{(1)}
\cr\cr
&=&[\Tnup\xcV,\Tnup\xcW]\con\txH_{(1)}\equiv\sfd^{(0)}_{\txH_{(1
)}}\bigl([\Tnup\xcV,\Tnup\xcW]\oplus 0\bigr)\,.
\qqq
\eroof \noindent The physical significance of the last result is
brought to the fore by the following simple consequence of
Propositions \ref{prop:sigmod-symm-def-junct} and
\ref{prop:2tw-bra-close}.
\becor\label{cor:full-GBra}
In the notation of Proposition \ref{prop:sigmod-symm-def-junct},
infinitesimal rigid symmetries of the two-dimensional non-linear
$\si$-model for network-field configurations $\,(X\,\vert\,\G)\,$ in
string background $\,\Bgt\,$ on world-sheet $\,(\Si,\g)\,$ with a
defect quiver $\,\G$, as described in Definition
\vref{def:sigmod}I.2.7, correspond to $(\iota_\a,\pi_n^{k,k+1}
)$-aligned sections of the restricted generalised tangent sheaves
$\,\widehat\sfE^{(1,2\sqcup 1\sqcup 0)}\xcF\,$ from $\,\ker\,
\pr_{\sfT^*_0 T_n}$,\ subject to the section descent equations with
\qq\nn
\txH_{(3)}\equiv\txH\,,\qquad\qquad\txH_{(2)}\equiv\om\,,\qquad
\qquad\txH_{(1)}\equiv 0\,,
\qqq
and such that their image under $\,\a_{\sfT M}\,$ is Killing for
$\,\txg$.\ Consequently, there exists a restricted $(\txH,\om,0;\D_Q
,\D_{T_n})$-twisted bracket structure $\,\widehat\Mgt_{(\iota_\a,
\pi_n^{k,k+1})}^{(2,1,0),(\txH,\om,0;\D_Q,\D_{T_n})}\,$ on the set
of these sections. \ecor We also readily establish, upon putting
together Proposition \ref{prop:Htw-Vbra-symm-tw} and Corollary
\ref{cor:full-GBra},
\becor\label{cor:Homtw-Vbra-symm-tw}
Adopt the notation of Definitions \ref{def:gen-tan-sheaf} and
\ref{def:tw-bra-str-sheaf}. The subspace within $\,\G(\sfT\xcF)\,$
given by $\,\a_{\sfT\xcF}\left(\G_{(\iota_\a,\pi_n^{k,k+1}),\sfd}
\left(\widehat\sfE^{(1,2\sqcup 1\sqcup 0)}\xcF\right)\right)\,$ is a
Lie subalgebra, to be denoted as $\,\ggt_\si$,\ within the Lie
algebra of vector fields on $\,\xcF=M \sqcup Q\sqcup T\,$ with a
Killing restriction to $\,M$.\ Fix a basis $\,\{\xcK_A\}_{A\in\ovl{1
,\dim\,\ggt_\si}}$,\ with restrictions $\,\xcK_A\vert_M=\Mup\xcK_A,\
\xcK_A\vert_Q=\Qup \xcK_A\,$ and $\,\xcK_A\vert_{T_n}=\Tnup\xcK_A\,$
such that the defining commutation relations
\qq\nn
[\xcK_A,\xcK_B]=f_{ABC}\,\xcK_C
\qqq
hold true for some structure constants $\,f_{ABC}$.\ The
corresponding $\si$-symmetric $(\iota_\a,\pi_n^{k,k+1})$-aligned
sections $\,\Kgt_A\,$ of the restricted generalised tangent sheaves
$\,\widehat\sfE^{(1,2\sqcup 1\sqcup 0)}\xcF\,$ with restrictions
\qq\nn
&\Kgt_A\vert_M=\Mup\xcK_A\oplus\kappa_A=:\Mup\Kgt_A\,,\qquad\qquad
\Kgt_A\vert_Q=\Qup\xcK_A\oplus k_A=:\Qup\Kgt_A\,,\qquad\qquad
\Kgt_A\vert_{T_n}=\Tnup\xcK_A\oplus 0=:\Tnup\Kgt_A\,,&\cr\cr\cr
&\left\{ \barr{l} \pLie{\Mup\xcK_A}\txg=0\cr\cr
\sfd_\txH\Mup\Kgt_A=0 \earr \right.\,,\qquad\qquad\left\{ \barr{l}
\iota_{\a\,*}\Qup\xcK_A=\Mup\xcK_A\vert_{\iota_\a(Q)}\cr\cr
\sfd_\om\Qup\Kgt_A+\D_Q\kappa_A=0 \earr \right.\,,\qquad\qquad
\left\{ \barr{l} \pi_{n\,*}^{k,k+1}\Tnup\xcK_A=\Qup\xcK_A
\vert_{\pi_n^{k,k+1}(T_n)}\cr\cr
\D_{T_n}k_A=0 \earr \right.&
\qqq
and the canonical contraction (with a trivial restriction to $\,Q
\sqcup T$)
\qq\nn
\txc_{(AB)}=\Vcon{\Kgt_A}{\Kgt_B}\vert_M
\qqq
satisfy the relations
\qq\label{eq:full-GBra-Ka-tw}
\GBra{\Kgt_A}{\Kgt_B}^{(\txH,\om,0;\D_Q,\D_{T_n})}=f_{ABC}\,\Kgt_C+
0\oplus\a_{AB}
\qqq
with
\qq\nn
\a_{AB}\vert_\xcM=\left\{ \barr{ll}
\pLie{\Mup\xcK_A}\kappa_B-f_{ABC}\,\kappa_C-\sfd \txc_{(AB)} \quad &
\tx{on} \quad \xcM=M \cr\cr \pLie{\Qup\xcK_A}k_B-f_{ABC} \,k_C+\D_Q
\txc_{(AB)} \quad & \tx{on} \quad \xcM=Q \cr\cr 0 \quad & \tx{on}
\quad \xcM=T_n \earr \right.\,.
\cr\cr
\qqq
\ecor

\subsection{The relative-cohomological interpretation}

Our derivation of the bracket structure on sections of the
restricted generalised tangent sheaves, while essentially devoid of
ambiguities, leaves us with a rather non-obvious definition of the
$(\txH,\om;\D_Q)$-twisted bracket, and hence also with an open
question as to the underlying algebraic structure. Below, we
reinterpret the definition in terms of the relative differential
geometry of the target space encoded in the sequence of smooth
(inter-)bi-brane maps
\qq\label{eq:rel-targeom}
\alxydim{@C=1.5cm@R=.05cm}{T\supset T_n \ar@<1ex>[r]^{\pi_n^{k,k+
1}}_{\ldots} \ar@<-.25ex>[r] \ar@<-1ex>[r] & Q
\ar@<.5ex>[r]^{\iota_\a} \ar@<-.5ex>[r] & M}
\qqq
subject to constraints \vref{eq:proto-simpl}(I.2.1). The latter
immediately suggests extending the standard de Rham complex of
$\,\xcF\,$ (resp.\ its dual) in the direction of structural
(\textit{e.g.}, categorial) descent indicated by the arrows in the
above diagram. This line of reasoning has found its application in
the cohomological discussion of gauge anomalies and inequivalent
gaugings presented in \Rxcite{Sec.\,11}{Gawedzki:2012fu}. Here, we
take it up anew with view to elucidating the bracket structure.

The naturalness of the appearance of relative (co)homology in a
rigorous description of target-space structures associated with
world-sheet defects of the two-dimensional $\si$-model was pointed
up and made clear already in \Rcite{Klimcik:1996hp} and,
subsequently, in Refs.\,\cite{Gawedzki:1999bq,Figueroa:2000kz},
where tensorial data of a boundary bi-brane were neatly organised
and classified in terms of relative cohomology of the pair
$\,(M,D)\,$ consisting of the target space $\,M\,$ and its
distinguished submanifold $\,\iota_D\ :\ D\emb M$,\ identified with
the world-volume of a D-brane, that supports a (global) primitive
$\,\om\in\Om^2(D)\,$ of the restricted Kalb--Ramond 3-form
$\,\txH\in\Om^3(M)\,$ (the gerbe curvature) and thus gives rise to a
$\iota_D^*$-relative de Rham 3-cocycle $\,\txH\oplus\om$,\
\qq\nn
\sfd^{(3)}_{\iota_D^*}(\txH\oplus\om):=\sfd\txH\oplus(-\sfd\om+
\iota_D^*\txH)=0\,.
\qqq
The approach pioneered by Klim\v c\'ik and \v Severa was later
adapted to the study of a distinguished class of non-boundary
bi-branes, with world-volumes $\,Q\subset M_1\x M_2\,$ embedded in
the cartesian product of the target spaces $\,M_\a,\ \a\in\{1,2\}\,$
assigned to the world-sheet patches on either side of the relevant
defect line, in \Rcite{Fuchs:2007fw} where a gerbe-theoretic
description of this class of defects was proposed. Below, we rework
the original argument of Fuchs \emph{et al.} in a manner that allows
for its generalisation to arbitrary string backgrounds, as
introduced in \Rcite{Runkel:2008gr}.\medskip

\subsubsection{\textbf{The cohomology for the target space in the
presence of defects}}

In what follows, we give a construction of a target-space
(co)homology underlying the definition of the two-dimensional
$\si$-model in the presence of defects admitting self-intersections.
To these ends, we extend to the more general setting of interest
(and in the spirit of \Rxcite{Sec.\,7}{Bott:1982}) the construction
advanced in \Rxcite{App.\,A}{Fuchs:2007fw}.

A natural point of departure in a systematic discussion of the
cohomology of the hierarchy of target-space geometries
\eqref{eq:rel-targeom} and its realisation in terms of differential
forms is the introduction of the relevant (singular) homology. Thus,
we begin with
\bedef\label{def:DQT-rel-hom}
Let $\,(M,Q,T_n),\ n\in\bN_{\geq 3}\,$ be a triple of smooth
manifolds, equipped with a collection of smooth maps $\,\iota_\a\ :\
Q\to M,\ \a\in\{1,2\}\,$ and $\,\pi_n^{k,k+1}\ :\ T_n\to Q,\ k\in\bZ
/n\bZ\,$ subject to the constraints
\qq\nn
\iota_2^{\vep_n^{k-1,k}}\circ\pi_n^{k-1,k}=\iota_1^{\vep_n^{k,k+1}}
\circ\pi_n^{k,k+1}\,,
\qqq
written, for some fixed choice of signs\footnote{Here, we are
abusing the original conventions of
\Rxcite{Sec.\,2.5}{Runkel:2008gr} slightly by denoting the component
of $\,T_n\,$ corresponding to the fixed choice of signs
$\,\vep_n^{k,k+1}\,$ with the same symbol.} $\,\vep_n^{k,k+1}$,\ in
the conventions of Definition \vref{def:bckgrnd}I.2.1. Moreover, let
$\,C_k(M),C_k(Q)\,$ and $\,C_k(T_n)\,$ be the respective chain
groups of the singular chain complexes $\,C_\bullet(M),C_\bullet(Q
)\,$ and $\,C_\bullet(T_n)$,\ equipped with the respective boundary
operators $\,\p_{(k)}^M,\p_{(k)}^Q\,$ and $\,\p_{(k)}^{T_n}$.\ Write
\qq\nn
\D^Q:=\iota_{2\,\sharp}-\iota_{1\,\sharp}\,,\qquad\qquad\D^{T_n}:=
\sum_{k=1}^n\,\vep_n^{k,k+1}\,\pi_{n\,\sharp}^{k,k+1}
\qqq
for the two combinations of pushforward maps $\,\iota_{\a\,
\sharp}\,$ and the $\,\pi_{n\,\sharp}^{k,k+1}\,$ on singular chains
induced by the $\,\iota_\a\,$ and the $\,\pi_n^{k,k+1}$,\
respectively. The \textbf{$k$-th $(\D^Q,\D^{T_n})$-relative chain
group} is defined as
\qq\label{eq:DQT-rel-chains}
C_k\left(M,Q,T_n\,\vert\,\D^Q,\D^{T_n}\right):=C_k(M)\oplus C_{k-1}(
Q)\oplus C_{k-2}(T_n)
\qqq
and the associated \textbf{$(\D^Q,\D^{T_n})$-relative boundary
operators} are given by
\qq\nn
\p^{(\D^Q,\D^{T_n})}_{(k)}\ &:&\ C_k\left(M,Q,T_n\,\vert\,\D^Q,
\D^{T_n}\right)\to C_{k-1}\left(M,Q,T_n\,\vert\,\D^Q,\D^{T_n}
\right)\cr\cr
&:&\ c_k^M\oplus c_{k-1}^Q\oplus c_{k-2}^{T_n}\mapsto\left(\p^M_{(k
)}c_k^M-\D^Q c_{k-1}^Q\right)\oplus\left(-\p^Q_{(k-1)}c_{k-1}^Q-
\D^{T_n}c_{k-2}^{T_n}\right)\oplus\p_{(k-2)}^{T_n}c_{k- 2}^{T_n}\,.
\qqq
They satisfy the fundamental relation
\qq\nn
\p^{(\D^Q,\D^{T_n})}_{(k)}\circ\p^{(\D^Q,\D^{T_n})}_{(k+1)}=0\,,
\qqq
and so they give rise to the \textbf{$(\D^Q,\D^{T_n})$-relative
(singular) chain complex}
\qq\nn
C_\bullet\left(M,Q,T_n\,\vert\,\D^Q,\D^{T_n}\right):=\bigoplus_{k
\geq 0}\,C_k\left(M,Q,T_n\,\vert\,\D^Q,\D^{T_n}\right)\,,
\qqq
\textit{i.e.}\ the total complex of the semi-bounded bicomplex
$\,C_\bullet( \xcM_\bullet)\,$ with
$\,(\xcM_1,\xcM_2,\xcM_3):=(T_n,Q,M)$,\ defined as
\qq\nn 
\alxydim{@C=1.cm@R=1.cm}{\cdots \ar[r]^{\p^{T_n}_{(4)}\quad} &
C_3(T_n) \ar[r]^{\p^{T_n}_{(3)}} \ar[d]^{\D^{T_n}} & C_2(T_n)
\ar[r]^{\p^{T_n}_{(2)}} \ar[d]^{\D^{T_n}} & C_1(T_n)
\ar[r]^{\p^{T_n}_{(1)}} \ar[d]^{\D^{T_n}} & C_0(T_n)
\ar[r]^{\quad\p^{T_n}_{(0)}} \ar[d]^{\D^{T_n}} & 0 \\ \cdots
\ar[r]^{\p^Q_{(4)}\quad} & C_3(Q) \ar[r]^{\p^Q_{(3)}} \ar[d]^{\D^Q}
& C_2(Q) \ar[r]^{\p^Q_{(2)}} \ar[d]^{\D^Q} & C_1(Q)
\ar[r]^{\p^Q_{(1)}} \ar[d]^{\D^Q} & C_0(Q) \ar[r]^{\quad\p^Q_{(0)}}
\ar[d]^{\D^Q} & 0 \\ \cdots \ar[r]^{\p^M_{(4)}\quad} & C_3(M)
\ar[r]^{\p^M_{(3)}} & C_2(M) \ar[r]^{\p^M_{(2)}} & C_1(M)
\ar[r]^{\p^M_{(1)}} & C_0(M) \ar[r]^{\quad\p^M_{(0)}} & 0}\,.
\qqq
Its $k$-th homology group
\qq\nn
H_k\left(M,Q,T_n\,\vert\,\D^Q,\D^{T_n}\right):=\frac{\ker\,
\p^{(\D^Q,\D^{T_n})}_{(k)}}{\im \,\p^{(\D^Q,\D^{T_n})}_{(k+1)}}
\qqq
will be termed the \textbf{$k$-th $(\D^Q,\D^{T_n})$-relative
homology group}. \exdef \noindent The dual structure is introduced
in
\bedef\label{def:DQT-rel-cohom}
In the notation of Definition \ref{def:DQT-rel-hom}, for $\,\D_Q=
\iota_2^*-\iota_1^*\,$ the dual of $\,\D^Q$,\ for $\,\D_{T_n}=
\sum_{k=1}^n\,\vep_n^{k,k+1}\,\pi_n^{k,k+1\,*}\,$ the dual of
$\,\D^{T_n}$,\ and for $\,R\,$ a ring, the \textbf{$k$-th
$(\D_Q,\D_{T_n})$-relative cochain group with values in $\,R\,$} is
defined as
\qq\nn
C^k\left(M,Q,T_n\,\vert\,\D_Q,\D_{T_n};R\right):=\mor_{R-\Mod}
\left(C_k \left(M,Q,T_n\,\vert\,\D^Q,\D^{T_n}\right),R\right)
\qqq
with $\,R-\Mod\,$ the category of $R$-modules. The attendant
\textbf{$(\D_Q,\D_{T_n})$-relative coboundary operators}
\qq\nn
\d_{(\D_Q,\D_{T_n})}^{(k)}\ :\ C^k\left(M,Q,T_n\,\vert\,\D_Q,
\D_{T_n};R\right)\to C^{k+1}\left(M,Q,T_n\,\vert\,\D_Q,\D_{T_n};R
\right)\,,
\qqq
defined by the duality relations (written for arbitrary $\,c^k\in
C^k\left(M,Q,T_n\,\vert\,\D_Q,\D_{T_n};R\right)\,$ and $\,c_{k+1}\in
C_{k+1}\left(M,Q,T_n\,\vert\,\D^Q,\D^{T_n}\right)$)
\qq\nn
\d_{(\D_Q,\D_{T_n})}^{(k)}c^k\left(c_{k+1}\right):=c^k\left(\p^{(
\D^Q,\D^{T_n})}_{(k+1)}c_{k+1}\right)\,,
\qqq
determine the \textbf{$(\D_Q,\D_{T_n})$-relative cochain complex
with values in $\,R$}
\qq\nn
C^\bullet\left(M,Q,T_n\,\vert\,\D_Q,\D_{T_n};R\right):=\bigoplus_{k
\geq 0}\,C^k\left(M,Q,T_n\,\vert\,\D_Q,\D_{T_n};R\right)
\qqq
and its $k$-th cohomology group
\qq\nn
H^k\left(M,Q,T_n\,\vert\,\D_Q,\D_{T_n};R\right):=\frac{\ker\,
\d_{(\D_Q,\D_{T_n})}^{(k)}}{\im\,\d_{(\D_Q,\D_{T_n})}^{(k-1)}}\,,
\qqq
to be termed the \textbf{$k$-th $(\D_Q,\D_{T_n})$-relative
(singular) cohomology group with values in $\,R$}. \exdef \noindent
It will prove useful, and -- indeed -- crucial for the discussion of
the situation without defect junctions, to consider also
\bedef\label{def:DQ-rel-co-hom}
Adopt the notation of Definitions \ref{def:DQT-rel-hom} and
\ref{def:DQT-rel-cohom}. The \textbf{$k$-th $\D^Q$-relative chain
group} is defined as
\qq\label{eq:DQ-rel-chains}
C_k\left(M,Q\,\vert\,\D^Q\right):=C_k(M)\oplus C_{k-1}(Q)
\qqq
and the associated \textbf{$\D^Q$-relative boundary operators} are
given by
\qq\nn
\p^{\D^Q}_{(k)}\ &:&\ C_k\left(M,Q\,\vert\,\D^Q\right)\to
C_{k-1}\left(M,Q\,\vert\,\D^Q\right)\cr\cr
&:&\ c_k^M\oplus c_{k-1}^Q\mapsto\left(\p^M_{(k)}c_k^M-\D^Q c_{k-
1}^Q\right)\oplus\left(-\p^Q_{(k-1)}c_{k-1}^Q\right)\,.
\qqq
They satisfy the relation
\qq\nn
\p^{\D^Q}_{(k)}\circ\p^{\D^Q}_{(k+1)}=0\,,
\qqq
and so they give rise to the \textbf{$\D^Q$-relative (singular)
chain complex}
\qq\nn
C_\bullet\left(M,Q\,\vert\,\D^Q\right):=\bigoplus_{k\geq 0}\,C_k
\left(M,Q\,\vert\,\D^Q\right)\,,
\qqq
\textit{i.e.}\ the total complex obtained by truncating the
bicomplex $\,C_\bullet(\xcM_\bullet)\,$ of Definition
\ref{def:DQT-rel-hom} as
\qq\nn 
\alxydim{@C=1.cm@R=1.cm}{\cdots \ar[r]^{\p^Q_{(4)}\quad} & C_3(Q)
\ar[r]^{\p^Q_{(3)}} \ar[d]^{\D^Q}
& C_2(Q) \ar[r]^{\p^Q_{(2)}} \ar[d]^{\D^Q} & C_1(Q)
\ar[r]^{\p^Q_{(1)}} \ar[d]^{\D^Q} & C_0(Q) \ar[r]^{\quad\p^Q_{(0)}}
\ar[d]^{\D^Q} & 0 \\ \cdots \ar[r]^{\p^M_{(4)}\quad} & C_3(M)
\ar[r]^{\p^M_{(3)}} & C_2(M) \ar[r]^{\p^M_{(2)}} & C_1(M)
\ar[r]^{\p^M_{(1)}} & C_0(M) \ar[r]^{\quad\p^M_{(0)}} & 0}\,.
\qqq
Its $k$-th homology group
\qq\nn
H_k\left(M,Q\,\vert\,\D^Q\right):=\frac{\ker\, \p^{\D^Q}_{(k)}}{\im
\,\p^{\D^Q}_{(k+1)}}
\qqq
will be termed the \textbf{$k$-th $\D^Q$-relative homology group}.

Analogously, the \textbf{$k$-th $\D_Q$-relative cochain group with
values in $\,R\,$} is defined as
\qq\nn
C^k\left(M,Q\,\vert\,\D_Q;R\right):=\mor_{R-\Mod}\left(C_k\left(M,Q
\,\vert\,\D^Q\right),R\right)\,.
\qqq
The attendant \textbf{$\D_Q$-relative coboundary operators}
\qq\nn
\d_{\D_Q}^{(k)}\ :\ C^k\left(M,Q\,\vert\,\D_Q;R\right)\to C^{k+1}
\left(M,Q\,\vert\,\D_Q;R\right)\,,
\qqq
defined by the duality relations (written for arbitrary $\,c^k\in
C^k\left(M,Q\,\vert\,\D_Q;R\right)\,$ and $\,c_{k+1}\in C_{k+1}
\left(M,Q\,\vert\,\D^Q\right)$)
\qq\nn
\d_{\D_Q}^{(k)}c^k\left(c_{k+1}\right):=c^k\left(\p^{\D^Q}_{(k+1)}
c_{k+1}\right)\,,
\qqq
determine the \textbf{$\D_Q$-relative cochain complex with values in
$\,R$}
\qq\nn
C^\bullet\left(M,Q\,\vert\,\D_Q;R\right):=\bigoplus_{k\geq 0}\,C^k
\left(M,Q\,\vert\,\D_Q;R\right)
\qqq
and its $k$-th cohomology group
\qq\nn
H^k\left(M,Q\,\vert\,\D_Q;R\right):=\frac{\ker\,\d_{\D_Q}^{(k
)}}{\im\,\d_{\D_Q}^{(k-1)}}\,,
\qqq
to be termed the \textbf{$k$-th $\D_Q$-relative (singular)
cohomology group with values in $\,R$}. \exdef \noindent The latter
cohomology is characterised in the following important
\berop\label{prop:lexact-DQ-rel-sing}
In the notation of Definition \ref{def:DQ-rel-co-hom}, the
(singular) cohomology groups $\,H^k(M;R)\,$ and $\,H^k(Q;R)$,\ and
the $\D_Q$-relative (singular) cohomology groups $\,H^k\left(M,Q\,
\vert\,\D_Q;R\right)$,\ all with values in ring $\,R$,\ fit into the
long exact sequence
\qq\nonumber\\
\begin{tikzpicture}[descr/.style={fill=white,inner sep=1.5pt}]
        \matrix (m) [
            matrix of math nodes,
            row sep=1.cm,
            column sep=1.cm,
            text height=1.5ex, text depth=0.25ex
        ]
        { \cdots & H^{k-1}(Q;R) & H^k\left(M,Q\,\vert\,\D_Q;R\right) & H^k(M;R) & \\
            & H^k(Q;R) & H^{k+1}\left(M,Q\,\vert\,\D_Q;R\right) & H^{k+1}(M;R) & \cdots \\
        };

        \path[overlay,->, font=\scriptsize,>=angle 45]
        (m-1-1) edge node[descr,yshift=2.3ex] {$B^{(k-1)}_{\D_Q;R}$} (m-1-2)
        (m-1-2) edge (m-1-3)
        (m-1-3) edge (m-1-4)
        (m-1-4) edge[out=355,in=175] node[descr,xshift=-15.ex,yshift=1.1ex] {$B^{(k)}_{\D_Q;R}$} (m-2-2)
        (m-2-2) edge (m-2-3)
        (m-2-3) edge (m-2-4)
        (m-2-4) edge node[descr,yshift=2.3ex] {$B^{(k+1)}_{\D_Q;R}$} (m-2-5);
\end{tikzpicture}\,, \label{eq:les-DQ-rel-sing-cohom}
\qqq
with the connecting homomorphisms
\qq\nn
B^{(k)}_{\D_Q;R}\ :\ H^k(M;R)\to H^k(Q;R)\ :\ [c^k_M]\mapsto[-\D^Q
\dagu c^k_M]
\qqq
defined as
\qq\nn
\D^Q\dagu c^k_M(c_k^Q):=c^k_M(\D^Q c_k^Q)
\qqq
for $\,c_k^Q\in C_k(Q)\,$ arbitrary.
\eerop
\beroof
The cohomology sequence is induced by the short exact sequence of
cochain groups
\qq\nn
0\to C^{k-1}(Q;R)\xrightarrow{\ \pr_{2\,(k)}^\dagger\ }C^k\left(M,Q
\,\vert\,\D_Q;R\right)\xrightarrow{\ \iota_{(k)}^\dagger\ }C^k(M;R)
\to 0\,,
\qqq
whose existence and properties stem from the fact that the (split)
short exact sequence of chain groups
\qq\nn
0\to C_k(M)\xrightarrow{\ \iota_{(k)}\ }C_k\left(M,Q\,\vert\,\D^Q
\right)\xrightarrow{\ \pr_{2\,(k)}\ }C_{k-1}(Q)\to 0\,,
\qqq
written in terms of the inclusion map $\,\iota_{(k)}\,$ and the
canonical projection map $\,\pr_{2\,(k)}\equiv\pr_2$,\ splits by
assumption, \textit{cf.}\ \Reqref{eq:DQ-rel-chains}. The former
sequence is obtained from the latter one through application of the
exact functor $\,\mor_{R-\Mod}(\cdot;R)$,\ and uses the dual maps
\qq\nn
\pr_{2\,(k)}^\dagger c^{k-1}_Q(c^M_k\oplus c^Q_{k-1})&:=&c^{k-1}_Q
\left(\pr_{2\,(k)}(c^M_k\oplus c^Q_{k-1})\right)=c^{k-1}_Q(c^Q_{k-
1})\,,\cr\cr \iota_{(k)}^\dagger c_{M,Q\vert\D_Q}^k(c_k^M)&:=&c_{M,Q
\vert\D_Q}^k\left(\iota_{(k)}(c_k^M)\right)=c_{M,Q\vert\D_Q}^k(c_k^M
\oplus 0)\,.
\qqq
Finally, the connecting (Bokshteyn) homomorphism is induced in the
usual manner upon noting that every $k$-cochain $\,c^k_M\,$ can be
written as
\qq\nn
c^k_M=\iota_{(k)}^\dagger(c^k_M\circ\pr_1)\,,
\qqq
and whenever it is co-closed, we find
\qq\nn
\d^{(k)}_{\D_Q}(c^k_M\circ\pr_1)=\pr_{2\,(k+1)}^\dagger(-\D^Q
{}^\dagger c^k_M)\,.
\qqq
\eroof \noindent Similarly, we establish
\berop\label{prop:lexact-DQT-rel-sing}
In the notation of Definitions \ref{def:DQT-rel-cohom} and
\ref{def:DQ-rel-co-hom}, the (singular) cohomology groups
$\,H^k(Q;R)$,\ the $\D_Q$-relative (singular) cohomology groups
$\,H^k\left(M,Q\,\vert\,\D_Q;R\right)\,$ and the
$(\D_Q,\D_{T_n})$-relative (singular) cohomology groups
$\,H^k\left(M,Q,T_n\,\vert\,\D_Q,\D_{T_n};R\right)$,\ all with
values in ring $\,R$,\ fit into the long exact sequence
\qq\nonumber\\
\begin{tikzpicture}[descr/.style={fill=white,inner sep=1.5pt}]
        \matrix (m) [
            matrix of math nodes,
            row sep=1.cm,
            column sep=.9cm,
            text height=1.5ex, text depth=0.25ex
        ]
        { \cdots & H^{k-2}(T_n;R) & H^k\left(M,Q,T_n\,\vert\,\D_Q,\D_{T_n};R\right) & H^k\left(M,Q\,\vert\,\D_Q;R\right) & \\
            & H^{k-1}(T_n;R) & H^{k+1}\left(M,Q,T_n\,\vert\,\D_Q,\D_{T_n};R\right) & H^{k+1}\left(M,Q\,\vert\,\D_Q;R\right) & \cdots \\
        };

        \path[overlay,->, font=\scriptsize,>=angle 45]
        (m-1-1) edge node[descr,xshift=-1.ex,yshift=2.5ex] {$B^{(k-1)}_{(\D_Q,\D_{T_n});R}$} (m-1-2)
        (m-1-2) edge (m-1-3)
        (m-1-3) edge (m-1-4)
        (m-1-4) edge[out=355,in=175] node[descr,xshift=-20.ex,yshift=1.3ex] {$B^{(k)}_{(\D_Q,\D_{T_n});R}$} (m-2-2)
        (m-2-2) edge (m-2-3)
        (m-2-3) edge (m-2-4)
        (m-2-4) edge node[descr,xshift=1.ex,yshift=2.5ex] {$B^{(k+1)}_{(\D_Q,\D_{T_n});R}$} (m-2-5);
\end{tikzpicture}\,,\nonumber\\\label{eq:les-DQT-rel-sing-cohom}
\qqq
with the connecting homomorphisms
\qq\nn
B^{(k)}_{(\D_Q,\D_{T_n});R}\ :\ H^k(M,Q\,\vert\,\D_Q;R)\to H^{k-1}(
T_n;R)\ :\ [c_{M,Q\vert\D_Q}^k]\mapsto[-\D^{T_n}_{(2)}{}^\dagger(
c^k_M\oplus c^{k-1}_Q)]
\qqq
defined as
\qq\nn
\D^{T_n}_{(2)}{}^\dagger c_{M,Q\vert\D_Q}^k(c_{k-1}^{T_n}):=c_{M,Q
\vert\D_Q}^k(0\oplus\D^{T_n}c_{k-1}^{T_n})
\qqq
for $\,c_{k-1}^{T_n}\in C_{k-1}(T_n)\,$ arbitrary.
\eerop
\beroof
The cohomology sequence is induced by the short exact sequence of
cochain groups
\qq\nn
0\to C^{k-2}(T_n;R)\xrightarrow{\ \widetilde\pr_{3\,(k)}^\dagger\ }
C^k\left(M,Q,T_n\,\vert\,\D_Q,\D_{T_n};R\right)\xrightarrow{\
\widetilde\iota_{(k)}^\dagger\ }C^k\left(M,Q\,\vert\,\D_Q;R\right)
\to 0\,,
\qqq
obtained, through application of the exact functor $\,\mor_{R-\Mod}(
\cdot;R)$,\ from the (split) short exact sequence of chain groups
\qq\nn
0\to C_k\left(M,Q\,\vert\,\D^Q\right)\xrightarrow{\ \widetilde
\iota_{(k)}\ }C_k\left(M,Q,T_n\,\vert\,\D^Q,\D^{T_n}\right)
\xrightarrow{\ \widetilde\pr_{3\,(k)}\ }C_{k-2}(T_n)\to 0\,,
\qqq
written in terms of the inclusion map $\,\widetilde\iota_{(k)}\,$
and the canonical projection map $\,\widetilde\pr_{3\,(k)}\equiv
\pr_3$.\ We have the dual maps
\qq\nn
\widetilde\pr_{3\,(k)}^\dagger c^{k-2}_{T_n}(c_k^M\oplus c_{k-1}^Q
\oplus c_{k-2}^{T_n})&:=&c^{k-2}_{T_n}(c_{k-2}^{T_n})\,,\cr\cr
\widetilde\iota_{(k)}^\dagger c_{M,Q,T_n\vert\D_Q,\D_{T_n}}^k(c_k^M
\oplus c_{k-1}^Q)&:=&c_{M,Q,T_n\vert\D_Q,\D_{T_n}}^k (c_k^M\oplus
c_{k-1}^Q\oplus 0)\,.
\qqq
The definition of the connecting homomorphism is, once again,
completely standard since every $\,c_{M,Q\vert\D_Q}^k\,$ can be
obtained as
\qq\nn
c_{M,Q\vert\D_Q}^k=\widetilde\iota_{(k)}^\dagger(c_{M,Q\vert\D_Q}^k
\circ\pr_{1,2})\,,
\qqq
and whenever it is ($\d^{(k)}_{\D_Q}$-)co-closed, we have
\qq\nn
\d^{(k)}_{(\D_Q,\D_{T_n})}(c_{M,Q\vert\D_Q}^k\circ\pr_{1,2})=
\widetilde\pr_{3\,(k)}^\dagger\left(-\D^{T_n}_{(2)}{}^\dagger
\right)c_{M ,Q\vert\D_Q}^k\,.
\qqq \eroof

It is convenient to have an explicit differential-geometric
realisation of the relative cohomologies defined above. The point of
departure towards establishing one will be the de Rham isomorphism
\qq\label{eq:deRham-iso}
[I^{(\bullet)}_\xcM]\ :\ H^\bullet_{\rm dR}(\xcM)\xrightarrow{\
\cong\ }H^\bullet(\xcM;\bR)
\qqq
between the de Rham cohomology and the $\bR$-valued singular
cohomology, induced from the cochain maps
\qq\nn
I^{(k)}_\xcM\ :\ \Om^k(\xcM)\to C^k(\xcM;\bR)\,,\qquad\qquad I^{(k
)}_\xcM(\om^k_\xcM)(c_k^\xcM):=\int_{c_k^\xcM}\,\om^k_\xcM\,.
\qqq
The only additional element needed for our construction is a
relative variant of the de Rham cohomology, to wit,
\bedef\label{def:DQT-rel-dR-cohom}
In the notation of Definitions \ref{def:DQT-rel-hom} and
\ref{def:DQT-rel-cohom}, and of \Reqref{eq:Del-comp}, the
\textbf{$k$-th $(\D_Q,\D_{T_n} )$-relative de Rham group} is the
vector space
\qq\nn
\Om^k_{\rm dR}\left(M,Q,T_n\,\vert\,\D_Q,\D_{T_n}\right)&:=&
\Om^k(M)\oplus\Om^{k-1}(Q)\oplus\Om^{k-2}(T_n)\,,\quad k\neq 0
\cr\cr
\Om^0_{\rm dR}\left(M,Q,T_n\,\vert\,\D_Q,\D_{T_n}\right)&:=&\Om^0(M
)\oplus\Om^{-1}(Q)
\qqq
with the additional convention that
\qq\nn
\Om^{-1}(\xcM):=\bR^{\pi_0(\xcM)}\,,\quad\xcM\in\{Q,T_n\}\,.
\qqq
The associated \textbf{$(\D_Q,\D_{T_n})$-relative coboundary
operators}
\qq\nn
\sfd_{(\D_Q,\D_{T_n})}^{(k)}\ &:&\ \Om^k_{\rm dR}\left(M,Q,T_n\,
\vert\,\D_Q\right)\to\Om^{k+1}_{\rm dR}\left(M,Q,T_n\,\vert\,\D_Q,
D_{T_n}\right)\cr\cr
&:&\ \om^k_M\oplus\om^{k-1}_Q\oplus\om^{k-2}_{T_n}\mapsto\sfd^{(k)}
\om^k_M\oplus(-\sfd^{(k-1)}\om^{k-1}_Q-\D_Q\om^k_M)\oplus(\sfd^{(k-
2)}\om^{k-2}_{T_n}-\D_{T_n}\om^{k-1}_Q)
\qqq
yield the \textbf{$(\D_Q,\D_{T_n})$-relative de Rham
complex}\footnote{We shall occasionally use the same name for the
pair $\,\left(\Om^k_{\rm dR}\left(M,Q,T_n\,\vert\,\D_Q,\D_{T_n}
\right),\sfd_{(\D_Q,\D_{T_n})}^{(k)}\right)$.}
\qq\nn
\Om^\bullet_{\rm dR}\left(M,Q,T_n\,\vert\,\D_Q,\D_{T_n}\right):=
\bigoplus_{k\geq 0}\,\Om^k_{\rm dR}\left(M,Q,T_n\,\vert\,\D_Q,
\D_{T_n}\right)\,,
\qqq
that is the total complex of the semi-bounded bicomplex
$\,\Om^\bullet(\xcM_\bullet)$,\ and its $k$-th cohomology group
\qq\nn
H^k_{\rm dR}\left(M,Q,T_n\,\vert\,\D_Q,\D_{T_n}\right):=\frac{\ker\,
\sfd_{(\D_Q,\D_{T_n})}^{(k)}}{\im\,\sfd_{(\D_Q,\D_{T_n})}^{(k-1)}}
\qqq
to be termed the \textbf{$k$-th $(\D_Q,\D_{T_n})$-relative de Rham
cohomology group}. \exdef \noindent Its truncated version is given
in
\bedef\label{def:DQ-rel-dR-cohom}
In the notation of Definitions \ref{def:DQ-rel-co-hom} and
\ref{def:DQT-rel-dR-cohom}, and of \Reqref{eq:Del-comp}, the
\textbf{$k$-th $\D_Q$-relative de Rham group} is the vector space
\qq\nn
\Om^k_{\rm dR}\left(M,Q\,\vert\,\D_Q\right):=\Om^k(M)\oplus\Om^{k-
1}(Q)\,.
\qqq
The associated \textbf{$\D_Q$-relative coboundary operators}
\qq\nn
\sfd_{\D_Q}^{(k)}\ &:&\ \Om^k_{\rm dR}\left(M,Q\,\vert\,\D_Q\right)
\to\Om^{k+1}_{\rm dR}\left(M,Q\,\vert\,\D_Q\right)\cr\cr
&:&\ \om^k_M\oplus\om^{k-1}_Q\mapsto\sfd^{(k)}\om^k_M\oplus(-
\sfd^{(k-1)}\om^{k-1}_Q-\D_Q\om^k_M)
\qqq
yield the \textbf{$\D_Q$-relative de Rham complex}\footnote{We shall
occasionally use the same name for the pair $\,\left(\Om^k_{\rm dR}
\left(M,Q\,\vert\,\D_Q\right),\sfd_{\D_Q}^{(k)}\right)$.}
\qq\nn
\Om^\bullet_{\rm dR}\left(M,Q\,\vert\,\D_Q\right):=\bigoplus_{k\geq
0}\,\Om^k_{\rm dR}\left(M,Q\,\vert\,\D_Q\right)
\qqq
and its $k$-th cohomology group
\qq\nn
H^k_{\rm dR}\left(M,Q\,\vert\,\D_Q\right):=\frac{\ker\,
\sfd_{\D_Q}^{(k)}}{\im\,\sfd_{\D_Q}^{(k-1)}}
\qqq
to be termed the \textbf{$k$-th $\D_Q$-relative de Rham cohomology
group}. \exdef We have the following relative counterpart of the de
Rham Theorem.
\bethe\cite[App.\,A]{Fuchs:2007fw}\label{thm:DQ-rel-deRham}
Adopt the notation of Definitions \ref{def:DQT-rel-hom},
\ref{def:DQT-rel-cohom}, \ref{def:DQ-rel-co-hom} and
\ref{def:DQ-rel-dR-cohom}. The $\bR$-linear map
\qq\nn
I^{(k)}_{\D_Q}\ :\ \Om^k_{\rm dR}\left(M,Q\,\vert\,\D_Q\right)\to
C^k\left(M,Q\,\vert\,\D_Q;\bR\right)
\qqq
defined by the formula
\qq\nn
I^{(k)}_{\D_Q}(\om^k_M\oplus\om^{k-1}_Q)(c_k^M\oplus c_{k-1}^Q):=
\int_{c_k^M}\,\om^k_M+\int_{c_{k-1}^Q}\,\om^{k-1}_Q\,,
\qqq
written for an arbitrary $\D^Q$-relative $k$-chain $\,c_k^M\oplus
c_{k-1}^Q\in C_k\left(M,Q\,\vert\,\D^Q\right)$,\ is a cochain map.
The induced homomorphism
\qq\label{eq:ind-DQ-rel-iso}
[I^{(k)}_{\D_Q}]\ :\ H^k_{\rm dR}\left(M,Q\,\vert\,\D_Q\right)\to
H^k\left(M,Q\,\vert\,\D_Q;\bR\right)
\qqq
is a group isomorphism. \ethe
\beroof
That $\,I^{(k)}_{\D_Q}\,$ is a cochain map readily follows from
direct computation,
\qq\nn
\d_{\D_Q}^{(k)}\left(I^{(k)}_{\D_Q}(\om^k_M\oplus\om^{k-1}_Q)
\right)(c_{k+1}^M\oplus c_k^Q)&\equiv&I^{(k)}_{\D_Q}(\om^k_M\oplus
\om^{k-1}_Q)\left(\left(\p^M_{(k+1)}c_{k+1}^M-\D^Q c_k^Q\right)
\oplus\left(-\p^Q_{(k)}c_k^Q\right)\right)\cr\cr
&=&\int_{\p^M_{(k+1)}c_{k+1}^M-\D^Q c_k^Q}\,\om^k_M-\int_{\p^Q_{(k
)}c_k^Q}\,\om^{k-1}_Q\cr\cr
&=&\int_{c_{k+1}^M}\,\sfd^{(k)}\om^k_M+\int_{c_k^Q}\,\left(-\sfd^{(
k-1)}\om^{k-1}_Q-\D_Q\om^k_M\right)\cr\cr
&\equiv&I^{(k)}_{\D_Q}\left(\sfd_{\D_Q}^{(k)}(\om^k_M\oplus\om^{k-
1}_Q)\right)(c_{k+1}^M\oplus c_k^Q)\,.
\qqq
We may, next, use the split exact sequence (existing by
construction)
\qq\nn
0\to\Om^{k-1}(Q)\xrightarrow{\ \iota_{\rm dR}^{(k)}\ }\Om^k_{\rm dR}
\left(M,Q\,\vert\,\D_Q\right)\xrightarrow{\ \pr_{1\,{\rm dR}}^{(k)}\
}\Om^k(M)\to 0\,,
\qqq
expressed in terms of the inclusion map $\,\iota_{\rm dR}^{(k )}\,$
and the canonical projection $\,\pr_{1\,{\rm dR}}^{(k)}\equiv
\pr_1$,\ to induce the long exact sequence
\qq\nonumber\\
\begin{tikzpicture}[descr/.style={fill=white,inner sep=1.5pt}]
        \matrix (m) [
            matrix of math nodes,
            row sep=1.cm,
            column sep=1.cm,
            text height=1.5ex, text depth=0.25ex
        ]
        { \cdots & H^{k-1}_{\rm dR}(Q) & H^k_{\rm dR}\left(M,Q\,\vert\,\D_Q\right) & H^k_{\rm dR}(M) & \\
            & H^k_{\rm dR}(Q) & H^{k+1}_{\rm dR}\left(M,Q\,\vert\,\D_Q\right) & H^{k+1}_{\rm dR}(M) & \cdots \\
        };

        \path[overlay,->, font=\scriptsize,>=angle 45]
        (m-1-1) edge node[descr,yshift=2.3ex] {$\b^{(k-1)}_{\D_Q}$} (m-1-2)
        (m-1-2) edge (m-1-3)
        (m-1-3) edge (m-1-4)
        (m-1-4) edge[out=355,in=175] node[descr,xshift=-15.ex,yshift=1.1ex] {$\b^{(k)}_{\D_Q}$} (m-2-2)
        (m-2-2) edge (m-2-3)
        (m-2-3) edge (m-2-4)
        (m-2-4) edge node[descr,yshift=2.3ex] {$\b^{(k+1)}_{\D_Q}$} (m-2-5);
\end{tikzpicture}\,, \label{eq:les-DQ-rel-dR-cohom}
\qqq
with the (standard) connecting homomorphisms
\qq\nn
\b^{(k)}_{\D_Q}\ :\ H^k_{\rm dR}(M)\to H^k_{\rm dR}(Q)\ :\
[\om^k_M]\mapsto[-\D_Q\om^k_M]\,.
\qqq
In conjunction with the long exact sequence of
\Reqref{eq:les-DQ-rel-sing-cohom}, it gives rise to the manifestly
commutative diagram with exact rows
\qq\nn
\alxydim{@C=1.cm@R=1.cm}{H^{k-1}_{\rm dR}(M) \ar[r]^{\b^{(k-1
)}_{\D_Q}} \ar[d]^{[I^{(k-1)}_M]} & H^{k-1}_{\rm dR}(Q) \ar[r]^{[
\iota_{\rm dR}^{(k)}]\qquad} \ar[d]^{[I^{(k-1)}_Q]} & H^k_{\rm
dR}\left(M, Q\,\vert\,\D_Q\right) \ar[r]^{\quad[\pr_{1\,{\rm
dR}}^{(k)}]} \ar[d]^{[ I^{(k)}_{\D_Q}]} & H^k_{\rm dR}(M)
\ar[r]^{\b^{(k)}_{\D_Q}}
\ar[d]^{[I^{(k)}_M]} & H^k_{\rm dR}(Q) \ar[d]^{[I^{(k)}_Q]} \\
H^{k-1}(M;\bR) \ar[r]_{B^{(k-1)}_{\D_Q;\bR}} & H^{k-1}(Q;\bR)
\ar[r]_{[\pr_{2\,(k)}^\dagger]\quad} & H^k\left(M,Q\,\vert\,\D_Q;\bR
\right) \ar[r]_{\qquad[\iota_{(k)}^\dagger]} & H^k(M;\bR)
\ar[r]_{B^{(k )}_{\D_Q;\bR}} & H^k(Q;\bR)}\,,
\qqq
in which the $\,[I^{(k)}_\xcM]\,$ are as in \Reqref{eq:deRham-iso},
and all maps in rectangular brackets are defined as the cohomology
lifts of the respective cochain maps, \textit{e.g.},
\qq\nn
[\pr_{1\,(k)}^\dagger][c^{k-1}_Q]:=[\pr_{1\,(k)}^\dagger c^{k-1}_Q]
\,.
\qqq
Since the $\,[I^{(k)}_\xcM]\,$ are isomorphisms, the commutativity
of the above diagram immediately implies, in virtue of the Five
Lemma of \Rxcite{Lemma I.3.3}{MacLane:1975}, that
$\,[I^{(k)}_{\D_Q}]\,$ is, indeed, an isomorphism.\eroof \noindent
The last theorem is instrumental in proving its own extended
version:
\bethe\label{thm:DQT-rel-deRham}
Adopt the notation of Definitions \ref{def:DQT-rel-hom},
\ref{def:DQT-rel-cohom} and \ref{def:DQT-rel-dR-cohom}. The
$\bR$-linear map
\qq\nn
I^{(k)}_{(\D_Q,\D_{T_n})}\ :\ \Om^k\left(M,Q,T_n\,\vert\,\D_Q,
\D_{T_n}\right)\to C^k\left(M,Q,T_n\,\vert\,\D_Q,\D_{T_n};\bR
\right)
\qqq
defined by the formula
\qq\nn
I^{(k)}_{(\D_Q,\D_{T_n})}(\om^k_M\oplus\om^{k-1}_Q\oplus\om^{k-
2}_{T_n})(c_k^M\oplus c_{k-1}^Q\oplus c_{k-2}^{T_n}):=\int_{c_k^M}
\,\om^k_M+\int_{c_{k-1}^Q}\,\om^{k-1}_Q+\int_{c_{k-2}^{T_n}}\,
\om^{k-2}_{T_n}\,,
\qqq
written for an arbitrary $(\D^Q,\D^{T_n})$-relative $k$-chain
$\,c_k^M\oplus c_{k-1}^Q\oplus c_{k-2}^{T_n}\in C_k\left(M,Q,T_n\,
\vert\,\D^Q,\D^{T_n}\right)$,\ is a cochain map. The induced
homomorphism
\qq\label{eq:ind-DQT-rel-iso}
[I^{(k)}_{(\D_Q,\D_{T_n})}]\ :\ H^k_{\rm dR}\left(M,Q,T_n\,\vert\,
\D_Q,\D_{T_n}\right)\to H^k\left(M,Q,T_n\,\vert\,\D_Q,\D_{T_n};\bR
\right)
\qqq
is a group isomorphism. \ethe
\beroof
That the $\,I^{(k)}_{(\D_Q,\D_{T_n})}\,$ are cochain maps follows
from a similar calculation as for the $\,I^{(k)}_{\D_Q}$.\ It
therefore remains to verify that the induced cohomology maps are
isomorphisms. Here, we consider the long exact sequence
\qq\nonumber\\
\begin{tikzpicture}[descr/.style={fill=white,inner sep=1.5pt}]
        \matrix (m) [
            matrix of math nodes,
            row sep=1.cm,
            column sep=1.cm,
            text height=1.5ex, text depth=0.25ex
        ]
        { \cdots & H^{k-2}_{\rm dR}(T_n) & H^k_{\rm dR}\left(M,Q,T_n\,\vert\,\D_Q,\D_{T_n}\right) & H^k_{\rm dR}(M,Q\,\vert\,\D_Q) & \\
            & H^{k-1}_{\rm dR}(T_n) & H^{k+1}_{\rm dR}\left(M,Q,T_n\,\vert\,\D_Q,\D_{T_n}\right) & H^{k+1}_{\rm dR}(M,Q\,\vert\,\D_Q) & \cdots \\
        };

        \path[overlay,->, font=\scriptsize,>=angle 45]
        (m-1-1) edge node[descr,xshift=-1.ex,yshift=2.3ex] {$\b^{(k-1)}_{(\D_Q,\D_{T_n})}$} (m-1-2)
        (m-1-2) edge (m-1-3)
        (m-1-3) edge (m-1-4)
        (m-1-4) edge[out=355,in=175] node[descr,xshift=-20.ex,yshift=1.1ex] {$\b^{(k)}_{(\D_Q,\D_{T_n})}$} (m-2-2)
        (m-2-2) edge (m-2-3)
        (m-2-3) edge (m-2-4)
        (m-2-4) edge node[descr,xshift=1.ex,yshift=2.3ex] {$\b^{(k+1)}_{(\D_Q,\D_{T_n})}$} (m-2-5);
\end{tikzpicture} \label{eq:les-DQT-rel-dR-cohom}
\qqq
induced by the split exact sequence
\qq\nn
0\to\Om^{k-2}(T_n)\xrightarrow{\ \widetilde\iota_{\rm dR}^{(k)}\ }
\Om^k_{\rm dR}\left(M,Q,T_n\,\vert\,\D_Q,\D_{T_n}\right)
\xrightarrow{\ \widetilde\pr_{1,2\,{\rm dR}}^{(k)}\ }\Om^k_{\rm dR}(
M,Q\,\vert\,\D_Q) \to 0
\qqq
in which $\,\widetilde\iota_{\rm dR}^{(k)}\,$ is the inclusion and
$\,\widetilde\pr_{1,2\,{\rm dR}}^{(k)}\equiv\pr_{1,2}\,$ is the
canonical projection. Above, $\,\b^{(k)}_{(\D_Q,\D_{T_n})}\,$ is the
(standard) connecting homomorphisms given by
\qq\nn
\b^{(k)}_{(\D_Q,\D_{T_n})}\ :\ H^k_{\rm dR}(M,Q\,\vert\,\D_Q)\to
H^{k-1}_{\rm dR}(T_n)\ :\ [\om^k_M\oplus\om^{k-1}_Q]\mapsto[-
\D_{T_n}\om^{k-1}_Q]\,.
\qqq
The long exact sequences \eqref{eq:les-DQT-rel-sing-cohom} and
\eqref{eq:les-DQT-rel-dR-cohom} altogether yield the manifestly
commutative diagram with exact columns
\qq\nn
\alxydim{@C=2.cm@R=1.cm}{H^{k-1}_{\rm dR}(M,Q\,\vert\,\D_Q)
\ar[r]^{[I^{(k-1)}_{\D_Q}]} \ar[d]_{\b^{(k-1)}_{(\D_Q,\D_{T_n})}} &
H^{k-1}(M,Q\,\vert\,\D_Q;\bR) \ar[d]^{B^{(k-1)}_{(\D_Q,\D_{T_n});
\bR}} \\ H^{k-2}_{\rm dR}(T_n) \ar[r]^{[I^{(k-2)}_{T_n}]}
\ar[d]_{[\widetilde\iota_{\rm dR}^{(k)}]} & H^{k-2}( T_n;\bR)
\ar[d]^{[\widetilde\pr_{3\,(k)}^\dagger]} \\ H^k_{\rm dR}\left(M,Q,
T_n\,\vert\,\D_Q,\D_{T_n}\right) \ar[r]^{[I^{(k)}_{(\D_Q,\D_{T_n})}
]} \ar[d]_{[\widetilde\pr_{1,2\,{\rm dR}}^{(k)}]} & H^k\left(M,Q,
T_n\,\vert\,\D_Q,\D_{T_n};\bR\right) \ar[d]^{[\widetilde\iota_{(k
)}^\dagger]} \\ H^k_{\rm dR}(M,Q\,\vert\,\D_Q) \ar[r]^{[I^{(k
)}_{\D_Q}]} \ar[d]_{\b^{(k)}_{(\D_Q,\D_{T_n})}} & H^k(M,Q\,\vert\,
\D_Q;\bR) \ar[d]^{B^{(k)}_{(\D_Q,\D_{T_n});\bR}} \\ H^{k-1}_{\rm dR}
(T_n) \ar[r]^{[I^{(k-1)}_{T_n}]} & H^{k-1}(T_n;\bR)}\,,
\qqq
in which both the $\,[I^{(k)}_{T_n}]\,$ and the $\,[I^{(k)}_{\D_Q}
]\,$ are isomorphisms, the latter by Theorem
\ref{thm:DQ-rel-deRham}, and all maps in rectangular brackets are
defined as the cohomology lifts of the respective cochain maps.
Adducing the Five Lemma once more, we conclude that the $\,[I^{(k
)}_{(\D_Q,\D_{T_n})}]\,$ are also isomorphisms, as claimed. \eroof

\subsubsection{\textbf{The relative Cartan calculus and the twisted
bracket}}\label{sub:rel-tw-Courant}

The replacement of the standard de Rham cohomology by its relative
counterpart in the presence of world-sheet defects and the
associated (inter-)bi-brane extension of the string background of
Definition \vref{}I.2.1 suggests that we reconsider the concept of a
twisted Courant bracket in the relative-geometric framework. Indeed,
the latter concept is based on two differential-geometric structures
present on the target space, namely the Lie algebra of vector fields
and the de Rham complex of forms that determines, through Cartan's
magic formula, the form component of the bracket. Taking as the
point of departure the geometry of the target space $\,\xcF\,$ of
the background with bi-branes and inter-bi-branes, the respective
structures on the target space $\,M$,\ on the bi-brane world-volume
$\,Q\,$ and on the (component) inter-bi-brane world-volumes
$\,T_n\,$ become related by the $(\iota_\a,\pi_n^{k,k+1})$-alignment
condition \eqref{eq:iotapi-align} and by the $(\D_Q,\D_{T_n})$-twist
in the de Rham complex. It therefore seems pertinent to enquire as
to a natural definition of the Courant bracket, with a twist now
determined by the pair $\,(\txH,\om)$,\ in this constrained setting.

We start by giving a relative variant of the Cartan calculus for the
coupled target-space geometries \eqref{eq:rel-targeom}.
\bedef\label{def:DQT-rel-Lie}
Adopt the notation of Definitions \ref{def:DQT-rel-hom},
\ref{def:DQT-rel-cohom} and \ref{def:DQT-rel-dR-cohom}. Denote the
space $\,\G(\sfT M\sqcup\sfT Q\sqcup\sfT T)\,$ of vector fields on
$\,M\sqcup Q\sqcup T,\ T=\bigsqcup_{n\geq 3}\,T_n\,$ with
restrictions $\,\xcV\vert_M=\Mup\xcV,\xcV\vert_Q=\Qup\xcV\,$ and
$\,\xcV\vert_{T_n}=\Tnup\xcV\,$ satisfying the $(\iota_\a,\pi_n^{k,k
+1})$-alignment condition \eqref{eq:iotapi-align} as $\,\Xgt_{(
\iota_\a,\pi_n^{k,k+1})}(M\sqcup Q\sqcup T)$.\ To every such vector
field $\,\xcV\in\Xgt_{(\iota_\a,\pi_n^{k,k+1})}(M\sqcup Q\sqcup T
)\,$ we associate a degree-$(-1)$ derivation of the $(\D_Q,\D_{T_n}
)$-relative de Rham complex
\qq\nn
\ic^{(\D_Q,\D_{T_n})}_\xcV\ &:&\ \Om^\bullet_{\rm dR}\left(M,Q,T_n\,
\vert\,\D_Q,\D_{T_n} \right)\to\Om^{\bullet-1}_{\rm dR}\left(M,Q,T_n
\,\vert\,\D_Q,\D_{T_n}\right)\cr\cr
&:&\ \om^k_M\oplus\om^{k-1}_Q\oplus\om^{k-2}_{T_n}\mapsto\left(\Mup
\xcV\con\om^k_M\right)\oplus\left(-\Qup\xcV\con\om^{k-1}_Q\right)
\oplus\left(\Tnup\xcV\con\om^{k-2}_{T_n}\right)\,,
\qqq
to be termed the \textbf{$(\D_Q,\D_{T_n})$-relative contraction}
henceforth.

The \textbf{$(\D_Q,\D_{T_n})$-relative Lie derivative on
$(\D_Q,\D_{T_n})$-relative de Rham complex along $(\iota_\a,\pi_n^{k
,k+1})$-aligned vector field $\,\xcV\,$} is defined by Cartan's
magic formula
\qq\nn
\pLie{\xcV}^{(\D_Q,\D_{T_n})}\vert_{\Om^k_{\rm dR}\left(M,Q,T_n\,
\vert\,\D_Q,\D_{T_n}\right)}:=\sfd_{(\D_Q,\D_{T_n})}^{(k-1)}\circ
\ic^{(\D_Q,\D_{T_n})\,(k)}_\xcV+\ic^{(\D_Q,\D_{T_n})\,(k+1)}_\xcV
\circ\sfd_{(\D_Q,\D_{T_n})}^{(k)}\,.
\qqq
\exdef

\brem The definition of the $(\D_Q,\D_{T_n})$-relative Lie
derivative is not only natural but also yields a simple object when
calculated explicitly,
\qq\nn
\pLie{\xcV}^{(\D_Q,\D_{T_n})}\left(\om^k_M\oplus\om^{k-1}_Q\oplus
\om^{k-2}_{T_n}\right)&=&\pLie{\Mup\xcV}\om^k_M\oplus\left(
\pLie{\Qup\xcV}\om^{k-1}-\D_Q\left(\Mup\xcV\con\om^k_M\right)+\Qup
\xcV\con\D_Q\om^k_M\right)\cr\cr
&&\oplus\left(\pLie{\Tnup\xcV}\om^{k-2}_{T_n}+\D_{T_n}\left(\Qup
\xcV\con\om^{k-1}_Q\right)-\Tnup\xcV\con\D_{T_n}\om^{k-1}_Q\right)
\cr\cr
&=&\pLie{\Mup\xcV}\om^k_M\oplus\pLie{\Qup\xcV}\om^{k-1}_Q\oplus
\pLie{\Tnup\xcV}\om^{k-2}_{T_n}\,.
\qqq
It ought to be emphasised that the form of the $(\D_Q,\D_{T_n}
)$-relative differential is essentially fixed by that of the $(\D_Q,
\D_{T_n})$-relative boundary operator, and so we may regard the
above observation as a rationale for the definition of the
$(\D_Q,\D_{T_n} )$-relative contraction and of the
$(\D_Q,\D_{T_n})$-relative Lie derivative. \erem

\noindent We readily establish
\berop\label{prop:Cartan-calc}
In the notation of Definitions \ref{def:DQT-rel-dR-cohom} and
\ref{def:DQT-rel-Lie}, the triple
\qq\nn
\left(\sfd_{(\D_Q,\D_{T_n})},\ic^{(\D_Q,\D_{T_n})}_\xcV,
\pLie{\xcV}^{(\D_Q,\D_{T_n})}\right)\,,
\qqq
with $\,\sfd_{(\D_Q,\D_{T_n})}:=(\sfd_{(\D_Q,\D_{T_n})}^{(k)})$,\
obeys the standard rules of Cartan's calculus:
\qq\nn
&\sfd_{(\D_Q,\D_{T_n})}^2=0\,,\qquad\qquad\left\{\ic^{(\D_Q,
\D_{T_n})}_\xcV,\ic^{(\D_Q,\D_{T_n})}_\xcW\right\}=0\,,\qquad\qquad
\left[\pLie{\xcV}^{(\D_Q,\D_{T_n})},\pLie{\xcW}^{(\D_Q,\D_{T_n})}
\right]=\pLie{[\xcV,\xcW]}^{(\D_Q,\D_{T_n})}\,,&\cr\cr
&\left\{\sfd_{(\D_Q,\D_{T_n})},\ic^{(\D_Q,\D_{T_n})}_\xcV
\right\}=\pLie{\xcV}^{(\D_Q,\D_{T_n})}\,,&\cr\cr
&\left[\sfd_{(\D_Q,\D_{T_n})},\pLie{\xcV}^{(\D_Q,\D_{T_n})}\right]=
0\,,\qquad\qquad\left[\pLie{\xcV}^{(\D_Q,\D_{T_n})},\ic^{(\D_Q,
\D_{T_n})}_\xcW\right]=\ic^{(\D_Q,\D_{T_n})}_{[\xcV,\xcW]}\,.&
\qqq
\eerop
\beroof Obvious, through inspection.\eroof
\noindent Thus, we may think of the triple $\,\left(\sfd_{(\D_Q,
\D_{T_n})},\ic^{(\D_Q,\D_{T_n})}_\xcV,\pLie{\xcV}^{(\D_Q,\D_{T_n})}
\right)\,$ as a natural counterpart of the standard triple $\,(\sfd,
\xcV\con,\pLie{\xcV})\,$ in the setting of coupled target-space
geometries. It is now straightforward to consider the corresponding
notion of a (twisted) Courant bracket.
\bedef\label{def:DQ-rel-Courant}
In the notation of Definitions \ref{def:DQT-rel-hom},
\ref{def:DQT-rel-cohom}, \ref{def:DQT-rel-dR-cohom} and
\ref{def:DQT-rel-Lie}, and for an arbitrary $\sfd_{(\D_Q,\D_{T_n}
)}$-closed $(\D_Q,\D_{T_n})$-relative 3-form $\,\eta\in\Om^3_{\rm
dR}\left(M,Q,T_n\,\vert\,\D_Q,\D_{T_n}\right)$,\ we define the
\textbf{$\eta$-twisted $(\D_Q,\D_{T_n})$-relative Courant bracket
on}
\qq\nn
\Egt_{(\iota_\a,\pi_n^{k,k+1})}(M\sqcup Q\sqcup T):=\Xgt_{(\iota_\a
,\pi_n^{k,k+1})}(M\sqcup Q\sqcup T)\oplus\Om^1_{\rm dR}\left(M,Q,
T_n\,\vert\,\D_Q,\D_{T_n}\right)
\qqq
by the formula
\qq\nn
[\xcV\oplus\upsilon,\xcW\oplus\varpi]_{\rm C}^\eta&:=&[\xcV,\xcW]
\oplus\bigg(\pLie{\xcV}^{(\D_Q,\D_{T_n})}\varpi-\pLie{\xcW}^{(\D_Q,
\D_{T_n})}\upsilon-\tfrac{1}{2}\,\sfd^{(1)}_{(\D_Q,\D_{T_n})}\left(
\ic^{(\D_Q,\D_{T_n})}_\xcV\varpi-\ic^{(\D_Q,\D_{T_n})}_\xcW\upsilon
\right)\cr\cr
&&\hspace{1.5cm}+\ic^{(\D_Q,\D_{T_n})}_\xcV\ic^{(\D_Q,\D_{T_n}
)}_\xcW\eta\bigg)\,.
\qqq
\exdef

The adequacy of the $(\D_Q,\D_{T_n})$-relative Cartan calculus in
the discussion of string backgrounds with bi-branes and
inter-bi-branes is illustrated amply by the following theorem, which
-- at the same time -- demystifies the previous definition of the
$(\txH,\om;\D_Q)$-twisted bracket structure on $(\iota_\a,
\pi_n^{k,k+1})$-paired restricted tangent sheaves.
\bethe\label{thm:brabra}
Adopt the notation of Definitions \ref{def:gen-tan-sheaf},
\ref{def:tw-bra-str-sheaf}, \ref{def:DQT-rel-dR-cohom},
\ref{def:DQT-rel-Lie} and \ref{def:DQ-rel-Courant}, and of Corollary
\ref{cor:full-GBra}. The $(\D_Q,\D_{T_n})$-relative 3-form
$\,\txH\oplus\om\oplus 0\,$ on $\,M\sqcup Q\sqcup T=\xcF\,$ is
$\sfd_{(\D_Q,\D_{T_n})}$-closed, and hence defines a
$(\txH\oplus\om\oplus 0)$-twisted $(\D_Q,\D_{T_n} )$-relative
Courant bracket on $\,\Egt_{(\iota_\a,\pi_n^{k,k+1})}( \xcF)$.\
Under the natural identification between the latter space and
$\,\G_{(\iota_\a,\pi_n^{k,k+1})}\bigl(\widehat\sfE^{(1,2\sqcup 1
\sqcup 0)}\xcF\bigr)\,$ (expressed in terms of the canonical
projections to the direct summands of $\,\Om^1_{\rm dR}\left(M,Q,T_n
\,\vert\,\D_Q,\D_{T_n}\right)=\Om^1(M)\oplus\Om^0(Q)\oplus\Om^{-1}(
T_n)$)
\qq\nn
\Psi\ &:&\ \Egt_{(\iota_\a,\pi_n^{k,k+1})}(\xcF)\xrightarrow{\ \cong
\ }\G_{(\iota_\a,\pi_n^{k,k+1})}\bigl(\widehat \sfE^{(1,2\sqcup 1
\sqcup 0)}\xcF\bigr)\cr\cr
&:&\ \xcV\oplus\upsilon\mapsto\left(\xcV\vert_M\oplus\pr_1(\upsilon
),\xcV\vert_Q\oplus\pr_2(\upsilon),\xcV\vert_{T_n}\oplus\pr_3(
\upsilon)\right)\,,
\qqq
we have
\qq\nn
\GBra{\cdot}{\cdot}^{(\txH,\om,0;\D_Q,\D_{T_n})}\circ(\Psi,\Psi)=
\Psi\circ[\cdot,\cdot]_{\rm C}^{\txH\oplus\om\oplus 0}\,.
\qqq
Furthermore, the $\si$-symmetric sections in $\,\G_{(\iota_\a,
\pi_n^{k,k+1})}\bigl(\widehat\sfE^{(1,2\sqcup 1\sqcup 0)}\xcF
\bigr)\,$ are identified with those elements of $\,\Egt_{(\iota_\a,
\pi_n^{k,k+1})}(\xcF)\,$ that satisfy the relation
\qq\label{eq:sigma-sym-new-def-rel}
\sfd_{(\D_Q,\D_{T_n}),\txH\oplus\om\oplus 0}(\xcV\oplus\upsilon):=
\sfd_{(\D_Q,\D_{T_n})}^{(1)}\upsilon+\ic^{(\D_Q,\D_{T_n})}_\xcV(
\txH\oplus\om\oplus 0)=0\,.
\qqq
\ethe
\beroof
Obvious, through inspection. \eroof

\brem It deserves to be emphasised that the above formalism
restricts in just the desired manner to the paired geometries $\,(M,
Q)$,\ \textit{i.e.}\ in the absence of inter-bi-branes. \erem

\subsubsection{\textbf{A proof of Proposition
\ref{prop:sympl-goes-ham-tw}}}\label{sub:proof} The preceding
considerations provide us with cohomological tools necessary for
verifying the thesis of Proposition \ref{prop:sympl-goes-ham-tw}
that will, in turn, prove central to the discussion of the small
gauge anomaly in the framework of generalised geometry in Section
\ref{sec:anomaly}.

Taking into account \Reqref{eq:Vinbra-Ka-tw} and identity
\eqref{eq:tiL-minus-DQ-tw}, we can explicitly write out
\Reqref{eq:Poiss-bra-ham-tw} in the basis $\,\Kgt_A\,$ as
\qq\nn
\{h^{\cB\,\vert\,\vep}_{\Kgt_A},h^{\cB\,\vert\,\vep}_{\Kgt_B}\,
\}_{\Om_{\si,\cB\,\vert\,(\pi,\vep)}}&=&f_{ABC}\,h^{\cB\,\vert\,
\vep}_{\Kgt_C}+\widetilde\sfL^{Q\vert(\pi,\vep)\,*}\bigl(\Mup\D_{A
B}-\sfd
\txc_{(AB)}\bigr)+\vep\,\widetilde{\unl\sfL}^{Q\vert(\pi,\vep)
\,*}\bigl(\Qup\D_{AB}+\D_Q \txc_{(AB)}\bigr)\cr\cr
&=&f_{ABC}\,h^{\cB\,\vert\,\vep}_{\Kgt_C}+\widetilde\sfL^{Q\vert(
\pi,\vep)\,*}\Mup\D_{AB}+\vep\,\widetilde{\unl\sfL}^{Q\vert(\pi,\vep
)\,*}\Qup\D_{AB}\,,
\qqq
where in the last line we used the notation of
\Reqref{eq:Vinbra-Ka}, with additional abbreviations
\qq\nn
\Mup\D_{AB}:=\pLie{\Mup\xcK_A}\kappa_B-f_{ABC}\,\kappa_C\,,
\qquad\qquad\Qup\D_{AB}:=\pLie{\Qup\xcK_A}k_B-f_{ABC}\,k_C\,.
\qqq
The realisation of $\,\ggt_\si\,$ is hamiltonian iff the identity
\qq\label{eq:ham-real-cond}
\left(\sfL^{Q\vert(\pi,\vep)\,*}\Mup\D_{AB}+\vep\,\unl\sfL^{Q\vert(
\pi,\vep)\,*}\Qup\D_{AB}\right)\left[(X,q)\right]=0
\qqq
obtains for every 1-twisted loop $\,(X,q)$.\ Write
\qq\nn
c_1^M(X,q):=X(\bS^1_\pi)\,,\qquad\qquad c_0^Q(X,q):=X(\pi)=q\,.
\qqq
The chain $\,(c_1^M\oplus\vep\,c_0^Q)(X,q)=:c_1^{\D^Q}(X,q)\,$
defines a $\D^Q$-relative 1-cycle, and -- clearly -- any such
1-cycle can be obtained within $\,\sfL_{Q\vert(\pi,\vep)}M$.\ We may
now rewrite condition \eqref{eq:ham-real-cond} as
\qq\nn
\langle [I^{(1)}_{\D_Q}][\Mup\D_{AB}\oplus\Qup\D_{AB}],[c_1^{\D^Q}(
X,q)]\rangle=0
\qqq
in terms of the induced isomorphism of \Reqref{eq:ind-DQ-rel-iso}
and of the standard pairing
\qq\nn
\langle\cdot,\cdot\rangle\ :\ H^1\left(M,Q\,\vert\,\D_Q;\bR\right)
\x H_1\left(M,Q\,\vert\,\D^Q\right)\to\bR\ :\ \left([c^1_{\D_Q}],[
c_1^{\D^Q}]\right)\mapsto c^1_{\D_Q}\left(c_1^{\D^Q}\right)\,.
\qqq
In view of the arbitrariness of $\,c_1^{\D^Q}(X,q)$,\ we conclude
that there must exist smooth functions $\,\Mup D_{AB}\in C^\infty(M,
\bR)\,$ and (local) constants $\,\Qup D_{AB}\,$ on $\,Q\,$ such that
\qq\nn
\Mup\D_{AB}\oplus\Qup\D_{AB}=\sfd_{\D_Q}^{(0)}(\Mup D_{AB}\oplus\Qup
D_{AB})\,,
\qqq
which reproduces \Reqref{eq:FM-2}.

We now obtain
\qq\nn
\GBra{\Kgt_A}{\Kgt_B}^{(\txH,\om;\D_Q)}=f_{ABC}\,\widetilde\Kgt_C+0
\oplus\bigl(\sfd(\Mup D_{AB}- \txc_{(AB)}),-\D_Q(\Mup
D_{AB}-\txc_{(AB)}) -\Qup D_{AB}\bigr)
\qqq
and, in the notation of the proof of \Reqref{eq:comm-preq-ham-tw},
\qq\nn
\Vbra{\widetilde\Kgt_{A\,\igt}}{\widetilde\Kgt_{B\,\igt}}=
\ee^{-\pr_{\sfL_{Q\vert(\pi,\vep)}M}^*E_{(\pi,\vep)\,\igt}}\lact
\widetilde\sfL^{Q\vert(\pi,\vep)}_{(1,1\sqcup 0)}
\GBra{\Kgt_A}{\Kgt_B}^{(\txH,\om;\D_Q)}=f_{ABC}\,\widetilde
\Kgt_{C\,\igt}
\qqq
which, indeed, yields Eqs.\,\eqref{eq:GBra-Ka-FM} and
\eqref{eq:tw-Vinbra-Kai-FM}. This concludes the proof.

\section{The gauge anomaly -- the sixfold way}\label{sec:anomaly}

In the preceding sections, we have identified a specific
target-space model of the algebraic structure on the set of charges
of a rigid symmetry of the multi-phase $\si$-model and elucidated
the underlying simple and universal
differential-geometric/cohomological scheme that is realised both in
the presence as well as in the absence of (symmetry-preserving)
world-sheet defects. In the course of our study, we have laid
considerable emphasis on the very fundamental gerbe-theoretic
aspects of the said structure, or -- to put it differently -- on the
naturalness of that structure in the setting of a target-space
geometry with the 2-category of bundle gerbes with connection over
it. This leaves us with a fairly complete understanding of
infinitesimal rigid symmetries of the two-dimensional field theory
of interest.

In this last section, we want to take our analysis of $\si$-model
symmetries to the next level by considering a local variant thereof.
A prerequisite for an in-depth treatment of the subject is a precise
identification and systematisation of potential obstructions to
rendering a global symmetry of the $\si$-model local. This task was
completed in a series of papers
\cite{Gawedzki:2003pm,Gawedzki:2004tu,Schreiber:2005mi,Gawedzki:2007uz,Gawedzki:2008um,Gawedzki:2010rn,Gawedzki:2012fu}
in which a cohomological classification scheme was worked out for
these so-called gauge anomalies and from which a universal Gauge
Principle has emerged.

In the intrinsically geometric context of the $\si$-model, the
field-theoretic gauging procedure admits a clear-cut interpretation:
It boils down to extending the target space by a principal
$\txG_\si$-bundle over the world-sheet and subsequently coupling the
string background to the attendant principal $\txG_\si$-connection
1-form in a manner that allows to descend the thus extended string
background, with its metric and gerbe-theoretic structure, to the
coset of the original target space with respect to the action of the
symmetry group $\,\txG_\si$.\ This yields the $\si$-model on the
coset of the original target space by the action of the group
whenever the latter coset exists within the smooth category.
Accordingly, gauge anomalies quantify obstructions to the existence
of equivalences between a given string background and the one
obtained through pullback of a string background from the
coset.\medskip

The rationale for taking up the issue of the gauge anomaly here is
twofold: First of all, we want to understand how gauge anomalies are
encoded in the algebroidal bracket structure introduced earlier, and
-- in so doing -- reassess the naturalness of the latter in the
context of the study of $\si$-model symmetries. As the bracket
captures infinitesimal features of the symmetries, we anticipate to
gain insights into the so-called small gauge anomaly in this manner.
The relation, established previously, between the Poisson algebra of
Noether charges of the rigid symmetry on the phase space of the
$\si$-model and the (relative) twisted Courant algebroid of the
corresponding $\si$-symmetric sections of the restricted generalised
tangent sheaves over the target space gives rise to an additional
expectation, to wit, that there is a canonical (\textit{i.e.}\
symplectic) interpretation of the small gauge anomaly. This
expectation will receive confirmation in the framework of a
canonical description of the gauged $\si$-model of
\Rxcite{Sect.\,10.2}{Gawedzki:2012fu} that we develop hereunder
along the lines of \Rxcite{Sect.\,3}{Suszek:2011hg}. The second
piece of motivation for studying gauge anomalies derives from the
findings of \Rcite{Suszek:2011hg} that establish an intimate
relation between dualities of the $\si$-model (including the
geometric symmetries studied in the present paper) and a
distinguished class of conformal defects. Thus, we shall demonstrate
how gauge anomalies, both large and small, obstruct the existence of
topological defect quivers implementing the action of the gauge
group $\,C^\infty(\Si,\txG_\si)\,$ on states of the gauged
$\si$-model in the manner detailed in
\Rxcite{Sec.\,4}{Suszek:2011hg}. As argued in
Refs.\,\cite[Sec.\,4]{Jureit:2006yf},\cite[Sec.\,2.9]{Runkel:2008gr}
and \cite[Sect.\,3]{Frohlich:2009gb}, defect quivers of this kind
give rise to a much intuitive world-sheet definition of the coset
$\si$-model. The definition, taking as the point of departure the
original string background (existing before the action of the
symmetry group $\,\txG_\si\,$ has been divided out), identifies
embeddings of the world-sheet in that background related by the
action of the gauge group and admits embeddings that are
continuously differentiable \emph{up to the action of the gauge
group}, the latter being realised by means of an arbitrarily fine
mesh of homotopically deformable (at no cost in the action
functional) $C^\infty(\Si, \txG_\si)$-jump defect lines, with defect
junctions that can be pulled through one another (once again, at no
cost in the action functional). Our construction of a full-fledged
duality background for the gauged $\si$-model in the presence of
defects transparent to the symmetries gauged will be seen to give
substance to some general claims of \Rxcite{Sec.\,4}{Suszek:2011hg}
concerning the duality-defect correspondence, and -- at the same
time -- will provide an explicit realisation of the somewhat
abstract Duality Scheme laid out in
\Rxcite{Rem.\,5.6}{Suszek:2011hg}. Clearly, the fundamental concept
that interrelates the various facets of the gauge anomaly outlined
above is the gerbe theory of the $\si$-model that serves to
characterise the anomaly itself, underlies the structure of the
Courant algebroid, and -- finally -- (co-)determines the world-sheet
definition of the coset $\si$-model.
\medskip

By way of preparation for the subsequent reinterpretations of the
gauge anomaly, let us recapitulate the relevant definitions and
results from \Rcite{Gawedzki:2012fu}. We begin with
\becon\label{conv:pullback}
In order to unclutter the notation, we fix a convention for
pullbacks. Thus, for any $p$-form $\,\eta\,$ on a smooth space
$\,\xcM:=\xcM_1\x\xcM_2\x\cdots\x\xcM_N\,$ equipped with canonical
projections
$\,\pr_{i_1,i_2,\ldots,i_n}:\xcM\to\xcM_{i_1}\x\xcM_{i_2}\x\cdots\x
\xcM_{i_n}\,$ given for $\,1\leq i_1<i_2<\ldots<i_n\leq N$,\ we
denote
\qq\nn
\eta_{[i_1,i_2,\ldots,i_n]^* }:=\pr_{i_1,i_2,\ldots,i_n}^*\eta\,.
\qqq
In particular,
\qq\nn
\eta_{i^*}\equiv\eta_{[i]^*}=\pr_i^*\eta
\qqq
for any $\,1\leq i\leq N$.\ Analogous convention will be used for
geometric objects such as bundles, gerbes \textit{etc.}

Given a triple of smooth manifolds $\,\xcM_1,\xcM_2,\xcN\,$ and a
pair of smooth maps $\,f_i:\xcM_i\to\xcN$,\ denote by
\qq\nn
\xcM_1\fibx{f_1}{f_2}\xcM_2:=\left\{\ (m_1,m_2)\in\xcM_1\x\xcM_2
\quad\vert\quad f_1(m_1)=f_2(m_2)\ \right\}
\qqq
the product of the $\,\xcM_i\,$ fibred over $\,\xcN$. \econ
\noindent We may now state
\bedef\label{def:gauged-sigmod}\cite[Cor.\,3.17]{Gawedzki:2012fu}
Adopt the notation of Definition \ref{def:sigmod-2d} and of
Corollary \ref{cor:Homtw-Vbra-symm-tw}. The \textbf{gauged
two-dimensional non-linear $\si$-model for network-field
configurations $\,(X\,\vert\,\G)\,$ in string background $\,\Bgt\,$
on world-sheet $\,(\Si,\g)\,$ with defect quiver $\,\G$} coupled to
a topologically trivial gauge field $\,\txA\in\Om^1(\Si)\ox
\ggt_\si\,$ is a theory of continuously differentiable maps $\,X\ :\
\Si\to\xcF\,$ determined by the principle of least action applied to
the action functional
\qq\label{eq:2d-gauge-sigma-def}
S_\si[(X\,\vert\,\G);\txA,\g]:=-\tfrac{1}{2}\,\int_\Si\,\txg_\txA(
\sfd\xi\overset{\wedge}{,}\star_\g\sfd\xi)-\sfi\,\log\Hol_{\cG_\txA
,\Phi_\txA,(\varphi_{n\,\txA})}(\xi\,\vert\,\G)\,,
\qqq
where $\,\xi=(\id_\Si,X)\ :\ \Si\to\Si\x\xcF\,$ is the extended
embedding field, and the extended string background $\,\Bgt_\txA:=(
\cM_\txA,\cB_\txA,\cJ_\txA)\,$ defining the action functional has
the following components:
\bit
\item the extended target $\,\cM_\txA\,$ composed of the target
space $\,\Si\x M\,$ with the metric
\qq\label{eq:gA-def}
\txg_\txA:=\txg_{2^*}-\txK_{A\,2^*}\ox\txA^A_{1^*}-\txA^A_{1^*}\ox
\txK_{A\,2^*}+\txh_{AB\,2^*}\,\left(\txA^A\ox\txA^B\right)_{1^*}\,,
\qqq
written in terms of the tensors
\qq\nn
\txK_A:=\txg(\Mup\xcK_A,\cdot)\,,\qquad\qquad\txh_{AB}:=\txg(\Mup
\xcK_A,\Mup\xcK_B)
\qqq
and implicitly understood to act on the second tensor factor in
\qq\nn
\sfd\xi(\si)=\left(\sfd\si^a\ox\p_a,\p_a X^\mu(\si)\,\sfd\si^a\ox
\p_\mu\vert_{X(\si)}\right)\,,
\qqq
and with the gerbe
\qq\label{eq:cGA}
\cG_\txA=\cG_{2^*}\ox I_{\rho_\txA}\,,
\qqq
containing in its definition a trivial gerbe $\,I_{\rho_\txA}\,$
with a global curving
\qq\label{eq:rhoA}
\rho_\txA:=\kappa_{A\,2^*}\wedge\txA_{1^*}^A-\tfrac{1}{2}\,\txc_{AB\,
2^*}\,\left(\txA^A\wedge\txA^B\right)_{1^*}\,\in\,\Om^2(\Si\x M) \,;
\qqq
\item the extended bi-brane $\,\cB_\txA\,$ with the world-volume
$\,\G\x Q$,\ bi-brane maps $\,\unl\iota_\a=\id_\G\x\iota_\a,\ \a\in
\{1,2\}$,\ the curvature
\qq\label{eq:omA}
\om_\txA=\om_{2^*}-\unl\D_Q\rho_\txA+\sfd\la_\txA\,,
\qqq
written in terms of the pullback operator $\,\unl\D_Q:=\unl\iota_2^*
-\unl\iota_1^*\,$ and
\qq\label{eq:laA}
\la_\txA:=-k_{A\,2^*}\,\txA_{1^*}^A\,\in\,\Om^1(\Si\x Q)\,,
\qqq
and with the 1-isomorphism
\qq\label{eq:PhiA}
\Phi_\txA=\Phi_{2^*}\ox J_{\la_\txA}\,,
\qqq
written in terms of a trivial 1-isomorphism (a trivial bundle)
$\,J_{\la_\txA}\,$ with a global connection 1-form $\,\la_\txA$;
\item the extended inter-bi-brane $\,\cJ_\txA\,$ with
component world-volumes $\,\Vgt_\G^{(n)}\x T_n,\ n\geq 3$,\ defined
in terms of the subsets $\,\Vgt_\G^{(n)}\subset\Vgt_\G\,$ composed
of vertices of valence $n$,\ with inter-bi-brane maps $\,\unl
\pi_n^{k,k+1}=\id_{\Vgt_\G^{(n)}}\x\pi_n^{k,k+1},\ k\in\ovl{1,n}$,\
and with 2-isomorphisms $\,\varphi_{n\,\txA}:=\varphi_{n\,2^*}$.
\eit
\exdef \noindent Fundamental invariance properties of the gauged
$\si$-model are expressed in the following two theorems in which we
also define the notions of the small and large gauge anomaly used in
the remainder of the paper.
\bethe\cite[Prop.\,3.1]{Gawedzki:2010rn}\cite[Cor.\,3.11]{Gawedzki:2012fu}\label{thm:small-ganom}
In the notation of Definition \ref{def:gauged-sigmod} and of
Corollary \ref{cor:Homtw-Vbra-symm-tw}, the action functional of
\Reqref{eq:2d-gauge-sigma-def} is invariant under infinitesimal
gauge transformations
\qq
X^\mu(\si)&\mapsto& X^\mu(\si)+\La^A(\si)\,\xcFup\xcK_A^\mu\left(X(
\si)\right)\,,\cr&&\label{eq:gauge-trafo}\\
\txA_a^A(\si)&\mapsto&\txA_a^A(\si)+\p_a\La^A(\si)-f_{ABC}\,\La^B(
\si)\,\txA_a^C(\si)\,,\nonumber
\qqq
written in terms of arbitrary functions $\,\La^A\in C^\infty(\Si,\bR
)$,\ iff the following \textbf{conditions for a consistent gauging}
are satisfied:
\qq\label{eq:gauge-constr}
\pLie{\xcK_A}\kappa_B=f_{ABC}\,\kappa_C\,,\qquad\qquad\pLie{\xcK_A}
k_B=f_{ABC}\,k_C\,,\qquad\qquad \txc_{(AB)}=0\,.
\qqq
\ethe
\bedef
In the notation of Corollary \ref{cor:Homtw-Vbra-symm-tw}, the
\textbf{small gauge anomaly} of the $\si$-model
\eqref{eq:2d-gauge-sigma-def} is the obstruction to the existence of
a choice of the objects $\,(\kappa_A,k_A),\ A\in\ovl{1,\dim\,
\ggt_\si}\,$ satisfying relations \eqref{eq:gauge-constr}. \exdef

The large gauge anomaly is most neatly described in the language of
Lie groupoids whose theory was developed in
Refs.\,\cite{MacKenzie:1987,Moerdijk:2003mm}. Below, we set up our
notation by way of preparation for the discussion to follow.
\bedef\label{def:grpd}
A \textbf{groupoid} is the septuple $\,\Gr=(\obj\,\Gr,\morf\,\Gr
,s,t,\Id,\Inv,\circ)\,$ composed of a pair of sets: the
\textbf{object set} $\,\obj\,\Gr\,$ and the \textbf{arrow set}
$\,\morf\,\Gr$,\ and a quintuple of \textbf{structure maps}: the
\textbf{source map} $\,s:\morf\,\Gr\to\obj\,\Gr\,$ and the
\textbf{target map} $\,t:\morf\,\Gr\to\obj\,\Gr$,\ the \textbf{unit
map} $\,\Id:\obj\,\Gr\to \morf\,\Gr:m\mapsto\Id_m$,\ the
\textbf{inverse map} $\,\Inv:\morf\,\Gr\to \morf\,\Gr:
\overrightarrow g\mapsto\overrightarrow g^{-1}\equiv\Inv(
\overrightarrow g)$,\ and the \textbf{multiplication map} $\,\circ:
\morf\,\Gr{}_s\hspace{-3pt}\x_t\hspace{-1pt}\morf\,\Gr\to\morf\,\Gr
:(\overrightarrow g,\overrightarrow h)\mapsto\overrightarrow g\circ
\overrightarrow h$.\ The structure maps satisfy the consistency
conditions (whenever the expressions are well-defined):
\bit
\item[(i)] $s(\overrightarrow g\circ\overrightarrow h)=s(
\overrightarrow h),\ t(\overrightarrow g\circ\overrightarrow h)=t(
\overrightarrow g)$;
\item[(ii)] $(\overrightarrow g\circ\overrightarrow h)\circ
\overrightarrow k=\overrightarrow g\circ(\overrightarrow h\circ
\overrightarrow k)$;
\item[(iii)] $\Id_{t(\overrightarrow g)}\circ\overrightarrow g=
\overrightarrow g=\overrightarrow g\circ\Id_{s(\overrightarrow g)}$
;
\item[(iv)] $s(\overrightarrow g^{-1})=t(\overrightarrow g),\ t(
\overrightarrow g^{-1})=s(\overrightarrow g),\ \overrightarrow g
\circ\overrightarrow g^{-1}=\Id_{t(\overrightarrow g)},\
\overrightarrow g^{-1}\circ\overrightarrow g=\Id_{s(\overrightarrow
g)}$.
\eit
Thus, a groupoid is a (small) category with all morphisms
invertible.

A \textbf{morphism} between two groupoids $\,\Gr_i,\ i\in\{1,2\}\,$
is a functor $\,\Phi:\Gr_1\to\Gr_2$.

A \textbf{Lie groupoid} is a groupoid whose object and arrow sets
are smooth manifolds, whose structure maps are smooth, and whose
source and target maps are surjective submersions. A morphism
between two Lie groupoids is a functor between them with smooth
object and morphism components. \exdef We may now proceed towards an
analysis of geometric symmetries of the $\si$-model to be gauged.
\bedef\label{def:G-act}
Let $\,\xcM\,$ be a smooth manifold, and let $\,\txG\,$ be a group.
A \textbf{left action of group $\,\txG\,$ on manifold $\,\xcM\,$} is
a smooth map
\qq\nn
\xcMup\ell\ :\ \txG\x\xcM\to\xcM\ :\ (g,m)\mapsto g.m\equiv\xcMup
\ell_g(m)\,.
\qqq
A manifold equipped with a left action of group $\,\txG\,$ is termed
a \textbf{$\txG$-space}.

The \textbf{action groupoid} associated to a $\txG$-space $\,\xcM\,$
is a Lie groupoid, usually denoted as
\qq\nn
\txG\lx\xcM\ :\ \alxydim{}{\txG\x\xcM \ar@<.5ex>[r]^{\quad s}
\ar@<-.5ex>[r]_{\quad t} & \xcM}\,,
\qqq
with the object and morphism sets
\qq\nn
\obj\,(\txG\lx\xcM)=\xcM\,,\qquad\qquad\morf\,(\txG\lx\xcM)=\txG\x
\xcM\,,
\qqq
with the source and target maps
\qq\nn
s(g,x):=x\,,\qquad\qquad t(g,x):=g.x\,,
\qqq
with the identity morphisms
\qq\nn
\Id_x:=(e,x)
\qqq
($e\,$ is the group unit), with the inversion map
\qq\nn
\Inv(g,m):=\bigl(g^{-1},g.m\bigr)\equiv(g,m)^{-1}\,,
\qqq
and, finally, with the composition of morphisms
\qq\nn
(g,h.x)\circ(h,x):=(g\cdot h,x)\,.
\qqq
The nerve of this category, termed the \textbf{nerve of action
groupoid $\,\txG\lx\xcM\,$} and denoted as
\qq\label{eq:nerve-cover}\qquad\qquad
\sfN^\bullet(\txG\lx\xcM)\quad :\quad \alxydim{}{ \cdots
\ar@<.75ex>[r] \ar@<.25ex>[r] \ar@<-.25ex>[r] \ar@<-.75ex>[r] &
\txG^2\times\xcM \ar@<.5ex>[r] \ar@<0.ex>[r] \ar@<-.5ex>[r] &
\txG\times\xcM \ar@<.5ex>[r] \ar@<-.5ex>[r] & \xcM}\,,
\qqq
is an incomplete simplicial object in the category of $\txG$-spaces
equipped with \textbf{face maps}
\qq\nn
\xcMup d_i^{(m)}\ :\ \txG^m\x\xcM\to\txG^{m-1}\x\xcM\,,\qquad
i\in\ovl{0,m-1}
\qqq
explicitly given by
\qq\nn
\xcMup d_0^{(m)}(g_{m},g_{m-1},\ldots,g_1,x)&=&(g_{m-1},g_{m-2},
\ldots,g_1,x)\,,\cr\cr \xcMup d_{m}^{(m)}(g_{m},g_{m-1},\ldots,g_1
,x)&=&(g_{m},g_{m-1},\ldots,g_2 ,g_1.x)\,,\cr\cr \xcMup
d_i^{(m)}(g_{m},g_{m-1},\ldots,g_1,x)&=&(g_{m},g_{m-1},\ldots,g_{m+2-i}
,g_{m+1-i}\cdot g_{m-i},g_{m-1-i},\ldots,g_1,x)\,.
\qqq
\exdef \noindent It is over the nerve of the action groupoid
$\,\txG_\si\lx\xcF\,$ that the construction of the gauged
$\si$-model is carried out. We start with the topologically trivial
case.
\berop\cite[Thm.\,4.12]{Gawedzki:2012fu}\label{prop:large-gauge-tt-sigmod}
Adopt the notation of Definitions \ref{def:sigmod-2d},
\ref{def:gauged-sigmod} and \ref{def:G-act}, of Corollary
\ref{cor:Homtw-Vbra-symm-tw}, and of Example \ref{eg:WZW-def}, and
denote by $\,\txG_\si\,$ the symmetry group of the $\si$-model of
Definition \ref{def:sigmod-2d} susceptible of gauging, with the Lie
algebra $\,\ggt_\si$.\ The action functional of
\Reqref{eq:2d-gauge-sigma-def} is invariant under gauge
transformations
\qq\nn
X(\si)\mapsto\xcFup\ell\left(\chi(\si),X(\si)\right)\equiv\chi.X(
\si)\,,\qquad\qquad\txA(\si)\mapsto\Ad_{\chi(\si)}\txA(\si)-\sfd
\chi\,\chi^{-1}(\si)\equiv\ups{\chi}\txA(\si)\,,
\qqq
written in terms of an arbitrary function $\,\chi\in C^\infty(\Si,
\txG_\si)$,\ iff there exist: a 1-isomorphism
\qq\label{eq:Ups-1-iso}
\Upsilon\ :\ \Mup d_1^{(1)\,*}\cG\xrightarrow{\cong}\Mup d_0^{(1)\,
*}\cG\ox I_\rho
\qqq
of gerbes over $\,\txG_\si\x M$,\ with
\qq\nn 
\rho:=\kappa_{A\,2^*}\wedge\theta_{L\,1^*}^A-\tfrac{1}{2}\,\txc_{AB\,
2^*}\,\left(\theta_L^A\wedge\theta_L^B\right)_{1^*}\,\in\,\Om^2(
\txG_\si\x M)\,,
\qqq
and a 2-isomorphism
\qq\label{eq:Xi}
\Xi\ :\ \Qup d_1^{(1)\,*}\Phi\xLongrightarrow{\cong}\bigl(\bigl(
\iota_2^{(1)\,*}\Upsilon^{-1}\ox\id\bigr)\circ\bigl(\Qup d_0^{(1)\,
*}\Phi\ox\id\bigr)\circ\iota_1^{(1)\,*}\Upsilon\bigr)\ox J_\la
\qqq
between the 1-isomorphisms over $\,\txG_\si\x Q$,\ with
$\,\iota_\a^{(1)}:=\id_{\txG_\si}\x \iota_\a\,$ and
\qq\nn 
\la:=-k_{A\,2^*}\,\theta_{L\,1^*}^A\,\in\,\Om^1(\txG_\si\x Q)\,,
\qqq
such that the identities
\qq
\Tnup d_1^{(1)\,*}\varphi_n=\psi_1\bullet\bigl(\id\circ\Tnup d_0^{(
1)\,*}\varphi_n\circ\id\bigr)\bullet\psi_2\bullet\bigl(\bigl(
\Xi_n^{n,1\,(1)}\ox\id\bigr)\circ\bigl(\Xi_n^{n-1,n\,(1)}\ox\id
\bigr)\circ\cdots\circ\Xi_n^{1,2\,(1)}\bigr)\qquad\qquad
\label{eq:varphi-equiv}
\qqq
hold true over $\,\txG_\si\x T_n\,$ (for all $\,n\geq 3$) for the
2-isomorphisms $\,\Xi_n^{k,k+1\,(1)}=\left(\id_{\txG_\si}\x\pi_n^{k
,k+1}\right)^*\Xi^{\vep_n^{k,k+1}}\,$ and for certain 2-isomorphisms
$\,\psi_\b,\ \b\in\{1,2\}\,$ canonically determined by components of
$\,\Bgt\,$ and by $\,\Upsilon$.
\eerop

Physical considerations presented in \Rcite{Gawedzki:2012fu} seem to
imply that a consistent quantum field theory of the gauged
$\si$-model requires incorporating topologically non-trivial gauge
fields into the lagrangean description. These are represented by
principal $\txG_\si$-connection 1-forms on arbitrary principal
$\txG_\si$-bundles over $\,\Si$,\ for which our conventions are
summarised in
\bedef\label{def:princ-g-bund}
Let $\,\txG\,$ be a (topological) group, and $\,\Si\,$ a topological
space\footnote{In the context of the present paper, the definition
will usually be restricted to the smooth category.}. A
\textbf{principal $\txG$-bundle over base $\,\Si\,$} is the
quadruple $\,\cP:=(\sfP,\Si ,\pi_\sfP,r_\sfP)\,$ composed of a fibre
bundle $\,\pi_\sfP:\sfP\to \Si\,$ with total space $\,\sfP\,$ and
base $\,\Si$,\ and of a continuous free and transitive fibrewise
right action
\qq\nn
r_\sfP\ :\ \sfP\x\txG\to\sfP\ :\ (p,g)\mapsto r(p,g)\equiv p.g\,.
\qqq
Under a local trivialisation
\qq\nn
\tau_i\ :\ \pi_\sfP^{-1}(\Si_i)\to\Si_i\x\txG
\qqq
associated with a choice $\,\{\Si_i\}_{i\in\xcI}\,$ of an open cover
of $\,\Si\,$ and defining a local section $\,\si_i:\Si_i\to
\pi_\sfP^{-1}(\Si_i)\,$ in the standard manner,
\qq\nn
\si_i(\si):=\tau_i^{-1}(\si,e)\,,
\qqq
the above action is related to the action of $\,\txG\,$ on itself by
right regular translations,
\qq\nn
\tau_i^{-1}(\si,g).h=\tau_i^{-1}(\si,g\cdot h)\,.
\qqq

For $\,\txG\,$ a Lie group with Lie algebra $\,\ggt$,\ the latter
having generators $\,t_A,\ A\in\ovl{1,\dim\,\ggt}\,$ subject to the
structure relations
\qq\label{eq:str-rel}
[t_A,t_B]=f_{ABC}\,t_C
\qqq
written in terms of structure constants $\,f_{ABC}$,\ a
\textbf{connection (1-form) on $\,\cP$},\ also termed the
\textbf{principal $\txG$-connection 1-form}, is a $\ggt$-valued
1-form $\,\cA\in\Om^1(\sfP)\ox\ggt\,$ with the defining properties
\qq\nn
\cA(p.g)=\Ad_{g^{-1}}\cA(p)\,,\qquad\qquad\ups{\sfP}\xcK_A\con\cA=
t_A\,,
\qqq
expressed in terms of the \textbf{fundamental vector fields}
$\,\ups{\sfP}\xcK_A\,$ on $\,\sfP\,$ determined by the formula
\qq\nn
(\ups{\sfP}\xcK_A f)(p):=\tfrac{\sfd\ }{\sfd t}\vert_{t=0}f\left(
\ee^{-tt_A}.p\right)\,,
\qqq
valid for any $\,f\in C^\infty(\sfP,\bR)$.

Finally, let $\,(\xcM,\xcMup\ell)\,$ be a (topological) $\txG$-space
in the sense of Definition \ref{def:grpd}. The \textbf{bundle
associated to $\,\cP\,$ by (left) action $\,\xcMup\ell$},\ also
termed the \textbf{associated bundle} for short whenever there is no
risk of confusion, is the fibre bundle
\qq\nn
\pi_{\sfP\x_\txG\xcM}\ :\ \sfP\x_\txG\xcM\equiv(\sfP\x\xcM)/\txG\to
\Si
\qqq
obtained as the smooth quotient of the product bundle $\,\sfP\x\xcM
\to\Si\,$ with respect to the right action $\,\widetilde r\,$ of
$\,\txG\,$ on the total space $\,\sfP\x\xcM\,$ given by
\qq\label{eq:right-diag-assoc}
\widetilde r\ :\ (\sfP\x\xcM)\x\txG\to\sfP\x\xcM\ :\ \bigl((p,m),g
\bigr)\mapsto\bigl(r_\sfP(p,g),\xcMup\ell(g^{-1},m)\bigr)\,,
\qqq
and equipped with the projection
\qq\nn
\pi_{\sfP\x_\txG\xcM}\left(\left[(p,m)\right]\right):=\pi_\sfP(p)
\,.
\qqq
\exdef \noindent We may now formulate
\bedef\label{def:P-ext-bgrnd}\cite[Def.\,10.1]{Gawedzki:2012fu}
Adopt the notation of Definitions \ref{def:sigmod-2d} and
\ref{def:gauged-sigmod}, and let $\,\pi_\sfP:\sfP\to\Si\,$ be a
principal $\txG_\si$-bundle over $\,\Si\,$ with a principal
connection 1-form $\,\cA\in\Om^1(\sfP)\ox\ggt_\si$.\ A
\textbf{$\sfP$-extended string background} is the string background
$\,\widetilde\Bgt_\cA:=(\widetilde\cM_\cA,\widetilde\cB_\cA,
\widetilde\cJ_\cA)\,$ with the following components
\bit
\item the \textbf{$\sfP$-extended target} $\,\widetilde\cM_\cA\,$
composed of the target space $\,\widetilde M=\sfP\x M\,$ with the
metric $\,\widetilde\txg_\cA\,$ and the gerbe
$\,\widetilde\cG_\cA\,$ defined as in Eqs.\,\eqref{eq:gA-def} and
\eqref{eq:cGA}, respectively, but with the global connection 1-form
$\,\txA\,$ on $\,\Si\,$ replaced by $\,\cA$;
\item the \textbf{$\sfP$-extended $\widetilde\cG_\cA$-bi-brane}
$\,\widetilde\cB_\cA\,$ with the world-volume $\,\widetilde
Q=\sfP\vert_\G\x Q$,\ with the bi-brane maps
$\,\widetilde\iota_\a=\id_\sfP\x\iota_\a,\ \a\in\{1,2\}$,\ and with
the curvature $\,\widetilde\om_\cA\,$ and the 1-isomorphism
$\,\widetilde\Phi_\cA\,$ defined as in Eqs.\,\eqref{eq:omA} and
\eqref{eq:PhiA}, respectively, but with the global connection 1-form
$\,\txA\,$ on $\,\Si\,$ replaced by $\,\cA$;
\item the \textbf{$\sfP$-extended $(\widetilde\cG_\cA,\widetilde\cB_\cA
)$-inter-bi-brane} $\,\widetilde\cJ_\cA\,$ with component
world-volumes $\,\widetilde T_n=\sfP \vert_{\Vgt_\G^{(n)}}\x T_n,\
n\geq 3$,\ with inter-bi-brane maps
$\,\widetilde\pi_n^{k,k+1}=\id_\sfP\x\pi_n^{k, k+1},\ k\in \bZ/n
\bZ$,\ and 2-isomorphisms $\,\widetilde\varphi_{n\,
\cA}=\varphi_{n\,2^*}$.\eit \exdef \noindent The idea behind the
introduction of the $\sfP$-extended string background is that it
permits to write the coupling between the original string background
and the topologically non-trivial gauge field in a manner that
generalises the treatment of the topologically trivial case. This
comes at a price: The target space $\,\widetilde\xcF=\sfP\x\xcF\,$
of the $\sfP$-extended string background is not the physical space
of the corresponding gauged $\si$-model. In order to keep the
original field content, we have to pass to the (smooth) quotient
$\,\widetilde\xcF/\txG_\si\cong\xcF\,$ with respect to the combined
(right) action
\qq\nn
\txcFup\ell\ :\ \widetilde\xcF\x\txG_\si\to\widetilde\xcF\ :\ \bigl(
(p,x),g\bigr)\mapsto\bigl(r_\sfP(p,g),\xcFup\ell\left(g^{-1},x
\right)\bigr)\,.
\qqq
We arrive thereby at associated bundles. Dividing out
$\,\txcFup\ell\,$ is straightforward on the level of the target
space, and the true challenge is to ensure that also the geometric
structure supported by $\,\widetilde\xcF\,$ descends to the quotient
space in the sense rendered rigorous in
\Rxcite{Sec.\,8}{Gawedzki:2012fu}. To describe such circumstances,
we need
\bedef\cite[Def.\,8.7]{Gawedzki:2012fu}\label{def:Gequiv-bgrnd}
Adopt the notation of Definitions \ref{def:sigmod-2d} and
\ref{def:G-act}, of Corollary \ref{cor:Homtw-Vbra-symm-tw}, and of
Proposition \ref{prop:large-gauge-tt-sigmod}. Let
$\,\{\tau^A\}^{A\in\ovl{1, \dim\,\ggt}}\,$ be the generators of
$\,\ggt_\si^*\,$ dual to the generators
$\,\{t_A\}_{A\in\ovl{1,\dim\,\ggt}}\,$ of $\,\ggt_\si\,$ satisfying
\Reqref{eq:str-rel}. A \textbf{$(\txG_\si,\rho,\la)$-equivariant
string background} is a triple
$\,\Bgt_{(\txG_\si,\rho,\la)}:=(\cM_{(\txG_\si,\rho)},\cB_{(
\txG_\si,\la)},\cJ_{\txG_\si})\,$ with the following components:
\bit
\item a \textbf{$(\txG_\si,\rho)$-equivariant target} $\,\cM_{(
\txG_\si,\rho)}:=\left(M,\txg,(\cG,\Upsilon,\g)\right)$,\ consisting
of a target space $\,M\,$ carrying the structure of a
$\txG_\si$-space with a $\txG_\si$-invariant metric $\,\txg$,\ and
of a gerbe $\,\cG\,$ with a $\txG_\si$-invariant curvature
$\,\txH\,$ admitting a $\ggt_\si$-equivariantly closed
$\txG_\si$-equivariant (Cartan-model) extension $\,\widehat\txH=\txH
-\kappa,\ \kappa=\kappa_A\ox\tau^A$,\ and endowed with a $(\txG_\si
,\rho)$-equivariant structure, \textit{i.e.}\ coming with a
1-isomorphism $\,\Upsilon\,$ of gerbes over $\,\txG_\si\x M$ as in
\Reqref{eq:Ups-1-iso} and with a 2-isomorphism
\qq\label{diag:2-iso-Gequiv}
\qquad\alxydim{@C=1cm@R=2cm}{ \bigl(\Mup d^{(1)}_1\circ\Mup d^{(2)}_1
\bigr)^*\cG \ar[r]^{\Mup d^{(2)\,*}_2\Upsilon\hspace{1cm}}
\ar[d]_{\Mup d^{(2)\,*}_1\Upsilon} & \bigl(\Mup d^{(1)}_1\circ\Mup
d^{(2)}_0\bigr)^*\cG\ox I_{\Mup d^{(2)\,*}_2\rho} \ar[d]^{\Mup
d^{(2) \,*}_0\Upsilon \ox\id} \ar@{=>}[dl]|{\,\g\ } \\ \bigl(\Mup
d^{(1)}_0\circ\Mup d^{(2)}_1\bigr)^*\cG\ox I_{\Mup d^{(2)\,*}_1\rho}
\ar@{=}[r] & \bigl(\Mup d^{(1)}_0\circ\Mup d^{(2)}_0\bigr)^*\cG\ox
I_{\Mup d^{(2)\,*}_0\rho+\Mup d^{(2)\,*}_2\rho}}
\qqq
between the 1-isomorphisms over $\,\txG_\si^2\x M$,\ satisfying,
over $\,\txG_\si^3\x M$,\ the coherence condition
\qq\label{eq:Gerbe-1iso-coh}
\Mup d_1^{(3)\,*}\g\bullet\bigl(\id\circ\Mup d_3^{(3)\,*}\g\bigr)=
\Mup d_2^{(3)\,*}\g\bullet\bigl(\bigl(\Mup d_0^{(3)\,*}\g\ox\id
\bigr)\circ\id\bigr)\,;
\qqq
\item a \textbf{$(\txG_\si,\la)$-equivariant $\cG$-bi-brane}
$\,\cB_{(\txG_\si,\la)}=\left(\cB,\Xi\right)$,\ consisting of a
bi-brane $\,\cB\,$ with a world-volume $\,Q\,$ carrying the
structure of a $\txG_\si$-space, with a $\txG_\si$-invariant
curvature $\,\om\,$ admitting a $\txG_\si$-equivariant
(Cartan-model) extension $\,\widehat\om=\om-k,\ k=k_A\ox\tau^A\,$
satisfying the relations
\qq\nn
\widehat\sfd\widehat\om=-\D_Q\widehat\txH\,,\qquad\qquad\D_{T_n}
\widehat\om=0\,,
\qqq
and endowed with a $(\txG_\si,\la)$-structure, \textit{i.e.}\ coming
with a 2-isomorphism $\,\Xi\,$ over $\,\txG_\si\x Q\,$ as in
\Reqref{eq:Xi}, subject to the coherence condition
\qq
\bigl(\bigl(\iota_2^{(1)\,*}\g^\sharp\ox\id\bigr)\circ\id\bigr)
\bullet\Qup d^{(2)\,*}_1\Xi=\bigl(\id\circ\iota_1^{(1)\,*}\g\bigr)
\bullet\bigl(\id\circ\bigl(\Qup d_0^{(2)\,*}\Xi\ox\id\bigr)\circ\id
\bigr)\bullet\Qup d_2^{(2)\,*}\Xi\cr\label{eq:Gequiv-bimod-coh}
\qqq
imposed over $\,\txG_\si^2\x Q$;
\item a \textbf{$\txG_\si$-equivariant $(\cG,\cB)$-inter-bi-brane}
$\,\cJ_{\txG_\si}=\cJ\,$ defined by a $(\cG,\cB)$-inter-bi-brane
$\,\cJ\,$ with component world-volumes $\,T_n\,$ each carrying the
structure of a $\txG_\si$-space, and with 2-isomorphisms
$\,\varphi_n\,$ satisfying relation \eqref{eq:varphi-equiv}.
\eit
\exdef \brem Above, the Cartan model of $\ggt_\si$-equivariant
cohomology of the $\txG_\si$-space $\,\xcF\,$ is taken with the
$\ggt_\si$-equivariant differential defined on
$\ggt_\si$-equivariant $p$-forms $\,\eta$, polynomially dependent on
elements of $\,\ggt_\si(\ni V)$,\ by the expression
\qq\nn
\widehat\sfd\eta(V)=\sfd\eta(V)-\ovl V\con\eta(V)\,,
\qqq
where $\,\ovl V\,$ is the vector field acting on smooth functions on
$\,\xcF\,$ according to the formula
\qq\label{eq:fund-vec}
(\ovl V f)(m):=\tfrac{\sfd\ }{\sfd t}\vert_{t=0}f\left(\ee^{-tV}.m
\right)\,.
\qqq
The imposition of the requirements that both $\,\txH\,$ and
$\,\om\,$ admit $\ggt_\si$-equivariant extensions, and that the
latter compose a $(\D_Q,\D_{T_n})$-relative $\ggt$-equivariant
3-cocycle guarantees that the small gauge anomaly vanishes.

Furthermore, $\,\circ\,$ and $\,\bullet\,$ are -- respectively --
the horizontal and the vertical composition of 1- and 2-isomorphisms
of the 2-category of bundle gerbes with connection (over the
relevant base $\,\txG_\si^m\x\xcF$), \textit{cf.}\
\Rxcite{Sect.\,1.2}{Waldorf:2007mm}, and $\,\psi^\sharp\ :\
\Psi_2^{-1}\xrightarrow{\ \cong\ }\Psi_1^{-1}\,$ is a 2-isomorphism
canonically induced by a given 2-isomorphism $\,\psi\ :\ \Psi_1
\xrightarrow{\ \cong\ }\Psi_2\,$ in a manner detailed in
\Rxcite{Sect.\,1.3}{Waldorf:2007mm}. \erem \noindent We may now
phrase the important
\bethe\cite[Thm.\,9.7, Cor.\,10.9 \& Def.\,10.10]{Gawedzki:2012fu}
In the notation of Definitions \ref{def:sigmod-2d} and
\ref{def:P-ext-bgrnd}, and of Proposition
\ref{prop:large-gauge-tt-sigmod}, a $\sfP$-extended string
background $\,\widetilde\Bgt_\cA\,$ with target space
$\,\widetilde\xcF\,$ descends to a unique string background
$\,\Bgt_\cA\,$ with target space $\,\widetilde\xcF/ \txG_\si\,$ if
the underlying string background $\,\Bgt\,$ carries a
$(\txG_\si,\rho,\la)$-equivariant structure of Definition
\ref{def:Gequiv-bgrnd}. The descendant string background
$\,\Bgt_\cA\,$ then defines the gauged $\si$-model coupled to a
gauge field $\,\cA$. \ethe \noindent The last theorem motivates
\bedef
In the notation of Proposition \ref{prop:large-gauge-tt-sigmod} and
of Definition \ref{def:Gequiv-bgrnd}, the \textbf{large gauge
anomaly} of the gauged $\si$-model is the obstruction to the
existence of a choice of a 1-isomorphism $\,\Upsilon\,$ and
2-isomorphisms $\,\g\,$ and $\,\Xi\,$ satisfying relations
\eqref{eq:Gerbe-1iso-coh}, \eqref{eq:Gequiv-bimod-coh} and
\eqref{eq:varphi-equiv}.\exdef

\brem A comment is due on the status of the gauging procedure
outlined, and -- consequently -- also on that of the gauge anomaly.
It ought to be kept in mind that the procedure involves choices,
such as, \textit{e.g.}, the choice of the coupling between the
string background and the gauge field (at most quadratic in the
latter, and in this sense `minimal') determining the form of the
small gauge anomaly, and that even for this distinguished choice of
the coupling the existence of a $\txG_\si$-equivariant structure on
the string background, tantamount to the vanishing of the large
gauge anomaly, is a \emph{sufficient} condition for a consistent
gauging of the global symmetry, and not an obviously
\emph{necessary} one (excepting the case of discrete symmetries,
treated in all generality in \Rcite{Gawedzki:2008um}, to which the
present procedure applies through reduction, and in which the answer
is known to be unique). It is, therefore, imperative to back up our
proposal for the universal gauge principle with additional
structural evidence attesting its naturalness and versatility as a
tool of description of the $\si$-model with a local symmetry. Steps
towards this end were taken already in \Rcite{Gawedzki:2012fu} where
the small gauge anomaly was given a simple interpretation in the
framework of $\ggt_\si$-equivariant cohomology of the target space
$\,\xcF\,$ (Sec.\,3.2, \emph{ib.}, but \textit{cf.}\ also
Refs.\,\cite{Jack:1989ne,Hull:1989,Figueroa:1994dj,Figueroa:1994ns,Witten:1991mm,Wu:1993iia,Gawedzki:2010rn}
for earlier results in this direction), where an infinitesimal
analogon of the $\txG_\si$-equivariant structure was extracted from
a local analysis of the string background with a vanishing small
gauge anomaly (Sec.\,7, \emph{ib.}), and where the necessity of the
existence of a full-fledged $\txG_\si$-equivariant structure was
demonstrated in the special situation in which the coset target
space $\,\xcF/\txG_\si\,$ is a smooth manifold (Sec.\,9,
\emph{ib.}). Below, we take up anew the pursuit of evidence in
favour of the proposal of \Rcite{Gawedzki:2012fu}, putting it in the
context of the canonical description of the $\si$-model in the
presence of conformal defects, with emphasis on their relation to
$\si$-model dualities, and that of the generalised geometry of rigid
symmetries of the two-dimensional field theory of interest. \erem

\subsection{The canonical description of the gauged $\si$-model}

In this first part of our discussion, we shall set up a canonical
description of the gauge anomaly. From this description, the
vanishing of the anomaly will be seen to emerge as a sufficient
condition for the existence of a hamiltonian realisation of the Lie
algebra $\,\ggt_\si\,$ of the symmetry group $\,\txG_\si\,$ on the
phase space of the $\si$-model, consistent with the
splitting-joining interactions and admitting a canonical extension
to a realisation of $\,C^\infty(\Si,\ggt_\si)\,$ as a gauge-symmetry
algebra on the phase space of the gauged $\si$-model coupled
`minimally' to a topologically trivial gauge field. Here, the term
``gauge symmetry'' indicates, in keeping with, \textit{e.g.},
\Rcite{Gawedzki:1972ms}, that the vector fields generating
infinitesimal gauge transformations on functions on the said phase
space belong to the characteristic distribution of the presymplectic
form of the gauged $\si$-model.

We commence by introducing the main elements of the subsequent
analysis.
\bedef\label{def:A-Psik}
Adopt the notation of Definitions \ref{def:sigmod-2d},
\ref{def:gauged-sigmod}, \vref{def:net-field}I.2.6,
\vref{def:cov-conf-bdle-si}I.3.5, \vref{def:untw-phspace}I.3.9 and
\vref{def:tw-phspace}I.3.10, and that of Propositions
\ref{prop:Htw-Vbra-symm} and \vref{prop:Cart-si-def}I.3.8. Let
$\,\pi_{\cF_\si}:\cF_\si\to\Si\,$ be the covariant configuration
bundles of the non-linear $\si$-model of Definition
\ref{def:sigmod-2d}. The \textbf{covariant configuration bundles of
the} corresponding \textbf{gauged non-linear $\si$-model} of
Definition \ref{def:gauged-sigmod} are given by the fibred product
\qq\nn
\pi_{\cF_\si}\circ\pr_1\ :\ \widetilde\cF_\si:=\cF_\si
\fibx{\pi_{\cF_\si}}{\pi_{\sfT^*\Si}}\left(\sfT^*\Si\ox\ggt_\si
\right)\to\Si\,.
\qqq
For the associated first-jet bundles, $\,\sfJ^1\widetilde\cF_\si\to
\Si$,\ we shall use the set of local coordinates from Definition
\vref{def:cov-conf-bdle-si}I.3.5, augmented by local coordinates
$\,(\txA^A_a,\z^A_{ab}),\ A\in\ovl{1,\dim\,\ggt_\si},\ a,b\in\{1,2
\}\,$ on the fibre of $\,\sfJ^1\sfT^*\Si\ox\ggt_\si$.

Similarly, we shall parameterise classical sections of $\,\sfJ^1
\widetilde\cF_\si\,$ (understood in the sense of an obvious
extension of Proposition \vref{prop:Cart-si-def}I.3.8 to the setting
of the gauged $\si$-model) by their Cauchy data localised on a
suitable space-like contour $\,\xcC\subset\Si$.\ The \textbf{state
space of the gauged non-linear $\si$-model} $\,\widetilde\sfP_\si
\subset\G\left(\sfJ^1\widetilde\cF_\si\right)\,$ composed of these
classical sections splits naturally into the untwisted and
$N$-twisted sectors (with $\,N\in\bN\setminus\{0\}$), \textit{cf.}
Remark \vref{rem:states}I.2.11, and the respective Cauchy data take
the following form:
\bit
\item in the untwisted sector, to be denoted as $\,\widetilde
\sfP_{\si,\emptyset}$,\ they are given by a quadruple $\,(X,\sfp,
\txA,\Pi)\,$ composed of smooth loops $\,X:\bS^1\to M\,$ and
$\,\txA:\bS^1\to\Om^1(\Si)\ox\ggt_\si$,\ and of the respective
normal covector fields $\,\sfp\,$ and $\,\Pi\,$ (\textit{cf.}\
Definition \vref{def:untw-phspace}I.3.9), where the latter pair,
$\,(\txA,\Pi)$,\ is implicitly determined by the former one,
$\,(X,\sfp )$,\ through the Euler--Lagrange equations for $\,\txA$,
\qq\label{eq:A-EL}
\bigg(\begin{smallmatrix} \txh_{AB} & -\txc_{[AB]} \\ \\
-\txc_{[AB]} & \txh_{AB} \end{smallmatrix}\bigg)\,
\bigg(\begin{smallmatrix} \left(\widehat n\con\txA^B\right) \\ \\
\left(\widehat t\con\txA^B\right) \end{smallmatrix}\bigg)=
\bigg(\begin{smallmatrix} \txK_{A\,\mu} & \kappa_{A\,\mu} \\ \\
\kappa_{A\,\mu} & \txK_{A\,\mu} \end{smallmatrix}\bigg)\,
\bigg(\begin{smallmatrix} (X_*\widehat n)^\mu \\ \\ (X_*\widehat
t)^\mu
\end{smallmatrix}\bigg)\,,\qquad\qquad \txc_{[AB]}:=\tfrac{1}{2}\,
\left(\txc_{AB}-\txc_{BA}\right)\,,
\qqq
obtained by varying the action functional of
\Reqref{eq:2d-gauge-sigma-def} in the direction of the world-sheet
gauge field\footnote{Recall that we have fixed a minkowskian gauge
for the world-sheet metric $\,\eta$.}, \textit{cf.}\
\Rxcite{Sec.\,9}{Gawedzki:2012fu}, and written in terms of the
vector field $\,\widehat t\,$ tangent to $\,\xcC\,$ and defining its
orientation, and of the vector field $\,\widehat
n=\eta^{-1}\left(\widehat t\con\Vol(\Si, \eta),\cdot\right)\,$
normal to it.
\item in the $N$-twisted sector, to be denoted as $\,\widetilde
\sfP_{\si,\cB\,\vert\,(P_k,\vep_k)}$,\ they are given by the $(2N+
4)$-tuple $\,(X,\sfp,q_k,V_k,\txA,\Pi\,\vert\,k\in\ovl{1,N})\,$
composed of smooth maps $\,X:\bS^1_{\{P_k\}}\to M\,$ and
$\,\txA:\bS^1\to\Om^1( \Si)\ox\ggt_\si\,$ (a loop), of the
respective normal covector fields $\,\sfp\,$ and $\,\Pi$,\ and of
$N$ points $\,(q_k,V_k)\in\sfT\unl Q$,\ related to $\,(X,\sfp)\,$ as
in Definition \vref{def:tw-phspace}I.3.10. Here, the submanifold
\qq\nn
\unl Q:=\bigcap_{A=1}^{\dim\,\ggt_\si}\,k_A^{-1}\bigl(\{0\}\bigr)
\subset Q
\qqq
is assumed smooth, \textit{cf.}\
\Rxcite{Eq.\,(9.5)}{Gawedzki:2012fu}, and the pair $\,(\txA,\Pi)\,$
is determined by the pair $\,(X,\sfp)\,$ through \Reqref{eq:A-EL}
taken in conjunction with the condition of continuity of the gauge
field along $\,\bS^1_{\{P_k\}}$.
\eit
\exdef \noindent We readily establish
\berop
Adopt the notation of Proposition \ref{prop:sympl-form-twuntw}, of
Corollary \ref{cor:Homtw-Vbra-symm-tw}, and of Definitions
\ref{def:gauged-sigmod} and \ref{def:A-Psik}. A (pre)symplectic form
on $\,\widetilde\sfP_{\si,\emptyset}$,
\qq\label{eq:A-sympl-form-untw}
\widetilde\Om_{\si,\emptyset}[(X,\sfp,\txA,\Pi)]=\Om_{\si,\emptyset}
[(X,\txp)]+\d\int_{\bS^1}\,\Vol\left(\bS^1\right)\wedge\D(X;\txA)\,,
\qqq
and that on $\,\widetilde\sfP_{\si,\cB\,\vert\,(\pi,\vep)}$,
\qq\label{eq:A-sympl-form-1tw}
\widetilde\Om_{\si,\cB\,\vert\,(\pi,\vep)}[(X,\sfp,q,V,\txA,\Pi)]=
\Om_{\si,\cB\,\vert\,(\pi,\vep)}[(X,\txp,q,V)]+\d\int_{\bS^1_{\{\pi
\}}}\,\Vol\left(\bS^1_{\{\pi\}}\right)\wedge\D(X;\txA)\,,
\qqq
differ from their counterparts from Proposition
\ref{prop:sympl-form-twuntw} by the 1-form
\qq\nn
\D(X;\txA):=\left(\widehat t\con\txA^A\right)\,X^*\kappa_A-\left(
\widehat n\con\txA^A\right)\,X^*\txK_A\,.
\qqq
\eerop
\beroof
An easy exercise using the basic methods of the first-order
formalism recapitulated in \Rxcite{Sec.\,3}{Suszek:2011hg}. \eroof

We are now fully prepared to give a canonical interpretation of the
small gauge anomaly. We begin with the unextended $\si$-model, prior
to the gauging. Putting together Propositions
\ref{prop:sympl-goes-ham-untw}, \ref{prop:ham-cont-across} and
\ref{prop:sympl-goes-ham-tw} and Theorems
\ref{thm:DJI-aug-intertw-untw} and \ref{thm:ham-cons-tw}, we obtain
\bethe\label{thm:sga-vs-ham}
If the small gauge anomaly of Theorem \ref{thm:small-ganom}
vanishes, then there exists a hamiltonian realisation of the
symmetry Lie algebra $\,\ggt_\si\,$ of Proposition
\ref{prop:Htw-Vbra-symm} on the full state space of the
two-dimensional non-linear $\si$-model of Definition
\ref{def:sigmod-2d}, and that realisation is continuous across the
defect quiver of the $\si$-model in the sense of Proposition
\ref{prop:ham-cont-across} and consistent with the splitting-joining
interactions in the sense of Theorems \ref{thm:DJI-aug-intertw-untw}
and \ref{thm:ham-cons-tw}. \ethe
\beroof
This is a simple corollary to the propositions and theorems listed
above. \eroof

The significance of the small gauge anomaly in the canonical
description of the gauged $\si$-model is emphasised by the following
\bethe\label{thm:sga-vs-gaugesym}
If the small gauge anomaly of Theorem \ref{thm:small-ganom}
vanishes, then there exists a canonical extension of each vector
field generating the action of the symmetry Lie algebra
$\,\ggt_\si\,$ of Proposition \ref{prop:Htw-Vbra-symm} on smooth
functions on the full state space of the two-dimensional non-linear
$\si$-model of Definition \ref{def:sigmod-2d} to a vector field
generating the action of $\,C^\infty(\Si,\ggt_\si)\,$ as a
gauge-symmetry algebra on the full state space of the gauged
two-dimensional non-linear $\si$-model of Definition
\ref{def:gauged-sigmod}. \ethe
\beroof
In the proof, we focus on the 1-twisted case exclusively. Clearly,
our result generalises straightforwardly to the $N$-twisted case
with $\,N\geq 1\,$ arbitrary. Furthermore, the claim for the
untwisted case can readily be recovered from what follows through a
trivialisation of the twist, \textit{i.e.}\ through setting
$\,\om=0,\ \g_A=0=k_A\,$ and $\,\iota_1=\id_M=\iota_2$.

Let $\,\La^A\,t_A\in\ggt_\si\,$ and $\,\ovl\La:=\La^A\,\xcK_A\in\G(
\sfT\xcF)$,\ and write the corresponding vector field on
$\,\sfP_{\si,\cB\,\vert\,(\pi,\vep)}\,$ explicitly as
\qq\nn
\widetilde\sfL_{\iota_\a\,*}^{Q\,\vert\,(\pi,\vep)}\ovl\La[(X,\sfp,
q,V)]&=&\int_{\bS^1_{\{\pi\}}}\,\Vol\left(\bS^1_{\{\pi\}}\right)\,
\La^A\,\left[\Mup\xcK^\mu_A\left(X(\cdot)\right)\,\tfrac{\d\quad\
}{\d X^\mu(\cdot)}-\sfp_\nu(\cdot)\,\p_\mu\Mup\xcK^\nu_A\bigl(X(
\cdot)\bigr)\,\tfrac{\d\quad\ }{\d\sfp_\mu(\cdot)}\right]\cr\cr
&&+\La^A\,\Qup\xcK^\mu_A(q)\,\tfrac{\d\quad\ }{\d X^\mu(q)}
\qqq
Let, next, $\,\La^A(\cdot)\,t_A\in C^\infty(\Si,\ggt_\si)$,\ and
abbreviate
\qq\nn
\txA^A_x:=\widehat x\con\txA^A\,,\qquad\qquad\La^A_{,x}:=\widehat x
\left(\La^A\right)\,,\qquad x\in\{t,n\}\,.
\qqq
Taking into account the transformation properties of the canonical
variables of the gauged $\si$-model under the infinitesimal gauge
transformation $\,\La^A(\cdot)\,t_A$,\ we may easily write out the
unique extension of $\,\widetilde\sfL_{\iota_\a
\,*}^{Q\,\vert\,(\pi,\vep)}\ovl\La\,$ in the form
\qq\nn
&&\widetilde\La[(X,\sfp,q,V,\txA,\Pi)]\cr\cr
&:=&\int_{\bS^1_{\{\pi\}}}\,\Vol\left(\bS^1_{\{\pi\}}\right)\,
\bigl\{\La^A(\cdot)\,\Mup\xcK^\mu_A\left(X(\cdot)\right)\,\tfrac{\d
\quad\ }{\d X^\mu(\cdot)}-\left[\La^A\,\sfp_\nu(\cdot)\,\p_\mu\Mup
\xcK^\nu_A\bigl(X(\cdot)\bigr)-\La^A_{,n}(\cdot)\,\txK_{A\,\mu}
\left(X(\cdot)\right)\right]\,\tfrac{\d\quad\ }{\d\sfp_\mu(\cdot)}
\cr\cr
&&+\left(\p_a\La^A-f_{ABC}\,\La^B\,\txA^C_a\right)(\cdot)\,
\tfrac{\d\quad\ }{\d\txA^A_a(\cdot)}\bigr\}+\La^A(\pi)\,\Qup
\xcK^\mu_A(q)\,\tfrac{\d\quad\ }{\d X^\mu(q)}\,.
\qqq
It now remains to check that $\,\widetilde\La\,$ is in the kernel of
$\,\widetilde\Om_{\si,\cB\,\vert\,(\pi,\vep)}$.\ Upon invoking the
defining formul\ae ~for the $\,\kappa_A, \txc_{AB}\,$ and $\,\txh_{A
B}$,\ we obtain
\qq\nn
&&\widetilde\La\con\widetilde\Om_{\si,\cB\,\vert\,(\pi,\vep)}[(X,
\sfp,q,V,\txA,\Pi)]\cr\cr
&=&\int_{\bS^1_{\{\pi\}}}\,\Vol\left(\bS^1_{\{\pi\}}\right)\,
\bigl\{-\La^A(\cdot)\,\bigl[\Mup\xcK^\mu_A\left(X(\cdot)\right)\,\d
\sfp_\mu(\cdot)-X_*\widehat t(\cdot)\con\d\kappa_A\left(X(\cdot)
\right)\cr\cr
&&+\txA^B_n(\cdot)\left(\pLie{\xcK_A}\txK_B-\d\txh_{AB}\right)
\left(X(\cdot)\right)-\txA^B_t(\cdot)\left(\pLie{\xcK_A}\kappa_B-\d
\txc_{AB}\right)\left(X(\cdot)\right)\cr\cr
&&+ \txc_{AB}\left(X(\cdot)\right)\,\d\txA^B_t(\cdot)-\txh_{AB}
\left(X(\cdot)\right)\,\d\txA^B_n(\cdot)+\sfp_\mu(\cdot)\,\d\Mup
\xcK_A^\mu\left(X(\cdot)\right)\bigr]+\La^A_{,n}(\cdot)\,\txK_A
\left(X(\cdot)\right)\cr\cr
&&+\left(\La^A_{,t}-f_{ABC}\,\La^B\,\txA^C_t\right)(\cdot)\,
\kappa_A\left(X(\cdot)\right)-\left(\La^A_{,n}-f_{ABC}\,\La^B\,
\txA^C_n\right)(\cdot)\,\txK_A\left(X(\cdot)\right)\bigr\}\cr\cr
&&+\vep\,\La^A(\pi)\,\Qup\xcK_A\con\om(q)\,.
\qqq
Further simplification of the last expression is achieved with the
help of the Killing equations for the $\,\Mup\xcK_A\,$ and the
defining formul\ae ~for the $\,k_A$.\ Altogether, we have
\qq\nn
&&\widetilde\La\con\widetilde\Om_{\si,\cB\,\vert\,(\pi,\vep)}[(X,
\sfp,q,V,\txA,\Pi)]\cr\cr
&=&\int_{\bS^1_{\{\pi\}}}\,\Vol\left(\bS^1_{\{\pi\}}\right)\,\left[
-\La^A(\cdot)\,\d\widetilde J_A(X,\sfp,\txA)(\cdot)+\La^A\,\txA^B_t(
\cdot)\,\left(\pLie{\xcK_A}\kappa_B-f_{ABC}\,\kappa_C\right)\left(X(\cdot
)\right)\right]\cr\cr
&&-\vep\,\La^A(\pi)\,\d k_A(q)\,,
\qqq
where
\qq\nn
\widetilde J_A(X,\txp;\txA):=\Mup\xcK_A(X)\con\txp+X_*\widehat t\con
\kappa_A(X)-\txh_{AB}(X)\,\txA^B_n+\txc_{AB}(X)\,\txA^B_t
\qqq
are extensions of the Noether currents $\,J_{\Kgt_A}\,$ of
\Reqref{eq:symm-curr-def}.

At this stage, we may start using the field equations\footnote{Note
that the last of them, $\,k_A=0$,\ implies the middle one of
equalities \eqref{eq:gauge-constr}.} for the gauge field alongside
conditions \eqref{eq:gauge-constr}, whereby the above immediately
reduces to the form
\qq\nn
\widetilde\La\con\widetilde\Om_{\si,\cB\,\vert\,(\pi,\vep)}[(X,\sfp
,q,V,\txA,\Pi)]=-\int_{\bS^1_{\{\pi\}}}\,\Vol\left(\bS^1_{\{\pi\}}
\right)\,\La^A(\cdot)\,\d\widetilde J_A(X,\sfp,\txA)(\cdot)\,.
\qqq
Taking into account the explicit formula
\qq\nn
\sfp=\txg_{\mu\nu}(X)\,(X_*\widehat n)^\mu\,\d X^\nu\,,
\qqq
we ultimately arrive at the expression
\qq\nn
\widetilde\La\con\widetilde\Om_{\si,\cB\,\vert\,(\pi,\vep)}[(X,\sfp
,q,V,\txA,\Pi)]=-\d\int_{\bS^1_{\{\pi\}}}\,\Vol\left(\bS^1_{\{\pi\}}
\right)\, \txc_{(AB)}\left(X(\cdot)\right)\,\La^A\,\txA^B_t(\cdot)
\qqq
which vanishes identically whenever the small gauge anomaly does.
\eroof

In the present section, we have identified the small gauge anomaly
as an obstruction to the existence of a canonical realisation of the
infinitesimal gauge symmetry of the gauged $\si$-model. This is to
be viewed as an alternative derivation of the corresponding results
obtained in the lagrangean picture in \Rcite{Gawedzki:2012fu}.
Taking guidance from the intuition developed in
\Rcite{Suszek:2011hg}, we are next led to expect that the vanishing
of the large gauge anomaly ensures, in turn, the existence of a lift
of the integrated (\textit{i.e.}\ finite) version of the symmetry to
an automorphism of the pre-quantum bundle of the gauged $\si$-model.
This expectation will be corroborated and -- indeed -- extended to a
proper world-sheet definition of the
$C^\infty(\Si,\txG_\si)$-reduction of the gauged $\si$-model in
Section \ref{sub:world-coset}. In the meantime, we pause to give a
very natural and purely geometric interpretation of the small gauge
anomaly, (apparently) in abstraction from the underlying
two-dimensional field theory.

\subsection{The generalised-geometric/groupoidal interpretation}

Below, we want to reexamine the small gauge anomaly from an
altogether different, intrinsically geometric perspective offered by
the previously introduced algebroidal structure on the set of
$\si$-symmetric sections of the restricted tangent sheaves over the
target space of the $\si$-model, \textit{cf.}\ Corollary
\ref{cor:Homtw-Vbra-symm-tw}. One of the key results of this section
was already announced in \Rxcite{Sec.\,6}{Gawedzki:2012fu}. However,
in view of its relevance to a more complete understanding of rigid
symmetries of the $\si$-model, as well as of the obvious structural
connection to the rest of the present paper, we have decided to
restate it in the language of Section \ref{sub:rel-tw-Courant}, in
this manner avoiding unnecessary repetitions. Thus, we shall
reinterpret conditions \eqref{eq:gauge-constr} for a consistent
gauging by establishing a straightforward link between the
$(\txH\oplus\om\oplus 0)$-twisted $(\D_Q,\D_{T_n})$-relative Courant
bracket on $\,\Egt_{(\iota_\a,\pi_n^{k,
k+1})}(\xcF)\cong\G_{(\iota_\a,\pi_n^{k,k+1})}\bigl(\widehat\sfE^{(
1,2\sqcup 1\sqcup 0)}\xcF\bigr)\,$ and the action groupoid
$\,\txG_\si\lx\xcF$.\ The latter emerges from the discussion of
large gauge transformations and the ensuing construction of a
$\txG_\si$-equivariant string background, and so its appearance in
the analysis of the gauge anomaly is not very surprising. That it is
actually quite natural will be demonstrated in the second part of
the present section in which we establish a correspondence between
the data of the gauged $\si$-model and the category of principal
bundles over $\,\Si\,$ with the structural groupoid $\,\txG_\si\lx
\xcF$.\ We shall elaborate the correspondence in Section
\ref{sub:world-coset}, where it will be shown to bridge the gap
between the infinitesimal description of the gauge symmetry and its
finite counterpart.

In what follows, we shall make frequent use of basic notions and
constructions of the theory of Lie groupoids and Lie algebroids, and
so we assume working knowledge thereof on the reader's part. For an
in-depth treatment of the theory, consult
Refs.\,\cite{MacKenzie:1987,Moerdijk:2003mm}.

In order to set the stage for subsequent considerations, we recall
\bedef\label{def:algbrd}
Let $\,\xcM\,$ be a smooth manifold. A \textbf{Lie algebroid} over
the \textbf{base} $\,\xcM\,$ is a quintuple $\,\gtGr=\bigl(V,\xcM,[
\cdot,\cdot],\pi_V,\a_{\sfT\xcM}\bigr)\,$ composed of a vector
bundle $\,\pi_V:V\to\xcM$,\ a Lie bracket $\,[\cdot,\cdot]\,$ on the
vector space $\,\G(V)\,$ of its sections, and a bundle map
$\,\a_{\sfT\xcM}:V\to \sfT\xcM\,$ termed the \textbf{anchor (map)}.
These are required to have the following properties:
\bit
\item[(i)] the induced map $\,\G(\a_{\sfT\xcM}):\G(V)\to\G(\sfT\xcM)$,
to be denoted by the same symbol $\,\a_{\sfT\xcM}\,$ in what
follows, is a Lie-algebra homomorphism (with respect to the standard
Lie-algebra structure on $\,\G(\sfT\xcM)\,$ defined by the Lie
bracket of vector fields);
\item[(ii)] $[\cdot,\cdot]\,$ obeys the \textbf{Leibniz
identity}
\qq\nn
[X,f\,Y]=f\,[X,Y]+\a_{\sfT\xcM}(X)(f)\,Y
\qqq
for all $\,X,Y\in\G(V)\,$ and any $\,f\in C^\infty(M,\bR)$.
\eit

A \textbf{morphism} between two Lie algebroids $\,\gtGr_i=\bigl(V_i,
\xcM,[\cdot,\cdot]_i,\pi_{V_i},\a_{\sfT\xcM\,i}\bigr),\ i\in\{1,2
\}\,$ is a bundle map $\,\phi:V_1\to V_2\,$ that satisfies the
relations\footnote{Again, we use the same symbol for the bundle map
and the induced map between spaces of sections.}
\qq\nn
\a_{\sfT\xcM\,1}=\a_{\sfT\xcM\,2}\circ\phi\,,\qquad\qquad \phi\circ
[\cdot,\cdot]_1=[\cdot,\cdot]_2\circ(\phi\x\phi)\,.
\qqq
\exdef \noindent We are now in a position to transcribe and quantify
in the setting in hand the old observation (\textit{cf.}\
\Rcite{Gualtieri:2003dx}) that a Courant bracket does not, in
general, respect the Leibniz rule or the Jacobi identity, and so the
associated Courant algebroid is \emph{not} a Lie algebroid. The
relevant objects are introduced in the following
\bedef\label{def:Courant-anomaly}
Let $\,\xcM\,$ be a smooth manifold, $\,\pi_\sfE:\sfE\to\xcM\,$ a
vector bundle equipped with a bundle map $\,\a_{\sfT\xcM}:\sfE\to
\sfT\xcM$,\ an antisymmetric bracket $\,[\cdot,\cdot]_{\rm C}\,$ and
a non-degenerate symmetric bilinear form $\,\corr{\cdot,\cdot}\,$ on
the set of smooth sections of $\,\sfE$.\ Assume that the quintuple
$\,(\sfE,\xcM,[\cdot,\cdot]_{\rm C},\pi_\sfE,\a_{\sfT\xcM})=:\Cgt\,$
satisfies the axioms of a Courant algebroid with base $\,\xcM$,\ as
stated, \textit{e.g.}, in \Rxcite{Def.\,2.1}{Liu:1997}. Take
arbitrary $\,\Xgt,\Ygt,\Zgt\in\G(\sfE)\,$ and $\,f\in
C^\infty(\xcM,\bR)$,\ and endow $\,\Cgt\,$ with the natural
structure of a $C^\infty(\xcM, \bR)$-module (with respect to
point-wise multiplication). The \textbf{Leibniz anomaly of
$\,\Cgt\,$} is defined as
\qq\nn
\xcL\ :\ \G(\sfE)^2\x C^\infty(\xcM,\bR)\to\G(\sfE)\ :\ (\Xgt,\Ygt,
f)\mapsto [\Xgt,f\cdot\Ygt]_{\rm C}-f\cdot[\Xgt,\Ygt]_{\rm
C}-\left(\pLie{\a_{\sfT\xcM}(\Xgt)}f\right)\cdot\Ygt
\qqq
and the \textbf{Jacobi anomaly} (or \textbf{Jacobiator}) \textbf{of
$\,\Cgt\,$} is given by the formula
\qq\nn
\xcJ\ :\ \G(\sfE)^3\to\G(\sfE)\ :\ (\Xgt,\Ygt,\Zgt)\mapsto[[\Xgt,
\Ygt]_{\rm C},\Zgt]_{\rm C}+[[\Zgt,\Xgt]_{\rm C},\Ygt ]_{\rm
C}+[[\Ygt,\Zgt]_{\rm C},\Xgt]_{\rm C}\,.
\qqq
The two anomalies determine the obstruction to $\,\Cgt\,$ becoming a
Lie algebroid. \exdef \noindent Upon specialisation of the above
general definition to the (relative-geometric) setting of interest,
we establish
\bethe\cite[Prop.\,6.2]{Gawedzki:2012fu}\label{thm:JacLie-sigma}
Adopt the notation of Corollary \ref{cor:Homtw-Vbra-symm-tw}, of
Theorem \ref{thm:brabra}, and of Definitions \ref{def:sigmod-2d},
\ref{def:tw-bra-str-sheaf}, \ref{def:DQT-rel-dR-cohom},
\ref{def:DQT-rel-Lie}, \ref{def:DQ-rel-Courant} and
\ref{def:Courant-anomaly}. Write
\qq\nn
K_A:=\kappa_A\oplus k_A\oplus 0\in\Om^1_{\rm
dR}\left(M,Q,T_n\,\vert\,\D_Q, \D_{T_n}\right)\,,\qquad\qquad
C_{AB}:=\txc_{AB}\oplus 0\in\Om^0_{\rm dR}
\left(M,Q,T_n\,\vert\,\D_Q,\D_{T_n}\right)\,,
\qqq
and
\qq\nn
\eta:=\txH\oplus\om\oplus 0\in\Om^3_{\rm
dR}\left(M,Q,T_n\,\vert\,\D_Q, \D_{T_n}\right)\,.
\qqq
In the basis $\,\{\Psi(\Kgt_A)=:\Psi_A\}_{A\in\ovl{1,\dim\,
\ggt_\si}}$,\ the obstruction to the involutivity of the
$\eta$-twisted $(\D_Q,\D_{T_n})$-relative Courant bracket on
$\,\Egt_{(\iota_\a,\pi_n^{k,k+1})}(\xcF)\,$ is given by the $(\D_Q,
\D_{T_n})$-relative 1-cycle
\qq\nn
\a_{AB}=\pLie{\Mup\xcK_A}^{(\D_Q,\D_{T_n})}K_B-f_{ABC}\,K_C-\sfd_{(
\D_Q,\D_{T_n})}^{(0)}C_{(AB)}\,,
\qqq
and the Leibniz and Jacobi anomalies are determined by the
expressions
\qq\nn
\xcL(\Psi_A,\Psi_B,f)=-0\oplus C_{(AB)}\cdot\sfd_{(\D_Q,\D_{T_n}
)}^{(0)}(f,0)\,,
\qqq
and
\qq\nn
\xcJ(\Psi_A,\Psi_B,\Psi_C)&=&(f_{ABD}\,\ic^{(\D_Q,\D_{T_n}
)}_{\xcK_C}+f_{CAD}\,\ic^{(\D_Q,\D_{T_n})}_{\xcK_B}+f_{BCD}\,\ic^{(
\D_Q,\D_{T_n})}_{\xcK_A})\,\ic^{(\D_Q,\D_{T_n})}_{\xcK_D}\eta\cr\cr
&&+\tfrac{1}{2}\,\sfd_{(\D_Q,\D_{T_n})}^{(0)}(\pLie{\Mup\xcK_A}^{(
\D_Q,\D_{T_n})}C_{[BC]}+\pLie{\Mup\xcK_C}^{(\D_Q,\D_{T_n})}C_{[AB]}
+\pLie{\Mup\xcK_B}^{(\D_Q,\D_{T_n})}C_{[CA]})\,,
\qqq
respectively.

Consequently, the triple
\qq\nn
\Sgt_\Bgt:=\bigl(\sfT_\Egt\xcF,[\cdot,\cdot]_{\rm C}^\eta,\a_{\sfT
\xcF}\bigr)\,,
\qqq
with $\,\sfT_\Egt\xcF\subset\sfT\xcF\,$ the subbundle whose space of
sections is defined as the $C^\infty(\xcF,\bR)$-linear span
\qq\nn
\G(\sfT_\Egt\xcF):=\oplus_{A=1}^{\dim\,\ggt_\si}\,C^\infty(\xcF,\bR
)\,\Psi_A\,,
\qqq
with the obvious bundle projection $\,\pi_\xcF$,\ and with the
$C^\infty(\xcF,\bR)$-linear map $\,\a_{\sfT\xcF}\,$ defined on the
base of $\,\G(\sfT_\Egt\xcF)\,$ as
\qq\nn
\a_{\sfT\xcF}(\Psi_A)=\xcFup\xcK_A\,,
\qqq
carries a canonical structure of a Lie algebroid over $\,\xcF\,$ iff
the small gauge anomaly, expressed in terms of the $\,\Kgt_A$,\
vanishes. The ensuing Lie algebroid is called the
\textbf{gauge-symmetry Lie algebroid of string background $\,\Bgt$}.
\ethe
\beroof
Obvious, through inspection. \eroof \brem It deserves to be
emphasised that the intrinsic ambiguity in the definition of the
$\si$-symmetric sections $\,\Kgt_A\,$ leaves room for nullifying the
small gauge anomaly (and, consequently, for the application of the
last part of the above theorem) even if the latter does not vanish
for the original choice of the $\,\Kgt_A$.\ Indeed, consider two
sets $\,\{\xcK_A\oplus K_A^i\}_{A\in\ovl{1,\dim\, \ggt_\si}},\
K_A^i:=\kappa_A^i\oplus k_A^i\oplus 0,\ i\in\{1,2\}\,$ of elements
of $\,\Egt_{(\iota_\a,\pi_n^{k,k+1})}(\xcF)\,$ satisfying the
defining relations \eqref{eq:sigma-sym-new-def-rel}. Write
\qq\nn
\widetilde\D_A:=K_A^2-K_A^1\,.
\qqq
We find
\qq\label{eq:ambig-in-ker}
\sfd_{(\D_Q,\D_{T_n})}^{(1)}\widetilde\D_A=0\,,
\qqq
and so the said ambiguity is parametrised by $\,\ker\,\sfd_{(\D_Q,
\D_{T_n})}^{(1)}$.\ In fact, it is easy to identify those $(\D_Q,
\D_{T_n})$-relative 1-cocycles whose contribution to the Courant
bracket and to the scalar product is trivial in the sense that it
cannot be used to cancel the anomalous terms $\,\a_{AB}\,$ and $\,
\txc_{AB}\,$ (obtained for the original sections $\,\xcK_A\oplus
K_A^1$). To this end, we calculate, using \Reqref{eq:ambig-in-ker},
\qq\nn
[\xcK_A\oplus(K_A^1+\widetilde\D_A),\xcK_B\oplus(K_B^1+\widetilde
\D_B)]&=&f_{ABC}\,\left(\xcK_C\oplus(K_C^1+\widetilde\D_C)\right)+0
\oplus\a_{AB}\cr\cr
&&+0\oplus\left(\tfrac{1}{2}\left(\pLie{\xcK_A}^{(\D_Q,\D_{T_n})}
\widetilde\D_B-\pLie{\xcK_B}^{(\D_Q,\D_{T_n})}\widetilde\D_A\right)
-f_{ABC}\,\widetilde\D_C\right)\,,\cr\cr\cr \Vcon{\xcK_A\oplus(K_A^1
+\widetilde\D_A)}{\xcK_B\oplus(K_B^1+\widetilde\D_B)}&=&
\txc_{(AB)}+
\tfrac{1}{2}\,(\ic^{(\D_Q,\D_{T_n})}_{\xcK_A}\widetilde\D_B+\ic^{(
\D_Q,\D_{T_n})}_{\xcK_B}\widetilde\D_A)\,.
\qqq
Thus, the conditions of triviality of the correction $\,\widetilde
\D_A\in\ker\,\sfd_{(\D_Q,\D_{T_n})}^{(1)}\,$ read
\qq\nn
\left\{\barr{l}
\tfrac{1}{2}\left(\pLie{\xcK_A}^{(\D_Q,\D_{T_n})}\widetilde\D_B-
\pLie{\xcK_B}^{(\D_Q,\D_{T_n})}\widetilde\D_A\right)-f_{ABC}\,
\widetilde\D_C=0\,,\cr\cr
\ic^{(\D_Q,\D_{T_n})}_{\xcK_A}\widetilde\D_B+\ic^{(\D_Q,\D_{T_n}
)}_{\xcK_B}\widetilde\D_A=0 \earr\right.\,,
\qqq
or -- equivalently --
\qq\nn
\left\{\barr{l}
\pLie{\xcK_A}\widetilde\D_B=f_{ABC}\,\widetilde\D_C\cr\cr
\Vcon{\xcK_A\oplus\widetilde\D_A}{\xcK_B\oplus\widetilde\D_B}=0
\earr\right.\,.
\qqq
We conclude that the freedom in the choice of the $\,K_A\,$ that can
be employed to set the small gauge anomaly to zero is
\emph{effectively} parametrised by those $(\D_Q,\D_{T_n})$-relative
1-cocycles that do not define, upon tensoring with the $\,\tau^A\in
\ggt_\si^*$,\ their own $(\D_Q,\D_{T_n})$-relatively
$\ggt_\si$-equivariantly closed ($\ggt_\si$-equivariant)
extensions.\erem

While the above result provides us with a neat quantitative
description of the small gauge anomaly of the multi-phase
$\si$-model, it leaves open questions concerning the nature of the
ensuing Lie algebroid (in particular, its integrability to a Lie
groupoid) and its intrinsic interpretation from the point of view of
the geometry of the target space $\,\xcF$.\ An answer to the former
question was given in \Rxcite{Sec.\,6}{Gawedzki:2012fu}, and we
recall it below, only to set up the context for the analysis of the
latter issue.

In order to be able to properly identify the gauge-symmetry Lie
algebroid, we need to introduce additional formal tools.
\bedef
Adopt the notation of Definitions \ref{def:grpd} and
\ref{def:algbrd}. Denote by
\qq\nn
R_{\overrightarrow g}\ :\ s^{-1}(\{t(\overrightarrow g)\})\to s^{-1}
(\{s(\overrightarrow g)\})\ :\ \overrightarrow h\mapsto
R_{\overrightarrow g}(\overrightarrow h):=\overrightarrow h\circ
\overrightarrow g
\qqq
the right multiplication map, written for an arbitrary
$\,\overrightarrow g\in\morf\,\Gr$,\ and let $\,\Xgt^s_{\rm
inv}(\morf\,\Gr)\,$ be the vector space of \textbf{right
$\Gr$-invariant vector fields} on $\,\morf\,\Gr$,\ given by
\qq\nn
\Xgt^s_{\rm inv}(\morf\,\Gr)=\{\ \xcV\in\G(\ker\,s_*)\quad\vert\quad
R_{\cdot\,*}(\xcV)=\xcV \ \}\,.
\qqq
The \textbf{tangent algebroid} of $\,\Gr\,$ is the Lie algebroid
$\,\gtgr=\bigl(\Id^*\ker\,s_*,\obj\,\Gr,[\cdot,\cdot],\pi_{\Id^*
\ker\,s_*},\a_{\sfT(\obj\,\Gr)}\bigr)\,$ with (the obvious bundle
projection $\,\pi_{\Id^* \ker\,s_*}\,$ and) the anchor
$\,\a_{\sfT(\obj\,\Gr )}\,$ inducing the map $\,t_*\circ i\,$
between spaces of sections, defined in terms of the canonical
vector-bundle isomorphism
\qq\label{eq:iso-for-tan-algbrd}
i\ :\ \Xgt^s_{\rm inv}(\morf\,\Gr)\xrightarrow{\ \cong\ }\G(\Id^*
\ker\,s_*)\,,
\qqq
and with the Lie bracket given by the unique bracket on $\,\G(\Id^*
\ker\,s_*)\,$ for which $\,i\,$ is an isomorphism of Lie algebras.
\exdef \noindent With hindsight, we next specialise the above
definition to the case of $\,\Gr=\txG\lx\xcM$,\ whereby we obtain
\berop\label{prop:act-algbrd}
Adopt the notation of Definitions \ref{def:grpd},
\ref{def:princ-g-bund} and \ref{def:algbrd}, and of Example
\ref{eg:WZW-def}. The tangent algebroid of the action groupoid
$\,\txG\lx\xcM\,$ is the quintuple
\qq\nn
\ggt\lx\xcM:=(V,\xcM,[\cdot,\cdot]_{\ggt\lx\xcM},\pi_V,\a_{\sfT
\xcM})
\qqq
composed of
\bit
\item the vector bundle $\,V\,$ with the space of sections
\qq\nn
\G(V):=\bigoplus_{A=1}^{\dim\,\ggt}\,C^\infty(\xcM,\bR)\,\xcR_A
\qqq
spanned by vector fields
\qq\nn
\xcR_A:=i(R_A\circ\pr_1)\in\G\left(\Id^*\ker\,\pr_{2\,*}\right)
\qqq
induced, through the isomorphism
\qq\nn
i\ :\ \Xgt^{\pr_2}_{\rm inv}(\txG\x\xcM)\xrightarrow{\ \cong\ }\G
\left(\Id^*\ker\,\pr_{2\,*}\right)
\qqq
of \Reqref{eq:iso-for-tan-algbrd}, from the right-invariant vector
fields $\,R_A\circ\pr_1\,$ on $\,\txG\x\xcM$,\ the latter being
defined in terms of the standard right-invariant vector fields
$\,R_A\,$ on $\,\txG\,$ dual to the right-invariant Maurer--Cartan
1-forms $\,\theta_R^A$;
\item the Lie bracket of smooth sections of $\,V$,
\qq\nn
[\la^A\,\xcR_A,\mu^B\,\xcR_B]_{\ggt\lx\xcM}:=f_{ABC}\,\la^A\,\mu^B
\,\xcR_C+(\pLie{\la^A\,\xcMup\xcK_A}\mu^B-\pLie{\mu^A\,\xcMup\xcK_A}
\la^B)\,\xcR_B\,,
\qqq
written, for arbitrary $\,\la^A,\mu^B\in C^\infty(\xcM,\bR)$,\ in
terms of the fundamental vector fields $\,\xcMup\xcK_A\equiv\ovl
t_A\,$ of \Reqref{eq:fund-vec};
\item the $C^\infty(\xcM,\bR)$-linear anchor map defined on the
basis by the formula
\qq\nn
\a_{\sfT\xcM}(\xcR_A):=\xcMup\xcK_A\,.
\qqq
\eit
The Lie algebroid thus defined is termed the \textbf{action
algebroid}.
\eerop
\beroof
A constructive proof of the proposition, based directly on
Definition \ref{def:algbrd}, can be found in
\Rxcite{App.\,D}{Gawedzki:2012fu}. \eroof \noindent We are now ready
to state the important identification result.
\bethe\cite[Thm.\,6.1]{Gawedzki:2012fu}\label{thm:gtAlgebroid}
Adopt the notation of Definitions \ref{def:sigmod-2d} and
\ref{def:DQ-rel-Courant}, of Corollary \ref{cor:Homtw-Vbra-symm-tw},
of Proposition \ref{prop:act-algbrd}, and of Theorems
\ref{thm:brabra} and \ref{thm:JacLie-sigma}. Whenever
$\,\Sgt_\Bgt\,$ is a Lie algebroid, it is canonically isomorphic
with the action algebroid $\,\ggt_\si\lx\xcF\,$ in the sense of
Definition \ref{def:algbrd}. \ethe

The emergence of the action groupoid $\,\txG_\si\lx\xcF\,$ as the
structure integrating those infinitesimal symmetries of the
$\si$-model that can be consistently gauged harmonises nicely with
its earlier appearance in the construction of a
$\txG_\si$-equivariant string background, but, at the same time, it
most definitely begs for elucidation, a task to which we turn next.

As mentioned earlier, physical consistency conditions appear to
necessitate coupling the lagrangean fields of the $\si$-model with a
target $\txG_\si$-space $\,\xcF\,$ to topologically non-trivial
gauge fields. This is tantamount to
\bit
\item introducing a (generic) principal $\txG_\si$-bundle
$\,\sfP_{\txG_\si}\,$ over the world-sheet, and subsequently
\item replacing the $\si$-model field $\,X\in C^1(\Si,\xcF)\,$ by a
\emph{global} section of the associated bundle $\,\sfP_{\txG_\si}
\x_{\txG_\si}\xcF$.\
\eit
It turns out that both constituents of the gauging algorithm listed
above find a most natural interpretation in the theory of principal
bundles with a structure Lie groupoid, as introduced (in the
geometric form) in \Rcite{Moerdijk:1991} (\textit{cf.} also
\Rcite{Haefliger:1984gh} for related work in the framework of the
theory of foliations), developed in \Rcite{Moerdijk:2003mm} and
reviewed in Refs.\,\cite{Rossi:2004dm,Rossi:2004lg} (from which we
borrow some of the proofs and most of the notation) in a much
accessible form in which the theory can be applied directly in the
context in hand. It is the last observation that plays a central
r\^ole in understanding the algebroidal interpretation of the small
gauge anomaly, and -- eventually -- also in a reinterpretation of
the large gauge anomaly. Therefore, with hindsight, we begin our
discussion by introducing a few more formal tools and results.

Let us first set up the scene by introducing the concept of a
principal bundle with a structure Lie groupoid. As it constitutes a
categorification of the concept of a principal $\txG$-bundle from
Definition \ref{def:princ-g-bund}, we start with
\bedef\label{def:gr-mod}
In the notation of Definition \ref{def:grpd}, a \textbf{right
$\Gr$-module space} is a triple $\,(\xcM,\mu,\rho_\xcM )\,$ composed
of a smooth manifold $\,\xcM$,\ a smooth map $\,\mu:
\xcM\to\obj\,\Gr\,$ called the \textbf{momentum (of the action)},
and a smooth map
\qq\nn
\rho_\xcM\ :\ \xcM{}_\mu\hspace{-3pt}\x_t\hspace{-1pt}\morf\,\Gr\to
\xcM\ :\ (m,\overrightarrow g)\mapsto\rho_\xcM(m,\overrightarrow g)
\equiv m.\overrightarrow g
\qqq
termed the \textbf{action (map)}. These satisfy the consistency
conditions (whenever the expressions are well-defined):
\bit
\item[(i)] $\mu(m.\overrightarrow g)=s(\overrightarrow g)$;
\item[(ii)] $m.\Id_{\mu(m)}=m$;
\item[(iii)] $(m.\overrightarrow g).\overrightarrow h=m.(
\overrightarrow g\circ\overrightarrow h)$.
\eit
A left $\Gr$-module space is defined similarly (with the r\^oles of
the source and target maps in the definition interchanged).

The (right) action $\,\rho_\xcM\,$ is termed \textbf{free} iff the
following implication obtains:
\qq\nn
m.\overrightarrow g=m\qquad\Longrightarrow\qquad\overrightarrow g=
\Id_{\mu(m)}\,,
\qqq
so that, in particular, the \textbf{isotropy group} $\,\Gr_x=s^{-1}
(\{x\})\cap t^{-1}(\{x\})\,$ of $\,x\in\obj\,\Gr\,$ acts freely (in
the usual sense) on the fibre $\mu^{-1}(\{x\})$.

The (right) action $\,\rho_\xcM\,$ is termed \textbf{transitive} iff
for any two points $\,m,m'\in\xcM\,$ there exists an arrow
$\,\overrightarrow g\in\morf\,\Gr\,$ such that $\,m'=m.
\overrightarrow g$.

Let $\,\Gr_i,\ i\in\{1,2\}\,$ be a pair of Lie groupoids and let
$\,(\xcM_i,\mu_i,\rho_{\xcM_i})\,$ be the respective
right-$\Gr_i$-module spaces. A \textbf{morphism} between the latter
is a pair $\,(\Theta,\Phi)\,$ consisting of a smooth manifold map
$\,\Theta:\xcM_1\to\xcM_2\,$ together with a functor $\,\Phi:\Gr_1
\to\Gr_2\,$ for which the following diagrams commute
\qq
&\alxydim{@C=1.5cm@R=1.cm}{\xcM_1 \ar[r]^{\Theta} \ar[d]_{\mu_1} &
\xcM_2 \ar[d]^{\mu_2}\cr \obj\,\Gr_1 \ar[r]^{\Phi} &
\obj\,\Gr_2}\,,& \label{diag:mu-Th-mu}\\\cr\cr
&\alxydim{@C=1.5cm@R=1.cm}{\xcM_1\fibx{\mu_1}{t_1}\morf\,\Gr_1
\ar[r]^{\Theta\x\Phi} \ar[d]_{\rho_{\xcM_1}} & \xcM_2
\fibx{\mu_2}{t_1}\morf\,\Gr_2 \ar[d]^{\rho_{\xcM_2}}\cr \xcM_1
\ar[r]^{\Theta} & \xcM_2}\,.& \label{diag:Th-intertw}
\qqq
\exdef \noindent We may now introduce the construct of immediate
interest, to wit,
\bedef\cite[Sec.\,1.2]{Moerdijk:1991}\label{def:princ-gr-bun}
In the notation of Definitions \ref{def:grpd} and \ref{def:gr-mod},
a \textbf{principal $\Gr$-bundle over base $\,\xcM\,$} is a
quintuple $\,\cP=(\sfP,\xcM,\pi_\sfP,\mu_\sfP,\rho_\sfP)\,$ composed
of a pair of smooth manifolds: the \textbf{total space} $\,\sfP\,$
of the bundle and its \textbf{base} $\,\xcM$,\ and a triple of
smooth maps: the surjective submersion $\,\pi_\sfP:\sfP\to\xcM$,\
termed the \textbf{bundle projection}, the \textbf{momentum (map)}
$\,\mu:\sfP\to\obj\,\Gr$,\ and the \textbf{action (map)}
$\,\rho_\sfP:\sfP{}_\mu\hspace{-3pt}\x_t\hspace{-1pt}\morf\,\Gr\to
\sfP\,$ with the following properties:
\bit
\item[(i)] $\,(\sfP,\mu_\sfP,\rho_\sfP)\,$ is a right $\Gr$-module;
\item[(ii)] $\,\pi_\sfP\,$ is $\Gr$-invariant in the sense made
precise by the commutative diagram (in which $\,\pr_1\,$ is the
canonical projection)
\qq\nn
\alxydim{@C=1.5cm@R=1.cm}{\sfP{}_\mu\hspace{-3pt}\x_t\hspace{-1pt}
\morf\,\Gr \ar[r]^{\hspace{1cm}\rho_\sfP} \ar[d]_{\pr_1} & \sfP
\ar[d]^{\pi_\sfP}\cr \sfP \ar[r]^{\pi_\sfP} & \xcM}\,;
\qqq
\item[(iii)] the map
\qq\nn
(\pr_1,\rho_\sfP)\ :\ \sfP{}_\mu\hspace{-3pt}\x_t\hspace{-1pt}\morf
\,\Gr\to\sfP\fibx{\pi_\sfP}{\pi_\sfP}\sfP\equiv\sfP^{[2]}\ :\ (p,
\overrightarrow g)\mapsto(p,p.\overrightarrow g)
\qqq
is a diffeomorphism, so that $\,\Gr\,$ acts freely and transitively
on $\pi_\sfP$-fibres. The smooth inverse of $\,(\pr_1,\rho_\sfP)\,$
takes the form
\qq\nn
(\pr_1,\rho_\sfP)^{-1}=:(\pr_1,\phi_\sfP)\,,\qquad\qquad\phi_\sfP\ :
\ \sfP^{[2]}\to\morf\,\Gr
\qqq
and $\,\phi_\sfP\,$ is called the \textbf{division map}.
\eit

Let $\,(\sfP_i,\xcM,\pi_{\sfP_i},\mu_i,\rho_{\sfP_i}),\ i\in\{1,2
\}\,$ be a pair of principal $\Gr$-bundles over a common base
$\,\xcM$.\ A \textbf{morphism}\footnote{In
\Rxcite{Sec.\,5.7}{Moerdijk:2003mm}, these morphisms were termed
``equivariant maps''.} between the two bundles is a fibre-preserving
morphism $\,(\Theta,\Id_\Gr)\,$ between the corresponding right
$\Gr$-modules $\,(\sfP_i,\mu_i,\rho_{\sfP_i})$.\ The category of
principal $\Gr$-bundles over a smooth manifold $\,\xcM\,$ shall be
denoted as $\,\Grbun{\xcM}$. \exdef \noindent Some useful properties
of the division map are summarised in the following
\berop\cite[Sec.\,5.7]{Moerdijk:2003mm}\label{prop:div-map}
In the notation of Definitions \ref{def:grpd} and
\ref{def:princ-gr-bun}, the division map $\,\phi_\sfP\,$ of
$\,\sfP\,$ has the following properties:
\bit
\item[(i)] it is determined uniquely by the relation
\qq\nn
q=p.\phi_\sfP(p,q)\,,
\qqq
valid for an arbitrary pair $\,(p,q)\in\sfP^{[2]}$;
\item[(ii)] $\,\phi_\sfP(p,q)\in\Gr_{\mu(q),\mu(p)}$,\ where
$\,\Gr_{x,y}:=s^{-1}\bigl(\{x\}\bigr)\cap t^{-1}\bigl(\{y\}
\bigr)\,$ for any $\,x,y\in\obj\,\Gr$;
\item[(iii)] $\,\phi_\sfP\circ(\Id_\sfP,\Id_\sfP)=\Id\circ\mu$;
\item[(iv)] $\,\phi_\sfP\circ\tau=\Inv\circ\phi_\sfP$,\ where
$\,\tau:\sfP^{[2]}\to\sfP^{[2]}:(p,q)\mapsto(q,p)$.
\eit
\eerop
\beroof
Obvious, though inspection. \textit{Cf.}\ also
\Rxcite{Sec.\,4.3}{Rossi:2004dm} for a simple proof.\eroof \noindent
We have a counterpart of the well-known result for the category
$\,\Gbun{\Si}\,$ of principal $\txG$-bundles over base $\,\Si$,\ to
wit,
\berop\cite[Sec.\,1.2]{Moerdijk:1991}\label{prop:morph-iso}
In the notation of Definition \ref{def:princ-gr-bun}, the category
$\,\Grbun{\xcM}\,$ is a groupoid.
\eerop
\beroof
Let $\,\cP_i,\ i\in\{1,2\}\,$ be a pair of objects of
$\,\Grbun{\xcM}$.\ Consider an arbitrary morphism
$\,(\Theta,\Id_\Gr) \in\morf_{\Grbun{\xcM}}(\cP_1,\cP_2)$.\ Assume
that $\,p_1,p_2\in \sfP_1\,$ satisfy the equation
\qq\label{eq:iso-inj}
\Theta(p_1)=\Theta(p_2)\,,
\qqq
whence also
\qq\nn
\pi_{\sfP_1}(p_1)=\pi_{\sfP_2}\bigl(\Theta(p_1)\bigr)=\pi_{\sfP_2}
\bigl(\Theta(p_2)\bigr)=\pi_{\sfP_1}(p_2)\qquad\Rightarrow\qquad(
p_1,p_2)\in\sfP_1^{[2]}\,.
\qqq
By property (iii) of Definition \ref{def:princ-gr-bun}, and in
virtue of point (i) of Proposition \ref{prop:div-map}, we then have
\qq\nn
p_2=p_1.\phi_{\sfP_1}(p_1,p_2)\,,
\qqq
so that the $\Gr$-equivariance of $\,\Theta$,\ expressed by diagram
\eqref{diag:Th-intertw}, implies
\qq\nn
\Theta(p_2)=\Theta(p_1).\phi_{\sfP_1}(p_1,p_2)\,.
\qqq
Taken in conjunction with the assumed equality \eqref{eq:iso-inj},
this yields
\qq\nn
(\pr_1,\rho_{\sfP_2})\bigl(\Theta(p_1),\Id_{\mu_{\sfP_2}\bigl(\Theta(p_1)
\bigr)}\bigr)&=&\bigl(\Theta(p_1),\Theta(p_1)\bigr)=\bigl(\Theta(
p_1),\Theta(p_2)\bigr)=\bigl(\Theta(p_1),\Theta(p_1).\phi_{\sfP_1}(
p_1,p_2)\bigr)\cr\cr
&=&(\pr_1,\rho_{\sfP_2})\bigl(\Theta(p_1),\phi_{\sfP_1}(p_1,p_2)
\bigr)\,,
\qqq
hence, owing to the invertibility of $\,(\pr_1,\rho_{\sfP_2})$,\
\qq\nn
\Id_{\mu_{\sfP_2}\bigl(\Theta(p_1)\bigr)}=\phi_{\sfP_1}(p_1,p_2)\,.
\qqq
Upon adducing the property of $\,\Theta\,$ encoded in diagram
\eqref{diag:mu-Th-mu}, we thus obtain
\qq\nn
\Id_{\mu_{\sfP_1}(p_1)}=\Id_{\mu_{\sfP_2}\bigl(\Theta(p_1)\bigr)}=
\phi_{\sfP_1}(p_1,p_2)\,,
\qqq
and so, finally,
\qq\nn
p_2=p_1.\phi_{\sfP_1}(p_1,p_2)=p_1.\Id_{\mu_{\sfP_1}(p_1)}=p_1\,,
\qqq
which proves the injectivity of $\,\Theta$.

Consider, next, an arbitrary point $\,p_2\in\sfP_2\,$ over
$\,\pi_{\sfP_2}(p_2)=:x\in\xcM$.\ Choose $\,q_1\in\pi_{\sfP_1}^{-1}
\bigl(\{x\}\bigr)$.\ Clearly, $\,\pi_{\sfP_2}\bigl(\Theta(q_1)\bigr)
=\pi_{\sfP_2}(p_2)$,\ and so
\qq\nn
p_2=\Theta(q_1).\phi_{\sfP_2}\bigl(\Theta(q_1),p_2\bigr)\,.
\qqq
Upon invoking point (ii) of Proposition \ref{prop:div-map} and, once
again, the property of $\,\Theta\,$ encoded in diagram
\eqref{diag:mu-Th-mu}, we establish
\qq\nn
t\bigl(\phi_{\sfP_2}\bigl(\Theta(q_1),p_2\bigr)\bigr)=\mu_{\sfP_2}
\bigl(\Theta(q_1)\bigr)=\mu_{\sfP_1}(q_1)\,.
\qqq
Define
\qq\nn
p_1:=q_1.\phi_{\sfP_2}\bigl(\Theta(q_1),p_2\bigr)\in\pi_{\sfP_1}^{-
1}\bigl(\{x\}\bigr)\,.
\qqq
We then find, using the $\Gr$-equivariance of $\,\Theta\,$ and
property (i) of Proposition \ref{prop:div-map},
\qq\nn
\Theta(p_1)=\Theta(q_1).\phi_{\sfP_2}\bigl(\Theta(q_1),p_2\bigr)=
p_2\,,
\qqq
which demonstrates the surjectivity of $\,\Theta\,$ and thus
concludes the proof. \eroof \noindent A particularly powerful tool
in the analysis of principal $\Gr$-bundles, instrumental also in our
subsequent discussion, is the local description, which we now set up
after Moerdijk and Mr\v cun. The first prerequisite is described in
\bedef\cite[Rem.\,5.34(2)]{Moerdijk:2003mm}\label{def:pull-grpd-bndle}
Adopt the notation of Definitions \ref{def:grpd} and
\ref{def:princ-gr-bun}. Let $\,\xcM,\xcN\,$ be a pair of smooth
manifolds, $\,f:\xcM\to\xcN\,$ a smooth map between them, and
$\,\cP\,$ a principal $\Gr$-bundle over $\,\xcN$.\ The
\textbf{pullback} of $\,\cP\,$ along $\,f\,$ is the principal
$\Gr$-bundle over $\,\xcM\,$ given by
\qq\nn
f^*\cP:=(f^* \sfP,\xcM,\pr_1,\mu_\sfP\circ\pr_2,\rho_{f^*\sfP})\,,
\qqq
where
\qq\nn
f^*\sfP:=\xcM{}_f\hspace{-3pt}\x_{\pi_\sfP}\hspace{-1pt}\sfP\,,
\qqq
and
\qq\nn
\rho_{f^*\sfP}\ :\ f^*\sfP\fibx{\mu_\sfP\circ\pr_2}{t}\morf\,\Gr\to
f^*\sfP\ :\ \bigl((m,p),\overrightarrow g\bigr)\mapsto(m,p.
\overrightarrow g)\,.
\qqq
\exdef \noindent Clearly, the above definition makes sense, that is
$\,f^*\cP\,$ is a principal $\Gr$-bundle. Indeed, properties (i) and
(ii) from Definition \ref{def:princ-gr-bun} are manifest. As for the
last property, we find, for any two points $\,(m_1,p_1),(m_2,p_2)\in
f^* \sfP\,$ from the same fibre,
\qq\nn
m_2=\pr_1(m_2,p_2)=\pr_1(m_1,p_1)=m_1\quad\Rightarrow\quad\pi_\sfP(
p_2)=f(m_2)=f(m_1)=\pi_\sfP(p_1)\,,
\qqq
and so, by virtue of Proposition \ref{prop:div-map},
\qq\nn
p_2=p_1.\phi_\sfP(p_1,p_2)\,.
\qqq
Hence, the smooth inverse of the map
\qq\nn
(\pr_1,\rho_{f^*\sfP})\ :\ f^*\sfP\fibx{\mu_\sfP\circ\pr_2}{t}\morf
\,\Gr\to f^*\sfP\fibx{\pr_1}{\pr_1}f^*\sfP\equiv f^* \sfP^{[2]}\ :\
\bigl((m,p),\overrightarrow g\bigr)\mapsto\bigl((m,p),(m,p.
\overrightarrow g)\bigr)
\qqq
reads
\qq\nn
(\pr_1,\phi_{f^*\sfP})\ :\ f^*\sfP^{[2]}\to f^*\sfP\fibx{\mu_\sfP
\circ\pr_2}{t}\morf\,\Gr\ :\ \bigl((m,p_1),(m,p_2)\bigr)\mapsto
\bigl((m,p_1),\phi_\sfP(p_1,p_2)\bigr)\,.
\qqq
In the next step, we consider
\bedef\cite[Rem.\,5.34(1)]{Moerdijk:2003mm}\label{def:triv-grpd-bndle}
In the notation of Definition \ref{def:grpd}, the \textbf{unit
bundle of $\,\Gr\,$} is the principal $\Gr$-bundle over $\,\obj\,
\Gr\,$ given by $\,\cU_\Gr:=(\morf\,\Gr,\obj\,\Gr,t,s,R)$,\ where
$\,R\,$ denotes right multiplication,
\qq\nn
R\ :\
\morf\,\Gr{}_s\hspace{-3pt}\x_t\hspace{-1pt}\morf\,\Gr\to\morf\,\Gr\
:\ ( \overrightarrow g,\overrightarrow h)\mapsto\overrightarrow g
\circ\overrightarrow h \,.
\qqq
\exdef \noindent Once more, properties (i) and (ii) from Definition
\ref{def:princ-gr-bun} are evident, and it remains to verify
property (iii). The map
\qq\nn
(\pr_1,R)\ :\ \morf\,\Gr{}_s\hspace{-3pt}\x_t\hspace{-1pt}\morf\,
\Gr\to\morf\,\Gr{}_t\hspace{-3pt}\x_t\hspace{-1pt}\morf\,\Gr\ :\ (
\overrightarrow g,\overrightarrow h)\mapsto(\overrightarrow g,
\overrightarrow g\circ\overrightarrow h)
\qqq
admits the smooth inverse
\qq\nn
(\pr_1,\phi_{\cU_\Gr})\ :\ \morf\,\Gr{}_t\hspace{-3pt}\x_t
\hspace{-1pt}\morf\,\Gr\to\morf\,\Gr{}_s\hspace{-3pt}\x_t
\hspace{-1pt}\morf\,\Gr\ :\ (\overrightarrow g,\overrightarrow h)
\mapsto\bigl(\overrightarrow g,\overrightarrow g^{-1}\circ
\overrightarrow h\bigr)\,.
\qqq
The last ingredient is
\bedef\cite[Rem.\,5.34(3)]{Moerdijk:2003mm}
In the notation of Definitions \ref{def:grpd},
\ref{def:pull-grpd-bndle} and \ref{def:triv-grpd-bndle}, and for
$\,\xcM\,$ a smooth manifold, a \textbf{trivial principal
$\Gr$-bundle} over $\,\xcM\,$ is the pullback $\,f^*\cU_\Gr\,$ of
the trivial bundle $\,\cU_\Gr\,$ along an arbitrary smooth map $\,f:
\xcM\to\obj\,\Gr$. \exdef

We are now ready to state the important
\berop\cite[Rem.\,5.34(4)]{Moerdijk:2003mm}\label{prop:loc-triv-Grbun}
In the notation of Definitions \ref{def:grpd} and
\ref{def:princ-gr-bun}, every principal $\Gr$-bundle $\,\cP\,$ is
locally trivialisable, \textit{i.e.}\ for every point $\,m\in\xcM\,$
of the base of $\,\cP$,\ there exists an open neighbourhood
$\,\cO\ni m\,$ and a smooth map $\,\mu_\cO:\cO\to\obj\,\Gr\,$ such
that $\,\cP \vert_\cO\,$ is isomorphic to the trivial $\Gr$-bundle
$\,\mu_\cO^* \cU_\Gr$.
\eerop
\beroof
Given a neighbourhood $\,\cO\,$ of a point $\,m\in\xcM$,\ choose a
smooth local section $\,\si_\cO:\cO\to\pi_\sfP^{-1}(\cO)\,$ of the
surjective submersion $\,\pi_\sfP:\sfP\to\xcM$,\ and define the
smooth map
\qq\nn
\mu_\cO:=\mu_\sfP\circ\si_\cO\,.
\qqq
The map
\qq\nn
\tau_\cO^{-1}:=\rho_\sfP\circ(\si_\cO\x\Id_{\morf\,\Gr})\ :\
\mu_\cO^*\morf\,\Gr\to\pi_\sfP^{-1}(\cO)\ :\ (m,\overrightarrow g)
\mapsto\si_\cO(m).\overrightarrow g
\qqq
is manifestly well-defined as
\qq\nn
(m,\overrightarrow g)\in\mu_\cO^*\morf\,\Gr\quad\Rightarrow\quad t(
\overrightarrow g)=\mu_\cO(m)\equiv\mu_\sfP\bigl(\si_\cO(m)\bigr)
\,,
\qqq
and smooth (as a composition of smooth maps). Viewed as a map
between the two bundles, it preserves the respective fibres as
\qq\nn
\pi_\sfP\circ\tau_\cO^{-1}(m,\overrightarrow g)=\pi_\sfP\bigl(
\si_\cO(m).\overrightarrow g\bigr)=\pi_\sfP\circ\si_\cO(m)=\id_\cO(m
)=m=\pr_1(m,\overrightarrow g)
\qqq
due to the $\Gr$-equivariance of $\,\pi_\sfP$.\ Moreover, it is
itself $\Gr$-equivariant, \textit{i.e.}\ it renders the
corresponding diagrams of Definition \ref{def:gr-mod} commutative.
Indeed, it preserves the momenta of the two bundles,
\qq\nn
\mu_\sfP\circ\tau_\cO^{-1}(m,\overrightarrow g)=\mu_\sfP\bigl(
\si_\cO(m).\overrightarrow g\bigr)=s(\overrightarrow g)=s\circ\pr_2(
m,\overrightarrow g)
\qqq
(owing to the defining property (i) of $\,\mu_\sfP$), and it
intertwines the respective right $\Gr$-actions,
\qq\nn
\rho_\sfP\circ(\tau_\cO^{-1}\x\Id_{\morf\,\Gr})\bigl((m,
\overrightarrow g),\overrightarrow h\bigr)&=&\rho_\sfP\bigl(\si_\cO
(m).\overrightarrow g,\overrightarrow h\bigr)=\bigl(\si_\cO(m).
\overrightarrow g\bigr).\overrightarrow h=\si_\cO(m).(
\overrightarrow g\circ\overrightarrow h)\cr\cr
&=&\tau_\cO^{-1}(m,\overrightarrow g\circ\overrightarrow h)=
\tau_\cO^{-1}\bigl(m,R(\overrightarrow g,\overrightarrow h)\bigr)=
\tau_\cO^{-1}\circ\rho_{\mu_\cO^*\cU_\Gr}\bigl((m,\overrightarrow
g),\overrightarrow h\bigr)\,.
\qqq
Thus, altogether, $\,\tau_\cO^{-1}\,$ is a morphism between the two
principal $\Gr$-bundles over $\,\cO$,\ and so -- by virtue of
Proposition \ref{prop:morph-iso} -- it is an isomorphism. It is the
inverse of the map
\qq\nn
\tau_\cO(p)=\bigl(\pi_\sfP(p),\phi_\sfP\bigl(\si_\cO\circ\pi_\sfP(p
),p\bigr)\bigr)\,.
\qqq
\eroof \noindent The above proposition paves the way to a local
description of principal $\Gr$-bundles that we shall find of great
use in the context in hand. An abstraction of the hitherto findings
yields
\bedef\cite[Sec.\,1.2]{Moerdijk:1991}\label{def:loc-triv-data}
Adopt the notation of Definitions \ref{def:grpd} and
\ref{def:princ-gr-bun}. Let $\,\cO_\xcM =\{\cO_i\}_{i\in\xcI}\,$ be
an open cover of $\,\xcM\,$ (with an index set $\,\xcI$), and let
$\,\si_i:\cO_i\to\pi_\sfP^{-1}(\cO_i)\,$ be the associated local
sections of the principal $\Gr$-bundle $\,\cP\,$ over $\,\xcM$.\
\textbf{Local (trivialising) data of $\,\cP\,$ (with values in
$\,\Gr$)} are given by the triple $\,(\cO_\xcM,\mu_i,\g_{ij}\ \vert\
i,j\in\xcI)\,$ composed of two collections of smooth maps:
\textbf{local momenta}
\qq\nn
\mu_i:=\mu_\sfP\circ\si_i\ :\ \cO_i\to\obj\,\Gr\,,
\qqq
and \textbf{transition maps}
\qq\nn
\g_{ij}\ :\ \cO_{ij}\to\morf\,\Gr\ :\ m\mapsto\phi_\sfP\left(\si_i(
m),\si_j(m)\right)\,,
\qqq
the latter being defined on non-empty double intersections $\,\cO_{i
j}=\cO_i\cap\cO_j\,$ and relating the respective restrictions of
\textbf{local trivialisations}
\qq\nn
\tau_i:=\left(\pi_\sfP,\phi_\sfP\circ(\si_i\circ\pi_\sfP,\id_\sfP)
\right)\ :\ \pi_\sfP^{-1}(\cO_i)\xrightarrow{\ \cong\ }\mu_i^*\morf
\,\Gr\,.
\qqq
\exdef \noindent Important properties of local trivialising data are
listed in the following proposition that, at the same time, provides
us with a key to understanding the geometry behind the data.
\berop\cite[Sec.\,1.2]{Moerdijk:1991}\cite[Lem.\,3.7]{Rossi:2004lg}
\label{prop:can-iso-bet-triv} Adopt the notation of Definitions
\ref{def:grpd}, \ref{def:pull-grpd-bndle}, \ref{def:triv-grpd-bndle}
and \ref{def:loc-triv-data}. Local trivialising data have the
defining properties:
\bit
\item[(i)] $\,t\circ\g_{ij}=\mu_i,\ s\circ\g_{ij}=\mu_j,\ \g_{ii}=
\Id\circ\mu_i$;
\item[(ii)] $\,\g_{ji}=\Inv\circ\g_{ij}$;
\item[(iii)] on a non-empty common intersection $\,\cO_{ijk}=\cO_i
\cap\cO_j\cap\cO_k\ni m\,$ of any three open sets $\,\cO_i,\cO_j,
\cO_k,\ i,j,k\in \xcI$,\ the \textbf{cocycle condition} $\,\g_{ik}(m
)=\g_{ij}(m)\circ\g_{jk}(m)\,$ obtains.
\eit
They canonically define an isomorphism between the trivial bundles
$\,\mu_i^*\cU_\Gr\vert_{\cO_{ij}}\,$ and $\,\mu_j^*\cU_\Gr
\vert_{\cO_{ij}}\,$ over every non-empty double intersection
$\,\cO_{ij}=\cO_i\cap\cO_j$.
\eerop
\beroof
Properties (i), (ii) and (iii) are straightforward consequences of
Proposition \ref{prop:div-map}. Thus, it remains to prove the
concluding statement. Consider the map
\qq\nn
\varphi_{ij}\ :\ \mu_j^*\cU_\Gr\vert_{\cO_{ij}}\to\mu_i^*\cU_\Gr
\vert_{\cO_{ij}}\ :\ (m,\overrightarrow g)\mapsto\bigl(m,\g_{ij}(m)
\circ\overrightarrow g\bigr)\,.
\qqq
The map is clearly well-defined as
\qq\nn
(m,\overrightarrow g)\in\mu_j^*\cU_\Gr\vert_{\cO_{ij}}\quad
\Rightarrow\quad t(\overrightarrow g)=\mu_j(m)=s\circ\g_{ij}(m)
\qqq
owing to property (i) of the local data, and
\qq\nn
t\bigl(\g_{ij}(m)\circ\overrightarrow g\bigr)=t\bigl(\g_{ij}(m)
\bigr)=\mu_i(m)\quad\Rightarrow\quad\varphi_{ij}(m,g)\in\mu_i^*
\cU_\Gr\,.
\qqq
Its surjectivity follows from the simple identity
\qq\nn
(m,\overrightarrow g)\in\mu_i^*\cU_\Gr\vert_{\cO_{ij}}\quad
\Rightarrow\quad(m,\overrightarrow g)&=&(m,\Id_{t(\overrightarrow g
)}\circ\overrightarrow g)=(m,\Id_{\mu_i(m)}\circ\overrightarrow g)=
\bigl(m,\g_{ii}(m)\circ\overrightarrow g\bigr)\cr\cr
&=&\bigl(m,\g_{ij}(m)\circ\g_{ji}(m)\circ\overrightarrow g\bigr)=
\varphi_{ij}\bigl(m,\g_{ji}(m)\circ\overrightarrow g\bigr)\,.
\qqq
The map is also manifestly fibre-preserving, and so it remains to
check that it also preserves the momenta, which follows from
\qq\nn
\mu_{\cO_{ij}\fibx{\mu_i}{t}\morf\,\Gr}\circ\varphi_{ij}(m,
\overrightarrow g)=s\circ\pr_2\bigl(m,\g_{ij}(m)\circ
\overrightarrow g\bigr)=s(\overrightarrow g)=s\circ\pr_2(m,
\overrightarrow g)=\mu_{\cO_{ij}\fibx{\mu_j}{t}\morf\,\Gr}(m,
\overrightarrow g)\,,
\qqq
and that it intertwines the two $\Gr$-actions,
\qq\nn
\varphi_{ij}\bigl((m,\overrightarrow g).\overrightarrow h\bigr)=
\varphi_{ij}(m,\overrightarrow g\circ\overrightarrow h)=\bigl(m,
\g_{ij}(m)\circ\overrightarrow g\circ\overrightarrow h\bigr)=\bigl(
m,\g_{ij}(m)\circ\overrightarrow g\bigr).\overrightarrow h=
\varphi_{ij}(m,\overrightarrow g).\overrightarrow h\,.
\qqq
The claim is now implied by Proposition \ref{prop:morph-iso}.\eroof
\noindent The dependence of a local trivialisation on the choice of
the local section is clarified by the following
\berop\cite[Lem.\,3.2]{Rossi:2004lg}\label{prop:Gr-loc-triv-amb}
Adopt the notation of Definitions \ref{def:grpd},
\ref{def:pull-grpd-bndle}, \ref{def:triv-grpd-bndle} and
\ref{def:loc-triv-data}. Let $\,\si_\cO^i:\cO\to\sfP,\ i\in\{1,2
\}\,$ be any two smooth local sections of a principal $\Gr$-bundle
$\,\cP\,$ over an open subset $\,\cO\subset\xcM\,$ of the base
$\,\xcM\,$ of $\,\cP$.\ The associated local trivialisations
$\,\tau_\cO^i\,$ of $\,\cP\,$ over $\,\cO\,$ are equivalent in the
sense that the corresponding trivial bundles $\,\mu_\cO^{i\,*}
\cU_\Gr$,\ defined in terms of the respective local momenta
$\,\mu_\cO^i\,$ associated with the $\,\si_\cO^i$,\ are isomorphic
as per
\qq\nn
\tau_\cO^{2,1}:=\tau_\cO^2\circ\bigl(\tau_\cO^1\bigr)^{-1}\ :\
\mu_\cO^{1\,*}\cU_\Gr\xrightarrow{\cong}\mu_\cO^{2\,*}\cU_\Gr\,.
\qqq
\eerop
\beroof
The claim follows from Proposition \ref{prop:morph-iso} and the
proven $\Gr$-equivariance of the local trivialisations. \eroof

We conclude our introductory presentation of principal Lie-groupoid
bundles by formulating a variant of the familiar clutching
construction.
\bethe\cite[Sec.\,1.2]{Moerdijk:1991}\cite[Thm.\,3.8]{Rossi:2004lg}
\label{thm:GrBun-loc} Adopt the notation of Definitions
\ref{def:grpd}, \ref{def:pull-grpd-bndle}, \ref{def:triv-grpd-bndle}
and \ref{def:loc-triv-data}. Define a manifold
\qq\nn
\vec\sfP_{\cO_\xcM}:=\bigsqcup_{i\in\xcI}\,\mu_i^*\cU_\Gr/\sim_{(
\g_{ij})}
\qqq
as the quotient with respect to the equivalence relation
\qq\label{eq:rel-equiv-loc}
(i,m_i,\overrightarrow g_i)\sim_{(\g_{ij})}(j,m_j,\overrightarrow
g_j)\quad\Leftrightarrow\quad\left(\ m_i=m_j\in\cO_{ij}\quad\land
\quad\overrightarrow g_i=\g_{ij}(m_i)\circ\overrightarrow g_j\
\right)\,,
\qqq
and maps
\qq\nn
\mu_{\vec\sfP_{\cO_\xcM}}\ &:&\ \vec\sfP_{\cO_\xcM}\to\obj\,\Gr\ :\
\bigl[(i,m,\overrightarrow g)\bigr]\mapsto s(\overrightarrow g)\,
\cr\cr
\pi_{\vec\sfP_{\cO_\xcM}}\ &:&\ \vec\sfP_{\cO_\xcM}\to\xcM\ :\
\bigl[(i,m,\overrightarrow g)\bigr]\mapsto m\,,\cr\cr
\rho_{\vec\sfP_{\cO_\xcM}}\ &:&\ \vec\sfP_{\cO_\xcM}
\fibx{\mu_{\vec\sfP_{\cO_\xcM}}}{t}\morf\,\Gr\to\vec\sfP_{\cO_\xcM}\
:\ \bigl(\bigl[(i,m,\overrightarrow g)\bigr],\overrightarrow h\bigr)
\mapsto\bigl[(i,m,\overrightarrow g\circ\overrightarrow h)\bigr]\,.
\qqq
The quintuple
\qq\nn
\vec\cP_{\cO_\xcM}:=\bigl(\vec\sfP_{\cO_\xcM},\xcM,\pi_{\vec
\sfP_{\cO_\xcM}},\mu_{\vec\sfP_{\cO_\xcM}},\rho_{\vec\sfP_{\cO_\xcM}}
\bigr)
\qqq
is a principal $\Gr$-bundle over $\,\xcM$. \ethe
\beroof
First of all, we convince ourselves that relation
\eqref{eq:rel-equiv-loc} is an equivalence relation using the
defining properties of the $\,\g_{ij}$.\ The space $\,\vec
\sfP_{\cO_\xcM}\,$ is then a smooth quotient of smooth spaces,
locally diffeomorphic to $\,\mu_i^*\cU_\Gr$,\ and the map
$\,\pi_{\vec\sfP_{\cO_\xcM}}\,$ is a surjective submersion. What has
to be shown is that the smooth map $\,\rho_{\vec\sfP_{\cO_\xcM}}\,$
endows the triple $\,(\vec\sfP_{\cO_\xcM},\mu_{\vec\sfP_{\cO_\xcM}},
\rho_{\vec\sfP_{\cO_\xcM}})\,$ with the structure of a right
$\Gr$-module, that $\,\pi_{\vec\sfP_{\cO_\xcM}}\,$ is invariant
under the $\Gr$-action, and that the map $\,(\pr_1,\rho_{\vec
\sfP_{\cO_\xcM}})\,$ is a diffeomorphism so that the $\Gr$-action is
free and transitive.

The former fact follows straightforwardly from the simple identities
\qq\nn
&\mu_{\vec\sfP_{\cO_\xcM}}\bigl(\bigl[(i,m,\overrightarrow g)\bigr].
\overrightarrow h\bigr)=\mu_{\vec\sfP_{\cO_\xcM}}\bigl(\bigl[(i,m,
\overrightarrow g\circ\overrightarrow h)\bigr]\bigr)=s(
\overrightarrow g\circ\overrightarrow h)=s(\overrightarrow h)\,,&
\cr\cr
&\bigl[(i,m,\overrightarrow g)\bigr].\Id_{\mu_{\vec\sfP_{\cO_\xcM}}(
[(i,m,\overrightarrow g)])}=\bigl[(i,m,\overrightarrow g\circ
\Id_{\mu_{\vec\sfP_{\cO_\xcM}}([(i,m,\overrightarrow g)])})\bigr]=
\bigl[(i,m,\overrightarrow g\circ\Id_{s(\overrightarrow g)})\bigr]=
\bigl[(i,m,\overrightarrow g)\bigr]\,,&\cr\cr
&\bigl(\bigl[(i,m,\overrightarrow g)\bigr].\overrightarrow h_1\bigr).
\overrightarrow h_2=\bigl[(i,m,\overrightarrow g\circ\overrightarrow
h_1)\bigr].\overrightarrow h_2=\bigl[(i,m,\overrightarrow g\circ
\overrightarrow h_1\circ\overrightarrow h_2)\bigr]=\bigl[(i,m,
\overrightarrow g)\bigr].(\overrightarrow h_1\circ\overrightarrow
h_2)\,,&
\qqq
and the $\Gr$-invariance of $\,\pi_{\vec\sfP_{\cO_\xcM}}\,$ is
self-evident.

In order to prove the latter fact, note that a pair $\,\bigl(\bigl[(
i,m_i,\overrightarrow g_i)\bigr],\bigl[(j,m_j,\overrightarrow
g_j)\bigr]\bigr)\,$ with a common projection to $\,\xcM$,\ that is
with $\,m_j=m_i=:m\in\cO_{ij}$,\ unambiguously defines a morphism
\qq\nn
\overrightarrow h(m):=\overrightarrow g_i^{-1}\circ\g_{ij}(m)\circ
\overrightarrow g_j\,.
\qqq
The definition makes sense as
\qq\nn
t\bigl(\g_{ij}(m)\bigr)=\mu_i(m)=t(\overrightarrow g_i)=s(
\overrightarrow g_i^{-1})\,,\qquad\qquad
s\bigl(\g_{ij}(m)\bigr)=\mu_j(m)=t(\overrightarrow g_j)\,,
\qqq
and it is independent of the choice of representatives of the two
equivalence classes. Indeed, for $\,\bigl[(k,m_k,\overrightarrow
g_k)\bigr]=\bigl[(i,m_i,\overrightarrow g_i)\bigr]\,$ and $\,[(l,
m_l,\overrightarrow g_l)\bigr]=\bigl[(j,m_j,\overrightarrow g_j)
\bigr]\,$ (with, necessarily, $\,m_k=m_l=m\in\cO_{ijkl}$),\ we find
\qq\nn
\overrightarrow g_k^{-1}\circ\g_{kl}(m)\circ\overrightarrow g_l&=&
\bigl(\g_{ki}(m)\circ\overrightarrow g_i\bigr)^{-1}\circ\g_{kl}(m)
\circ\bigl(\g_{lj}(m)\circ\overrightarrow g_j\bigr)=\overrightarrow
g_i^{-1}\circ\g_{ik}(m)\circ\g_{kl}(m)\circ\g_{lj}(m)\circ
\overrightarrow g_j\cr\cr
&=&\overrightarrow g_i^{-1}\circ\g_{ij}(m)\circ \overrightarrow g_j
\,.
\qqq
We may then write down the smooth inverse of $\,(\pr_1,\rho_{\vec
\sfP_{\cO_\xcM}})\,$ in the form
\qq\nn
(\pr_1,\rho_{\vec\sfP_{\cO_\xcM}})^{-1}\bigl(\bigl[(i,m_i,
\overrightarrow g_i)\bigr],\bigl[(j,m_j,\overrightarrow g_j)\bigr]
\bigr):=\bigl(\bigl[(i,m_i,\overrightarrow g_i)\bigr],
\overrightarrow h(m_i)\bigr)
\qqq
and check its desired properties:
\qq\nn
(\pr_1,\rho_{\vec\sfP_{\cO_\xcM}})\circ(\pr_1,\rho_{\vec
\sfP_{\cO_\xcM}})^{-1}\bigl(\bigl[(i,m_i,\overrightarrow g_i)\bigr],
\bigl[(j,m_j,\overrightarrow g_j)\bigr]\bigr)=(\pr_1,\rho_{\vec
\sfP_{\cO_\xcM}})\bigl(\bigl[(i,m_i,\overrightarrow g_i)\bigr],
\overrightarrow h(m_i)\bigr)\cr\cr
=\bigl(\bigl[(i,m_i,\overrightarrow
g_i)\bigr],\bigl[(i,m_i,\overrightarrow g_i\circ\overrightarrow
g_i^{-1}\circ\g_{ij}(m)\circ\overrightarrow g_j)\bigr]\bigr)=\bigl(
\bigl[(i,m_i,\overrightarrow g_i)\bigr],\bigl[(i,m_i,\g_{ij}(m)\circ
\overrightarrow g_j)\bigr]\bigr)\cr\cr
=\bigl(\bigl[(i,m_i,\overrightarrow g_i)\bigr],\bigl[(i,m_j,
\overrightarrow g_j)\bigr]\bigr)
\qqq
and
\qq\nn
(\pr_1,\rho_{\vec\sfP_{\cO_\xcM}})^{-1}\circ(\pr_1,\rho_{\vec
\sfP_{\cO_\xcM}})\bigl(\bigl[(i,m_i,\overrightarrow g_i)\bigr],
\overrightarrow
g\bigr)=(\pr_1,\rho_{\vec\sfP_{\cO_\xcM}})^{-1}\bigl(
\bigl[(i,m_i,\overrightarrow g_i)\bigr],\bigl[(i,m_i,\overrightarrow
g_i\circ\overrightarrow g)\bigr]\bigr)\cr\cr
=\bigl(\bigl[(i,m_i,\overrightarrow g_i)\bigr],\overrightarrow
g_i^{-1}\circ\g_{ii}(m_i)\circ(\overrightarrow g_i\circ
\overrightarrow g)\bigr)=\bigl(\bigl[(i,m_i,\overrightarrow g_i)
\bigr],\overrightarrow g\bigr)\,.
\qqq
\eroof

The local language can equally well be developed for morphisms
between principal Lie-groupoid bundles. We begin with
\berop\label{prop:GrBun-morf-loc}
Adopt the notation of Definitions \ref{def:grpd} and
\ref{def:loc-triv-data}. Let $\,\cP_A, A\in\{1,2\}\,$ be a pair of
principal $\Gr$-bundles with the respective local bundle data
$\,(\cO_\xcM,\mu_i^A,\g_{ij}^A\,\vert\,i,j\in\xcI)$.\ The existence
of a morphism $\,\Th\in\mor_{\Grbun{\xcM}}(\cP_1,\cP_2)\,$ is
equivalent to the existence of a collection of locally smooth maps
$\,\theta_i:\cO_i\to\morf\,\Gr\,$ with the defining properties:
\bit
\item[(i)] $\,s\circ\theta_i=\mu_i^1,\ t\circ\theta_i=\mu_i^2$;
\item[(ii)] on a non-empty common intersection $\,\cO_{ij}=\cO_i
\cap\cO_j\ni m\,$ of any two open sets $\,\cO_i,\cO_j,\ i,j\in
\xcI$,\ the \textbf{intertwiner condition} $\,\theta_i(m)\circ\g_{i
j}^1(m)=\g_{ij}^2(m)\circ\theta_j(m)\,$ obtains.
\eit
\eerop
\beroof
First, consider local trivialisations $\,\tau_i^A:\pi_{\sfP_A}^{-1}
(\cO_i)\to\mu_i^{A\,*}\morf\,\Gr\,$ associated with the local data
$\,(\cO_\xcM,\mu_i^A,\g_{ij}^A\ \vert\ i,j\in\xcI)$,\ and the
corresponding local sections $\,\si_i^A:\cO_i\to\sfP_A$.\ Denote by
$\,\phi_{\sfP_A}\,$ the respective division maps. As $\,\Theta\,$
preserves fibres, we necessarily find, for any $\,m\in\cO_i$,
\qq\nn
\tau_i^2\circ\Theta\circ\si_i^1(m)=\left(m,\phi_{\sfP_2}\left(
\si_i^2(m),\Theta\circ\si_i^1(m)\right)\right)\,,
\qqq
and so
\qq\nn
\Theta\circ\si_i^1(m)=\si_i^2(m).\phi_{\sfP_2}\left(\si_i^2(m),
\Theta\circ\si_i^1(m)\right)\,.
\qqq
Define
\qq\nn
\theta_i\ :\ \cO_i\to\morf\,\Gr\ :\ m\mapsto\phi_{\sfP_2}\left(
\si_i^2(m),\Theta\circ\si_i^1(m)\right)\,.
\qqq
Adducing point (ii) of Proposition \ref{prop:div-map} and
subsequently using the defining property of a $\Gr$-bundle morphism
encoded in diagram \eqref{diag:mu-Th-mu}, we readily find the
desired identity
\qq\nn
s\circ\theta_i(m)=\mu_{\sfP_2}\circ\Theta\circ\si_i^1(m)=
\mu_{\sfP_1}\circ\si_i^1(m)\equiv\mu_i^1(m)\,.
\qqq
Next, once more with the help of point (ii) of Proposition
\ref{prop:div-map}, we obtain
\qq\nn
t\circ\theta_i(m)=\mu_{\sfP_2}\circ\si_i^2(m)\equiv\mu_i^2(m)\,.
\qqq
Finally, the intertwiner property of $\,\Theta\,$ captured by
diagram \eqref{diag:Th-intertw}, enables us to demonstrate the
validity of point (ii),
\qq\nn
\si_j^2(m).\left(\theta_j(m)\circ\g_{ji}^1(m)\right)&\equiv&\left(
\tau_j^{2}\right)^{-1}(m,\Id_{\mu_j^2(m)}).\left(\theta_j(m)\circ
\g_{ji}^1(m)\right)=\left(\left(\tau_j^2\right)^{-1}(m,\Id_{\mu_j^2
(m)}).\theta_j(m)\right).\g_{ji}^1(m)\cr\cr
&=&\Theta\circ\left(\tau_j^1\right)^{-1}(m,\Id_{\mu_j^1(m)}).\g_{j
i}^1(m)=\Theta\circ\left(\tau_j^1\right)^{-1}\left(m,\g_{ji}^1(m)
\right)\cr\cr
&=&\Theta\circ\left(\tau_j^1\right)^{-1}\circ\tau_j^1\circ\left(
\tau_i^1\right)^{-1}(m,\Id_{\mu_i^1(m)})\equiv\Theta\circ\left(
\tau_i^1\right)^{-1}(m,\Id_{\mu_i^1(m)})\cr\cr
&=&\left(\tau_i^2\right)^{-1}\left(m,\theta_i(m)\right)=\left(
\tau_j^2\right)^{-1}\circ\tau_j^2\circ\left(\tau_i^2\right)^{-1}(m,
\Id_{\mu_i^2(m)}).\theta_i(m)\cr\cr
&=&\si_j^2(m).\left(\g_{ji}^2(m)\circ\theta_j(m)\right)\,.
\qqq
Here, we are using the fact that the defining $\Gr$-action on
$\,\sfP_2\,$ is free.

Conversely, let $\,(\theta_i)\,$ be a collection of locally smooth
maps satisfying conditions (i) and (ii). We shall demonstrate that
they induce $\Gr$-bundle (iso)morphisms $\,\widetilde\theta_i:
\mu_i^{1\,*}\cU_\Gr\xrightarrow{\cong}\mu_i^{2\,*}\cU_\Gr\,$ between
the local trivialisations of the principal $\Gr$-bundles $\,\cP_1\,$
and $\,\cP_2$,\ respectively. Define a smooth map
\qq\nn
\widetilde\theta_i\ :\ \xcM\fibx{\mu_i^1}{t}\morf\,\Gr\to\xcM
\fibx{\mu_i^2}{t}\morf\,\Gr\ :\ (m,\vec g)\mapsto\left(m,\theta_i(m)
\circ\vec g\right)\,.
\qqq
The definition makes sense as for $\,(m,\vec g)\,$ such that
$\,\mu_i^1(m)=t(\vec g)\,$ we have
\qq\nn
s\left(\theta_i(m)\right)=\mu_i^1(m)=t(\vec g)
\qqq
and
\qq\nn
t\left(\theta_i(m)\circ\vec g\right)=t\circ\theta_i(m)=\mu_i^2(m)
\,.
\qqq
The map is surjective,
\qq\nn
(m,\vec g)\in\xcM\fibx{\mu_i^2}{t}\morf\,\Gr\quad\Rightarrow\quad
(m,\vec g)=\widetilde\theta_i\left(m,\theta_i(m)^{-1}\circ\vec g
\right)\,,
\qqq
preserves fibres,
\qq\nn
\pi_{\mu_i^{2\,*}\morf\,\Gr}\circ\widetilde\theta_i(m,\vec g)\equiv
\pr_1\left(m,\theta_i(m)\circ\vec g\right)=m=\pr_1(m,\vec g)\equiv
\pi_{\mu_i^{1\,*}\morf\,\Gr}(m,\vec g)\,,
\qqq
intertwines the momenta,
\qq\nn
\mu_{\mu_i^{2\,*}\morf\,\Gr}\circ\widetilde\theta_i(m,\vec g)\equiv
s\circ\pr_2\left(m,\theta_i(m)\circ\vec g\right)=s(\vec g)=s\circ
\pr_2(m,\vec g)\equiv\mu_{\mu_i^{1\,*}\morf\,\Gr}(m,\vec g)\,,
\qqq
and is manifestly (right-)$\Gr$-equivariant,
\qq\nn
\widetilde\theta_i\left((m,\vec g).\vec h\right)\equiv\widetilde
\theta_i(m,\vec g \circ\vec h)=\left(m,\theta_i(m)\circ\vec
g\circ\vec h\right)=\left( m,\theta_i(m)\circ\vec g\right).\vec
h\equiv\widetilde\theta_i(m,\vec g).\vec h\,,
\qqq
which altogether means that it is a $\Gr$-bundle morphism, and hence
an isomorphism. \eroof

Theorem \ref{thm:GrBun-loc} and Proposition
\ref{prop:GrBun-morf-loc} permit to reduce the analysis of the
category $\,\Grbun{\xcM}\,$ to that of local data for its objects
and morphisms. We shall use this fact below in an explicit
discussion of the case of interest, which is that of the action
groupoid $\,\Gr=\txG\lx M\,$ introduced earlier. By way of
preparation, we formulate
\bedef\label{def:princ-Gbun-sec}
Adopt the notation of Definitions \ref{def:grpd} and
\ref{def:princ-g-bund}. Let $\,\Si\,$ be a smooth space, $\,\txG\,$
a Lie group, and $\,\xcM\,$ a smooth $\txG$-space. Denote by
$\,\Gbun{\Si}\,$ the groupoid of principal $\txG$-bundles with base
$\,\Si$.\ The \textbf{groupoid $\,\Gbun{\Si\,\Vert\,\xcM}\,$ of
principal $\txG$-bundles with base $\,\Si\,$ gauging $\,\xcM\,$} is
the subgroupoid of $\,\Gbun{\Si}\,$ composed of the objects
$\,\cP_\txG=(\sfP_\txG,\Si,\pi_{\sfP_\txG},\rho_{\sfP_\txG})\,$ of
the latter category (and all morphisms between them) with the
property that the corresponding associated bundles
$\,\sfP_\txG\x_\txG\xcM\to\Si\,$ admit a global section. \exdef
\brem\label{rem:G-princ-iso-loc} The definition makes sense as every
isomorphism $\,\chi\in\mor_{\Gbun{\Si}}(\cP_\txG^1,\cP_\txG^2)\,$
between bundles $\,\cP_\txG^1,\cP_\txG^2\in\obj\,\Gbun{\Si\,\Vert\,
\xcM}\,$ canonically induces an isomorphism $\,\widetilde\chi\in
\mor_{\Gbun{\Si\,\Vert\,\xcM}}(\cP_\txG^1,\cP_\txG^2)$.\ This is
readily verified in the local description associated with a choice
$\,\cO_\Si:=\{\Si_i\}_{i\in\xcI}\,$ of an open cover of the common
base $\,\Si\,$ of the two bundles in which $\,\chi$,\ being an
invertible fibre-preserving $\txG$-map $\,\chi:\sfP_\txG^1\to
\sfP_\txG^2$,\ is described by a collection of locally smooth maps
$\,\chi_i:\Si_i\to\txG\,$ defined by the formul\ae
\qq\label{eq:chii-def}
\tau_i^2\circ\chi\circ\si_i^1(\si)=:\left(\si,\chi_i(\si)\right)\,,
\qqq
written for $\,\si\in\Si_i\,$ and for a local section
$\,\si_i^1(\si)= \left(\tau_i^1\right)^{-1}(\si,e)\,$ determined by
a local trivialisation
$\,\tau_i^1:\pi_{\sfP_\txG^1}^{-1}(\Si_i)\to\Si_i\x\txG$,\ and
satisfying the identity
\qq\label{eq:g2g1-Piso}
g_{ij}^2(\si)=\chi_i(\si)\cdot g_{ij}^1(\si)\cdot\chi_j(\si)^{-1}
\qqq
written for $\,\si\in\Si_{ij}\,$ and the transition maps $\,g_{i
j}^A:\Si_{ij}\to\txG\,$ of $\,\sfP_\txG^A$,\ the latter being
defined by the relation
\qq\nn
\tau_i^A\circ\left(\tau_j^A\right)^{-1}(\si,e)=:\left(\si,g_{ij}^A
(\si)\right)\,.
\qqq
This follows from a specialisation of Proposition
\ref{prop:GrBun-morf-loc} to the case of the Lie group $\,\txG\,$
viewed as a groupoid with the object manifold given by a singleton
$\,\{\bullet\}$.

Consider, now, a global section $\,\eta^1\in\G(\sfP_\txG^1\x_\txG
\xcM)\,$ with
\qq\nn
\eta^1\ :\ \Si_i\to\sfP_\txG^1\x_\txG\xcM\ :\ \si\mapsto\left[\left(
\si_i^1(\si),m_i(\si)\right)\right]
\qqq
such that, for any $\,\si\in\Si_{ij}$,\ we obtain
\qq\nn
\left[\left(\si_j^1(\si),m_j(\si)\right)\right]=\left[\left(\si_i^1
(\si),m_i(\si)\right)\right]\,.
\qqq
The left-hand side equals
\qq\nn
\left[\left(\si_j^1(\si),m_j(\si)\right)\right]=\left[\left(\si_i^1
(\si).g_{ij}^1(\si),m_j(\si)\right)\right]=\left[\left(\si_i^1(\si)
,g_{ij}^1(\si).m_j(\si)\right)\right]\,,
\qqq
and so we must require that the gluing condition
\qq\nn
m_i(\si)=g_{ij}^1(\si).m_j(\si)
\qqq
hold true over $\,\Si_{ij}$.\ The existence of the locally smooth
maps $\,m_i:\Si_i\to\xcM\,$ is thus tantamount to the existence of a
global section of the associated bundle $\,\sfP_\txG^1\x_\txG\xcM$.\
Define
\qq\nn
\widetilde\chi(\eta^1)(\si):=\left[\left(\si_i^2(\si),\chi_i(\si).
m_i(\si)\right)\right]\,,
\qqq
where, as usual, the local section $\,\si_i^2(\si)=\left(\tau_i^2
\right)^{-1}(\si,e)\,$ is defined in terms of the very same local
trivialisation $\,\tau_i^2\,$ as the one entering the definition of
the $\,\chi_i$.\ Clearly, $\,\left[\left(\si_i^2(\cdot),\chi_i(\cdot
).m_i(\cdot)\right)\right]\,$ is a local section of $\,\sfP_\txG^2
\x_\txG\xcM\,$ over $\,\Si_i$,\ from which it follows that
$\,\widetilde\chi(\eta^1)\,$ is a collection of local sections of
the associated bundle $\,\sfP_\txG^2\x_\txG\xcM$.\ We readily
convince ourselves that this last section is global,
\qq\nn
\left[\left(\si_j^2(\si),\chi_j(\si).m_j(\si)\right)\right]&=&\left[
\left(\left(\tau_i^2\right)^{-1}\circ\tau_i^2\circ\left(\tau_j^2
\right)^{-1}(\si,e),\left(\chi_j(\si)\cdot g_{ji}^1(\si)\right).m_i(
\si)\right)\right]\cr\cr
&=&\left[\left(\si_i^2(\si).g_{ij}^2(\si),\left(\chi_j(\si)\cdot
g_{ji}^1(\si)\right).m_i(\si)\right)\right]\cr\cr
&=&\left[\left(\si_i^2(\si),\left(g_{ij}^2(\si)\cdot\chi_j(\si)\cdot
g_{ji}^1(\si)\right).m_i(\si)\right)\right]\cr\cr
&=&\left[\left(\si_i^2(\si),\chi_i(\si).m_i(\si)\right)\right]\,,
\qqq
as stipulated by the definition. \erem \noindent We come to the main
result of our considerations.
\bethe\label{thm:Gbun-vs-Grbun} Adopt the notation of Definitions
\ref{def:grpd}, \ref{def:princ-gr-bun} and \ref{def:princ-Gbun-sec}.
There exists an isomorphism of groupoids
\qq\nn
\Gbun{\Si\,\Vert\,\xcM}\cong\GMbun{\Si}\,.
\qqq
\ethe
\beroof
By way of a proof, we give an explicit construction of a essentially
surjective fully faithful functor
\qq\nn
Gr\ :\ \Gbun{\Si\,\Vert\,\xcM}\to\GMbun{\Si}
\qqq
in the local description of both (small) categories. Thus, as the
point of departure of our construction we take an open cover $\,\{
\Si_i\}_{i\in\xcI}=:\cO_\Si\,$ of $\,\Si$,\ to which we associate
local data of principal $\txG$-bundles, bundles associated to them,
principal $\txG\lx \xcM$-bundles, and (iso)morphisms between them.

Take a principal $\txG$-bundle $\,\cP_\txG=(\sfP_\txG,\Si,
\pi_{\sfP_\txG},\rho_{\sfP_\txG})\,$ with local trivialisations
\qq\nn
\tau_i\ :\ \pi_{\sfP_\txG}^{-1}(\Si_i)\to\Si_i\x\txG
\qqq
and transition maps
\qq\nn
\tau_{ij}(\si,e)=\tau_i\circ\tau_j^{-1}(\si,e)=\left(\si,g_{ij}(\si
)\right)\,,
\qqq
written in terms of a \v Cech 1-cocycle $\,g_{ij}:\Si_{ij}\to
\txG$.\ Form the associated bundle $\,\sfP_\txG\x_\txG\xcM\to\Si\,$
and assume the existence of a global section
\qq\nn
\eta\ :\ \Si\to\sfP_\txG\x_\txG\xcM
\qqq
with restrictions
\qq\nn
\eta\ :\ \Si_i\to\sfP_\txG\x_\txG\xcM\ :\ \si\mapsto\bigl[\bigl(
\tau_i^{-1}(\si,e),m_i(\si)\bigr)\bigr]\,,
\qqq
written in terms of some locally smooth maps $\,m_i:\Si_i\to\xcM\,$
that satisfy the relation
\qq\nn
m_i(\si)=g_{ij}(\si).m_j(\si)
\qqq
over double intersections $\,\Si_{ij}\ni\si$.

To $\,\cP_\txG$,\ we associate a principal $\txG\lx\xcM$-bundle as
follows: Define locally smooth maps
\qq\nn
\mu_i\ :\ \Si_i\to\xcM\ :\ \si\mapsto m_i(\si)\,,\qquad\g_{ij}\ :\
\Si_{ij}\to\txG\x\xcM\ :\ \si\mapsto\left(g_{ij}(\si),m_j(\si)
\right)\,.
\qqq
These satisfy the identities
\qq\nn
&t\circ\g_{ij}(\si)=g_{ij}(\si).m_j(\si)=m_i(\si)\equiv\mu_i(\si)
\,,\qquad\qquad s\circ\g_{ij}(\si)=m_j(\si)\equiv\mu_j(\si)\,,&\cr
\cr
&\g_{ii}(\si)=\left(e,m_i(\si)\right)\equiv\Id_{m_i(\si)}\equiv\Id
\circ\mu_i(\si)\,,&\cr\cr
&\g_{ji}(\si)=\left(g_{ij}(\si)^{-1},m_i(\si)\right)=\left(g_{ij}(\si)^{-1},
g_{ij}(\si).m_j(\si)\right)\equiv\Inv\circ\g_{ij}(\si)\,,&\cr\cr
&\,\g_{ij}(\si)\circ\g_{jk}(\si)=\left(g_{ij}(\si),g_{jk}(\si).m_k(
\si)\right)\circ\left(g_{jk}(\si),m_k(\si)\right)=\left(g_{ij}(\si)
\cdot g_{jk}(\si),m_k(\si)\right)=\g_{ik}(\si)\,,&
\qqq
and so we conclude that the collection $\,\bigl(\cO_\Si,m_i,(g_{ij},
m_j)\ \vert\ i,j\in\xcI\bigr)\,$ defines local data of a principal
$\txG\lx\xcM$-bundle over $\,\Si$.\ Upon applying the clutching
construction of Theorem \ref{thm:GrBun-loc}, we thus obtain the
total space
\qq\nn
\vec\sfP_{\cO_\Si}:=\bigsqcup_{i\in\xcI}\,m_i^*\cU_{\txG\lx\xcM}/
\sim_{(g_{ij},m_j)}
\qqq
of a principal $\txG\lx\xcM$-bundle which we declare to be the
$Gr$-image of $\,\cP_\txG$,
\qq\nn
Gr(\cP_\txG):=\left(\vec\sfP_{\cO_\Si},\Si,\pi_{\vec\sfP_{\cO_\Si}}
,\mu_{\vec\sfP_{\cO_\Si}},\rho_{\vec\sfP_{\cO_\Si}}\right)\,.
\qqq

Next, we shall verify that the above assignment is functorial by
associating morphisms between $\txG\lx\xcM$-bundles to those between
$\txG$-bundles gauging $\,\xcM$.\ To this end, consider a pair
$\,\cP_\txG^1,\cP_\txG^2\in\obj\,\Gbun{\Si\,\Vert\,\xcM}\,$ and
assume given local data $\,(\chi_i)\,$ (associated with $\,\cO_\Si$)
of an isomorphism $\,\chi:\cP_\txG^1\xrightarrow{\cong}\cP_\txG^2\,$
determined by the relations
\qq\nn
\tau_i^2\circ\chi\circ\left(\tau_i^1\right)^{-1}(\si,e)=\left(\si,
\chi_i(\si)\right)\,,
\qqq
written, for $\,\si\in\Si_i$,\ in terms of local trivialisations
$\,\tau_i^A:\pi_{\sfP_\txG^A}^{-1}(\Si_i)\to\Si_i\x\txG,\ A\in\{1,2
\}$.\ The corresponding global sections of the associated bundles
are related as described in Remark \ref{rem:G-princ-iso-loc}, that
is
\qq\nn
\eta^1\ :\ \Si_i\to\pi_{\sfP_\txG^1\x_\txG\xcM}^{-1}(\Si_i)\ :\ \si
\mapsto\left[\left(\left(\tau_i^1\right)^{-1}(\si,e),m_i(\si)
\right)\right]
\qqq
is mapped to
\qq\nn
\widetilde\chi(\eta^1)\ :\ \Si_i\to\pi_{\sfP_\txG^2\x_\txG\xcM}^{-
1}(\Si_i)\ :\ \si\mapsto\left[\left(\left(\tau_i^2\right)^{-1}(\si,e
),\chi_i(\si).m_i(\si)\right)\right]
\qqq
by the induced isomorphism $\,\widetilde\chi$.\ Accordingly, we find
\qq\nn
\vec\sfP_{\cO_\Si}^1=\bigsqcup_{i\in\xcI}\,m_i^*\cU_{\txG\lx\xcM}/
\sim_{(g_{ij}^1,m_j)}
\qqq
and
\qq\nn
\vec\sfP_{\cO_\Si}^2=\bigsqcup_{i\in\xcI}\,(\chi_i.m_i)^*\cU_{\txG
\lx\xcM}/\sim_{(\chi_i.g_{ij}^1.(\Inv\circ\chi_j),\chi_j.m_j)}
\qqq
as total spaces of $\,Gr(\cP_\txG^1)\,$ and $\,Gr(\cP_\txG^2)$,\
respectively, and so we conclude that the desired isomorphism
\qq\nn
Gr(\chi)\ :\ Gr(\cP_\txG^1)\xrightarrow{\ \cong\ }Gr(\cP_\txG^2)
\qqq
is determined by local data
\qq\nn
\theta_i:=(\chi_i,m_i)\equiv(\chi_i,\mu_i^1)\,.
\qqq
Indeed, we obtain
\qq\nn
s\circ\theta_i(\si)=m_i(\si)\equiv\mu_i^1(\si)\,,\qquad\qquad t
\circ\theta_i(\si)=\chi_i(\si).m_i(\si)\equiv\mu_i^2(\si)
\qqq
and
\qq\nn
\theta_i(\si)\circ\g_{ij}^1(\si)\circ\theta_j(\si)^{-1}&\equiv&
\left(\chi_i(\si),m_i(\si)\right)\circ\left(g_{ij}^1(\si),m_j(\si)
\right)\circ\left(\chi_j(\si)^{-1},\chi_j(\si).m_j(\si)\right)\cr
\cr
&=&\left(\chi_i(\si),g_{ij}^1(\si).m_j(\si)\right)\circ\left(g_{i
j}^1(\si),m_j(\si)\right)\circ\left(\chi_j(\si)^{-1},\chi_j(\si).
m_j(\si)\right)\cr\cr
&=&\left(\chi_i(\si)\circ g_{ij}^1(\si)\circ\chi_j(\si)^{-1},
\chi_j(\si).m_j(\si)\right)\cr\cr
&=&\g_{ij}^2(\si)\,,
\qqq
in conformity with Proposition \ref{prop:GrBun-morf-loc}. It is
clear from the very definition of the mapping $\,Gr\,$ that its
morphism component preserves composition of morphisms as for
\qq\nn
\cP_\txG^1\xrightarrow{\ \chi\ }\cP_\txG^2\xrightarrow{\ \chi'\ }
\cP_\txG^3
\qqq
we get
\qq\nn
Gr(\chi'\circ\chi)=(\chi_i'\cdot\chi_i,\mu_i^1)=(\chi_i',\chi_i.
\mu_i^1)\circ(\chi_i,\mu_i^1)\equiv(\chi_i',\mu_i^2)\circ(\chi_i,
\mu_i^1)=Gr(\chi')\circ Gr(\chi)\,.
\qqq
Moreover, the $Gr$-image of the identity $\txG$-bundle morphism is
the identity $\txG\lx\xcM$-bundle morphism,
\qq\nn
Gr(\Id_{\cP_\txG})=(e,\mu_i^1)\equiv(\Id_{\mu_i^1})=\Id_{Gr(
\cP_\txG)}\,.
\qqq
Thus, the mapping $\,Gr\,$ does, indeed, define a functor
\qq\nn
Gr\ :\ \Gbun{\Si\,\Vert\,\xcM}\to\GMbun{\Si}\,.
\qqq
We shall next demonstrate that the latter functor is an equivalence.

We begin by showing that $\,Gr\,$ is essentially surjective. Take a
$\txG\lx\xcM$-bundle $\,\cP_{\txG\lx\xcM}=(\sfP_{\txG\lx\xcM},\Si,
\pi_{\sfP_{\txG\lx\xcM}},\mu_{\sfP_{\txG\lx\xcM}},\rho_{\sfP_{\txG
\lx\xcM}})\,$ with local momenta $\,\mu_i:\Si_i\to\xcM\,$ and
transition maps $\,\g_{ij}:\Si_{ij}\to\txG\x\xcM$.\ The latter
decompose as
\qq\label{eq:Gr-trans-fun-decomp}
\g_{ij}=(\g_{ij}^\txG,\g_{ij}^\xcM)\,,\qquad\qquad\g_{ij}^\xcX\ :\
\Si_{i j}\to\xcX\,,\quad\xcX\in\{\xcM,\txG\}\,,
\qqq
and identities (i) of Definition \ref{def:loc-triv-data} yield, for
$\,\si\in\Si_{ij}$,
\qq\nn
&\g_{ij}^\xcM(\si)=s\circ\g_{ij}(\si)=\mu_j(\si)\,,\qquad\qquad
\g_{ij}^\txG(\si).\mu_j(\si)=\g_{ij}^\txG(\si).\g_{ij}^\xcM(\si)=t
\circ\g_{ij}(\si)=\mu_i(\si)\,,&\cr\cr
&\bigl(\g_{ii}^\txG(\si),\mu_i(\si)\bigr)=\bigl(\g_{ii}^\txG(\si),
\g_{ii}^\xcM(\si)\bigr)\equiv\g_{ii}(\si)=\Id_{\mu_i(\si)}=(e,\mu_i
(\si))\,.&
\qqq
Identity (ii) of the same definition now implies
\qq\nn
\bigl(\g_{ji}^\txG(\si),\mu_i(\si)\bigr)=\g_{ji}(\si)=\bigl(\g_{i
j}^\txG(\si),\mu_j(\si)\bigr)^{-1}=\bigl(\g_{ij}^\txG(\si)^{-1},
\g_{ij}^\txG(\si).\mu_j(\si)\bigr)=\bigl(\g_{ij}^\txG(\si)^{-1},
\mu_i(\si)\bigr)\,,
\qqq
and identity (iii) transcribes as
\qq\nn
\bigl(\g_{ik}^\txG(\si),\mu_k(\si)\bigr)&=&\g_{ik}(\si)=\g_{ij}(\si
)\circ\g_{jk}(\si)=\bigl(\g_{ij}^\txG(\si),\mu_j(\si)\bigr)\circ
\bigl(\g_{jk}^\txG(\si),\mu_k(\si)\bigr)\cr\cr
&=&\bigl(\g_{ij}^\txG(\si),\g_{jk}^\txG(\si).\mu_k(\si)\bigr)\circ
\bigl(\g_{jk}^\txG(\si),\mu_k(\si)\bigr)=\bigl(\g_{ij}^\txG(\si)
\cdot\g_{jk}^\txG(\si),\mu_k(\si)\bigr)
\qqq
for any $\,\si\in\Si_{ijk}$.\ Thus, altogether, the local bundle
data consist of the smooth functions
\qq\nn
\mu_i\ :\ \Si\to\xcM\,,\qquad\qquad g_{ij}:=\g_{ij}^\txG\ :\
\Si_{ij}\to\txG
\qqq
with the following properties
\qq\nn
&\mu_i(\si)=g_{ij}(\si).\mu_j(\si)\,,&\cr\cr
&g_{ik}(\si)=g_{ij}(\si)\cdot g_{jk}(\si)\,,\qquad\qquad g_{ji}(\si
)=g_{ij}(\si)^{-1}\,,\qquad\qquad g_{ii}(\si)=e\,.&
\qqq

Using the data $\,g_{ij}$,\ we obtain a principal $\txG$-bundle
$\,\cP_{\cO_\Si}=(\sfP_{\cO_\Si},\Si,\pi_{\sfP_{\cO_\Si}},
\rho_{\sfP_{\cO_\Si}})\,$ via the standard clutching construction.
Its total space is
\qq\nn
\sfP_{\cO_\Si}:=\bigsqcup_{i\in\xcI}\,(\Si_i\x\txG)/\sim_{(g_{ij})}
\,,
\qqq
with the equivalence relation defined as
\qq\nn
(i,\si_i,g_i)\sim_{(g_{ij})}(j,\si_j,g_j)\quad\Leftrightarrow\quad
\left(\ \si_j=\si_i\in\Si_{ij}\quad\land\quad g_i=g_{ij}(\si_i)\cdot
g_j\ \right)\,.
\qqq
The projection to the base $\,\Si\,$ reads
\qq\nn
\pi_{\sfP_{\cO_\Si}}\ :\ \sfP_{\cO_\Si}\to\Si\ :\
\bigl[(i,\si,g)\bigr] \mapsto\si\,,
\qqq
and the right $\txG$-action is given by the formula
\qq\nn
\rho_{\sfP_{\cO_\Si}}\ :\ \sfP_{\cO_\Si}\x\txG\to\sfP_\txG\ :\
\bigl([(i,\si,g)],h\bigr)\mapsto[(i,\si,g\cdot h)]\,.
\qqq
The latter is manifestly fibre-preserving, free and transitive.

Local trivialisations of $\,\cP_{\cO_\Si}\,$ are given by the maps
\qq\nn
\tau_i\ :\ \pi_{\sfP_{\cO_\Si}}^{-1}(\Si_i)\to\Si_i\x\txG\ :\ [(i,
\si,g)]\mapsto(\si,g)
\qqq
with inverses
\qq\nn
\tau_i^{-1}\ :\ \Si_i\x\txG\to\pi_{\sfP_{\cO_\Si}}^{-1}(\Si_i)\ :\ (
\si,g)\mapsto[(i,\si,g)]
\qqq
that have the desired $\txG$-equivariance property
\qq\nn
\tau_i^{-1}(\si,g)=\left[(i,\si,e\cdot g)\right]=\left[(i,\si,e)
\right].g=\tau_i^{-1}(\si,e).g
\qqq
and hence, in particular, satisfy the gluing relations
\qq\nn
\tau_i^{-1}(\si,e)=[(i,\si,e)]=\bigl[\bigl(j,\si,g_{ji}(
\si)\bigl)\bigr]=[(j,\si,e)].g_{ji}(\si)=\tau_j^{-1}(\si,
e).g_{ji}(\si)\,.
\qqq
The associated transition maps read
\qq\nn
\tau_{ij}:=\tau_i\circ\tau_j^{-1}\ :\ (\si,g)\mapsto[(j,\si,g)]=
\bigl[\bigl(i,\si,g_{ij}(\si)\cdot g\bigr)\bigr]\mapsto\bigl(
\si,g_{ij}(\si)\cdot g\bigr)\,.
\qqq

The $\,\tau_i\,$ in conjunction with the $\,\mu_i\,$ give rise to
\emph{global} sections of the associated bundle $\,\sfP_{\cO_\Si}
\x_\txG\xcM\to\Si$,\ with the total space given by the smooth
quotient $\,(\sfP_{\cO_\Si}\x\xcM)/\txG\,$ with respect to the
(right) diagonal $\txG$-action of \Reqref{eq:right-diag-assoc}.
Indeed, write
\qq\nn
\eta_i\ :\ \Si_i\to(\sfP_{\cO_\Si}\x\xcM)/\txG\ :\ \si\mapsto\bigl[
\bigl(\tau_i^{-1}(\si,e),\mu_i(\si)\bigr)\bigr]\,.
\qqq
We readily check that the $\,\eta_i\,$ compose a global section as
for an arbitrary $\,\si\in\Si_{ij}$,
\qq\nn
\eta_j(\si)=\bigl[\bigl(\tau_j^{-1}(\si,e),\mu_j(\si)\bigr)\bigr]
&=&\bigl[\bigl(\tau_i^{-1}(\si,e).g_{ij}(\si),\mu_j(\si)\bigr)
\bigr]=\bigl[\bigl(\tau_i^{-1}(\si,e),g_{ij}(\si).\mu_j(\si)\bigr)
\bigr]\cr\cr
&=&\bigl[\bigl(\tau_i^{-1}(\si,e),\mu_i(\si)\bigr)\bigr]=\eta_i(\si
)\,.
\qqq
The above construction yields a map
\qq\nn
\gimel\ :\ \obj\,\GMbun{\Si}\to\obj\,\Gbun{\Si\,\Vert\,\xcM}\ : \
\cP_{\txG\lx\xcM}\mapsto\cP_{\cO_\Si}\,,
\qqq
and we readily check that
\qq\nn
Gr\circ\gimel(\cP_{\txG\lx\xcM})\cong\cP_{\txG\lx\xcM}\,,
\qqq
with the isomorphism determined by local trivialisations of
$\,\cP_{\txG\lx\xcM}\,$ as described in Proposition
\ref{prop:Gr-loc-triv-amb}. This proves that $\,Gr\,$ is essentially
surjective, as claimed.

In the next step, we show that $\,Gr\,$ is full by explicitly
constructing a counterpart of $\,\gimel\,$ acting on morphisms, to
be denoted by the same symbol. Here, we consider a pair of principal
$\txG\lx\xcM$-bundles $\,\cP_{\txG\lx\xcM}^A,\ A\in\{1,2\}\,$ over
$\,\Si\,$ with the respective local data $\,( \cO_\Si,\mu_i^A,\g_{i
j}^A\ \vert\ i,j\in\xcI)\,$ associated with an open cover introduced
before. Following Proposition \ref{prop:GrBun-morf-loc}, we take the
data to be related as
\qq\nn
\mu_i^1=s\circ\theta_i\,,\qquad\qquad\mu_i^2=t\circ\theta_i\,,
\qqq
and, for any $\,\si\in\Si_{ij}$,
\qq\nn
\g_{ij}^2(\si)=\theta_i(\si)\circ\g_{ij}^1(\si)\circ\theta_j(\si
)^{-1}
\qqq
by a collection $\,(\theta_i)_{i\in\xcI}\,$ of locally smooth maps
$\,\theta_i:\Si_i\to\morf\,(\txG\lx\xcM)$.\ Taking into account the
specific form of the source and target maps of $\,\txG\lx\xcM$,\ we
may write the $\,\theta_i\,$ in the component form
\qq\nn
\theta_i=(\xi_i,\mu_i^1)\,,
\qqq
with $\,\xi_i:\Si_i\to\txG\,$ chosen such that, for all $\,\si\in
\Si_i$,
\qq\nn
\xi_i(\si).\mu_i^1(\si)=t\circ\theta_i(\si)=\mu_i^2(\si)\,.
\qqq
Writing out the $\,\g_{ij}^A\,$ in components as in
\Reqref{eq:Gr-trans-fun-decomp}, we then find over $\,\Si_{ij}\ni
\si$,
\qq\nn
\left(g_{ij}^2(\si),\mu_j^2(\si)\right)&\equiv&\g_{ij}^2(\si)=
\theta_i(\si)\circ\g_{ij}^1(\si)\circ\theta_j(\si)^{-1}\cr\cr
&=&\left(\xi_i(\si),\mu_i^1(\si)\right)\circ\left(g_{ij}^1(\si),
\mu_j^1(\si)\right)\circ\left(\xi_j(\si)^{-1},\xi_j(\si).\mu_j^1(
\si)\right)\cr\cr
&=&\left(\xi_i(\si)\cdot g_{ij}^1(\si)\cdot\xi_j(\si)^{-1},\xi_j(
\si).\mu_j^1(\si)\right)\,,
\qqq
whence we infer that the local data of the bundles $\,\gimel(
\cP_{\txG\lx\xcM}^1)\,$ and $\,\gimel(\cP_{\txG\lx\xcM}^2)\,$ are
related by a $\txG$-bundle (iso)morphism
\qq\nn
\gimel(\Theta)\ :\ \gimel(\cP_{\txG\lx\xcM}^1)\xrightarrow{\ \cong\
}\gimel(\cP_{\txG\lx\xcM}^2)
\qqq
with local data $\,(\xi_i)\,$ (that automatically belongs to
$\,\mor_{\Gbun{\Si}}\left(\gimel(\cP_{\txG\lx\xcM}^1),\gimel(
\cP_{\txG\lx\xcM}^2)\right)$). It is now a matter of a simple check
to see that
\qq\nn
Gr\circ\gimel(\Theta)=\Theta\,.
\qqq

The last property of $\,Gr\,$ to be substantiated is its
faithfulness. This one follows immediately from the construction of
the functor. Indeed, different $\txG$-bundle isomorphisms would
necessarily have different local data that would -- in turn -- yield
different $\txG \lx\xcM$-bundle isomorphisms under $\,Gr$.\ Thus, we
have established that the functor $\,Gr\,$ is an equivalence of
categories, which concludes the proof of the theorem. \eroof

\brem The above theorem is an extension of the statement of a
one-to-one correspondence between objects of the two categories
worked out in \Rxcite{Sec.\,3.3.3}{Rossi:2004lg}. \erem

\begin{figure}[hbt]~\\[50pt]

$$
 \raisebox{0pt}{\begin{picture}(50,50)
  \put(-170,-14){\scalebox{0.15}{\includegraphics{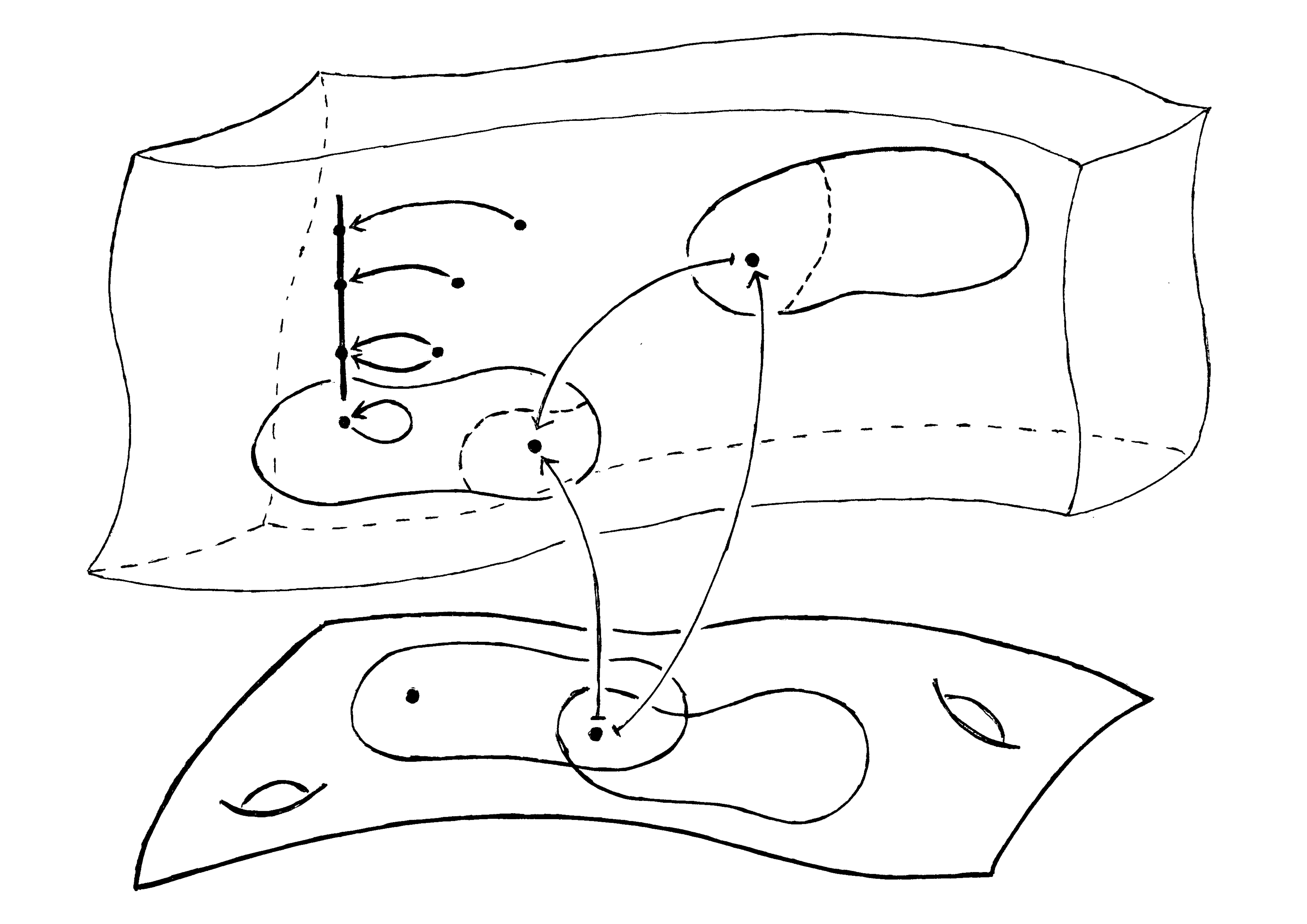}}}
  \end{picture}
  \put(0,0){
     \setlength{\unitlength}{.60pt}\put(-28,-16){
     \put(-252,90)     { $\Si$  }
     \put(-155,71)     { $\Si_i$  }
     \put(110,39)     { $\Si_j$  }
     \put(-34,82)     { {\small $\si$}  }
     \put(-135,116)     { {\small $\si'$}  }
     \put(-41,167)     { $\mu_i$  }
     \put(49,187)     { $\mu_j$  }
     \put(272,213)     { $M$  }
     \put(-22,307)     { $g_{ij}(\si).$  }
     \put(-252,220)     { $\mu_i(\Si_i)$  }
     \put(-91,234)     { {\small $\mu_i(\si)$}  }
     \put(-200,245)     { {\small $\mu_i(\si')$}  }
     \put(-125,272)     { {\scriptsize $\left(e,\mu_i(\si')\right)$}  }
     \put(140,316)     { $\mu_j(\Si_j)$  }
     \put(54,360)     { {\small $\mu_j(\si)$}  }
     \put(-92,335)     { {\tiny $h^{-1}.\mu_i(\si')$}  }
     \put(-58,367)     { {\tiny $g^{-1}.\mu_i(\si')$}  }
     \put(-98,312)     { {\tiny $h_1^{-1}.\mu_i(\si')$}  }
     \put(-104,297)     { {\tiny $=\hspace{-2pt}h_2^{-1}.\mu_i(\si')$}  }
     \put(-150,387)     { {\scriptsize $\left(g,g^{-1}.\mu_i(\si')\right)$}  }
           }\setlength{\unitlength}{1pt}}}
$$

\caption{The principal $\txG_\si\lx M$-bundle
$\,Gr(\cP_{\txG_\si})\,$ over $\,\Si\,$ in the (dual) local
description of the gauged mono-phase $\si$-model. Open
neighbourhoods $\,\Si_i\subset\Si\,$ are mapped into $\,M\,$ by
local momenta $\,\mu_i\,$ extracted from the definition of a global
section of the associated bundle $\,\sfP_{\txG_\si}\x_{\txG_\si}
M$.\ Points in the image of a double intersection $\,\Si_{ij}\,$ of
the neighbourhoods are related by the action of the transition map
$\,g_{ij}\,$ of the principal $\txG_\si$-bundle $\,\cP_{\txG_\si}$.\
Over each point $\,\mu_i(\si')$,\ there is an entire fibre of arrows
from $\,\txG_\si\x M\,$ ending at $\,\mu_i(\si')$.\ In addition to
the complete information about the $\txG_\si$-orbit
$\,\txG_\si.\mu_i(\si)$,\ the fibre encodes information on the
isotropy subgroup $\,\txG_{\si\,\mu_i(\si)}\,$ (\textit{cf.}\ the
pair of arrows with a common source and target).}
\label{fig:Gr-bundle}
\end{figure}

The last theorem offers a most natural explanation of the appearance
of the tangent algebroid of the action groupoid $\,\txG_\si\lx
\xcF\,$ in the analysis of the rigid symmetries of the $\si$-model
that admit gauging. On top of that, it gives rise to a simple local
presentation of field configurations of the gauged $\si$-model on
the purely geometric level, \textit{i.e.}\ in the setting in which
the presence of the gauge field on the world-sheet and that of the
metric and gerbe-theoretic structure on the target space has been
forgotten. In this presentation, sketched in Figure
\ref{fig:Gr-bundle} in the simplest case of a mono-phase
$\si$-model, elements of an open cover of the world-sheet are
embedded smoothly into the target space using local momenta of the
principal $\txG_\si\lx\xcF$-bundle $Gr$-dual to the principal
$\txG_\si$-bundle of the gauged $\si$-model. This is done in such a
manner that images, under the respective local momenta, of points
from double intersections of elements of the open cover are related
by arrows from the morphism set $\,\txG_\si\x\xcF\,$ of the action
groupoid determined by the appropriate transition maps of the
principal $\txG_\si$-bundle. The picture thus obtained is largely
reminiscent of the well-established idea of realising field
configurations of the $\si$-model on the orbit space of the action
of a group on the target space of a parent $\si$-model through
patchwise smooth field configurations of the parent $\si$-model, in
which field discontinuities that occur upon passing between
neighbouring patches are determined by the action of the group that
is being gauged. Of course, for this idea to be applicable, one
would have to take into account the extra structure, both on the
world-sheet\footnote{An extension of the equivalence between the
category of principal $\txG_\si$-bundles gauging the target space of
the $\si$-model and the category of principal bundles with the
corresponding action groupoid to the setting \emph{with connection}
should be possible and relatively straightforward within the
differential-geometric framework developed in
Refs.\,\cite{MacKenzie:1987} and \cite{Schreiber:2007pt}. We hope to
return to this issue in future work.} and on the target space, that
enters the definition of the gauged $\si$-model. Nevertheless, even
in its present over-simplified form, it does provide us with
qualitative insights into the local structure of the gauged
$\si$-model, and that with direct reference to the algebroidal
structure discovered earlier on the set of infinitesimal symmetries
under gauging. We shall take up this newly established intuition in
the next section and combine it, along the lines of
Refs.\,\cite[Sec.\,2]{Runkel:2008gr} and
\cite[Sec.\,3]{Frohlich:2009gb}, with the concept of a duality
defect of \Rxcite{Sec.\,3}{Suszek:2011hg} with view to obtaining a
world-sheet definition of a field configuration of the gauged
$\si$-model (in the presence of the full-fledged
differential-geometric structure on the world-sheet and on the
target space) locally twisted by the symmetry group under gauging.
Remarkably enough, as a byproduct of our analysis, we find a novel
field-theoretic interpretation of the large gauge anomaly. But even
prior to such refinement, the theorem clearly demonstrates, on
purely geometric grounds\footnote{In
Refs.\,\cite{Gawedzki:2010rn,Gawedzki:2012fu}, the incorporation of
topologically non-trivial gauge fields into a unified framework was
motivated by purely field-theoretic arguments relying on
inconclusive (in this respect) analyses of
Refs.\,\cite{Schellekens:1989am,Schellekens:1990xy,Hori:1994nc,Fuchs:1995tq}.},
the necessity of having gauge fields of \emph{arbitrary} topology
coupled to the string background of the parent $\si$-model (with the
target space $\,\xcF$) for a \emph{complete} formulation of the
gauged resp.\ coset $\si$-model, taking into account the existence
of the $\txG_\si$-twisted sector.

\subsection{Topological gauge-symmetry defect networks and
$\txG_\si$-equivariance}\label{sub:world-coset}

Our hitherto careful investigation of the algebraic aspects of the
passage from global symmetries of the multi-phase $\si$-model to
their local counterparts has brought to the fore the r\^ole of the
action groupoid $\,\txG_\si\lx\xcF\,$ as the structure underlying
symmetries of the gauged $\si$-model. Furthermore, it has led to the
emergence of a suggestive local geometric picture of the latter
field theory. In the remainder of this section, we want to formalise
these observations in a manner consistent with the extra structure
present on the world-sheet (the gauge field) and over the target
space (the string background). The findings of the previous section
suggest two directions in which we can develop the discussion of the
gauged $\si$-model, to wit,
\bit
\item a local implementation of the gauge symmetry through patchwise
smooth network-field configurations with $C^\infty(\Si,\txG_\si
)$-jump discontinuities localised along (topological-)defect lines,
forming an arbitrarily dense mesh as in
Refs.\,\cite[Sec.\,2]{Runkel:2008gr} and
\cite[Sec.\,3]{Frohlich:2009gb};
\item a systematic reconstruction of network-field configurations in
the background of a \emph{topologically non-trivial} gauge field
through local trivialisation of the gauge bundle and subsequent
application of the clutching construction using local transition
maps.
\eit
Technically speaking, the two constructions are intimately related:
Both entail splitting $\,\Si\,$ into a collection of patches
$\,\Si_i\,$ through the embedding of an oriented graph $\,\G\,$ (a
defect graph resp.\ a graph defining the triangulation of $\,\Si\,$
subordinate to the open cover used in the local trivialisation) and
pulling back data of local trivialisations of the geometric
structure over the world-sheet (\textit{i.e.}\ the gauge field
coupled to the string background) to the patches, and data of local
morphisms relating the trivialisations to the edges and vertices of
the graph. Both impose consistency conditions on the data pulled
back to the multi-valent vertices of the graph (associativity
\textit{etc.}). Finally, both require (local) extendibility of the
local data (to ensure topologicality, a distinctive feature of a
duality defect network, resp.\ to ensure independence of the
construction of the arbitrary choices made in the trivialisation
procedure). The sole formal difference between the two constructions
consists in the choice of the gluing maps
$\,\chi:\Egt_\G\sqcup\Vgt_\G\to\txG_\si\,$ (\textit{cf.}\ Definition
\vref{def:net-field}I.2.6), but that is readily accounted for: In
the former case, one uses restrictions of globally defined (smooth)
maps $\,\chi\in C^\infty(\Si,\txG_\si)$;\ in the latter case, the
construction of topologically non-trivial gauge bundles necessitates
the use of locally smooth maps $\,\chi_{ij}\in C^\infty(\Si_{ij},
\txG_\si)\,$ without global extensions, \textit{cf.}\ the discussion
closing the previous section. In the light of the structural
affinity between the two constructions, and with view to keeping the
discourse less cluttered with technical notation, we choose to
present in detail only the first construction. Incidentally, this
will enable us to give an explicit realisation of the abstract ideas
outlined in Remark \vref{rem:duality-scheme}I.5.6, and -- in so
doing -- will provide us with a new interpretation of the large
gauge anomaly. Upon completing the presentation, we comment briefly
on the application of the methods developed along the way in the
second construction.
\medskip

The embedding in the world-sheet of a defect network implementing
the action of the gauge group on fields of the gauged
\emph{multi-phase} $\si$-model divides naturally into three stages.
The first stage is restricted to a single phase of the theory. It
consists in defining the $C^\infty(\Si,\txG_\si)$-jump bi-brane and
ensuring that the associated (component) $C^\infty(\Si,\txG_\si
)$-jump defects are topological and can be fused in an associative
manner, leading to the emergence of a topological $C^\infty(\Si,
\txG_\si)$-jump defect network. In the second stage, one ensures
compatibility of the former definition with the structure of a
conformal defect $\,\xcD_\txA\,$ between phases of the gauged
$\si$-model (assuming $\,\xcD_\txA\,$ to be $\txG_\si$-symmetric) by
defining a junction between $\,\xcD_\txA\,$ and an arbitrary
$C^\infty(\Si,\txG_\si )$-jump defect, and by requiring subsequently
that the presence of $\,\xcD_\txA\,$ do not destroy the crucial
feature of topologicality of the $C^\infty(\Si,\txG_\si)$-jump
defect network. The third and final stage of the construction boils
down to securing topologicality in the presence of
self-intersections of $\,\xcD_\txA$.\ We shall now go step by step
through the successive stages.

\begin{figure}[hbt]~\\[5pt]

$$
 \raisebox{-50pt}{\begin{picture}(50,50)
  \put(-79,-4){\scalebox{0.25}{\includegraphics{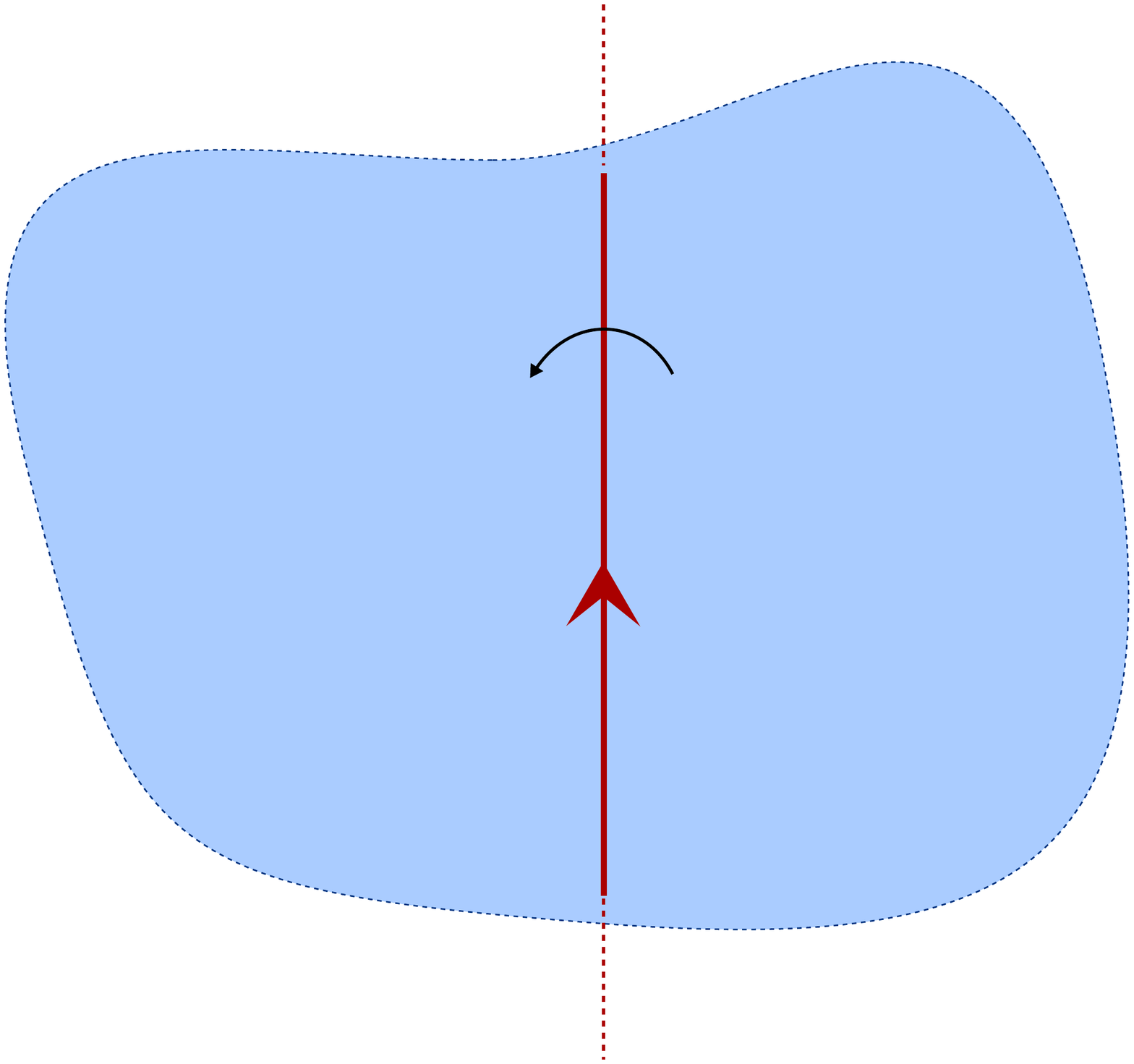}}}
  \end{picture}
  \put(0,0){
     \setlength{\unitlength}{.60pt}\put(-28,-16){
     \put(-33,30)     { $\cB_\chi$  }
     \put(-10,240)     { $\xcD_\chi$  }
     \put(-125,155)     { $(X_{|1},\txA_{|1})$ }
     \put(-130,135)     { $=(\chi.X_{|2},\ups{\chi}\txA_{|2})$ }
     \put(10,145)      { $(X_{|2},\txA_{|2})$ }
     \put(-163,200)   { $U_1$   }
     \put(93,200)    { $U_2$   }
     \put(0,190)    { $L_\chi$ }
            }\setlength{\unitlength}{1pt}}}
$$

\caption{The $C^\infty(\Si,\txG_\si)$-jump defect $\,\xcD_\chi\,$
associated with the mapping $\,\chi\in C^\infty(\Si,\txG_\si)\,$ and
carrying the data of the $C^\infty(\Si,\txG_\si)$-jump bi-brane
$\,\cB_\chi$.} \label{fig:gauge-def}
\end{figure}

Let us start by taking into consideration a single phase of the
gauged $\si$-model. The point of departure in our discussion is the
following
\bedef\label{def:gauge-jump-defect}
Adopt the notation of Definitions \vref{def:net-field}I.2.6 and
\ref{def:gauged-sigmod}, and of Proposition
\ref{prop:large-gauge-tt-sigmod}. Given an arbitrary map $\,\chi\in
C^\infty(\Si,\txG_\si)$,\ the associated \textbf{$C^\infty(\Si,
\txG_\si)$-jump defect $\,\xcD_\chi\,$} for the gauged $\si$-model
of \Reqref{eq:2d-gauge-sigma-def} is the one-dimensional locus
$\,\ell\subset\Si\,$ (of the topology of a line segment or that of a
circle) of discontinuity of the lagrangean fields of the theory of
the form
\qq\label{eq:LG-jump}
X_{|1}(p)=\chi(p).X_{|2}(p)\,,\qquad\qquad\txA_{|1}(p)=\ups{\chi}
\txA_{|2}(p)\,,\qquad p\in\ell\,,
\qqq
\textit{cf.}\ Figure \ref{fig:gauge-def}, carrying the data of the
distinguished \textbf{(component) $C^\infty(\Si,\txG_\si)$-jump
bi-brane}, that is the $(L_\chi^*\cG_{\ups{\chi}\txA},\cG_\txA
)$-bi-brane
\qq\nn
\cB_\chi:=\left(\{\chi\}\x\Si\x M\equiv\Si\x M,L_\chi,\id_{\Si\x M},
\Upsilon_\chi,0\right)\,,
\qqq
written in terms of the gerbe 1-isomorphism
\qq\nn
\Upsilon_\chi:=(\chi\x\id_M)^*\Upsilon\ :\ L_\chi^*\cG_{\ups{\chi}
\txA}\xrightarrow{\ \cong\ }\cG_\txA\,,
\qqq
with
\qq\nn
L_\chi\ :\ \Si\x\xcF\to\Si\x\xcF\ :\ (\si,m)\mapsto\left(\si,\chi(
\si).m\right)\,.
\qqq
The pair $\,\left(\xi_{|2},\txA_{|2}\right)$,\ with $\,\xi_{|2}
\equiv(\id_\Si,X_{|2})\,$ are taken as the restriction of the field
configuration of the (gauged) $\si$-model in the presence of the
$C^\infty(\Si,\txG_\si)$-jump defect $\,\xcD_\chi\,$ to the defect
line $\,\ell$.\ In keeping with the original notation of
\Rxcite{Sec.\,2}{Runkel:2008gr}, they are to be denoted as
$\,(X,\txA)\vert_\ell$.

Taking a disjoint union over the gauge group of component
$C^\infty(\Si,\txG_\si)$-jump bi-branes associated with various maps
$\,\chi\in C^\infty(\Si,\txG_\si)$,\ we obtain the \textbf{(total)
$C^\infty(\Si,\txG_\si)$-jump bi-brane}
\qq\nn
\cB_{C^\infty(\Si,\txG_\si)}=\bigsqcup_{\chi\in C^\infty(\Si,
\txG_\si)}\,\cB_\chi\,.
\qqq
\exdef

The physical relevance of the above definition stems from the
following
\berop
The $C^\infty(\Si,\txG_\si)$-jump defect $\,\xcD_\chi\,$ of
Definition \ref{def:gauge-jump-defect} is conformal in the sense of
\Rxcite{Sec.\,2.9}{Runkel:2008gr}.
\eerop
\beroof
As shown in \Rxcite{Sec.\,2.9}{Runkel:2008gr}, conformality of a
world-sheet defect is implied by the Defect Gluing Condition
\vref{eq:DGC}(I.2.8) being satisfied by the corresponding
circle-field configuration. Hence, it suffices to verify the
appropriate DGC for $\,\xcD_\chi$,\ obtained as the term in the
variation of the action functional \eqref{eq:2d-gauge-sigma-def}
localised at the defect line. In order to simplify matters
further\footnote{A lengthy but otherwise completely straightforward
analysis free of such simplifying assumptions can readily be carried
out along the lines of \Rxcite{App.\,A.2}{Runkel:2008gr}.}, take a
cohomologically trivial target
\qq\label{eq:cohom-triv-target}
\cM=(M,\txg,\cG)\,,\qquad\cG:=I_\txB\,,
\qqq
endowed with a cohomologically trivial $\txG_\si$-equivariant
structure (\textit{cf.}\ \Rxcite{Def.\,8.1}{Gawedzki:2012fu})
\qq\label{eq:cohom-triv-Gequiv-str-target}
(\Upsilon,\gamma):=(J_E,f)\,,\qquad(E,f)\in\Om^1(\txG_\si \x
M,\bR)\x C^\infty\left(\txG_\si^2\x M,\bR\right)\,.
\qqq
Use the adapted world-sheet coordinates $\,(\si^1,\si^2)\equiv(t,
\varphi)\,$ in which the defect line $\,\ell\,$ is the locus of the
equation $\,t=0\,$ and the right-handed basis of $\,\sfT_p\Si$
considered in Definition \vref{def:net-field}I.2.6 is given by
$\,(\p_t,\p_\varphi)$.\ Given a variation $\,X^\mu\mapsto
X^\mu+\xcV^\mu,\ \xcV\in\G(\sfT M)\,$ the DGC reads
\qq\nn
{\rm DGC}(X;\txA)(\varphi)&:=&\txg_{\mu\nu}\left(\chi.X_{|2}(0,
\varphi)\right)\,\left(D_{\ups{\chi}\txA_{|2}}(\chi.X_{|2})^\mu
\right)_t(0,\varphi)\,(\ell_{\chi(0,\varphi)\,*}\xcV)^\nu\left(X_{|2}(
0,\varphi)\right)\cr\cr
&&-\txg_{\mu\nu}\left(X_{|2}(0,\varphi)\right)\,\left(D_{\txA_{|2}}
(X_{|2})^\mu\right)_t(0,\varphi)\,\xcV^\nu\left(X_{|2}(0,\varphi)
\right)\cr\cr
&&+\kappa_{A\,\mu}\left(\chi.X_{|2}(0,\varphi)\right)\,\left(
\ups{\chi}\txA_{|2}^A\right)_\varphi\,(\ell_{\chi(0,\varphi)\,*}\xcV
)^\mu\left(X_{|2}(0,\varphi)\right)\cr\cr
&&-\kappa_{A\,\mu}\left(X_{|2}(0,\varphi)\right)\,\left(\txA_{|
2}^A\right)_\varphi\,\xcV^\mu\left(X_{|2}(0,\varphi)\right)\cr\cr
&&+2\txB_{\mu\nu}\left(\chi.X_{|2}(0,\varphi)\right)\,(\ell_{\chi(
0,\varphi)\,*}\xcV)^\mu\left(X_{|2}(0,\varphi)\right)\,\p_\varphi
\left(\chi.X_{|2}\right)^\nu(0,\varphi)\cr\cr
&&-2\txB_{\mu\nu}\left(X_{|2}(0,\varphi)\right)\,\xcV^\mu\left(X_{|2}(
0,\varphi)\right)\,\p_\varphi X_{|2}^\nu(0,\varphi)\cr\cr
&&+\xi_{|2\,*}\p_\varphi\con \xcV\left(X_{|2}(0,\varphi)\right)\con\sfd
E_\chi\left((0,\varphi),X_{|2}(0,\varphi)\right)
\qqq
with
\qq\nn
E_\chi:=(\chi\x\id_M)^*E\,.
\qqq
It is our task to show that the DGC vanishes identically. Its first
two terms cancel out due to the assumed $\txG_\si$-invariance of the
target-space metric (recall the tensorial transformation law for the
covariant derivative, \textit{cf.}\
\Rxcite{Eq.\,(3.13)}{Gawedzki:2012fu}). Taking into account the
$\txG_\si$-equivariance of $\,\kappa$,\ \textit{cf.}\
\Reqref{eq:gauge-constr}, in the integrated form
\qq\nn
\Mup\ell_{\chi(\si)}^*\kappa(X)=\kappa\left(\Ad_{\chi(\si)^{-1}}X
\right)\,,\qquad X\in\ggt_\si\,,
\qqq
in conjunction with the defining formula
\qq\nn
\sfd E_\chi(\si,m)=\txB(m)-\ee^{-\ovl{\chi^*\theta_L(\si)}}\cdot
\Mup\ell_{\chi(\si)}^*\txB(m)+\rho_{\chi^*\theta_L}(\si,m)
\qqq
that uses the notation of \Rxcite{Conv.\,2.11}{Gawedzki:2012fu}, we
reduce the DGC to the form
\qq\nn
{\rm DGC}(X;\txA)(\varphi)&=&-\kappa_{A\,\mu}\left(X_{|2}(0,\varphi
)\right)\,\left(\chi^{-1}\,\p_\varphi\chi\right)^A(0,\varphi)\,
\xcV^\mu\left(X_{|2}(0,\varphi)\right)\cr\cr
&&+2\txB_{\mu\nu}\left(\chi.X_{|2}(0,\varphi)\right)\,(\ell_{\chi(
0,\varphi)\,*}\xcV)^\mu\left(X_{|2}(0,\varphi)\right)\,\p_\varphi
\left(\chi.X_{|2}\right)^\nu(0,\varphi)\cr\cr
&&-2\txB_{\mu\nu}\left(X_{|2}(0,\varphi)\right)\,\xcV^\mu\left(X_{|2}(
0,\varphi)\right)\,\p_\varphi X_{|2}^\nu(0,\varphi)\cr\cr
&&+X_{|2\,*}\p_\varphi\con \xcV\con\left(\txB-\Mup\ell_{\chi(0,\varphi
)}^* \txB\right)\left(X_{|2}(0,\varphi)\right)\cr\cr
&&-\left(\chi^{-1}\,\p_\varphi\chi\right)^A(0,\varphi)\,\left(\xcV\con
\Mup\xcK_A\con\Mup\ell_{\chi(0,\varphi)}^*\txB-\xcV\con\kappa_A\right)
\left(X_{|2}(0,\varphi)\right)
\qqq
It is now evident that the DGC vanishes identically, \textit{cf.}\
\Rxcite{Eq.\,(2.36)}{Gawedzki:2012fu}.

We conclude that the existence of $\,\Upsilon\,$ ensures the
existence of an \emph{element-wise} realisation of $\,C^\infty(\Si,
\txG_\si)\,$ on extended targets through equivalences
(\textit{i.e.}\ it maps a target to a physically equivalent
one).\eroof

The study, initiated in \Rcite{Runkel:2008gr} and carried out at
length in \Rcite{Suszek:2011hg}, of the correspondence between
conformal defects and dualities of the $\si$-model (the latter being
understood in the sense of Definition \vref{def:pqsymm}I.4.7) has
singled out the topological defects of Definition
\vref{def:def-top}I.4.3 as natural candidates for world-sheet
representatives of the said dualities. This conforms with
predictions of various alternative approaches to the CFT of the
$\si$-model, including those of the categorial quantisation scheme
reported in Refs.\,\cite{Frohlich:2004ef,Frohlich:2006ch}. We are
thus led to enquire as to the topologicality of the $C^\infty(\Si,
\txG_\si)$-jump defect. The answer to this question is given in the
following
\berop
The circle-field configuration (understood in the sense of
\Rxcite{Sec.\,2.4}{Runkel:2008gr}) for the $C^\infty(\Si,\txG_\si
)$-jump defect $\,\xcD_\chi\,$ of Definition
\ref{def:gauge-jump-defect} is extendible, and so the defect is
topological in the sense of \Rxcite{Sec.\,2.9}{Runkel:2008gr}.
\eerop
\beroof
An extension
\qq\nn
\widehat\xi:=(\id_U,\widehat X)\ :\ U\to U\x M
\qqq
of a circle-field configuration $\,(\xi\,\vert\,\G)\,$ on a
world-sheet $\,\Si\,$ with an embedded (oriented) circular defect
line $\,\G\cong\bS^1\,$ that carries the data of $\,\xcD_\chi\,$ to
a tubular neighbourhood $\,U\,$ of $\,\G\,$ within $\,\Si\,$ takes
the form
\qq\nn
\widehat X(\si)=\left\{ \barr{cl} L_{\Inv\circ\chi}\circ X(\si) &
\tx{if } \si\in U_1 \cr\cr X(\si) & \tx{if } \si\in U_2 \earr
\right.\,.
\qqq
The extension of the original configuration $\,\txA\,$ assigned to
$\,\xcD_\chi\,$ as in Definition \ref{def:gauge-jump-defect} reads
\qq\nn
\widehat\txA(\si)=\left\{ \barr{cl} \ups{\Inv\circ\chi}\txA(\si)
& \tx{if } \si\in U_1 \cr\cr \txA(\si) & \tx{if } \si\in U_2 \earr
\right.\,.
\qqq
Adducing the very same arguments as in the proof of the vanishing of
$\,{\rm DGC}(X;\txA)$,\ we convince ourselves that the above
extension of the defect field configuration $\,(X,\txA)
\vert_\ell\,$ satisfies Eq.\,(2.113) of \Rcite{Runkel:2008gr}, from
which we infer that the $C^\infty(\Si,\txG_\si)$-jump defect is
extendible, and hence -- by the arguments of
\Rxcite{Sec.\,2.9}{Runkel:2008gr} -- (off-shell) topological.\eroof

It is to be stressed that the existence of a $C^\infty(\Si,
\txG_\si)$-jump defect is a straightforward consequence of the
assumed $C^\infty(\Si,\txG_\si)$-invariance of the gauged
$\si$-model in the presence of the topologically trivial gauge
field, by which we mean that it does not call for any additional
structure beyond the one required for the $C^\infty(\Si,\txG_\si
)$-invariance, \textit{cf.}\ Proposition
\ref{prop:large-gauge-tt-sigmod}. On the other hand, from the
arguments presented in the Introduction to \Rcite{Runkel:2008gr}, we
infer that the presence of defects in a self-consistent quantum CFT
unavoidably leads to the emergence of defect junctions at which the
convergent defects undergo fusion. As seen from the world-sheet
perspective, the latter is to be understood as a relation between
the limiting values attained by the defect embedding maps together
with a 2-isomorphism trivialising a (horizontal) composition of the
pullbacks of the defect 1-isomorphisms to the inter-bi-brane
world-volume in which the defect junction is embedded, both
following the scheme detailed in \Rxcite{Sec.\,2.5}{Runkel:2008gr}.
Thus, internal consistency of the field theory in hand is contingent
upon the existence of the above-mentioned fusion 2-isomorphism.

\begin{figure}[hbt]~\\[5pt]

$$
 \raisebox{-50pt}{\begin{picture}(50,50)
  \put(-79,-4){\scalebox{0.25}{\includegraphics{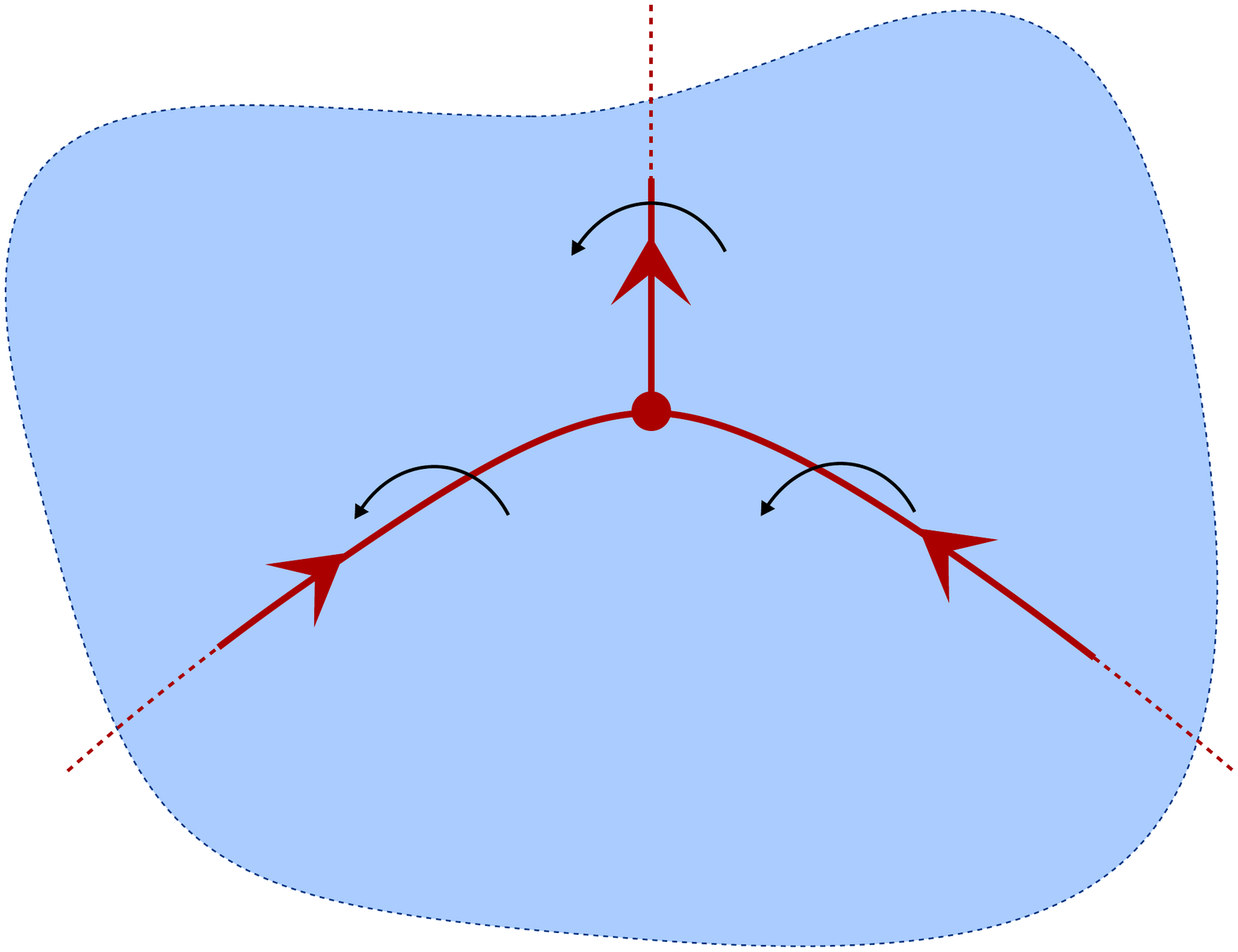}}}
  \end{picture}
  \put(0,0){
     \setlength{\unitlength}{.60pt}\put(-28,-16){
     \put(-10,245)     { $\xcD_{\chi_1\cdot\chi_2}$  }
     \put(-200,180)    { $\left((\chi_1\cdot\chi_2).X,\ups{\chi_1\cdot\chi_2}\txA\right)$ }
     \put(30,180)      { $(X,\txA)$ }
     \put(-10,210)    { $L_{\chi_1\cdot\chi_2}$ }
     \put(-95,152)    { $L_{\chi_1}$ }
     \put(38,152)    { $L_{\chi_2}$ }
     \put(-55,100)    { $(\chi_2.X,\ups{\chi_2}\txA)$ }
     \put(-150,70)     { $\xcD_{\chi_1}$  }
     \put(100,70)     { $\xcD_{\chi_2}$  }
     \put(-42,140)    { $\xcJ_{\chi_1,\chi_2}$ }
            }\setlength{\unitlength}{1pt}}}
$$

\caption{A trivalent junction $\,\xcJ_{\chi_1,\chi_2}$ of the
$C^\infty(\Si,\txG_\si)$-jump defects: the two incoming ones,
$\,\xcD_{\chi_1}\,$ and $\,\xcD_{\chi_2}$,\ and the outgoing product
defect $\,\xcD_{\chi_1\cdot\chi_2}$.} \label{fig:gauge-def-fus}
\end{figure}

While there is no \textit{a priori} relation between the
inter-bi-brane world-volume and the components of the bi-brane
world-volume into which the convergent defect lines are mapped, or
between world-volumes of inter-bi-branes corresponding to junctions
of different valence, the study of the structure of inter-bi-branes
in specific situations in which the relevant defects implement the
action of a symmetry group of the $\si$-model (such as,
\textit{e.g.}, the $Z(\txG)$-jump defects of the WZW model dealt
with in \Rcite{Runkel:2008gr}, or the more general maximally
symmetric defects of the same model analysed in
Refs.\,\cite{Runkel:2009sp,Runkel:2010} and
\cite[Sec.\,5]{Gawedzki:2012fu}) indicates that a distinguished form
of a string background is favoured in such circumstances, to wit, a
string background with induction. This concept was introduced in
\Rxcite{Sec.\,2.8}{Runkel:2008gr} and further elaborated in
\Rxcite{Rem.\,5.6}{Suszek:2011hg}. Its basis is the reconstruction
of an elementary (trivalent) inter-bi-brane that we give in
\bedef\label{def:gauge-jump-defect-junct}
Adopt the notation of Definitions \vref{def:net-field}I.2.6 and
\ref{def:gauge-jump-defect}, and of Propositions
\ref{prop:large-gauge-tt-sigmod} and \ref{def:Gequiv-bgrnd}. Given
arbitrary maps $\,\chi_1,\chi_2\in C^\infty(\Si,\txG_\si)$,\ the
associated \textbf{elementary $C^\infty(\Si,\txG_\si)$-jump defect
junction $\,\xcJ_{\chi_1,\chi_2}\,$} for the gauged $\si$-model of
\Reqref{eq:2d-gauge-sigma-def} is the point $\,\jmath_{(3)}\subset
\Si\,$ of convergence of a triple of $C^\infty(\Si,\txG_\si)$-jump
defects $\,\xcD_{\chi_1},\xcD_{\chi_2}\,$ and $\,\xcD_{\chi_1\cdot
\chi_2}\,$ of the type depicted in Figure \ref{fig:gauge-def-fus},
carrying the data of the \textbf{(component) elementary
$C^\infty(\Si,\txG_\si )$-jump inter-bi-brane}
\qq\nn
\cJ_{\chi_1,\chi_2}:=\cB_{\chi_1}\left(\{(\chi_1,\chi_2)\}\x\Si\x M
\equiv\Si\x M;\pi_3^{1,2},\pi_3^{2,3},\pi_3^{3,1};\g_{\chi_1,\chi_2}
\right)\,,
\qqq
including the inter-bi-brane maps
\qq\nn
\pi_3^{1,2}\ &:&\ \{(\chi_1,\chi_2)\}\x\Si\x M\to\{\chi_1\}\x\Si\x M
\ :\ (\chi_1,\chi_2,\si,m)\mapsto\left(\chi_1,\si,\chi_2(\si).m
\right)\,,\cr\cr \pi_3^{2,3}\ &:&\ \{(\chi_1,\chi_2)\}\x\Si\x M\to\{
\chi_2\}\x\Si\x M\ :\ (\chi_1,\chi_2,\si,m)\mapsto(\chi_2,\si,m)\,,
\cr\cr \pi_3^{3,1}\ &:&\ \{(\chi_1,\chi_2)\}\x\Si\x M\to\{\chi_1
\cdot\chi_2\} \x\Si\x M\ :\ (\chi_1,\chi_2,\si,m)\mapsto(\chi_1\cdot
\chi_2,\si,m)
\qqq
and the 2-isomorphism
\qq\nn
\g_{\chi_1,\chi_2}:=\left((\chi_1,\chi_2)\x\id_M\right)^*\g\ :\
\left(\Upsilon_{\chi_2}\ox\Id\right)\circ L_{\chi_2}^*
\Upsilon_{\chi_1}\xLongrightarrow{\ \cong\ }\Upsilon_{\chi_1\cdot
\chi_2}\,.
\qqq
Taking a disjoint union over the gauge group of component elementary
$C^\infty(\Si,\txG_\si)$-jump inter-bi-branes associated with
various maps $\,\chi_1,\chi_2\in C^\infty(\Si,\txG_\si)$,\ we obtain
the \textbf{(total) elementary $C^\infty(\Si,\txG_\si)$-jump
inter-bi-brane}
\qq\nn
\cJ_{C^\infty(\Si,\txG_\si)}^{++-}:=\bigsqcup_{\chi_1,\chi_2\in
C^\infty(\Si,\txG_\si)}\, \cJ_{\chi_1,\chi_2}\,.
\qqq
\exdef

\brem Note that the component elementary $C^\infty(\Si,\txG_\si
)$-jump inter-bi-brane can be identified, by a slight abuse of the
notation, with the product of bi-branes fibred over the target space
in terms of the bi-brane maps,
\qq\nn
\cJ_{\chi_1,\chi_2}=\cB_{\chi_1}\fibx{\id_{\Si\x M}}{L_{\chi_2}}
\cB_{\chi_2}\,.
\qqq
Under this identification, the inter-bi-brane maps become the
canonical projections $\,\pr_1,\ \pr_2\,$ and $\,\left(\txm\circ(
\pr_1\circ\pr_1,\pr_1\circ\pr_2),(\pr_2,\pr_3)\circ\pr_2\right)$,\
respectively. \erem

\begin{figure}[hbt]~\\[40pt]
$$
   \raisebox{-50pt}{\begin{picture}(150,45)
   \put(-74,-37){\scalebox{0.65}{\includegraphics{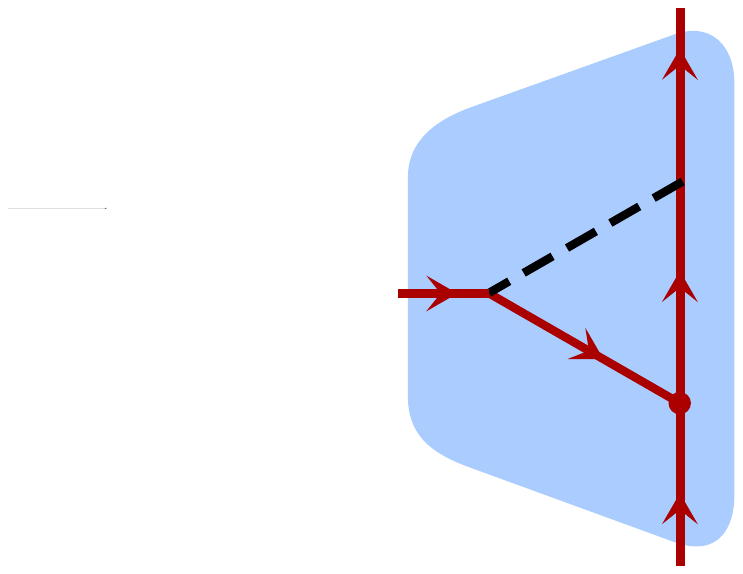}}}
   \put(0,0){
      \setlength{\unitlength}{.65pt}\put(-112,-87){
      \put(262,198)   {\scriptsize $ \upsilon_2 $}
      \put(262,134)   {\scriptsize $ \upsilon_1 $}
      \put(228,138)   {\scriptsize $ \ell_1 $}
      \put(228,194)   {\scriptsize $ \ell_2 $}
      \put(265,165)   {\scriptsize $ \ell_3 $}
      \put(235,165)   {\scriptsize $ \triangle $}
      \put(158,162)   {\scriptsize $ \xcD_{\chi_1} $}
      \put(252,253)   {\scriptsize $ \xcD_{\chi_1\cdot\chi_2} $}
      \put(252,78)   {\scriptsize $ \xcD_{\chi_2} $}
      }\setlength{\unitlength}{1pt}}
   \end{picture}}
   ~~ \longrightarrow ~~
   \raisebox{-50pt}{\begin{picture}(150,45)
   \put(-74,-37){\scalebox{0.65}{\includegraphics{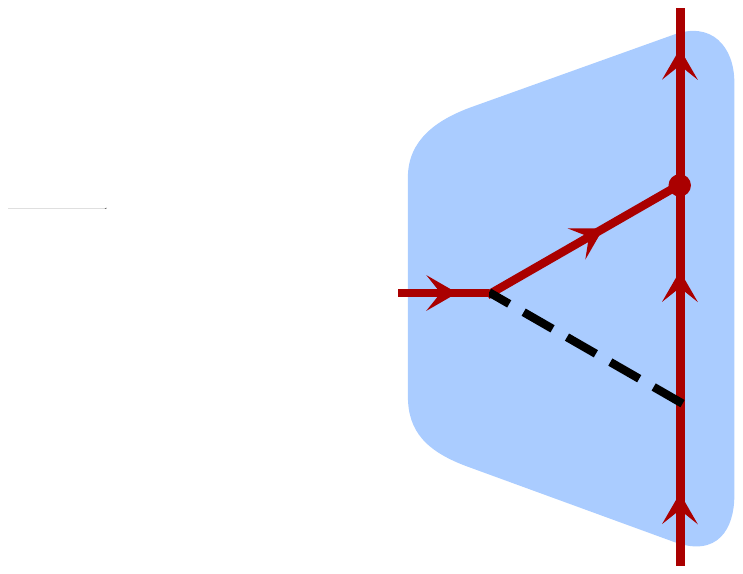}}}
   \put(0,0){
      \setlength{\unitlength}{.65pt}\put(-112,-87){
      \put(262,198)   {\scriptsize $ \upsilon_2 $}
      \put(262,134)   {\scriptsize $ \upsilon_1 $}
      \put(228,138)   {\scriptsize $ \ell_1 $}
      \put(228,194)   {\scriptsize $ \ell_2 $}
      \put(265,165)   {\scriptsize $ \ell_3 $}
      \put(235,165)   {\scriptsize $ \triangle $}
      \put(158,162)   {\scriptsize $ \xcD_{\chi_1} $}
      \put(252,253)   {\scriptsize $ \xcD_{\chi_1\cdot\chi_2} $}
      \put(252,78)   {\scriptsize $ \xcD_{\chi_2} $}
      }\setlength{\unitlength}{1pt}}
   \end{picture}}
$$

\caption{A homotopic displacement of a trivalent vertex of a
$C^\infty(\Si,\txG_\si)$-jump defect network along a defect line.}
\label{fig:homotopy-move}
\end{figure}

A distinctive feature of string backgrounds with induction is the
extendibility of the associated network-field configurations
\emph{in the presence of defect junctions}. The feature allows to
translate defect junctions along defect lines without changing the
value of the $\si$-model action functional. In the present setting,
we find
\berop
The network-field configuration (understood in the sense of
\Rxcite{Sec.\,2.6}{Runkel:2008gr}) for a graph of the $C^\infty(\Si,
\txG_\si)$-jump defects of Definition \ref{def:gauge-jump-defect}
with at most trivalent junctions, as described in Definition
\ref{def:gauge-jump-defect-junct}, is extendible, and so the defect
defined by the graph is topological in the sense of
\Rxcite{Sec.\,2.9}{Runkel:2008gr}.
\eerop
\beroof
We describe in full detail the extension of the network-field
configuration in the vicinity of the defect junction drawn on the
left-hand side of Figure \ref{fig:homotopy-move}, and study the
effect of the local homotopic deformation of the defect quiver,
using the extension, on the value of the $\si$-model action
functional, \textit{cf.}\ \Rxcite{App.\,A.3}{Runkel:2008gr}.
Extension of our considerations to generic homotopy moves of
trivalent $C^\infty(\Si,\txG_\si)$-jump defect junctions within the
world-sheet is straightforward and therefore left as an exercise to
the reader.

\noindent We define
\qq\nn
\widehat\xi_{\ell_1}:=\left(\id_\Si,\widehat X_{\ell_1}\right)\ &:&\
\triangle\to\{\chi_1\}\x\triangle\x M\ :\ \si\mapsto
\left\{\barr{cl} \left(\si,\chi_1 (\si)^{-1}.X(\si)\right) & \tx{if
} \si\not\in\ell_1 \cr\cr \left(\si,X(\si)\right) & \tx{if }
\si\in\ell_1 \earr\right.\,,\cr\cr\cr
\widehat\xi_{\upsilon_1}:=\left(\id_\Si,\widehat X_{\upsilon_1}
\right)\ &:&\ \ell_3\to\{(\chi_1,\chi_2)\}\x\ell_3\x M\ :\ \si
\mapsto\left(\si,X(\si)\right)\,.
\qqq
Assuming, as previously, cohomological triviality of the various
gerbe-theoretic structures involved, we can calculate the difference
between the values attained by the action functional on the two
network-field configurations from Figure \ref{fig:homotopy-move}, of
which the right one, $\,(\widetilde X\,\vert\,\widetilde\G)$,\
defined for $\,\widetilde\G\,$ resulting from the homotopic
displacement of the defect junction, is determined by the above
extension as
\qq\nn
\widetilde X\vert_{\Si\setminus\triangle}=X\vert_{\Si\setminus
\triangle}\,,\qquad\qquad\widetilde X\vert_{\triangle\setminus
\ell_3}=\widehat X_{\ell_1}\vert_{\triangle
\setminus\ell_3}\,,\qquad\qquad\widetilde X\vert_{\ell_3}=\widehat
X_{\upsilon_1}\,.&
\qqq
Completing the definition of the new configuration by redefining the
gauge field in an obvious manner (it is understood that the limiting
values attained by the gauge field on either side of a defect line
are determined by the field's smooth functional dependence on the
point in the bulk, as specified below),
\qq\nn
\widetilde\txA\vert_{\Si\setminus\triangle}=\txA\vert_{\Si\setminus
\triangle}\,,\qquad\qquad\widetilde\txA\vert_{\triangle\setminus\p
\triangle}=\ups{\Inv\circ\chi_1}\txA\vert_{\triangle\setminus\p
\triangle}\,,
\qqq
we find
\qq\nn
&&S_\si[(\widetilde X\,\vert\,\widetilde\G);\widetilde\txA,\g]-
S_\si[(X\,\vert\,\G);\txA,\g]\cr\cr
&=&-\tfrac{1}{2}\,\int_\triangle\,\left[\txg\left(\widetilde X(
\cdot)\right)\left(D_{\widetilde\txA}\widetilde X\overset{\wedge}{,}
\star_\g D_{\widetilde\txA}\widetilde X\right)(\cdot)-\txg\left(
\chi_1.\widetilde X(\cdot)\right)\left(D_{\ups{\chi_1}\widetilde
\txA}(\chi_1.\widetilde X)\overset{\wedge}{,}\star_\g
D_{\ups{\chi_1}\widetilde\txA}(\chi_1.\widetilde X\right)(\cdot)
\right]\cr\cr
&&+\int_\triangle\,\left[\txB\left(\widetilde X(\cdot)\right)-\Mup
\ell^*\txB(\chi_1,\widetilde X)(\cdot)+\kappa_A \left(\widetilde
X(\cdot)\right)\wedge\widetilde\txA^A(\cdot)-\Mup
\ell^*\kappa_A(\chi_1,\widetilde X)(\cdot)\wedge
\ups{\chi_1}\widetilde\txA^A(\cdot)\right]\cr\cr
&&-\tfrac{1}{2}\,\int_\triangle\,\left[ \txc_{AB}\left(\widetilde X(
\cdot)\right)\,\left(\widetilde\txA^A\wedge\widetilde\txA^B\right)(
\cdot)-\Mup\ell^* \txc_{AB}(\chi_1,\widetilde
X)(\cdot)\,\left(\ups{\chi_1}\widetilde\txA^A\wedge\ups{\chi_1}
\widetilde\txA^B\right)(\cdot)\right]\cr\cr
&&+\int_{\ell_3}\,(E_{\chi_2}-E_{\chi_1\cdot\chi_2})\left(\cdot,
\widetilde X(\cdot)\right)+\int_{\ell_2}\,E_{\chi_1}\left(\cdot,
\widetilde X(\cdot)\right)-\int_{\ell_1}\,E_{\chi_1}\left(\cdot,
\widetilde X(\cdot)\right)\cr\cr
&&+f_{\chi_1,\chi_2}\left(\widetilde X(v_2)\right)-f_{\chi_1,
\chi_2}\left(\widetilde X(v_1)\right)\,,
\qqq
with
\qq\nn
f_{\chi_1,\chi_2}:=\left((\chi_1,\chi_2)\x\id_M\right)^*f\,,
\qqq
the latter satisfying the defining relation
\qq\label{eq:df-def}
\sfd f_{\chi_1,\chi_2}(\si,m)=E_{\chi_1\cdot\chi_2}(\si,m)-
E_{\chi_2}(\si,m)-\ee^{-\ovl{\chi_2^*\theta_L(\si)}}\cdot\left(
\id_\Si\x\Mup\ell_{\chi_2(\si)}\right)^*E_{\chi_1}(\si,m) \,.
\qqq
Reasoning as in the discussion of the DGC for the
$C^\infty(\Si,\txG_\si)$-jump defect, we reduce the above expression
to the form
\qq\nn
&&S_\si[(\widetilde X\,\vert\,\widetilde\G);\widetilde\txA,\g]-
S_\si[(X\,\vert\,\G);\txA,\g]\cr\cr
&=&\int_\triangle\,\left[\txB\left(\widetilde X(\cdot)\right)-\Mup
\ell^*\txB(\chi_1,\widetilde X)(\cdot)+\rho_{\chi_1^*\theta_L}\left(
\cdot,\widetilde X(\cdot)\right)\right]+\int_{\ell_3}\,(E_{\chi_2}-
E_{\chi_1\cdot\chi_2})\left(\cdot,\widetilde X(\cdot)\right)\cr\cr
&&+\int_{\ell_2}\,E_{\chi_1}\left(\cdot,\widetilde X(\cdot)\right)-
\int_{\ell_1}\,E_{\chi_1}\left(\cdot,\widetilde X(\cdot)\right)+
f_{\chi_1,\chi_2}\left(\widetilde X(v_2)\right)-f_{\chi_1,\chi_2}
\left(\widetilde X(v_1)\right)\cr\cr
&=&\int_\triangle\,\sfd E_{\chi_1}\left(\cdot,\widetilde X(\cdot)
\right)+\int_{\ell_3}\,(E_{\chi_2}-E_{\chi_1\cdot\chi_2})\left(
\cdot,\widetilde X(\cdot)\right)+\int_{\ell_2}\,E_{\chi_1}\left(
\cdot,\widetilde X(\cdot)\right)-\int_{\ell_1}\,E_{\chi_1}\left(
\cdot,\widetilde X( \cdot)\right)\cr\cr
&&+f_{\chi_1,\chi_2}\left(\widetilde X(v_2)\right)-f_{\chi_1,
\chi_2}\left(\widetilde X(v_1)\right)\cr\cr
&=&\int_{\ell_3}\,\left[(E_{\chi_2}-E_{\chi_1\cdot\chi_2})\left(
\cdot,\widetilde X(\cdot)\right)+\left(\id_\Si\x\Mup\ell\right)^*
E_{\chi_1}\left(\cdot,\chi_2(\cdot),\widetilde X(\cdot)\right)
\right]+f_{\chi_1,\chi_2}\left(\widetilde X(v_2)\right)-f_{\chi_1,
\chi_2}\left(\widetilde X(v_1)\right)\,.
\qqq
Taking \Reqref{eq:df-def} into account, we finally arrive at the
desired equality
\qq\nn
S_\si[(\widetilde X\,\vert\,\widetilde\G);\widetilde\txA,\g]=S_\si[
(X\,\vert\,\G);\txA,\g]\,.
\qqq

We conclude that the existence of $\,\g\,$ ensures the existence of
a \emph{homomorphic} realisation of $\,C^\infty(\Si,\txG_\si)\,$ on
extended targets through equivalences. \eroof

\begin{figure}[hbt]~\\[5pt]

\qq\nonumber
\Si_L \hspace{4cm} & \hspace{3cm} \Si_{L|R} & \hspace{4.2cm}
\Si_R\cr
   \raisebox{-40pt}{\begin{picture}(85,80)
   \put(-10,-3){\scalebox{0.55}{\includegraphics{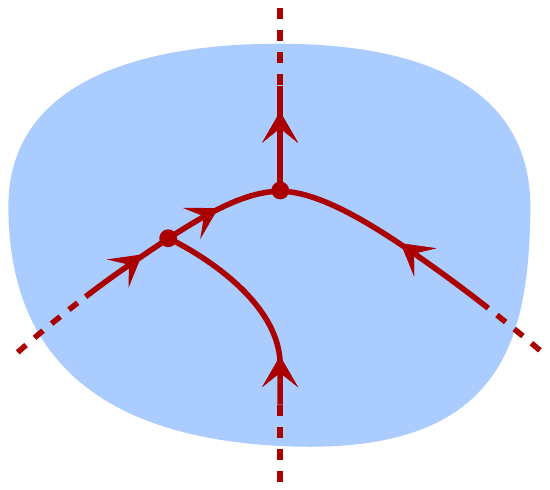}}}
   \put(0,0){
      \setlength{\unitlength}{.55pt}\put(-18,-4){
      \put(85,95)     {\scriptsize $ \upsilon $}
      \put(80,75)     {\scriptsize $ \vep_L $}
      \put(5,33)      {\scriptsize $ \xcD_{\chi_1} $}
      \put(85,-10)    {\scriptsize $ \xcD_{\chi_2} $}
      \put(165,33)    {\scriptsize $ \xcD_{\chi_3} $}
      \put(70,152)    {\scriptsize $ \xcD_{\chi_1\cdot\chi_2\cdot\chi_3} $}
      \put(36,97)     {\scriptsize $ \xcD_{\chi_1\cdot\chi_2} $}
      \put(102,98)    {\scriptsize $ \g_{\chi_1\cdot\chi_2,\chi_3} $}
      \put(-5,80)     {\scriptsize $ \chi_3.\g_{\chi_1,\chi_2} $}
      }\setlength{\unitlength}{1pt}}
   \end{picture}}
   ~~~~~~ \xrightarrow{~~\vep_L\rightarrow 0~~}
   &\raisebox{-40pt}{\begin{picture}(85,80)
   \put(-10,-3){\scalebox{0.55}{\includegraphics{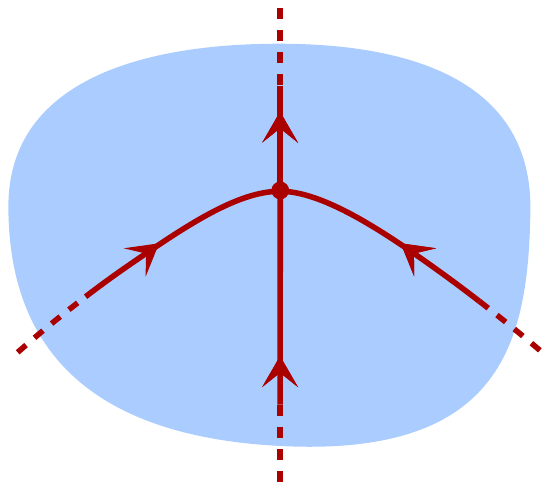}}}
   \put(0,0){
      \setlength{\unitlength}{.55pt}\put(-18,-4){
      \put(102,95)    {\scriptsize $ \upsilon $}
      \put(5,33)      {\scriptsize $ \xcD_{\chi_1} $}
      \put(85,-10)    {\scriptsize $ \xcD_{\chi_2} $}
      \put(165,33)    {\scriptsize $ \xcD_{\chi_3} $}
      \put(70,152)    {\scriptsize $ \xcD_{\chi_1\cdot\chi_2\cdot\chi_3} $}
      \put(9,99)     {\scriptsize $ \Mup u^3_{\chi_1,\chi_2,\chi_3;\txA} $}
      }\setlength{\unitlength}{1pt}}
   \end{picture}}&
    \xleftarrow{~~\vep_R\rightarrow 0~~} ~~~~~~
   \raisebox{-40pt}{\begin{picture}(85,80)
   \put(-10,-3){\scalebox{0.55}{\includegraphics{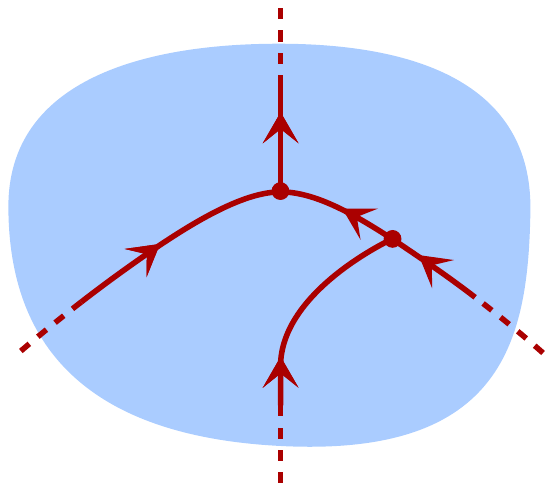}}}
   \put(0,0){
      \setlength{\unitlength}{.55pt}\put(-18,-4){
      \put(102,95)    {\scriptsize $ \upsilon $}
      \put(100,75)    {\scriptsize $ \vep_R $}
      \put(5,33)      {\scriptsize $ \xcD_{\chi_1} $}
      \put(85,-10)    {\scriptsize $ \xcD_{\chi_2} $}
      \put(165,33)    {\scriptsize $ \xcD_{\chi_3} $}
      \put(70,152)    {\scriptsize $ \xcD_{\chi_1\cdot\chi_2\cdot\chi_3} $}
      \put(117,97)    {\scriptsize $ \xcD_{\chi_2\cdot\chi_3} $}
      \put(35,98)     {\scriptsize $ \g_{\chi_1,\chi_2\cdot\chi_3} $}
      \put(137,80)    {\scriptsize $ \g_{\chi_2,\chi_3} $}
      }\setlength{\unitlength}{1pt}}
   \end{picture}}
\qqq

\caption{A four-valent defect vertex of a
$C^\infty(\Si,\txG_\si)$-jump defect network in $\,\Si_{L|R}\,$
obtained as a result of collapsing a pair of three-valent vertices
in two inequivalent ways, whereby the two 2-morphisms
$\,\varphi^L\,$ and $\,\varphi^R\,$ are induced at the vertex. These
differ by the anomaly 3-cocycle (associator) $\,\Mup
u^3_{\chi_1,\chi_2,\chi_2;\txA}$.} \label{fig:associator}
\end{figure}

Prepared by the foregoing considerations, we now come to discuss the
main point of the induction scheme. The latter is founded on the
premise that defect junctions of valence greater than 3 can be fixed
(up to an irremovable ambiguity quantified in the discussion
surrounding Eqs.\,(2.77)-(2.79) in
\Rxcite{Sec.\,2.8}{Runkel:2008gr}) in terms of the data of the
elementary (\textit{i.e.}\ trivalent) defect junctions in a limiting
procedure applied to a defect graph obtained from the original one
by an arbitrary resolution of its vertices of valence greater that
or equal to 4 into trees of trivalent vertices. The resolution,
first proposed in \Rxcite{Sec.\,2.8}{Runkel:2008gr} and elaborated
in \Rxcite{Rem.\,5.6}{Suszek:2011hg}, is effected through
introduction of intermediate defect lines whose length vanishes in
the limit taken, \textit{cf.}\ Figure \ref{fig:associator}. As
discussed in the papers cited, internal consistency of the induction
scheme, automatically inherited by defect junctions of valence
greater than 4 from the 4-valent ones, is ensured by a cohomological
(cocycle) constraint imposed upon the 2-isomorphism data carried by
the trivalent defect junction. In the context of the gauged
$\si$-model, we establish
\berop\label{prop:simpl-ind-jump-def-net}
Adopt the notation of Definition \ref{def:sigmod-2d}. The data
carried by $C^\infty(\Si,\txG_\si)$-jump defects of Definition
\ref{def:gauge-jump-defect} and by the elementary $C^\infty(\Si,
\txG_\si)$-jump defect junctions of Definition
\ref{def:gauge-jump-defect-junct} give rise to a simplicial string
background in the sense of \Rxcite{Rem.\,5.6}{Suszek:2011hg}, with
data of (component) $C^\infty(\Si,\txG_\si)$-jump inter-bi-branes of
valence greater than 3 induced, in the manner described
\textit{ibidem}, from those of the elementary $C^\infty(\Si,\txG_\si
)$-jump inter-bi-branes of the latter definition iff the
$\txG_\si$-equivariant structure on the target $\,\cM\,$ is coherent
in the sense of Definition \ref{def:Gequiv-bgrnd}.
\eerop
\beroof
Through a simple computation carried out along the lines of
\Rxcite{Sec.\,2.9}{Runkel:2008gr}, the validity of the induction
scheme in the present setting is readily shown to be tantamount to
the triviality of the $C^\infty(\Si,\txG_\si)$-valued
\textbf{associator 3-cocycle}
\qq\nn
\Mup u^3_{\chi_1,\chi_2,\chi_3;\txA}:=\g_{\chi_1,\chi_2\cdot\chi_3}
\bullet\left[(\g_{\chi_2,\chi_3}\ox\Id)\circ\Id\right]\bullet\left(
\Id\circ L_{\chi_3}^*\g_{\chi_1,\chi_2}\right)^{-1}\bullet\g_{\chi_1
\cdot\chi_2,\chi_3}^{-1}
\qqq
whose value at the image of the four-valent defect junction under
the embedding map $\,\xi\,$ measures the difference between the
respective contributions of the two alternative sets of four-valent
inter-bi-brane data to the action functional, obtained in the two
limiting procedures shown in Figure \ref{fig:associator}.

The associator 3-cocycle is next identified with the pullback, along
the map $\,(\chi_1,\chi_2,\chi_3)\x\id_M\ :\ \Si\x M \to\txG_\si^3\x
M$,\ of the anomaly 3-cocycle\footnote{Strictly speaking, the
associator 3-cocycle is the pullback of the preimage of the said
$\uj^{\pi_0(M)}$-valued 3-cocycle on $\,\pi_0(\txG_\si)\,$ with
respect to the identification between the groups $\,H^0\left(
\txG_\si^3\x M,\uj\right)\,$ and $\,C^3\left(\pi_0(\txG_\si),
\uj^{\pi_0(M)}\right)\,$ discussed in the papers cited. We shall
keep this identification implicit in what follows.} $\,\Mup u^3\in
Z^3\left(\pi_0(\txG_\si),\uj^{\pi_0(M)}\right)\,$ of
\Rxcite{Cor.\,6.7}{Gawedzki:2010rn} and
\Rxcite{Cor.\,11.7}{Gawedzki:2012fu} whose class measures the
obstruction to the existence of a \emph{coherent}
$\txG_\si$-equivariant structure on the bulk gerbe $\,\cG$,
\qq\nn
\Mup u^3_{\chi_1,\chi_2,\chi_3;\txA}=\left((\chi_1,\chi_2,\chi_3)\x
\id_M\right)^*\Mup u^3\,.
\qqq
The vanishing of the said class is a sufficient and necessary
condition for the coherence condition of \Reqref{eq:Gerbe-1iso-coh}
to be satisfied by the 2-isomorphism $\,\g\,$ entering the
definition of the trivalent $C^\infty(\Si,\txG_\si)$-jump defect
junction.

We conclude that the existence of a full-fledged
$\txG_\si$-equivariant structure on the string background $\,\cM\,$
of the mono-phase $\si$-model ensures the existence of an
\emph{associative} realisation of $\,C^\infty(\Si,\txG_\si)\,$ on
extended targets through equivalences, associated with topological
world-sheet defects. \eroof

Prior to taking up to the multi-phase case, we pause to emphasise
the distinct status of the various elements of the construction of
the $C^\infty(\Si,\txG_\si)$-jump defect network detailed above.
Thus, while the existence of the topological $C^\infty(\Si,\txG_\si
)$-jump defect is ensured by the assumed $C^\infty(\Si,\txG_\si
)$-invariance of the gauged $\si$-model in the presence of a
topologically trivial gauge field, the existence of the elementary
$C^\infty(\Si,\txG_\si)$-jump defect junction and the validity of
the induction scheme should be regarded as conditions necessary and
sufficient for the existence of an extension of the construction of
the gauge-symmetry defect to a consistent quantum field theory with
the factorisation property, admitting a natural -- from the physical
point of view -- induction scheme for multi-valent defect junctions.
In view of the cohomological significance of the conditions, our
findings mark the first step towards an explanation of the
full-fledged large gauge anomaly in abstraction from the topological
properties of the world-sheet gauge field, and in conformity with
the infinitesimal symmetry structure of the $\si$-model captured by
the small gauge anomaly.\medskip

\begin{figure}[hbt]~\\[5pt]

$$
 \raisebox{-50pt}{\begin{picture}(50,50)
  \put(-79,-4){\scalebox{0.25}{\includegraphics{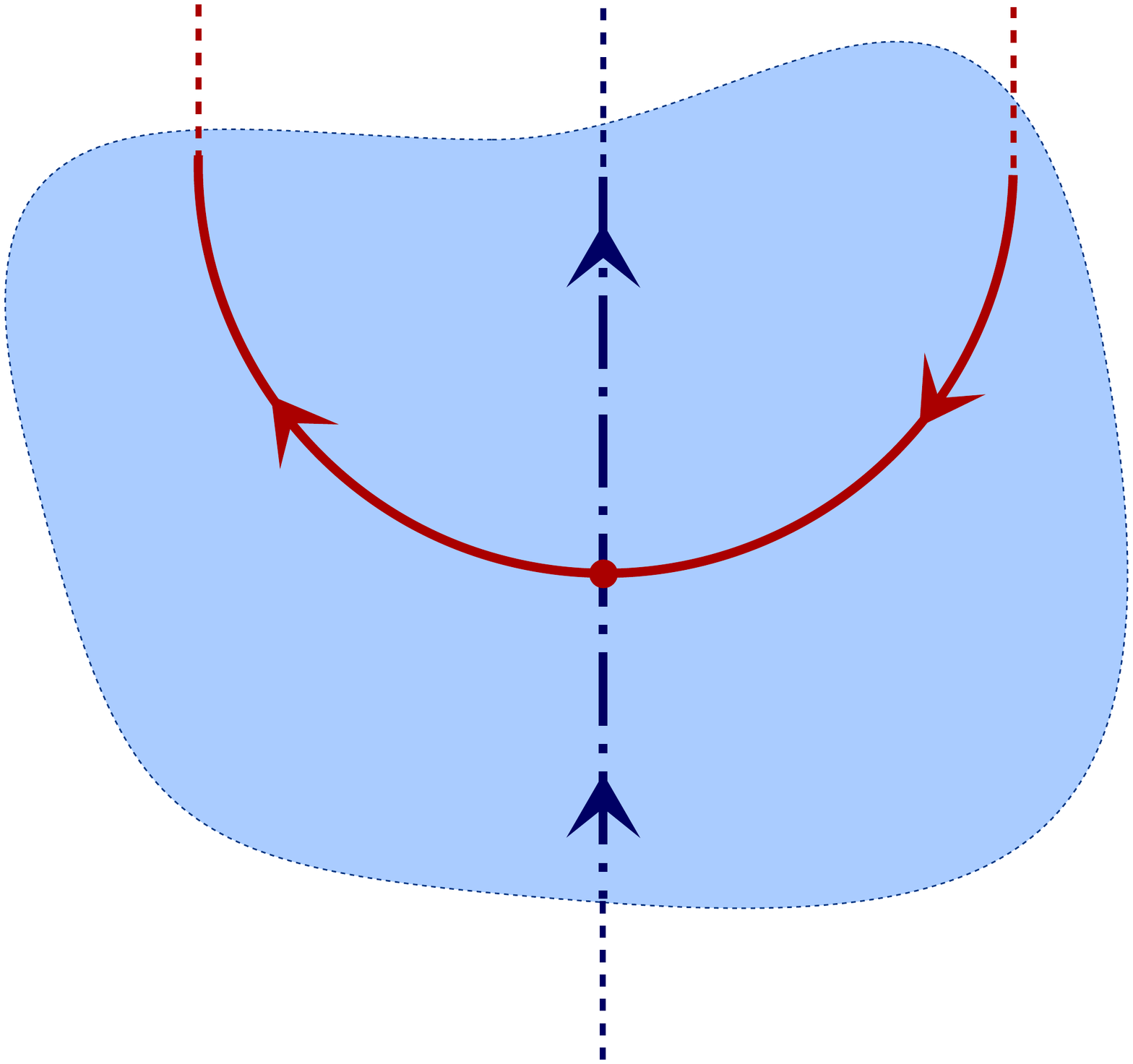}}}
  \end{picture}
  \put(0,0){
     \setlength{\unitlength}{.60pt}\put(-28,-16){
     \put(-10,245)     { $\xcD_\txA$  }
     \put(-88,185)    { $(X_{|1},\txA)$ }
     \put(-115,105)    { $(\chi.X_{|1},\ups{\chi}\txA)$ }
     \put(7,185)      { $(X_{|2},\txA)$ }
     \put(15,105)    { $(\chi.X_{|2},\ups{\chi}\txA)$ }
     \put(-125,245)     { $\xcD_\chi$  }
     \put(-35,135)    { $\Xi_\chi$ }
            }\setlength{\unitlength}{1pt}}}
$$

\caption{A four-valent crossing between a
$C^\infty(\Si,\txG_\si)$-jump defect $\,\xcD_\chi\,$ (red) and a
generic $\txG_\si$-transparent defect of the gauged $\si$-model
$\,\xcD_\txA\,$ (dark blue). The crossing carries the data of the
2-isomorphism $\,\Xi_\chi$.} \label{fig:gauge-trans-def-fus}
\end{figure}

Having completed the first stage of our construction for the gauged
multi-phase $\si$-model, we may next pass to the investigation of
conditions of coexistence of the conformal defect associated with
the bi-brane of the original (gauged) $\si$-model and the newly
introduced topological $C^\infty(\Si,\txG_\si)$-jump defect network.
The analysis that follows splits into two steps: First of all, we
set up a world-sheet description of a crossing between a $C^\infty(
\Si,\txG_\si)$-jump defect and a generic
$\txG_\si$-transparent\footnote{By $\txG_\si$-transparency we mean
preservation of the Noether charges of the global symmetry under
gauging across the domain wall. As argued earlier, it is only in the
presence of such distinguished defect lines that we can consistently
gauge the symmetry $\,\txG_\si$.} domain wall that separates phases
of the gauged $\si$-model. Secondly, we demand that the $C^\infty(
\Si,\txG_\si)$-jump defect network constructed previously remain
topological in the presence of the domain wall. Accordingly, we
begin with
\bedef\label{def:gauge-jump-trans-defect}
Adopt the notation of Definitions \vref{def:net-field}I.2.6,
\ref{def:gauged-sigmod} and \ref{def:gauge-jump-defect}, and of
Proposition \ref{prop:large-gauge-tt-sigmod}. Denote by
\qq\nn
\cB_{C^\infty(\Si,\txG_\si);\txA}:=\cB_{C^\infty(\Si,\txG_\si)}
\sqcup\cB_\txA
\qqq
the composite bi-brane of the gauged multi-phase $\si$-model with an
embedded $C^\infty(\Si,\txG_\si)$-jump defect network. Given an
arbitrary map $\,\chi\in C^\infty(\Si,\txG_\si)\,$ and a
$\txG_\si$-transparent conformal defect $\,\xcD_\txA\,$ carrying the
data of the extended bi-brane $\,\cB_\txA$,\ the associated
\textbf{elementary $C^\infty(\Si,\txG_\si)$-jump trans-defect
junction $\,\cJ_{\chi;\txA}\,$} for the gauged $\si$-model of
\Reqref{eq:2d-gauge-sigma-def} is the point $\,\jmath_{(4)}\subset
\Si\,$ of intersection of a $C^\infty(\Si,\txG_\si)$-jump defect
$\,\xcD_\chi\,$ with $\,\xcD_\txA\,$ of the type depicted in Figure
\ref{fig:gauge-trans-def-fus}, carrying the data of the
\textbf{(component) elementary $C^\infty(\Si,\txG_\si)$-jump
crossing inter-bi-brane} (the labelling of the defect lines
converging at the junction starts with the bottom half of the
vertical line in the figure, to which we assign label $(1,2)$,\ and
continues -- as usual -- in the counter-clockwise direction)
\qq\nn
\cJ_{\chi;\txA}:=(\{\chi\}\x\Si\x Q\equiv\Si\x Q;\pi_4^{1,2},
\pi_4^{2,3},\pi_4^{3,4},\pi_4^{4,1};\Xi_\chi)\,,
\qqq
with inter-bi-brane maps
\qq\nn
\pi_4^{1,2}\ &:&\ \{\chi\}\x\Si\x Q\to\Si\x Q\ :\ (\chi,\si,q)
\mapsto\left(\si,\chi(\si).q\right)\,,\cr\cr \pi_4^{2,3}\ &:&\ \{
\chi\}\x\Si\x Q\to\{\chi\}\x\Si\x M\ :\ (\chi,\si,q)\mapsto\left(
\chi,\si,\iota_2(q)\right)\,,\cr\cr \pi_4^{3,4}\ &:&\ \{\chi\}\x\Si
\x Q\to\Si\x Q\ :\ (\chi,\si,q)\mapsto(\si,q)\,,\cr\cr \pi_4^{4,1}\
&:&\ \{\chi\}\x\Si\x Q\to\{\chi\}\x\Si\x M\ :\ (\chi,\si,q)\mapsto
\left(\chi,\si,\iota_1(q)\right)
\qqq
and the 2-isomorphism
\qq\nn
\Xi_\chi:=(\chi\x\id_Q)^*\Xi\ :\ L_\chi^*\Phi_{\ups{\chi}\txA}
\xLongrightarrow{\ \cong\ }\left(\unl\iota_2^*\Upsilon_\chi^{-1}\ox
\Id\right)\circ(\Phi_\txA\ox\Id)\circ\unl \iota_1^*\Upsilon_\chi\,,
\qqq
where $\,\unl\iota_\a:=\id_\Si\x\iota_\a$.

Taking a disjoint union over the gauge group of component elementary
$C^\infty(\Si,\txG_\si)$-jump crossing inter-bi-branes associated
with various maps $\,\chi\in C^\infty(\Si,\txG_\si)$,\ we obtain the
\textbf{(total) elementary $C^\infty(\Si,\txG_\si)$-jump crossing
inter-bi-brane}
\qq\nn
\cJ_{C^\infty(\Si,\txG_\si);\txA}=\bigsqcup_{\chi\in C^\infty(\Si,
\txG_\si)}\,\cJ_{\chi;\txA}\,.
\qqq
\exdef

\brem Clearly, the consistency conditions of
Eq.\,\vref{eq:proto-simpl}(I.2.1) for the inter-bi-brane maps are
satisfied owing to the assumed $\txG_\si$-equivariance of the
bi-brane maps. \erem

\begin{figure}[hbt]~\\[40pt]
$$
   \raisebox{-50pt}{\begin{picture}(150,45)
   \put(-45,-12){\scalebox{0.2}{\includegraphics{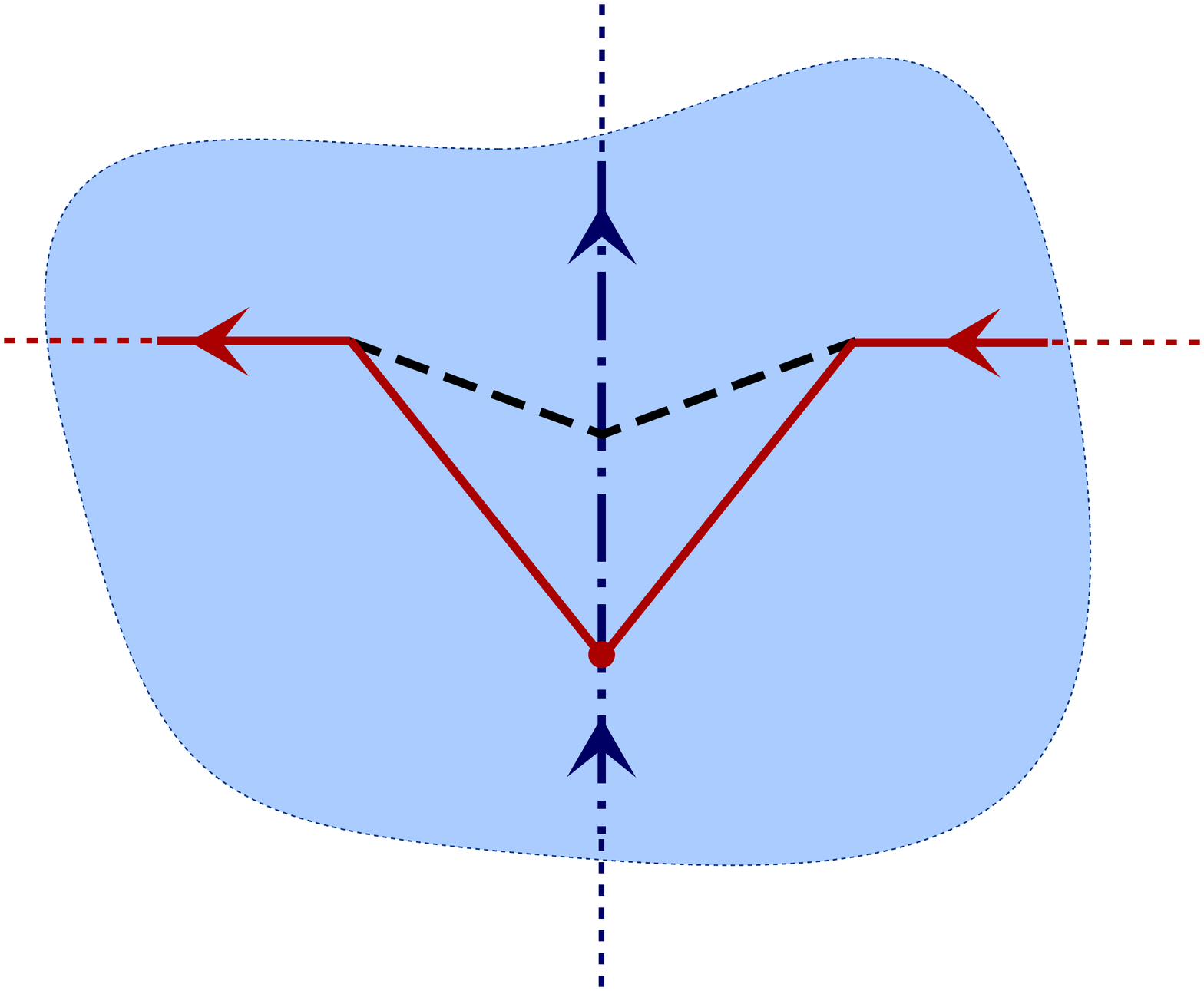}}}
   \put(0,0){
      \setlength{\unitlength}{.65pt}\put(-112,-87){
      \put(178,169)   {\scriptsize $ \upsilon_2 $}
      \put(162,134)   {\scriptsize $ \upsilon_1 $}
      \put(139,156)   {\scriptsize $ \ell_{1|1} $}
      \put(144,190)   {\scriptsize $ \ell_{2|1} $}
      \put(198,156)   {\scriptsize $ \ell_{1|2} $}
      \put(189,190)   {\scriptsize $ \ell_{2|2} $}
      \put(178,150)   {\scriptsize $ \ell $}
      \put(160,165)   {\scriptsize $ \nabla $}
      \put(60,198)   {\scriptsize $ \xcD_\chi $}
      \put(179,245)   {\scriptsize $ \xcD_\txA $}
      }\setlength{\unitlength}{1pt}}
   \end{picture}}
   ~~ \longrightarrow ~~
   \raisebox{-50pt}{\begin{picture}(150,45)
   \put(25,-12){\scalebox{0.2}{\includegraphics{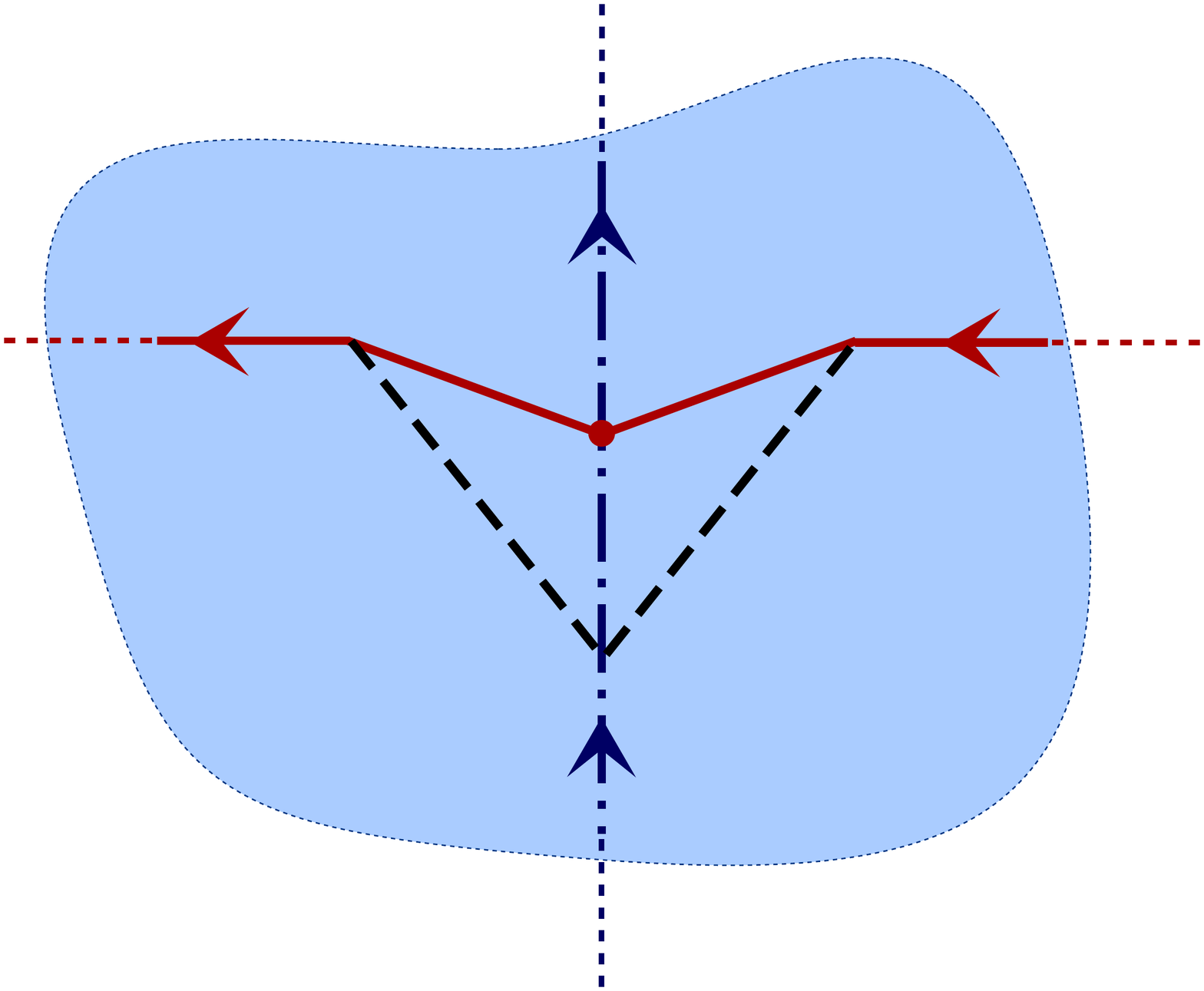}}}
   \put(0,0){
      \setlength{\unitlength}{.65pt}\put(-112,-87){
      \put(285,169)   {\scriptsize $ \upsilon_2 $}
      \put(269,134)   {\scriptsize $ \upsilon_1 $}
      \put(246,156)   {\scriptsize $ \ell_{1|1} $}
      \put(251,190)   {\scriptsize $ \ell_{2|1} $}
      \put(306,156)   {\scriptsize $ \ell_{1|2} $}
      \put(296,190)   {\scriptsize $ \ell_{2|2} $}
      \put(285,150)   {\scriptsize $ \ell $}
      \put(267,165)   {\scriptsize $ \nabla $}
      \put(167,198)   {\scriptsize $ \xcD_\chi $}
      \put(286,245)   {\scriptsize $ \xcD_\txA $}
      }\setlength{\unitlength}{1pt}}
   \end{picture}}
$$

\caption{A homotopic displacement of a four-valent crossing between
a $C^\infty(\Si,\txG_\si)$-jump defect (red) and a generic
$\txG_\si$-transparent defect of the gauged $\si$-model (dark blue)
along the defect line of the latter.}
\label{fig:homotopy-trans-move}
\end{figure}

As in the mono-phase setting, it is imperative for the
interpretation of the $C^\infty(\Si,\txG_\si)$-jump defect as a
world-sheet realisation of the local symmetry of the gauged
$\si$-model to ensure that the presence of the defect does not
affect the action of the conformal group on field configurations.
Consistently with the earlier discussion, this is amenable to direct
verification which consists in determining a suitable (local)
extension of the network-field configuration for the left-hand side
of Figure \ref{fig:homotopy-trans-move} and checking that the value
of the action functional does not change upon translating the defect
junction along the defect $\,\xcD_\txA\,$ to its new position as in
the right-hand side of the same figure\footnote{Since the defect
$\,\xcD_\txA\,$ is not, \emph{a priori}, topological, we should only
insist on invariance of the action functional under homotopic
deformations of the $C^\infty(\Si,\txG_\si)$-jump defect inducing
translations of the $C^\infty(\Si,\txG_\si)$-jump trans-defect
junction along the defect line of $\,\xcD_\txA$.}.
In this way, we establish
\berop
The network-field configuration for the defect $\,\xcD_\chi\sqcup
\xcD_\txA\,$ composed of the $C^\infty(\Si,\txG_\si)$-jump defect
$\,\xcD_\chi\,$ of Definition \ref{def:gauge-jump-defect} and of an
arbitrary $\txG_\si$-transparent conformal defect $\,\xcD_\txA\,$ is
extendible in a neighbourhood of the elementary $C^\infty(\Si,
\txG_\si)$-jump trans-defect junction $\,\cJ_{\chi;\txA}\,$ of
Definition \ref{def:gauge-jump-trans-defect} in such a manner as to
ensure that the $C^\infty(\Si,\txG_\si)$-jump defect remains
topological, in the sense of \Rxcite{Sec.\,2.9}{Runkel:2008gr}, also
in the presence of $\,\xcD_\txA$.
\eerop
\beroof
For the sake of simplicity, we shall only consider homotopic
deformations of the $C^\infty(\Si,\txG_\si)$-jump defect of the sort
depicted in Figure \ref{fig:homotopy-trans-move}, leaving a
verification of the claim in the case of a topological defect
$\,\xcD_\txA\,$ (in which also the latter could be deformed) to the
reader. We shall also assume a cohomologically trivial
$\txG_\si$-equivariant string background, with a target as in
Eqs.\,\eqref{eq:cohom-triv-target} and
\eqref{eq:cohom-triv-Gequiv-str-target}, and with a bi-brane
\qq\nn
\cB=(Q,\iota_1,\iota_2,\om,\Phi)\,,\qquad\qquad\Phi:=J_P
\qqq
endowed with a $\txG_\si$-equivariant structure
\qq\nn
\Xi=\z\in C^\infty\left(\txG_\si\x Q,\bR\right)\,,
\qqq
so that we end up with a smooth function
\qq\nn
\Xi_\chi=(\chi\x\id_Q)^*\z=:\z_\chi\in C^\infty\left(\Si\x Q,\bR
\right)
\qqq
satisfying the defining equation
\qq\nn
\sfd\z_\chi(\si,q)=P(\si,q)-\left(\id_\Si\x\Qup\ell\right)^*P(\si,
\chi,q)+E_\chi\left(\si,\iota_1(q)\right)-E_\chi\left(\si,\iota_2(q
)\right)+\la_{\chi^*\theta_L}(\si,q)\,.
\qqq
Define an extension of the network-field configuration
$\,(X\,\vert\,\G)\,$ for the left-hand side of Figure
\ref{fig:homotopy-trans-move} by the following formul\ae:
\qq\nn
\widehat\xi_{\ell_1}:=\left(\id_\Si,\widehat X_{\ell_1}\right)\ &:&\
\nabla\setminus\ell\to\{\chi\}\x(\nabla \setminus\ell)\x M\ :\ \si
\mapsto\left(\chi,\si,X(\si)\right)\,,\cr\cr
\widehat\xi_{\upsilon_1}:=\left(\id_\Si,\widehat X_{\upsilon_1}
\right)\ &:&\ \ell\to\{\chi\}\x \ell\x Q\ :\ \si\mapsto\left(\chi,
\si,X(\si)\right)\,,
\qqq
and subsequently use it to write a network-field configuration
$\,(\widetilde X\,\vert\,\widetilde\G)\,$ for the right-hand side of
the same figure as
\qq\nn
&\widetilde X\vert_{\Si\setminus\nabla}=X\vert_{\Si\setminus
\nabla}\,,\qquad\qquad\widetilde X\vert_{\nabla\setminus(\ell_{2|
1}\cup\ell_{2|2}\cup\ell)}=\chi.\widehat X_{\ell_1}
\vert_{\nabla\setminus(\ell_{2|1}\cup\ell_{2|2}\cup\ell)}\,,&\cr
\cr
&\widetilde X\vert_{(\ell_{2|1}\cup\ell_{2|2})\setminus\{\upsilon_2
\}}=\widehat X_{\ell_1}\vert_{(\ell_{2|1}\cup\ell_{2|2})\setminus\{
\upsilon_2\}}\,,\qquad\qquad\widetilde X\vert_{\ell\setminus\{
\upsilon_2\}}=\chi.\widehat X_{\upsilon_1}\vert_{\ell\setminus
\{\upsilon_2\}}\,,\qquad\qquad\widetilde X\vert_{\{\upsilon_2\}}=
\widehat X_{\upsilon_1}\vert_{\{\upsilon_2\}}\,.&
\qqq
This is to be augmented by the definition of the new gauge field,
\qq\nn
\widetilde\txA\vert_{\Si\setminus\nabla}=\txA\vert_{\Si\setminus
\nabla}\,,\qquad\qquad\widetilde\txA\vert_{\nabla\setminus\p
\nabla}=\ups{\chi}\txA\vert_{\nabla\setminus\p\nabla}\,.
\qqq
Repeating previous arguments, we readily establish
\qq\nn
&&S_\si[(\widetilde X\,\vert\,\widetilde\G);\widetilde\txA,\g]-
S_\si[(X\,\vert\,\G);\txA,\g]\cr\cr
&=&\int_\nabla\,\left[\Mup\ell^*\txB(\chi,X)(\cdot)-\txB\left(X(
\cdot)\right)+\Mup\ell^*\kappa_A(\chi,X)(\cdot)\wedge\ups{\chi}
\txA^A(\cdot)-\kappa_A\left(X(\cdot)\right)\wedge\txA^A(\cdot)
\right]\cr\cr
&&-\tfrac{1}{2}\,\int_\nabla\,\left[\Mup\ell^* \txc_{AB}(\chi,X)(
\cdot)\,\left(\ups{\chi}\txA^A\wedge\ups{\chi}\txA^B\right)(\cdot)-
 \txc_{AB}\left(X(\cdot)\right)\,\left(\txA^A\wedge\txA^B\right)(
\cdot)\right]\cr\cr
&&+\int_\ell\,\left[\left(\id_\Si\x\Qup\ell\right)^*P\left(\cdot,
\chi(\cdot),X(\cdot)\right)-P\left(\cdot,X(\cdot)\right)-\Qup\ell^*
k_A(\chi,X)(\cdot)\,\ups{\chi}\txA^A(\cdot)+k_A\left(X(\cdot)\right)
\,\txA^A(\cdot)\right]\cr\cr
&&+\int_{\ell_{2|1}\cup\ell_{2|2}\cup(-\ell_{1|1})\cup(-\ell_{1|2}
)}\,E_\chi\left(\cdot,X(\cdot)\right)+\z_\chi\left(v_2,X(v_2)
\right)-\z_\chi\left(v_1,X(v_1)\right)\cr\cr
&=&-\int_{\ell_{2|1}\cup\ell_{2|2}\cup(-\ell_{1|1})\cup(-\ell_{1|2}
)}\,E_\chi\left(\cdot,X(\cdot)\right)-\int_\ell\,\left[E_\chi\left(
\cdot,\iota_1\circ X(\cdot)\right)-E_\chi\left(\cdot,\iota_2\circ X
(\cdot)\right)\right]\cr\cr
&&+\int_\ell\,\left[\left(\id_\Si\x\Qup\ell\right)^*P\left(\cdot,
\chi(\cdot),X(\cdot)\right)-P\left(\cdot,X(\cdot)\right)-
\la_{\chi^*\theta_L}\left(\cdot,X(\cdot)\right)\right]\cr\cr
&&+\int_{\ell_{2|1}\cup\ell_{2|2}\cup(-\ell_{1|1})\cup(-\ell_{1|2}
)}\,E_\chi\left(\cdot,X(\cdot)\right)+\z_\chi\left(v_2,X(v_2)
\right)-\z_\chi\left(v_1,X(v_1)\right)\cr\cr
&=&-\int_\ell\,\sfd\z_\chi\left(\cdot,X(\cdot)\right)+\z_\chi\left(
v_2,X(v_2)\right)-\z_\chi\left(v_1,X(v_1)\right)=0\,,
\qqq
which is the desired result. \eroof

\begin{figure}[hbt]~\\[20pt]

\qq\nonumber
\raisebox{-50pt}{\begin{picture}(85,80)
   \put(-25,7){\scalebox{0.15}{\includegraphics{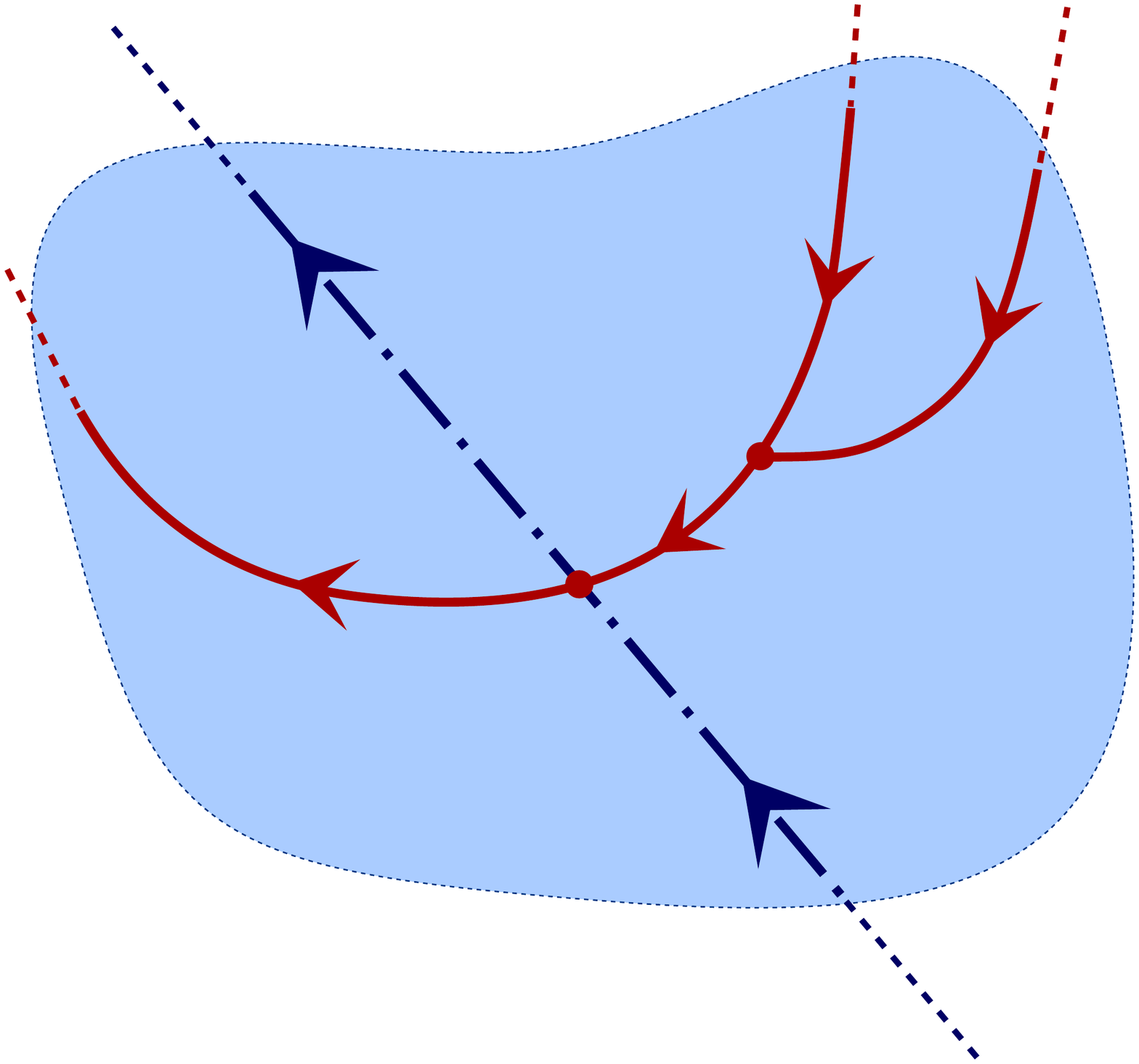}}}
   \put(0,0){
      \setlength{\unitlength}{.55pt}\put(-18,-4){
      \put(-37,146)    {\scriptsize $ \xcD_{\chi_1\cdot\chi_2} $}
      \put(99,180)    {\scriptsize $ \xcD_{\chi_2} $}
      \put(158,175)    {\scriptsize $ \xcD_{\chi_1} $}
      \put(19,175)    {\scriptsize $ \xcD_\txA $}
      \put(83,103)    {\scriptsize $ \vep_L $}
      \put(106,95)    {\scriptsize $ \unl\iota_2^{(2)\,*}\g_{\chi_1,\chi_2}^\sharp $}
      \put(36,99)    {\scriptsize $ \Xi_{\chi_1\cdot\chi_2} $}
      }\setlength{\unitlength}{1pt}}
   \end{picture}}\hspace{.5cm}
   \xrightarrow{~~\vep_L\rightarrow 0~~}\hspace{.5cm}
   \raisebox{-40pt}{\begin{picture}(85,80)
   \put(-10,-3){\scalebox{0.15}{\includegraphics{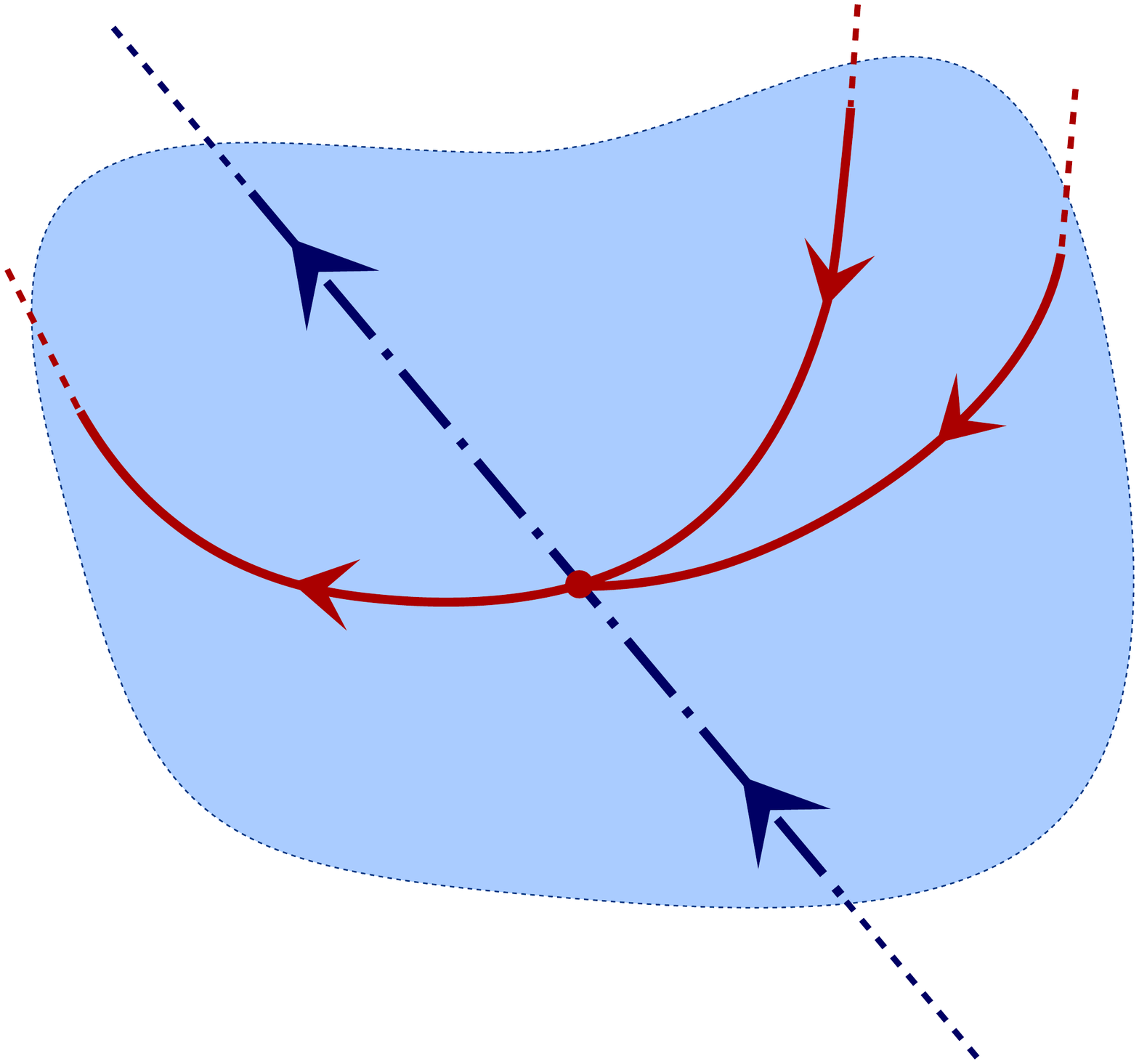}}}
   \put(0,0){
      \setlength{\unitlength}{.55pt}\put(-18,-4){
      \put(-11,127)    {\scriptsize $ \xcD_{\chi_1\cdot\chi_2} $}
      \put(125,161)    {\scriptsize $ \xcD_{\chi_2} $}
      \put(187,146)    {\scriptsize $ \xcD_{\chi_1} $}
      \put(46,157)    {\scriptsize $ \xcD_\txA $}
      \put(119,60)    {\scriptsize $ \Qup u^2_{\chi_1,\chi_2;\txA} $}
      }\setlength{\unitlength}{1pt}}
   \end{picture}}\hspace{1.2cm}
   \xleftarrow{~~\vep_R^i\rightarrow 0~~}
   \raisebox{-40pt}{\begin{picture}(85,80)
   \put(10,-3){\scalebox{0.15}{\includegraphics{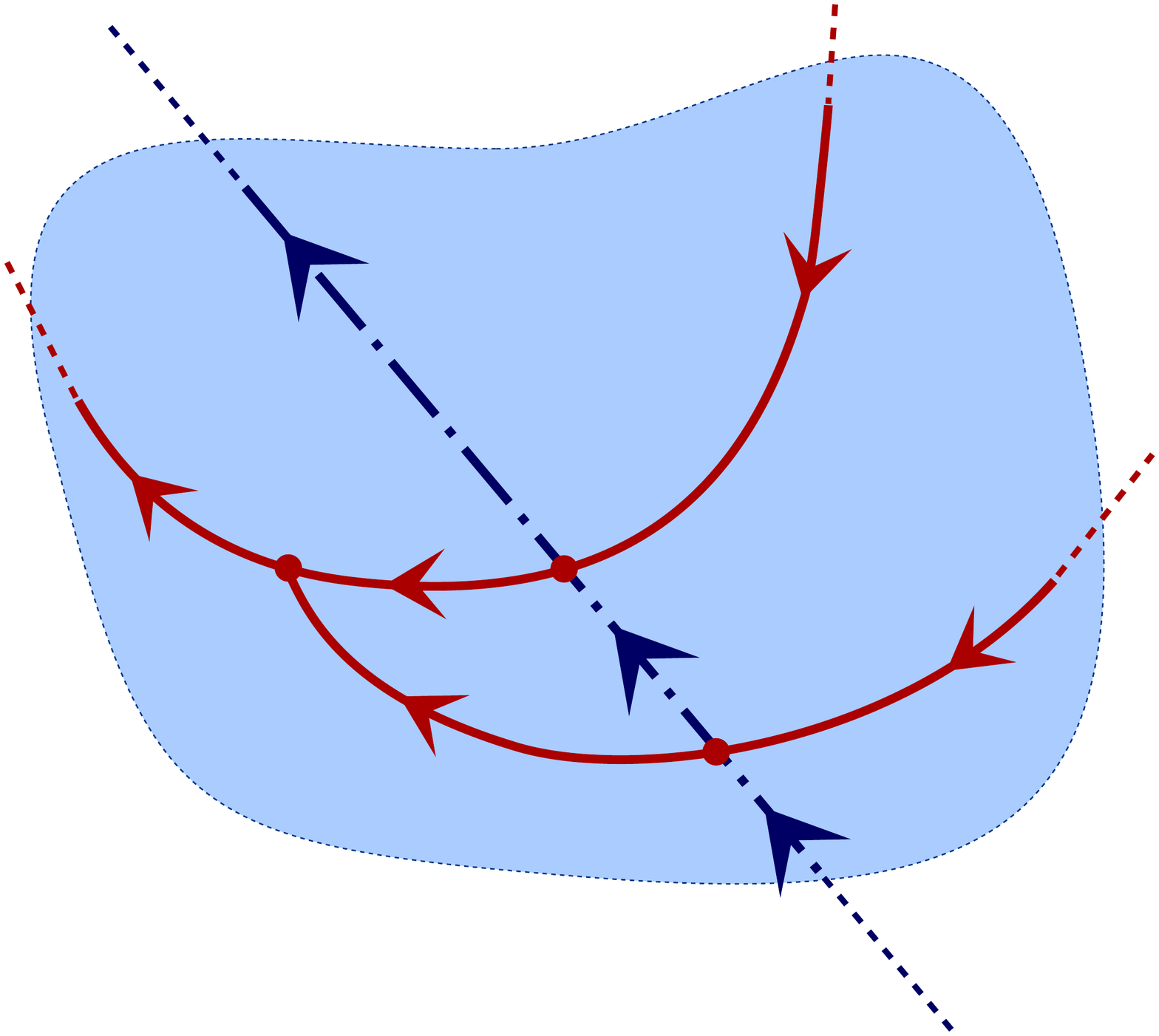}}}
   \put(0,0){
      \setlength{\unitlength}{.55pt}\put(-18,-4){
      \put(26,127)    {\scriptsize $ \xcD_{\chi_1\cdot\chi_2} $}
      \put(162,161)    {\scriptsize $ \xcD_{\chi_2} $}
      \put(238,90)    {\scriptsize $ \xcD_{\chi_1} $}
      \put(83,157)    {\scriptsize $ \xcD_\txA $}
      \put(33,72)    {\scriptsize $ \unl\iota_1^{(2)\,*}\g_{\chi_1,\chi_2} $}
      \put(155,67)    {\scriptsize $ \Xi_{\chi_2} $}
      \put(179,41)    {\scriptsize $ \chi_2.\Xi_{\chi_1} $}
      \put(111,79)    {\scriptsize $ \vep_R^1 $}
      \put(114,38)    {\scriptsize $ \vep_R^2 $}
      \put(141,53)    {\scriptsize $ \vep_R^3 $}
      }\setlength{\unitlength}{1pt}}
   \end{picture}}
\qqq

\caption{Pulling a tri-valent defect vertex of a
$C^\infty(\Si,\txG_\si)$-jump defect network across a generic
$\txG_\si$-transparent defect line yields the anomaly 3-cocycle
$\,\Qup u^2_{\chi_1,\chi_2;\txA}$.} \label{fig:anomaly-cycle-bib}
\end{figure}

So far, no structure beyond that which is required for the $C^\infty
(\Si,\txG_\si)$-invariance of the gauged $\si$-model in the presence
of $\txG_\si$-transparent defects was necessary. However, in order
to ensure topologicality of the $C^\infty(\Si,\txG_\si)$-jump defect
network, we should also demand invariance of the action functional
under homotopies of the network that pull the $C^\infty(\Si,\txG_\si
)$-jump defect junction across $\,\xcD_\txA\,$ as in Figure
\ref{fig:anomaly-cycle-bib}.
This imposes familiar constraints upon the data carried by the two
crossing defect networks.
\berop\label{prop:pull-jump-across}
Adopt the notation of Definition \ref{def:sigmod-2d} and assume that
the conditions stated in Proposition
\ref{prop:simpl-ind-jump-def-net} are satisfied. The network-field
configuration for an arbitrary graph of the $C^\infty(\Si,\txG_\si
)$-jump defects of Definition \ref{def:gauge-jump-defect} crossing a
$\txG_\si$-transparent conformal defect $\,\xcD_\txA\,$ is
extendible (as long as there are no topological obstructions within
the world-sheet) in a neighbourhood of the crossing in such a manner
as to ensure that the $C^\infty(\Si,\txG_\si)$-jump defect network
remains topological, in the sense of
\Rxcite{Sec.\,2.9}{Runkel:2008gr}, also in the presence of
$\,\xcD_\txA\,$ iff the $\txG_\si$-equivariant structure on the
bi-brane $\,\cB\,$ is coherent in the sense of Definition
\ref{def:Gequiv-bgrnd}.
\eerop
\beroof
Reasoning along the same lines as in the case of Figure
\ref{fig:associator}, we find out that the elementary $C^\infty(\Si,
\txG_\si)$-jump trans-defect junction can be pulled through
$\,\xcD_\txA\,$ iff the \textbf{cross-multiplication 2-cocycle}
\qq\nn
\Qup u^2_{\chi_1,\chi_2;\txA}:=\left(\Id\circ\unl\iota_1^*\g_{\chi_1
,\chi_2}\right)\bullet\left(\Id\circ\left(\Xi_{\chi_2}\ox\Id\right)
\circ\Id\right)\bullet L_{\chi_2}^*\Xi_{\chi_1}\bullet\Xi_{\chi_1
\cdot\chi_2}^{-1}\bullet\left(\left(\unl\iota_1^*\g_{\chi_1,
\chi_2}^\sharp\ox\Id\right)\circ\Id\right)^{-1}
\qqq
trivialises in cohomology.

The latter is the pullback, along the map $\,(\chi_1,\chi_2)\x\id_Q
\ :\ \Si\x Q\to\txG_\si^2\x Q$,\ of the anomaly 2-cocycle $\,\Qup
u^2\in Z^2\left(\pi_0(\txG_\si),\uj^{\pi_0(Q)}\right)\,$ of
\Rxcite{Cor.\,11.21}{Gawedzki:2012fu} whose class measures the
obstruction to the existence of a \emph{coherent}
$\txG_\si$-equivariant structure on the bi-brane $\,\cB$,
\qq\nn
\Qup u^2_{\chi_1,\chi_2;\txA}=\left((\chi_1,\chi_2)\x\id_Q\right)^*
\Qup u^2\,.
\qqq
Triviality of $\,\Qup u^2_{\chi_1,\chi_2;\txA}\,$ is a sufficient
and necessary condition for the coherence condition of
\Reqref{eq:Gequiv-bimod-coh} to be satisfied by the 2-isomorphism
$\,\Xi\,$ entering the definition of the elementary $C^\infty(\Si,
\txG_\si)$-jump trans-defect junction.

We conclude that the existence of a full-fledged
$\txG_\si$-equivariant structure on the string background of a
multi-phase $\si$-model on a world-sheet $\,\Si\,$ with circular
(\textit{i.e.}\ non-intersecting) $\txG_\si$-transparent defect
lines ensures that the associative realisation of $\,C^\infty(\Si,
\txG_\si)\,$ mentioned in the proof of Proposition
\ref{prop:simpl-ind-jump-def-net} extends to the multi-phase
setting.\eroof

\begin{figure}[hbt]~\\[20pt]

\qq\nonumber
\raisebox{-50pt}{\begin{picture}(85,80)
   \put(-35,7){\scalebox{0.15}{\includegraphics{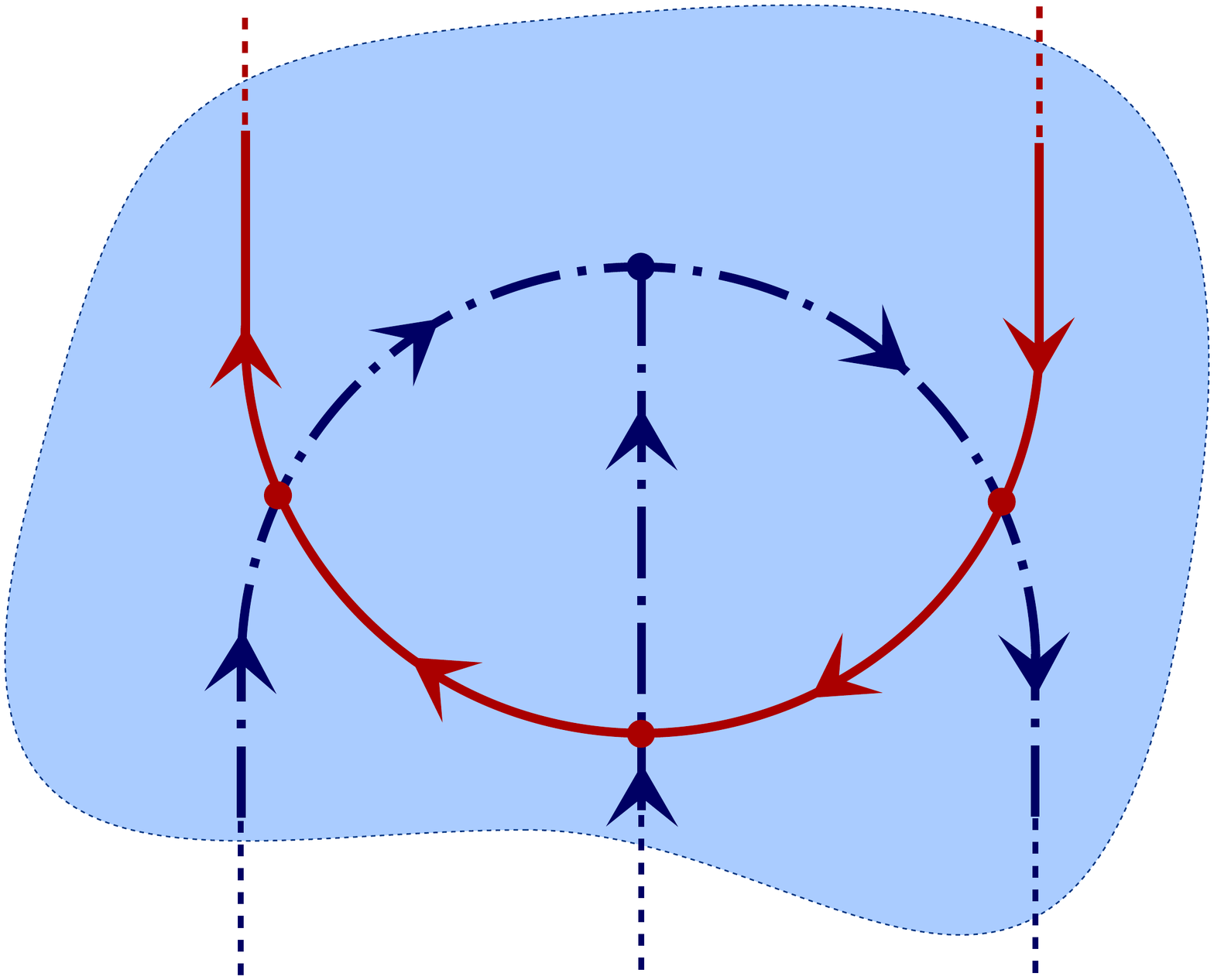}}}
   \put(0,0){
      \setlength{\unitlength}{.55pt}\put(-18,-4){
      \put(134,168)    {\scriptsize $ \xcD_\chi $}
      \put(5,12)    {\scriptsize $ \xcD_\txA^{(1)} $}
      \put(64,12)    {\scriptsize $ \xcD_\txA^{(2)} $}
      \put(122,12)    {\scriptsize $ \xcD_\txA^{(3)} $}
      \put(30,126)    {\scriptsize $ \vep_L^1 $}
      \put(76,105)    {\scriptsize $ \vep_L^2 $}
      \put(104,126)    {\scriptsize $ \vep_L^3 $}
      \put(-52,95)    {\scriptsize $ \pi_3^{1,2\,(1)\,*}\Xi_\chi $}
      \put(40,80)    {\scriptsize $ \pi_3^{2,3\,(1)\,*}\Xi_\chi $}
      \put(130,95)    {\scriptsize $ \pi_3^{3,1\,(1)\,*}\Xi_\chi^{\sharp\,-1} $}
      \put(69,135)    {\scriptsize $ \varphi_3 $}
      \put(36,52)    {\scriptsize $ \vep_L^4 $}
      \put(93,52)    {\scriptsize $ \vep_L^5 $}
      }\setlength{\unitlength}{1pt}}
   \end{picture}}\hspace{.5cm}
   \xrightarrow{~~\vep_L^i\rightarrow 0~~}\hspace{.5cm}
   \raisebox{-40pt}{\begin{picture}(85,80)
   \put(-10,-3){\scalebox{0.15}{\includegraphics{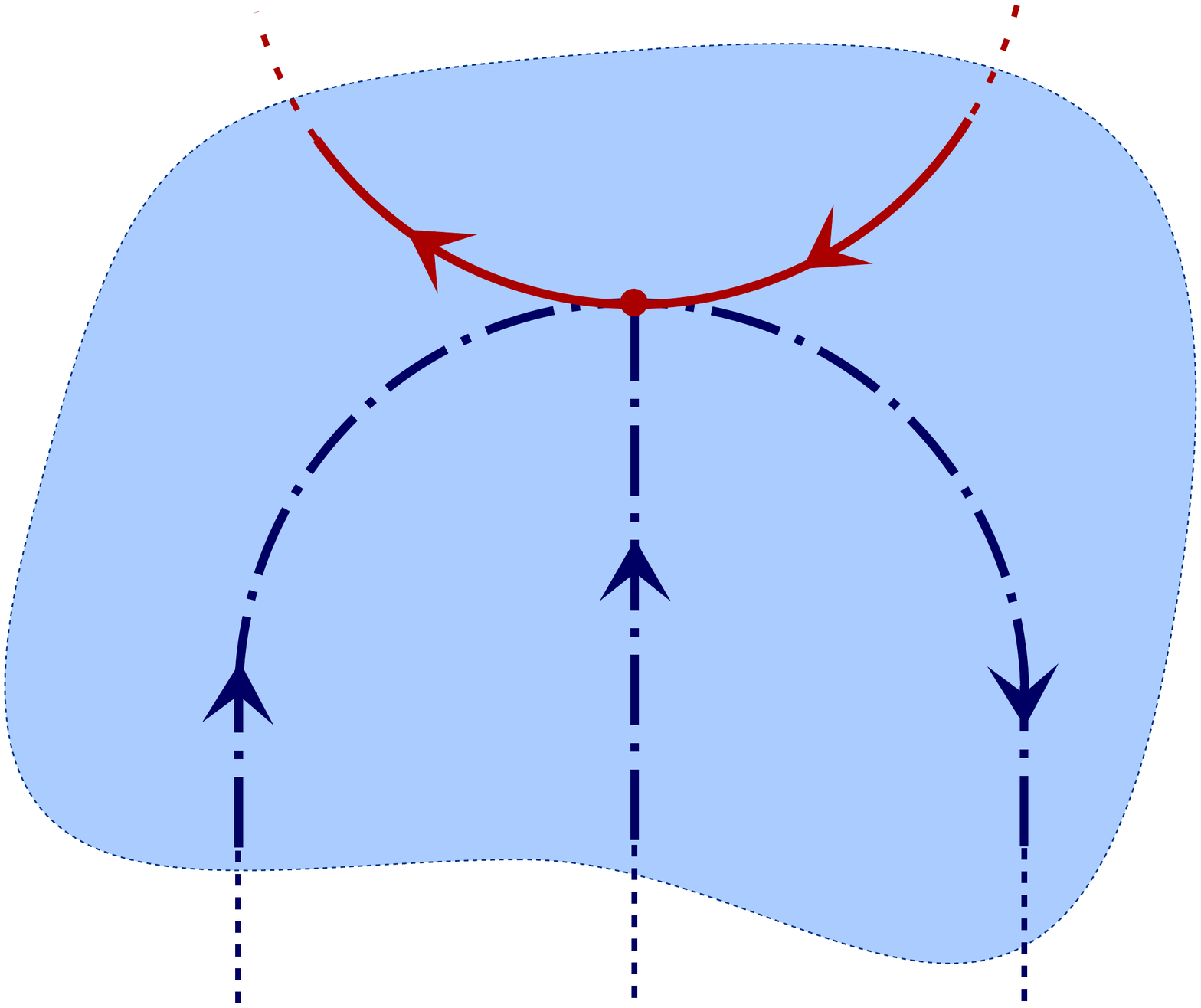}}}
   \put(0,0){
      \setlength{\unitlength}{.55pt}\put(-18,-4){
      \put(178,154)    {\scriptsize $ \xcD_\chi $}
      \put(50,-6)    {\scriptsize $ \xcD_\txA^{(1)} $}
      \put(109,-6)    {\scriptsize $ \xcD_\txA^{(2)} $}
      \put(167,-6)    {\scriptsize $ \xcD_\txA^{(3)} $}
      \put(97,121)    {\scriptsize $ \ups{T_3}u^1_{\chi;\txA} $}
      }\setlength{\unitlength}{1pt}}
   \end{picture}}\hspace{1.2cm}
   \xleftarrow{~~\vep_R\rightarrow 0~~}
   \raisebox{-40pt}{\begin{picture}(85,80)
   \put(5,-3){\scalebox{0.15}{\includegraphics{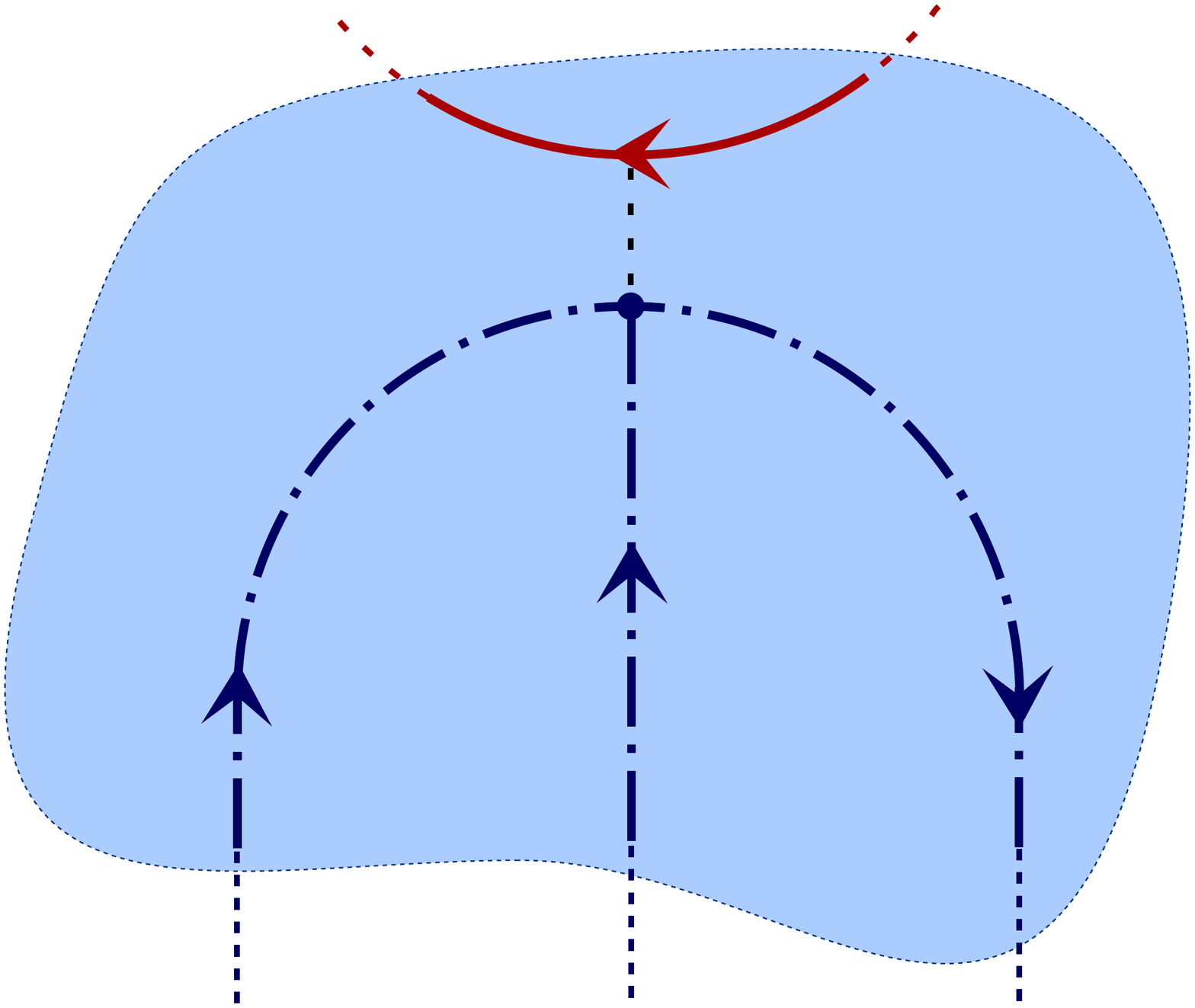}}}
   \put(0,0){
      \setlength{\unitlength}{.55pt}\put(-18,-4){
      \put(194,155)    {\scriptsize $ \xcD_\chi $}
      \put(77,-6)    {\scriptsize $ \xcD_\txA^{(1)} $}
      \put(136,-6)    {\scriptsize $ \xcD_\txA^{(2)} $}
      \put(194,-6)    {\scriptsize $ \xcD_\txA^{(3)} $}
      \put(148,117)    {\scriptsize $ \chi.\varphi_3 $}
      \put(128,124)    {\scriptsize $ \vep_R $}
      }\setlength{\unitlength}{1pt}}
   \end{picture}}
\qqq

\caption{Pulling a $C^\infty(\Si,\txG_\si)$-jump defect past a
three-valent defect junction of a generic $\txG_\si$-transparent
defect network yields the anomaly 3-cocycle
$\,\ups{T_3}u^1_{\chi;\txA}$.} \label{fig:anomaly-cycle-ibb}
\end{figure}

In the last stage of the construction, it remains to examine the
fate of the topologicality of the $C^\infty(\Si,\txG_\si)$-jump
defect network in the presence of arbitrary defect junctions of
$\txG_\si$-transparent defects of the gauged multi-phase
$\si$-model. This boils down to calculating the correction to the
action functional induced in the process of pulling the $C^\infty(
\Si,\txG_\si)$-jump defect past a vertex of the
$\txG_\si$-transparent defect network as, \textit{e.g.}, in Figure
\ref{fig:anomaly-cycle-ibb}.
As expected, we obtain
\berop\label{prop:lga-vs-top-def-net}
Adopt the notation of Definition \ref{def:sigmod-2d} and assume that
the conditions stated in Proposition \ref{prop:pull-jump-across} are
satisfied. The network-field configuration for an arbitrary graph of
the $C^\infty(\Si,\txG_\si)$-jump defects of Definition
\ref{def:gauge-jump-defect} crossing a network of
$\txG_\si$-transparent conformal defects is extendible (as long as
there are no topological obstructions within the world-sheet) in a
neighbourhood of every vertex of the latter network in such a manner
as to ensure that the $C^\infty(\Si,\txG_\si)$-jump defect network
remains topological, in the sense of
\Rxcite{Sec.\,2.9}{Runkel:2008gr}, also in the presence of junctions
of the $\txG_\si$-transparent defects.
\eerop
\beroof
The claim of the proposition follows directly from cohomological
triviality of the \textbf{intertwiner 1-cocycle} expressible as the
pullback, along the map $\,\chi\x\id_{T_n}\ :\ \Si\x T_N\to\txG_\si
\x T_N$,\ of the anomaly 1-cocycle $\,\Tnup u^2\in Z^2\left(\pi_0(
\txG_\si),\uj^{\pi_0(T_n)}\right)\,$ of
\Rxcite{Cor.\,11.26}{Gawedzki:2012fu},
\qq\nn
\Tnup u^1_{\chi;\txA}=(\chi\x\id_Q)^*\Tnup u^1\,.
\qqq
The class of the latter measures the obstruction to the existence of
a \emph{coherent} $\txG_\si$-equivariant structure on the complete
string background, as expressed by \Reqref{eq:varphi-equiv}.

\begin{figure}[hbt]~\\[30pt]

$$
 \raisebox{-50pt}{\begin{picture}(50,50)
  \put(-169,0){\scalebox{0.45}{\includegraphics{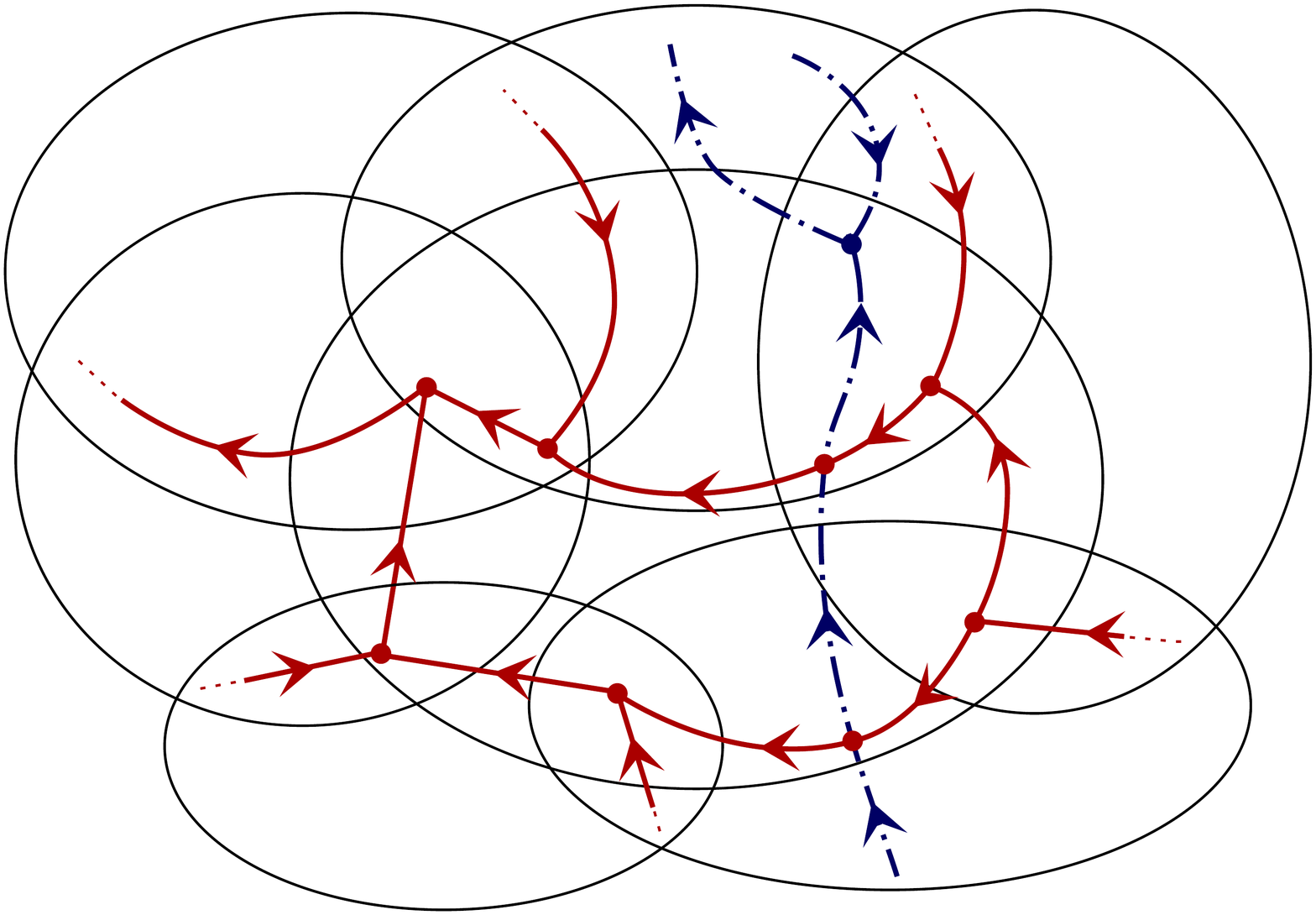}}}
  \end{picture}
  \put(10,0){
     \setlength{\unitlength}{.60pt}\put(-28,-16){
     \put(-330,400)    {$ \Si_i $}
     \put(-327,150)    {$ \Si_j $}
     \put(-251,40)    {$ \Si_k $}
     \put(176,40)    {$ \Si_l $}
     \put(232,400)    {$ \Si_m $}
     \put(-20,455)    {$ \Si_n $}
     \put(172,225)    {$ \Si_o $}
     \put(16,429)    {$ \xcD_{\txA_n} $}
     \put(-63,429)    {$ \xcD_{\txA_n} $}
     \put(11,321)    {$ \xcJ_{\txA_n}^{++-} $}
     \put(77,51)    {$ \xcD_{\txA_l} $}
     \put(14,140)    {$ \xcD_{\txA_o} $}
     \put(115,168)    {$ \xcJ_{g_{lm},g_{mo}} $}
     \put(23,290)    {$ \xcD_{\txA_n} $}
     \put(106,352)    {$ \xcD_{g_{mn}} $}
     \put(131,220)    {$ \xcD_{g_{om}} $}
     \put(94,276)    {$ \xcJ_{g_{om},g_{mn}} $}
     \put(68,147)    {$ \xcD_{g_{lo}} $}
     \put(-21,232)    {$ \xcD_{g_{on}} $}
     \put(78,242)    {$ \xcD_{g_{on}} $}
     \put(40,228)    {$ \xcJ_{g_{on},\txA_n} $}
     \put(158,138)    {$ \xcD_{g_{lm}} $}
     \put(-8,114)    {$ \xcD_{g_{lo}} $}
     \put(57,99)    {$ \xcJ_{g_{lo},\txA_o} $}
     \put(-83,101)    {$ \xcD_{g_{kl}} $}
     \put(-90,140)    {$ \xcJ_{g_{kl},g_{lo}} $}
     \put(-131,123)    {$ \xcD_{g_{ko}} $}
     \put(-210,122)    {$ \xcD_{g_{jk}} $}
     \put(-154,190)    {$ \xcD_{g_{jo}} $}
     \put(-165,155)    {$ \xcJ_{g_{jk},g_{ko}} $}
     \put(-110,262)    {$ \xcD_{g_{oi}} $}
     \put(-96,339)    {$ \xcD_{g_{ni}} $}
     \put(-85,245)    {$ \xcJ_{g_{on},g_{ni}} $}
     \put(-160,280)    {$ \xcJ_{g_{jo},g_{oi}} $}
     \put(-237,251)    {$ \xcD_{g_{ji}} $}
     \put(-287,244)    {$ e_{ji} $}
     \put(112,148)    {$ v_{lmo} $}
     \put(-250,180)    {$ p_j $}
     \put(-250,300)    {$ p_i $}
     \put(15,241)    {$ c_{on} $}
     \put(58,334)    {$ \jmath_n $}
     \put(62,302)    {$ s_n $}
     \put(51,270)    {$ v_{omn} $}
                 }\setlength{\unitlength}{1pt}}}
$$

\caption{A reconstruction of an arbitrary principal
$\txG_\si$-bundle with connection over the world-sheet through local
trivialisation. The local description uses data of a topological
$C^\infty(\sfN^1\cO_\Si,\txG_\si)$-jump defect network.}
\label{fig:princ-bndle-constr}
\end{figure}

We conclude that the existence of a full-fledged
$\txG_\si$-equivariant structure on the string background of a
multi-phase $\si$-model on a world-sheet $\,\Si\,$ with arbitrary
$\txG_\si$-transparent defect lines ensures that the associative
realisation of $\,C^\infty(\Si, \txG_\si)\,$ mentioned in the proof
of Proposition \ref{prop:simpl-ind-jump-def-net} extends to the
multi-phase setting. \eroof \brem\label{rem:lga-vs-tnt-prin-g-bun}
We close the present section with a comment on how the hitherto
findings can be employed to reconstruct the gauged $\si$-model
coupled to an arbitrary gauge bundle $\,\cP_{\txG_\si}\,$ with
connection over $\,\Si\,$ out of data of the trivial gauge bundle
and those of a topological $C^\infty(\sfN^1\cO_\Si,\txG_\si)$-jump
defect network, defined over the nerve $\,\sfN^\bullet\cO_\Si\,$ of
an open cover $\,\cO_\Si\,$ of the world-sheet $\,\Si$.\ Rather than
formalising our discussion, we illustrate the general idea by
referring to the generic local world-sheet situation depicted in
Figure \ref{fig:princ-bndle-constr}.
The latter shows a piece of the world-sheet covered by a number of
elements of an open cover $\,\cO_\Si=\{\Si_i\}_{i\in\xcI}\,$ over
which the principal $\txG_\si$-bundle is assumed to trivialise, so
that, \textit{e.g.}, $\,\sfP_{\txG_\si}\vert_{\Si_i}\cong\Si_i\x
\txG_\si\,$ and the principal $\txG_\si$-connection 1-form induces a
locally smooth 1-form on the base
$\,\txA_i\in\Om^1(\Si_i)\ox\ggt_\si$.

We set up the local description as follows: Assume given an embedded
defect quiver $\,\G\subset\Si$.\ Consider an oriented trivalent
graph $\,\G_{\cO_\Si}\subset\Si\,$ that is \textbf{$\G$-transversal
and $\G$-simple, and subordinate to $\,\cO_\Si\,$} in the sense that
$\,\G \cap\G_{\cO_\Si}\,$ is discrete (\textit{i.e.}\ composed of a
finite number of intersection points) and does not contain vertices
of $\,\G$,\ and is such that for every edge of the graph there
exists a pair $\,(i,j)\in\xcI^2\,$ of indices of the cover with the
property that the edge is contained in $\,\Si_{ij}=\Si_i\cap\Si_j$.\
Clearly, to every vertex of the graph, we may associate a triple
$\,(i,j,k)\in \xcI^3\,$ of indices corresponding to the three edges
converging at the vertex. The graph splits $\,\Si\,$ into a
collection of disjoint plaquettes. Each of them will be labelled by
the index of the element of the open cover in which it is contained,
\textit{e.g.}, $\,p_i\subset\Si_i$.\ We label the edge separating
plaquettes $\,p_i\,$ and $\,p_j\,$ with the two indices $i$ and $j$
written in the order determined by the orientation of the edge as in
the case of the edge $\,e_{ji}\,$ in the figure. Furthermore, we
label each vertex of the graph by an arbitrary cyclic permutation of
the three indices associated to the edges meeting at the vertex,
read off anti-clockwise around the vertex, as in the case of the
vertex $\,v_{lmo}\,$ in the figure. Finally, each junction and each
(segment of a) defect line of $\,\G\,$ is labelled by the index of
the plaquette in which it sits, and every (4-valent) crossing
between an edge of $\,\G_{\cO_\Si}\,$ and an edge of $\,\G\,$ by the
pair of indices assigned to the edge of $\,\G_{\cO_\Si}\,$ going
through it, as the junction $\,\jmath_n$,\ the segment $\,s_n\,$ and
the crossing $\,c_{on}\,$ in the figure, respectively.

Once the assignment of labels has been accomplished, and given a
collection of local trivialisations $\,\tau_i:
\pi_{\cP_{\txG_\si}}^{-1}(\Si_i)\to\Si_i\x\txG_\si\,$ of
$\,\cP_{\txG_\si}$,\ the attendant local connection 1-forms
$\,\txA_i\,$ and transition maps $\,g_{ij}: \Si_{ij}\to\txG_\si$,\
and a global section $\,[(g_{ij},X_i)]\,$ of the associated bundle
$\,\sfP_{\txG_\si}\x_{\txG_\si}\xcF$,\ it is straightforward to
associate (local) geometric objects to elements of the decorated
triangulation of the world-sheet defined (as above) by
$\,\G_{\cO_\Si}$.\ Thus, to a plaquette $\,p_i\,$ we pull back,
along the local section $\,(\id_{p_i},X_i)$,\ the data of the
extended background $\,\Bgt_{\txA_i}\,$ (including the extended
metric $\,\txg_{\txA_i}$,\ the extended gerbe $\,\cG_{\txA_i}\,$
\textit{etc}.). In particular, we pull back the 1-isomorphism
$\,\Phi_{\txA_n}\,$ to the segment $\,s_n$,\ and the 2-isomorphism
$\,\varphi_{\txA_n}^{++-}\,$ to the junction $\,\jmath_n$.\ Based on
our previous considerations, we endow the edge $\,e_{ji}\,$ with the
structure of a \textbf{local transition defect} $\,\xcD_{g_{ji}}\,$
(denoted, by a slight abuse of the notation, by the same symbol as
and) differing from the component $C^\infty(\Si,\txG_\si)$-jump
defect $\,\xcD_\chi\,$ of Definition \ref{def:gauge-jump-defect}
\emph{exclusively} in the choice of the associated $\txG_\si$-valued
map ($\,g_{ji}\,$ \textit{versus} $\,\chi$), required to be smooth
only \emph{locally} in the present case. Similarly, we put over the
crossing $\,c_{on}\,$ the \textbf{local trans-defect transition
junction $\,\xcJ_{g_{on},\txA_n}$},\ which is none other than the
elementary $C^\infty(\Si,\txG_\si)$-jump trans-defect junction
$\,\xcJ_{\chi,\txA}\,$ of Definition
\ref{def:gauge-jump-trans-defect} with locally smooth data $\,(g_{o
n},\txA_n)\,$ instead of the globally smooth ones $\,(\chi,\txA)$.\
Up to now, we have been using only the structure necessitated by the
$C^\infty(\Si,\txG_\si)$-invariant gauged $\si$-model coupled to the
topologically trivial gauge field. The existence of the latter also
ensures that local transition defects can be deformed homotopically
(\textit{i.e.}\ drawn arbitrarily) within the domains of definition
of the respective transition maps without changing the value of the
action functional, also along defect lines of $\,\G\,$ and past its
junctions, so that, \textit{e.g.}, we may pull $\,e_{on}\,$ up past
$\,\jmath_n$.\ This renders the field theory defined in terms of the
local data introduced above independent of some of the arbitrary
choices made in the trivialisation procedure. It is through the
imposition of requirements of internal consistency of the ensuing
field theory with local transition defects, and arguments of
independence of the latter theory of the arbitrary choices made that
we shall next rediscover the remaining components of the
$\txG_\si$-equivariant structure on $\,\Bgt\,$ as necessary
ingredients of the local construction of a topologically non-trivial
gauge bundle over $\,\Si\,$ coupled to $\,\Bgt$.

Given the assignment of geometric data to edges of
$\,\G_{\cO_\Si}$,\ detailed arguments of
\Rxcite{Sec.\,2.7}{Runkel:2008gr} force us to pull back to its
vertices data of appropriate trivialising gerbe 2-isomorphisms,
\textit{cf.}\ Definition \vref{def:bckgrnd}I.2.1. Taking into
account the cocycle condition satisfied by the transition maps of
$\,\cP_{\txG_\si}$,\ we are thus led to require the existence of the
2-isomorphism $\,\g_{g_{lm},g_{mo}}\,$ for $\,v_{lmo}\,$ (determined
by the transition maps in the very same manner as the 2-isomorphisms
$\,\g_{\chi_1,\chi_2}\,$ of Definition
\ref{def:gauge-jump-defect-junct} are determined by the gauge maps
$\,\chi_1,\chi_2$), and so -- ultimately, in view of the
arbitrariness of the transition functions -- the existence of the
underlying 2-isomorphism $\,\g\,$ of Definition
\ref{def:Gequiv-bgrnd}. The \textbf{local transition defect
junctions $\,\xcJ_{g_{lm},g_{m o}}\,$} thus obtained can be moved
around within the domain $\,\Si_{lm o}\,$ of their definition at no
cost in the value of the action functional as long as they do not
cross a defect line of $\,\G$,\ and so it remains to ensure that
this feature prevails also in the presence of the defect lines (so
that we may, \textit{e.g.}, pull $\,v_{omn}\,$ across $\,s_n\,$ in
the figure), and that the field theory determined by the local data
of the trivialisation of $\,\cP_{\txG_\si}\,$ including the local
transition defect junctions does not suffer from any ambiguities
under refinement of a given open cover or a simple change of the
choice of indices in quadruple and higher-order intersections of
elements of $\,\cO_\Si\,$ (as, \textit{e.g.}, in $\,\Si_{ijon}\,$ in
the figure). As the discussion conveyed in the context of the
implementation of the $C^\infty(\Si ,\txG_\si)$-action through
defects indicates, we need the coherence condition
\eqref{eq:Gequiv-bimod-coh} for the former, and the standard
argument for the quadruple intersection (used previously in the
context of the associativity of the world-sheet realisation of
$\,C^\infty(\Si,\txG_\si)$) demonstrates the necessity (and
sufficiency) of imposing the coherence condition
\eqref{eq:Gerbe-1iso-coh}. It is now clear that the systematic
procedure leads to a reconstruction of a consistent coupling of the
non-trivial principal $\txG_\si$-bundle $\,\cP_{\txG_\si}\,$ over
the world-sheet to the original string background, and yields a
gauged $\si$-model manifestly independent of the arbitrary choices
entering its local description. Our analysis shows, once again, that
the passage from trivial to non-trivial gauge bundles coupled to the
string background $\,\Bgt\,$ of the $\si$-model with a global
$\txG_\si$-symmetry (with a vanishing small gauge anomaly) does
\emph{necessitate} the existence of a full-fledged
$\txG_\si$-equivariant structure on $\,\Bgt$. \erem

The findings of the last section (and those of the previous one) are
summarised in
\bethe\label{thm:LGA-vs-coset}
The gauged non-linear two-dimensional $\si$-model coupled to gauge
fields of an arbitrary topology, whose incorporation is necessary to
account for the existence of $\txG_\si$-twisted (network-)field
configurations in the non-linear two-dimensional $\si$-model with
the target space given by the orbit space of the parent $\si$-model
with respect to the action of a group $\,\txG_\si\,$ of rigid
symmetries of the latter $\si$-model, exists iff the string
background of the parent $\si$-model can be endowed with a
$\txG_\si$-equivariant structure. \ethe

\section{Conclusions and outlook}\label{sec:con-&-out}

The paper gives an account of a comprehensive treatment of algebraic
and differential-geometric aspects of rigid symmetries of the
multi-phase two-dimensional non-linear $\si$-model and of their
gauging, laying due emphasis on the underlying gerbe theory and --
also in this latter context -- exploiting the interplay between
$\si$-model dualities and conformal defects. It develops a scheme of
description of the said symmetries based on the concept of the
(relative) generalised geometry and thus naturally adapted to the
setting of the target space of the $\si$-model endowed with the
structure of the 2-category of bundle gerbes with connection over
it, discusses the transgression of that scheme to the phase space of
the $\si$-model, and -- finally -- extracts from it a simple
geometric measure of the gauge anomaly that obstructs an attempt at
rendering the rigid symmetries local. The naturalness of this
measure is subsequently corroborated in the framework of the theory
of principal bundles with a structural action groupoid over the
world-sheet of the $\si$-model, leading to a systematic construction
of topological defect networks implementing the action of the gauge
group as well as those realising a local (world-sheet)
trivialisation of a gauge bundle of an arbitrary topology in the
gauged multi-phase $\si$-model. The latter construction demonstrates
the necessity of the existence of a full-fledged equivariant
structure on the string background of the $\si$-model for a
consistent gauging of its rigid symmetries.

For the sake of concreteness, and by way of a concise summary, we
list the main results of our work hereunder.
\ben
\item The definition of an algebroidal target-space model of the
Poisson algebra of Noether charges of a rigid symmetry, inspired by
earlier work of Alekseev and Strobl, and that of Hitchin and
Gualtieri, and formulated in terms of a twisted bracket structure on
the space of sections of generalised tangent bundles over the target
space. It is based on the following:
\bit
\item a reconstruction of the model through the study of
infinitesimal lagrangean symmetries (Propositions
\ref{prop:sigmod-n-symm}, \ref{prop:sisym-iotalign} and
\ref{prop:sigmod-symm-def-junct}; Corollary
\ref{cor:sigmod-symm-E11});
\item a classification of its automorphisms (Propositions
\ref{prop:Vin-str-auts} and \ref{prop:aut-pair-tw-bra-str});
\item a reformulation in terms of local data of the 2-category of
abelian bundle gerbes with connection over the target space of the
$\si$-model, \textit{via} Hitchin-type isomorphisms (Corollary
\ref{cor:tw-gen-tan-vs-ngerb}; Proposition
\ref{prop:pair-tw-gen-tan-bun}), compatible with the action of gerbe
morphisms (Theorem \ref{thm:bib-as-morph});
\item a homomorphic transgression to the state space of the
$\si$-model (Theorems \ref{thm:ind-quasi-morph-glob-Vin} and
\ref{thm:ind-quasi-morph-glob-Vin-tw}; Propositions
\ref{prop:ham-vs-sisymsec} and \ref{prop:ham-vs-sisymsec-tw}),
consistent with the defect-duality correspondence (Proposition
\ref{prop:dual-vs-morf}).
\eit
\item A description of the relative (co)homology of the hierarchical
target space of the $\si$-model and a simple reinterpretation of the
physically motivated bracket structure from the previous point.
These include
\bit
\item an elementary characterisation of the relative singular
(co)homology (Proposition \ref{prop:lexact-DQT-rel-sing}), alongside
its realisation in terms of differential forms (Theorem
\ref{thm:DQT-rel-deRham});
\item introduction of a relative variant of the Cartan calculus and
identification of the associated (relative) twisted Courant
algebroid as the aforementioned bracket structure (Proposition
\ref{prop:Cartan-calc}; Theorem \ref{thm:brabra}).
\eit
\item A canonical description of rigid and gauged symmetries of the
$\si$-model in the first-order formalism of \Rcite{Suszek:2011hg}.
Here, we present
\bit
\item an investigation of conditions of continuity of Noether
charges across conformal defects and of their additive conservation
in trans-defect (resp.\ twisted-sector) splitting-joining
interactions (Propositions \ref{prop:ham-cont-across} and
\ref{prop:ham-simil-across}; Theorems \ref{thm:DJI-aug-intertw-untw}
and \ref{thm:ham-cons-tw});
\item a reinterpretation of the small gauge anomaly as an
obstruction to the existence of a hamiltonian realisation of the
symmetry algebra on states of the $\si$-model, consistent with
interactions (Theorem \ref{thm:sga-vs-ham}), resp.\ to a canonical
realisation of the (infinitesimal) gauge symmetry on the state space
of the gauged $\si$-model through elements of the characteristic
distribution of the relevant presymplectic form (Theorem
\ref{thm:sga-vs-gaugesym}).
\eit
\item An investigation of the Lie-groupoidal geometry of the gauge
anomaly. It yields
\bit
\item a reinterpretation of the small gauge anomaly in the
algebroidal framework introduced (as a combined Leibniz, Jacobi and
involutivity anomaly obstructing the existence of a Lie algebroid
within the relative twisted Courant algebroid), from which there
emerges the tangent algebroid of the action groupoid associated with
the action of the symmetry group $\,\txG_\si\,$ on the target space
$\,\xcF\,$ of the $\si$-model (Theorems \ref{thm:JacLie-sigma} and
\ref{thm:gtAlgebroid});
\item an elucidation of the latter phenomenon through a categorial
equivalence between -- on the one hand -- a category formed from
fundamental structures of a consistent gauged $\si$-model (a
principal $\txG_\si$-bundle over the world-sheet with the property
that the bundle associated to it through a $\txG_\si$-action on
$\,\xcF\,$ admits a global section) and morphisms between them and
-- on the other hand -- the groupoid of principal bundles over the
world-sheet with the structural action groupoid
$\,\txG_\si\lx\xcF\,$ whose appearance is the first hint of a local
world-sheet description of an orbit space of $\,\xcF\,$ with respect
to the action of $\,\txG_\si\,$ (Theorem \ref{thm:Gbun-vs-Grbun});
\item an extension of the said equivalence to the setting with
connection through an explicit construction of a topological defect
network implementing gauge transformations (defined globally or only
locally) on states of the gauged $\si$-model, giving rise to a
hands-on realisation of the concept of a simplicial duality
background of \Rxcite{Rem.\,5.6}{Suszek:2011hg} (Section
\ref{sub:world-coset});
\item a reinterpretation of the large gauge anomaly as an
obstruction to the existence of a consistent quantum CFT of the
gauged $\si$-model with topological gauge-symmetry defects resp.\ to
the existence of the gauged $\si$-model coupled to a gauge field of
an arbitrary topology whose indispensable incorporation in a unified
field-theoretic framework is understood from a purely geometric
point of view (Proposition \ref{prop:lga-vs-top-def-net}; Remark
\ref{rem:lga-vs-tnt-prin-g-bun}; Theorem \ref{thm:LGA-vs-coset}).
\eit
\een

The study reported in the present paper, taken in conjunction with
the earlier works on the subject, and in particular
Refs.\,\cite{Gawedzki:2008um,Runkel:2008gr,Suszek:2011hg,Gawedzki:2010rn,Gawedzki:2012fu}
(\textit{cf.}\ also the references listed there), of which it
constitutes a natural completion, leaves us with a fairly good
understanding of the deeper nature of rigid symmetries of the
multi-phase $\si$-model. It also motivates and lays the groundwork
for a number of new lines of research, of which we mention the
following:
\ben
\item A systematic construction of \emph{all} bi-branes of the coset
(resp.\ gauged) $\si$-model (upon relaxing, in particular, the
restrictive assumption of abelianness of the associated gerbe
bimodules, {\it cf.}\ \Rcite{Gawedzki:2004tu}), and comparison of
its results with predictions of the categorial quantisation
scheme\footnote{The author is grateful to Ingo Runkel for raising
this point in a private discussion.}.
\item Application of the gauge principle in the study of T-duality
in the context of the gerbe theory of the $\si$-model.
\item A world-sheet construction of the $\si$-model on an orbit space
of the target space with respect to the action of \textit{bona fide}
dualities, based on the defect-duality correspondence
(\textit{e.g.}, T-folds).
\item Incorporation of world-sheet/target-space supersymmetry into
the gerbe-theoretic framework of description of the $\si$-model,
with direct reference to the concept of a pure spinor but also with
view to deriving a 2-categorially-twisted (relative) extension of
Gualtieri's generalised complex geometry of its target space.
\item Study of relations between gauged multi-phase $\si$-models and
Poisson $\si$-models in the context of the underlying algebroidal
structure over the target space (drawing inspiration from but also
going beyond the long-known correspondence between gauged WZW models
and certain distinguished Poisson $\si$-models).
\een
We hope to return to these problems in near future.

\bibliographystyle{amsalpha}

\end{document}